\documentclass[10pt,preprint]{aastex}
\usepackage{natbib}
\def\etal{et al.~}

\def\ang{\AA}

\def\gapprox{\lower.4ex\hbox{$\;\buildrel >\over{\scriptstyle\sim}\;$}}
\def\lapprox{\lower.4ex\hbox{$\;\buildrel <\over{\scriptstyle\sim}\;$}}
\def\captio#1{\caption{\small {#1} \normalsize}}

\shortauthors{ASCHWANDEN ET AL. 2014}
\shorttitle{25 YEARS SELF-ORGANIZED CRITICALITY}

\begin{document}
\renewcommand{\topfraction}{0.95}
\renewcommand{\bottomfraction}{0.95}
\renewcommand{\textfraction}{0.05}
\renewcommand{\floatpagefraction}{0.95}
\renewcommand{\dbltopfraction}{0.95}
\renewcommand{\dblfloatpagefraction}{0.95}

%{\sl  Manuscript, version ....; accepted ... }

\title{25 Years of Self-Organized Criticality: Solar and Astrophysics}

\author{Markus J. Aschwanden}
\affil{Lockheed Martin, Solar and Astrophysics Laboratory (LMSAL),
       Advanced Technology Center (ATC),
       A021S, Bldg.252, 3251 Hanover St.,
       Palo Alto, CA 94304, USA;
       e-mail: aschwanden@lmsal.com}

\author{Norma Crosby}
\affil{Belgian Institute for Space Aeronomy, Ringlaan-3-Avenue 
	Circulaire, B-1180 Brussels, Belgium}

\author{Michaila Dimitropoulou}
\affil{Kapodistrian University of Athens, Dept. Physics, 15483, 
	Athens, Greece}

\author{Manolis K. Georgoulis}
\affil{Research Center Astronomy and Applied Mathematics,
	Academy of Athens, 4 Soranou Efesiou St., Athens, Greece,
	GR-11527.}

\author{Stefan Hergarten}
\affil{Institute f\"ur Geo- und Umweltnaturwissenschaften,
 	Albert-Ludwigs-Universit\"at Freiburg, 
	Albertstr. 23B, 79104 Freiburg, Germany}

\author{James McAteer}
\affil{Dept. Astronomy, P.O.Box 30001, New Mexico State University, 
	MSC 4500, Las Cruces, USA} 

\author{Alexander V. Milovanov}
\affil{Associazione EURATOM-ENEA sulla Fusione, Italian National 
	Agency for New Technologies, Frascati Research Centre,
	Via E. Fermi 45, C.P.-65, I-00044 Frascati, Rome, Italy; }
\affil{Space Research Institute, Russian Academy of Sciences,
	Profsoyuznaya str. 84/32, 117997 Moscow, Russia;}
\affil{Max Planck Institute for the Physics of Complex Systems,
	 Noethnitzer Str. 38, 01187 Dresden, Germany}

\author{Shin Mineshige}
\affil{Dept. Astronomy, Kyoto University, Kyoto 606-8602, Japan}

\author{Laura Morales}
\affil{Canadian Space Agency, Space Science and Technology Branch,  
	6767 Route de l'Aeroport, Saint Hubert, Quebec, J3Y8Y9, Canada}

\author{Naoto Nishizuka}
\affil{Inst. Space and Astronautical Science, Japan Aerospace Exploration
	Agency, Sagamihara, Kanagawa, 252-5210, Japan}

\author{Gunnar Pruessner}
\affil{Dept. Mathematics, Imperial College London, 180 Queen's
	Gate, London SW7 2AZ, United Kingdom}

\author{Raul Sanchez}
\affil{Dept. Fisica, Universidad Carlos III de Madrid, Avda. de la
	Universidad 30, 28911 Leganes, Madrid, Spain}

\author{Surja Sharma}
\affil{Dept. Astronomy, University of Maryland, College Park, MD 20740, USA}

\author{Antoine Strugarek}
\affil{Dept. de Physique, University of Montreal, C.P. 6128 Succ.~Centre-Ville,
	Montreal, Quebec, H3C-3J7, Canada}

\author{Vadim Uritsky}
\affil{NASA Goddard Space Flight Center, Code 671.0, Greenbelt,
	MD 20771, USA}

\begin{abstract}
Shortly after the seminal paper {\sl ``Self-Organized 
Criticality: An explanation of 1/f noise''} by Bak, Tang, and Wiesenfeld
(1987), the idea has been applied to solar physics, in {\sl ``Avalanches 
and the Distribution of Solar Flares''} by Lu and Hamilton (1991). 
In the following years, an inspiring cross-fertilization from complexity 
theory to solar and astrophysics took place, where the SOC concept was 
initially applied to solar flares, stellar flares, and magnetospheric 
substorms, and later extended to the radiation belt, the heliosphere, 
lunar craters, the asteroid belt, the Saturn ring, 
pulsar glitches, soft X-ray repeaters, blazars, black-hole 
objects, cosmic rays, and boson clouds. The application of SOC concepts 
has been performed by numerical cellular automaton simulations, by 
analytical calculations of statistical (powerlaw-like) distributions 
based on physical scaling laws, and by observational tests of theoretically
predicted size distributions and waiting time distributions. Attempts have been 
undertaken to import physical models into the numerical SOC toy models,
such as the discretization of magneto-hydrodynamics (MHD) processes.
The novel applications stimulated also vigorous debates about the 
discrimination between SOC models, SOC-like, and non-SOC processes,
such as phase transitions, turbulence, random-walk diffusion, 
percolation, branching processes, network theory, chaos theory, 
fractality, multi-scale, and other complexity phenomena. 
We review SOC studies from the last 25 years and highlight new trends,
open questions, and future challenges, as discussed during two recent 
ISSI workshops on this theme.
\end{abstract}


.
\clearpage
 
\tableofcontents
\clearpage

\section{INTRODUCTION}

About 25 years ago, the concept of {self-organized criticality (SOC)}
emerged (Bak et al.~1987), initially envisioned to explain the ubiquitous
1/f-power spectra, which can be characterized by a powerlaw function
$P(\nu) \propto \nu^{-1}$. The term {\sl 1/f power spectra} or
{\sl flicker noise} should actually be understood in broader terms, including
power spectra with pink noise ($P(\nu) \propto \nu^{-1}$), 
red noise ($P(\nu) \propto \nu^{-2}$), and black noise
($P(\nu) \propto \nu^{-3}$), essentially everything except white noise
($P(\nu) \propto \nu^{0}$). While white noise represents traditional
random processes with uncorrelated fluctuations, 1/f power spectra 
are a synonym for time series with non-random structures that exhibit
long-range correlations. These non-random time structures represent the
avalanches in Bak's paradigm of sandpiles. Consequently, Bak's seminal
paper in 1987 triggered a host of numerical simulations of sandpile 
avalanches, which all exhibit powerlaw-like size distributions of
avalanche sizes and durations. These numerical simulations were, most
commonly, cellular automata in the language of complexity theory, 
which are able to produce complex spatio-temporal patterns by iterative
application of a simple mathematical redistribution rule. The numerical
algorithms of cellular automata are extremely simple, basically a
one-liner that defines the redistribution rule, with an iterative loop 
around it, but can produce the most complex dynamical patterns, similar to 
the beautiful geometric patterns created by Mandelbrot's fractal algorithms  
(Mandelbrot 1977, 1983, 1985). An introduction and exhaustive description
of cellular automaton models that simulate SOC systems is given in
Pruessner (2012, 2013), and a review of cellular automaton
models applied to solar physics is given in Charbonneau et al.~(2001). 

Four years after introduction, Bak's SOC concept was applied to solar flares,
which were known to exhibit similar powerlaw size distributions for hard X-ray
peak fluxes, total fluxes, and durations as the cellular automaton
simulations produced for avalanche sizes and durations (Lu and Hamilton
1991). This discovery enabled a host of new applications of
the SOC concept to astrophysical phenomena, such as solar and stellar
flare statistics, magnetospheric substorms, X-ray pulses from accretion 
disks, pulsar glitches, and so forth. A compilation of SOC applications 
to astrophysical phenomena is given in a recent textbook (Aschwanden 2011a), 
as well as in recent review articles (Aschwanden 2013; Crosby 2011).
The successful spreading of the SOC concept in astrophysics mirrored the
explosive trend in other scientific domains, such as the application
of SOC in magnetospheric physics (auroras, substorms; see
review by Sharma \etal~(2014), in geophysics (earthquakes, 
mountain and rock slides, snow avalanches, forest fires; see Hergarten 
2002 and review by Hergarten in this volume), in biophysics (evolution 
and extinctions, neuron firing, spread of diseases), in laboratory 
physics (Barkhausen effect, magnetic domain patterns, Ising model, 
tokamak plasmas; Jensen 1998), financial physics (stock market 
crashes; Sornette 2003), and social sciences (urban growth, traffic, 
global networks, internet) or sociophysics (Galam 2012). 
This wide range of applications 
elevated the SOC concept to a truly interdisciplinary research area, 
which inspired Bak's vision to explain ``how nature works'' (Bak 1996). 
What is common to all these systems is the statistics of nonlinear
processes, which often ends up in powerlaw-like size distributions.
Other aspects that are in common among the diverse applications are 
complexity, contingency, and criticality (Bak and Paczuski 1995), 
which play a grand role in complexity theory and systems theory. 

What became clear over the last 25 years of SOC applications is the
duality of (1) a universal statistical aspect, and (2) a special 
physical system aspect. The universal aspect is a statistical
argument that can be formulated in terms of the scale-free probability
conjecture (Aschwanden 2012a), which explains the powerlaw function and
the values of the powerlaw slopes of most occurrence frequency distributions 
of spatio-temporal parameters in avalanching systems. This statistical 
argument for the probability distributions of nonlinear systems is 
as common as the statistical argument for binomial or Gaussian
distributions in linear or random systems. In this sense,
solar flares, earthquakes, and stockmarket systems have a 
statistical commonality (e.g., de Arcangelis et al.~2006).
On the other hand, each SOC system 
may be governed by different physical principles unique 
to each observed SOC phenomenon, such as plasma magnetic reconnection
physics in solar flares, mechanical stressing of tectonic plates
in earthquakes, or the networking of brokers in stock market crashes.
So, one should always be aware of this duality of model components when
creating a new SOC model. There is no need to re-invent the universal 
statistical aspects or powerlaw probability distributions each time,
while the modeling of physical systems may be improved with more
accurate measurements and model parameterizations in every new SOC
application. 

There is another duality in the application of SOC:
the numerical world of lattice simulation toy models, and the
real world of quantitative observations governed by physical laws.
The world of lattice simulations has its own beauty in producing
complexity with mathematical simplicity, but it cannot
capture the physics of a SOC system. It can be easily designed,
controlled, modified, and visualized. It allows us to perform Monte-Carlo
simulations of SOC models and may give us insights about the universal
statistical aspects of SOC. Real world phenomena, in contrast, 
need to be observed and measured with large statistics and reliable 
parameters that have been cleaned from systematic bias effects,
incomplete sampling, and unresolved spatial and temporal scales,
which is often hard to achieve. However, computer power has increased
drastically over the last 25 years, exponentially according to Gordon
Moore's law, so that enormous databases with up to $\approx 10^9$ events
have been gathered per data set from some SOC phenomena, such as from
solar small-scale phenomena for instance (McIntosh and Gurman 2005).

We organize this review by describing first some basics of SOC systems 
(Section 2), concerning SOC definitions, elements of a SOC system,
the probability concept, geometric scaling laws, transport process,
derivation of occurrence frequency distributions, waiting time 
distributions, separation of time scales, and the application of 
cellular automata. Then we deliver an overview on astrophysical
applications (Section 3), grouped by observational results and
theoretical models in solar physics, magnetospheres, planets,
stars, galaxies, and cosmology. In Section 4 we capture some
discussions, open issues and challenges, critiques, limitations, 
and new trends on the SOC subject, including also discussions of SOC-related 
processes, such as turbulence and percolation. The latter section
mostly results from discussions during two weeks of dedicated
workshops on ``Self-organized Criticality and Turbulence'', held
at the International Space Science Institute (ISSI) Bern during 2012 
and 2013, attended by participants who have contributed 
to this review.
 
\clearpage

\section{BASICS OF SELF-ORGANIZED CRITICALITY SYSTEMS}

\subsection{SOC Definitions}

The original definition of the term {\sl self-organized criticality (SOC)}
was inspired by a numerical lattice simulation of a dynamical system 
with spatially complex patterns, mimicking avalanches of a sandpile,
which became the BTW model (Bak, Tang, and Wiesenfeld 1987), and
demonstrated that:

\begin{itemize}
\item	{\sl Dynamical systems with extended spatial degrees of freedom
	naturally evolve into self-organized critical structures of
	states which are barely stable. Flicker noise, or 1/f noise,
	can be identified with the dynamics of the critical state.
	This picture also yields insight into the origin of fractal
	objects.} (Bak et al.~1987)
\end{itemize}

In this first seminal paper, the authors had already fractal structures
like cosmic strings, mountain landscapes, and coastal lines as potential
applications in mind and concluded: {\sl We believe that the new concept
of self-organized criticality can be taken much further and might be
{\bf the} underlying concept of dissipative systems with extended degrees
of freedom} (Bak et al.~1987). In this spirit, the application of the SOC 
concept has been broadened substantially over the last 25 years.

If we read a modern definition of SOC, we find:

\begin{itemize}
\item   {\sl In physics, self-organized criticality (SOC) is a property of
	(classes of) dynamical systems which have a critical point as an
	attractor. The macroscopic behavior thus displays the spatial 
	and/or temporal scale-invariance characteristic of the critical
	point of a phase transition, but without the need to tune control
	parameters to precise values} (Wikipedia).
\end{itemize}

In the same vein, it is stated in the original paper of the SOC creators: 
{\sl The criticality in our theory is fundamentally different from the
critical point at phase transitions in equilibrium statistical mechanics
which can be reached by tuning of a parameter, for instance the
temperature} (Bak et al.~1987). The aspect of self-tuning in SOC systems
is the most crucial difference to (second-order) phase transitions, where 
fine-tuning is necessary and is not automatically arranged by nature.
The implications and theoretical details of this peculiar feature are 
discussed in Watkins et al.~(2014). However, whenever
there is a threshold for instabilities, the threshold value itself
could be called a ``critical point'' that decides whether an instability,
also called a nonlinear energy dissipation event, or avalanche,
happens or not. Over the past 25 years, a lot of applications of the SOC 
concept have been made to slowly-driven systems with a critical threshold,
especially in solar and astrophysics, as reviewed in this article.
We therefore like to use a more pragmatic and physics-based 
definition of a SOC system:

\begin{figure}[t]
\centerline{\includegraphics[width=0.8\textwidth]{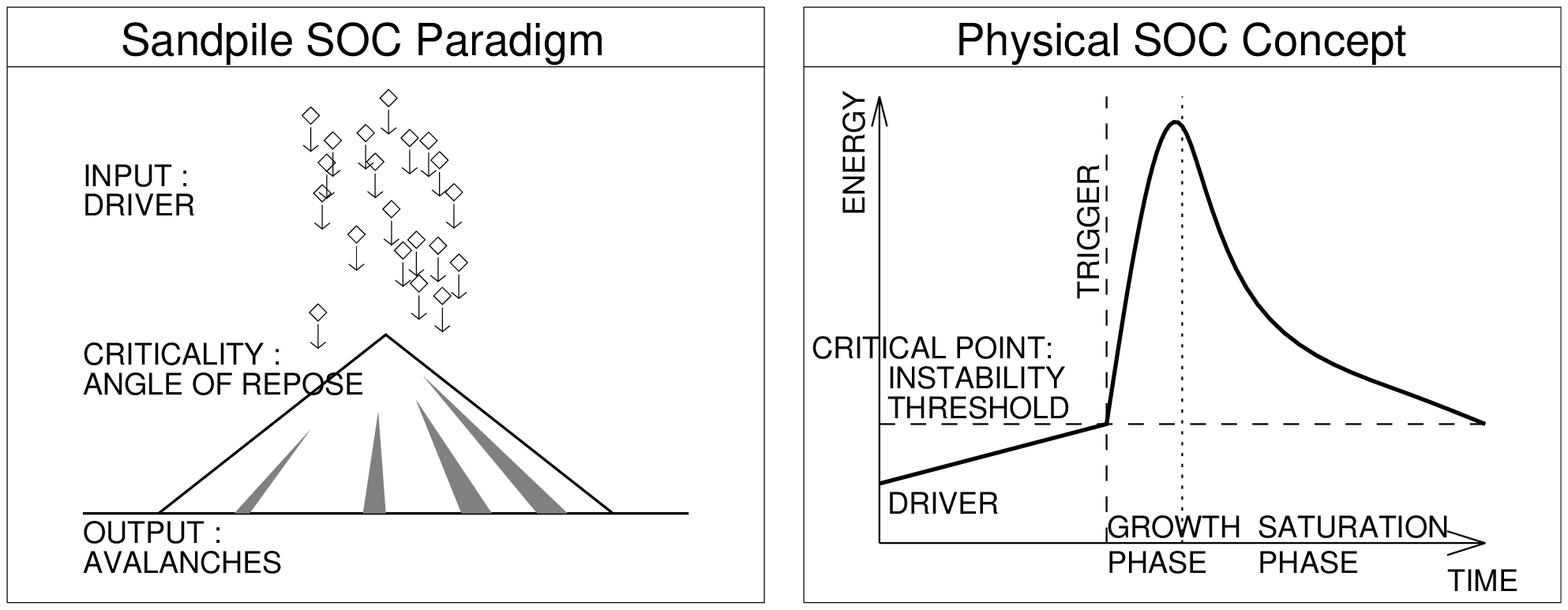}}
\captio{{\sl Left:} The original sandpile SOC paradigm, consisting
of the (input) driver, the self-organized criticality mechanism
(self-tunig angle of repose), and the (output) avalanches.
{\sl Right:} In a physical SOC concept, the driver is a slow and
continuous energy input rate, the criticality mechanism is replaced
by a critical point in form of an instability threshold, where
an avalanche is triggered, usually consisting of a nonlinear
growth phase and a subsequent saturation phase.}  
\end{figure}

\begin{itemize}
\item   {\sl SOC is a critical state of a nonlinear energy dissipation
	system that is slowly and continuously driven towards a critical value
	of a system-wide instability threshold, producing scale-free,
	fractal-diffusive, and intermittent avalanches with powerlaw-like 
	size distributions} (Aschwanden 2014).
\end{itemize}
 
With this definition we broaden the meaning of the term ``criticality'' 
to a more general meaning of a ``critical point'', which includes almost
any nonlinear system with a (global) instability threshold (Fig.~1). 
In addition,
a SOC system has to be self-organizing or self-tuning without external
control parameter, which is accomplished by a {\sl slow and continuous
driver}, which brings the system back to the critical point after each
avalanche. Thus, we can say that a SOC system has energy balance 
between the slowly-driven input and the (spontaneous) avalanching output, 
and thus energy is conserved in the system (in the time average). 
 
\subsection{The Driver}

The driver is the input part of a SOC system. Without a driver, avalanching
would die out and the system becomes subcritical and static. On the other
side, the driver must be slowly and continuous, so that the critical
state is restored in the asymptotic limit, while a strong driver would
lead the system into a catastrophic collapse and may destroy the system.
In the classical BTW model, sand grains are dripped 
under the action of gravity at a slow rate, at random locations of the 
sandpile, which re-fill and restore dents from previous avalanches 
towards the critical angle of repose. In astrophysical systems, the driver 
or energy input of a SOC system may be gravity (in galaxy formation,
star formation, black holes, planet formation, asteroid formation), 
gravitational disturbances (in Saturn ring), or creation and stressing 
of magnetic flux (in solar flares, stellar flares, neutron stars,
pulsars). The driver must bring the system back to the critical point
after each major avalanche, which means that the system is locally pushed
towards the instability threshold again, so that further avalanching
can occur. In the slowly-driven limit, the time duration of an avalanche
is much longer than the (waiting) time intervals between two
subsequent events, which warrants a separation of time scales.
In some natural systems the driver may temporarily or permanently stop,
such as the solar dynamo during the Maunder minimum that stopped
solar flaring, or the final stage of the sweep-up of debris left over 
from the formation of the solar system 4.0 billion years ago that 
stopped lunar cratering.

\subsection{Instability and Criticality}

We broaden the meaning of ``criticality'' in the original BTW model to
a system-wide ``instability threshold'', which does not need to be
tuned by external parameters, since an ``instability threshold'' is
established by common physical conditions throughout a system. For instance,
an earthquake is triggered at a critical stressing brake point that may have
a similar threshold in different tectonic plates around the globe, 
due to similar geophysical conditions (i.e., the gravity force at 
the same distance from Earth center, similar continental drift rates, 
rock constitutions, and crust fracturing conditions).
In analogy, a magnetic instability leading to magnetic reconnection
is caused by similar physical threshold conditions in solar active regions
(such as the kink instability, the torus instability, or the tearing
mode instability), and thus solar or stellar flares occur whenever
such global instability thresholds are exceeded locally. 
Such instabilities occur naturally because the driver continuously
brings the system back to the instability threshold. In sandpiles,
the dripping of additional sand grains rises the angle of repose
wherever it is subcritical. In earthquakes, the continental drift is
continuously driven by forces that are rooted deeper below the Earth
crust. In solar flares, differential rotation, emergence of magnetic
flux, and braiding of magnetic fields by random motion in the
subphotospheric magneto-convection layer continuously build up
nonpotential free magnetic energy that can be released in
subsequent avalanches. The analogy of unstable coherent structures 
in a near-critical state in sandpiles and solar flares is visualized
in Fig.~(2).

\begin{figure}[t]
\centerline{\includegraphics[width=0.8\textwidth]{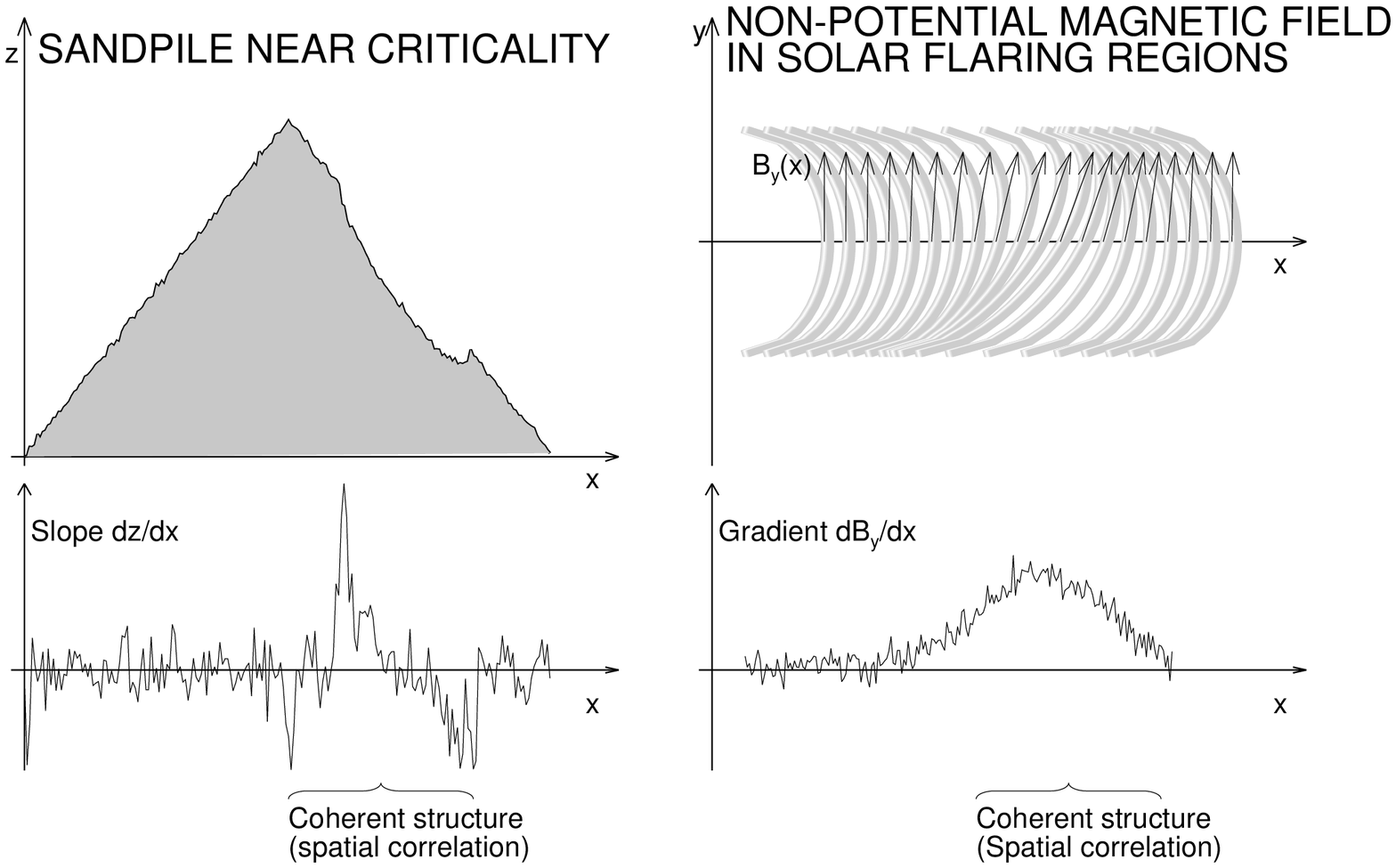}}
\captio{{\sl Left:} A sandpile in a state in the vicinity of criticality
is shown with a vertical cross-section $z(x)$, with the the slope
(or repose angle) $\textrm{d}z/\textrm{d}x$
(bottom), exhibiting short-range fluctuations
due to noise and long-range correlations due to local deviations
from the mean critical slope. {\sl Right:} The solar analogy of a flaring
region is visualized in terms of a loop arcade straddling along a neutral line
in $x$-direction, consisting of loops with various shear angles that
are proportional to the gradient of the field direction $B_x/B_y$,
showing some local (non-potential) deviations from the potential 
magnetic field (bottom).}
\end{figure}

\subsection{Avalanches}

Avalanches are defined as nonlinear energy dissipation events,
which occur in our generalized SOC definition whenever and
wherever a local instability threshold is exceeded. Avalanches
are the output part of a SOC system, which balance the energy
input rate in the time average for conservative SOC systems.
Avalanches are detectable events, which can be obtained in
astrophysical observations with large statistics, such as 
length scales ($L$), time scales or durations ($T$), fluxes ($F$),
fluences or energies ($E$). The occurrence frequency distributions of
these observables tend to be powerlaw-like functions, a
hallmark of SOC systems, but deviations from powerlaw functions
can be explained by measurement bias effects (such as
incomplete sampling, finite system-size effects, truncations
of distributions), or could reflect multiple physical
processes. Unnecessary to say that these observables and
their size distributions and underlying scaling laws provide 
the most important evidence and tests of SOC models.  

The time evolution of avalanches contain essential information
on the underlying spatio-temporal transport process (i.e.,
diffusion, fractal diffusion, percolation, turbulence, etc.). 
A generic time evolution is an initially nonlinear (i.e., 
exponential) growth phase, followed by a quenching or saturation 
phase (as expressed in the popular saying  
``No trees grow to the sky!''). In solar flares, for instance,
the initial growth phase is called ``impulsive phase'', and the
subsequent saturation phase is called ``postflare phase''.
In earthquakes, the terms ``precursors'' and ``after shocks''
are common. 

\begin{figure}[t]
\centerline{\includegraphics[width=1.0\textwidth]{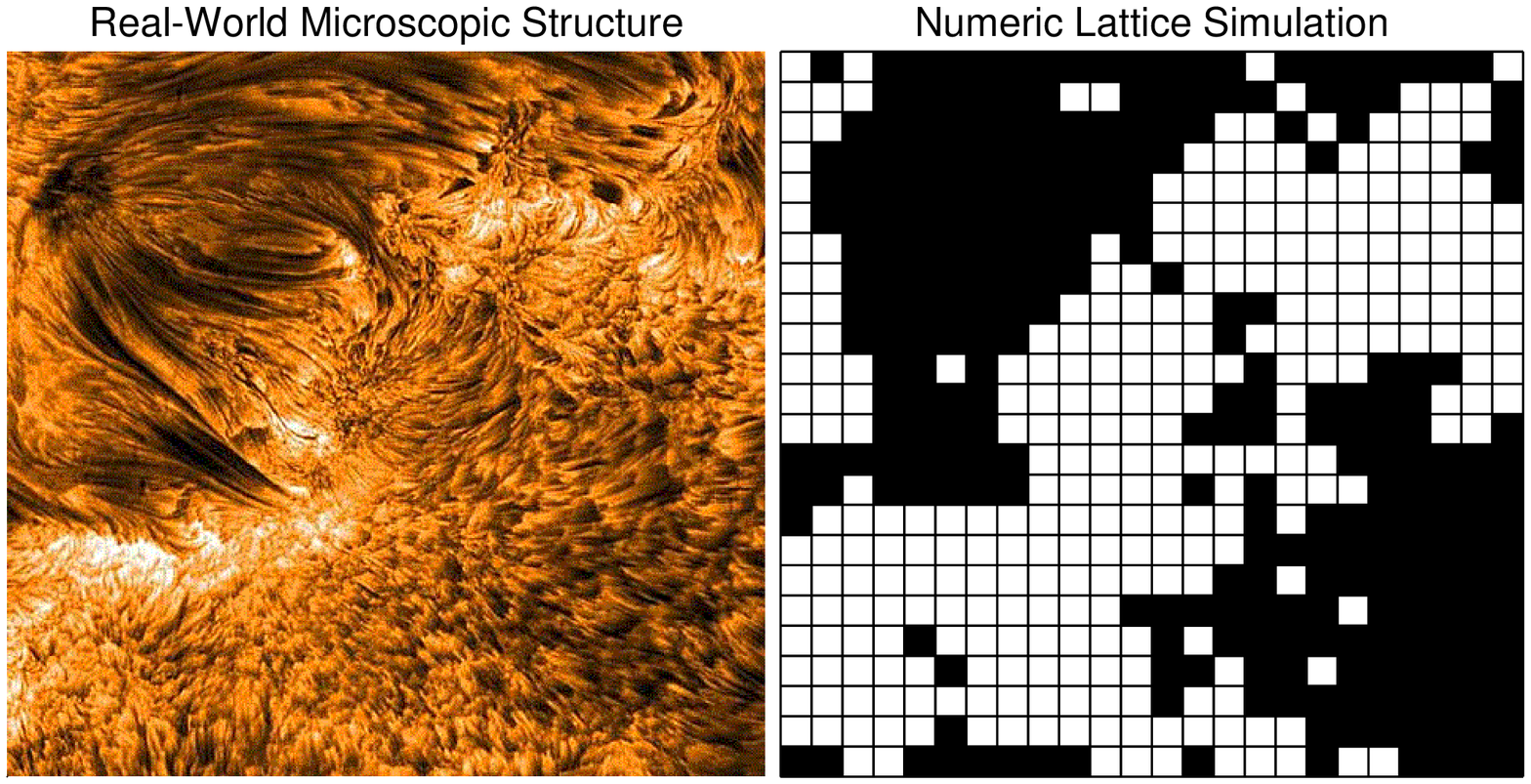}}
\captio{{\sl Left:} A high-resolution image (480 $\times$ 480 pixel)
of chromospheric spiculae 
in solar active region 10380, observed on 2003 June 16 with the 
Swedish 1-m Solar Telescope (SST) on La Palma, Spain, using a tunable 
filter, tuned to the blue-shifted line wing of the H$\alpha$ 6536 \ang\ 
line (Courtesy of Bart DePontieu). 
{\sl Right:} A digitized binary version of the left solar image, 
using a lattice grid with a size of 24 $\times$ 24 nodes. The left
image shows the microscopic structure of real-world data, while
the right image shows the rendering of numerical lattice simulations
used in SOC models.}
\end{figure}

\subsection{Microscopic Structure and Complexity}

SOC systems are a means to study complexity, systems with extended 
degrees of freedom. Ultimately, a real-world object consists of
atoms that has as many degrees of freedom as the Avogadro number
of atoms per mol quantifies, i.e., $6.0 \times 10^{23}$. Such large numbers
prevent us from modeling complex nonlinear systems in a deterministic 
way. In order to deal with SOC systems, we have to resort to numerical 
simulations with far fewer degrees of freedom, and we have to 
approximate the complexity of microscopic structures by macroscopic 
parameters and statistical probability distributions. For example,
the complex microscopic structure of the solar chromosphere (Fig.~3,
left panel) can be rendered with a binary lattice on a much coarser 
scale (Fig.~3, right). The question is, whether the basic physics
that governs the dynamics of a real-world system can also be
adequately represented by numerical lattice simulations.
In the example shown in Fig.~3, one binary node of a lattice
corresponds to a cube with 1000 km length scale on the solar surface,
where the complex plasma dynamics driven by magneto-hydrodynamic 
processes exceeds the information content of a binary lattice node
by far, so that it appears to be hopeless to mimic the dynamics
of a SOC system with numerical cellular automaton simulations.
Interestingly however, numerical lattice simulations do reproduce
the emergent complex behavior in physical systems to some extent,
regardless of the vaste discrepancy of spatial scales and
information content. For instance, the statistical size distribution
of solar flares can be reproduced with cellular automata for
various physical parameters (spatial, temporal scales, flux, and energy),
as demonstrated by Lu and Hamilton (1991). Therefore, SOC models
have the powerful ability to give us insight into system dynamics
in complex systems, regardless of the intricate details of 
real-world microscopic fine structure. On the other side, the
mathematical world of numerical lattice simulations created a
whole new cosmos of complex spatial patterns (i.e., Wolfram 2002)
and cellular automaton toy models (i.e., Pruessner 2012), which 
appear to have nothing in common with real-world microscopic fine structure,
except that they provide practical means to simulate the same dynamic 
behavior of complex nonlinear systems. Consequently, in this review
on solar and astrophysical SOC applications, the emphasis is not on
mathematical and numerical SOC models (except when they were
specifically designed for astrophysical applications), although 
they make up for more than half of the extant SOC literature.

\subsection{The Scale-Free Probability Conjecture}

Common characterizations of SOC systems are statistical distributions 
of SOC parameters (also called ``size distributions'', ``occurrence 
frequency distributions'', or ``log(N)-log(S) plots''). How do we
derive a statistical {\sl probability distribution function (PDF)} for SOC
systems? This question has been answered in the original SOC papers
(Bak et al.~1987, 1988) in an empirical way, by performing numerical 
Monte-Carlo simulations of avalanches in cartesian lattice grids,
according to the well-known algorithm with next-neighbor interactions
(BTW model). Several theoretical attempts have been made to derive 
statistical probabilities, by considering avalanches as a branching
process (Harris 1963; Christensen and Olami 1993), by exact solutions
of the Abelian sandpile (Dhar and Ramaswamy 1989; Dhar 1990, 1999;
Dhar and Majumdar 1990), by considering the BTW cellular automaton
as a discretized diffusion process using the Langevin equations
(Wiesenfeld 1989; Zhang 1989; Forster et al.~1977;
Medina et al.~1989), or by renormalization group theory
(Medina et al.~1989; Pietronero and Schneider 1991; Pietronero et al.~1994;
Vespignani et al.~1995; Loreto et al.~1995, 1996). Most of these 
analytical theories represent special solutions to a particular
set of mathematical redistribution rules, but predict different 
powerlaw exponents for the probability distribution functions obtained
with each method, and thus lack the generality to interpret the ubiquitous
and omnipresent SOC phenomena observed in nature.

A simple approach to estimate the size distributions of SOC avalanche 
sizes has recently been proposed by making a simple statistical
probability argument, called the {\sl scale-free probability conjecture} 
(Aschwanden 2012a, 2014), which predicts the functional form of powerlaws 
for most observable SOC parameters, and predicts specific values for 
their powerlaw slopes (or exponents). The derivation goes as follows.
If we consider the derivation of a normal or Gaussian distribution 
function, we can toss a number of dice and enumerate all possible 
statistical outcomes, ending up with a binomial distribution function,
which converges to a Gaussian distribution function for a large number
of dice, and thus characterizes a maximum likelihood distribution. 
Similarly, we can enumerate all statistically possible sizes $L$
of avalanches in a system bound by a finite size $L_{max}$, which is simply
a number density that is reciprocal to the volume $V=L^d$ of
avalanches with size $L$, i.e.,
\begin{equation}
	N(L) dL \propto L^{-d} dL \qquad {\rm for}\  L \le L_{max} \ , 
\end{equation}
where $d$ is the Euclidean dimension of the SOC system. This distribution
function is based on the principle of statistical maximum likelihood,
which follows from braking up a finite system volume into smaller pieces.
This distribution function is also related to {\sl packing rules}
(e.g., sphere packing) in geometric aggregation problems. A similar
approach using geometric scaling laws was also applied to earthquakes
(Main and Burton 1984). Of course, for slowly-driven SOC systems,
only one avalanche happens at a time, and thus the whole SOC system
is not fully ``packed'' with avalanches occurring at once, but the statistical
likelihood probability for an avalanche of a given size is nevertheless proportional
to the packing density, for a statistically representative subset of 
all possible avalanche sizes (in a system with $L \le L_{max}$). 
This basic scale-free probability conjecture (Eq.~1)
straightforwardly predicts the size distribution of length 
scales of SOC avalanches, namely $N(L) \propto L^{-3}$ in 3D space,
and can be used to derive the size distributions of other geometric 
parameters. 

\subsection{Geometric Scaling Laws}

Other geometric parameters are the Euclidean area $A$ or the
Euclidean volume $V$. The simplest definition of an area $A$
as a function of a length scale $L$ is the square-dependence,
\begin{equation}
        A \propto L^2 \ .
\end{equation}
A direct consequence of this
simple geometric scaling law is that the statistical probability
distribution of avalanche areas is directly coupled to the
scale-free probability distribution of length scales (Eq.~1),
and can be computed by substitution of $L(A) \propto A^{1/2}$ (Eq.~2),
into the distribution of Eq.~(1),
$N(L)=N(L[A])=L[A]^{-d}=(A^{1/2})^{-d}=A^{-d/2}$,
and by inserting the derivative $dL/dA \propto A^{-1/2}$,
\begin{equation}
        N(A) dA \propto N(L[A]) \left| {dL \over dA} \right| dA
                \propto A^{-(1+d)/2}\ dA\ .
\end{equation}
Thus we expect an area distribution of $N(A) \propto A^{-2}$ in
3D-space.

Similarly to the area, we can derive the geometric scaling for 
volumes $V$, which simply scales with the cubic power in 3D space
($d=3$), or generally as, 
\begin{equation}
        V \propto L^d \ .
\end{equation}
Consequently, we can also derive
the probability distribution $N(V) dV$ of volumes $V$ directly from
the scale-free probability conjecture (Eq.~1). 
Substituting $L \propto V^{1/d}$ into
$N(L[V]) \propto L[V]^d \propto V^{-1}$, and inserting the derivative
$dL/dV=V^{1/d-1}$, we obtain,
\begin{equation}
        N(V) dV \propto N(L[V]) \left| {dL \over dV} \right| dV
                \propto V^{-(2-1/d)} dV \  
                \propto V^{-\alpha_V} dV \ .
\end{equation}
Thus, a powerlaw slope of $\alpha_V=2-1/d=5/3\approx 1.67$ is predicted in 3D
space $(d=3)$. Since all the assumptions made so far are universal, such as the
scale-free probability conjecture (Eq.~1) and the geometric scaling
laws $A \propto L^2$ (Eq.~2) and $V \propto L^3$ (Eq.~4), the resulting 
predicted occurrence frequency distributions of $N(A) \propto A^{-2}$ 
(Eq.~3) and $N(V) \propto V^{-5/3}$ (Eq.~5) are universal too, 
and thus powerlaw functions are predicted from this derivation from first 
principles, which is consistent with the property of {\sl universality} 
in theoretical SOC definitions.

\begin{figure}[t]
\centerline{\includegraphics[width=0.9\textwidth]{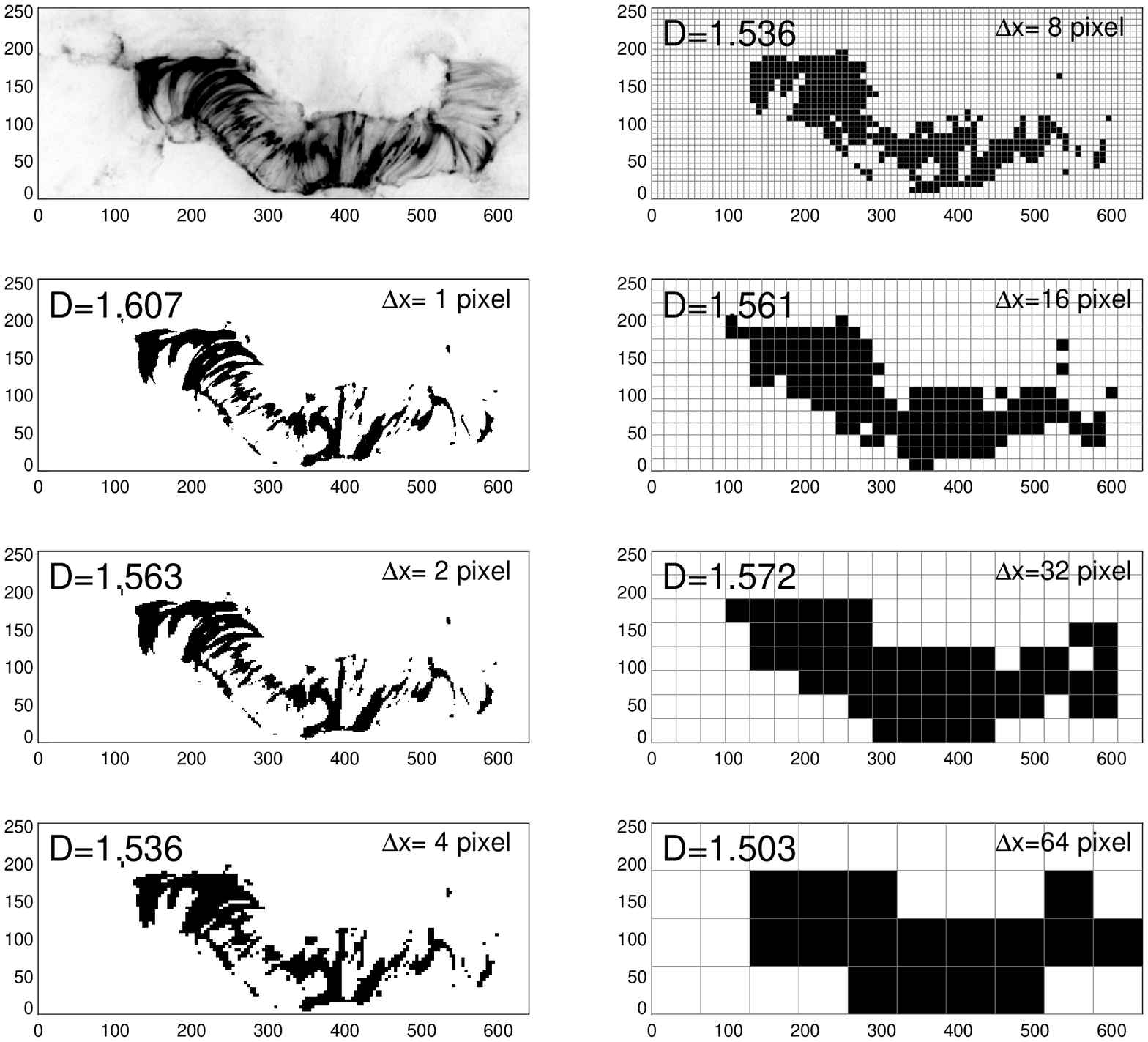}}
\captio{Measurement of the fractal area of a solar flare, observed
by TRACE 171 \ang\ on 2000-Jul-14, 10:59:32 UT. The Hausdorff dimension is
evaluated with a box-counting algorithm for pixels above a threshold of
20\% of the peak flux value, yielding a mean of $D_2=1.55\pm0.03$ for the 7
different spatial scales ($\Delta x=1, 2, 4, ..., 64$ pixels) shown here 
(Aschwanden and Aschwanden 2008a).}  
\end{figure}

\subsection{Fractal Geometry}

{\sl ``Fractals in nature originate from self-organized critical
dynamical processes''} (Bak and Chen 1989). The fractal geometry
has been postulated for SOC processes by the first proponents of SOC.
However, the geometry of fractals has been explored at least a decade 
before the SOC concept existed (Mandelbrot 1977, 1983, 1985).
An extensive discussion of measuring the fractal geometry in SOC systems
associated with solar and planetary data
is given in Aschwanden (2011a, chapter 8) and McAteer (2013a).

The simplest fractal is the Hausdorff dimension
$D_d$, which is a monofractal and depends on the Euclidean space dimension
$d=1,2,3$. The Hausdorff dimension $D_3$ for the 3D Euclidean space ($d=3$) is
\begin{equation}
        D_3 = {\log{V_f(t)} \over \log{(L)} } \ ,
\end{equation}
and analogously for the 2D Euclidean space ($d=2$),
\begin{equation}
        D_2 = {\log{A_f(t)} \over \log{(L)} } \ ,
\end{equation}
with $A_f(t)$ and $V_f(t)$ being the fractal area and volume of a SOC
avalanche during an instant of time $t$. These fractal dimensions can be
determined by a box-counting method, where the area fractal $D_2$ can
readily be obtained from images from the real world (e.g., for a solar
flare as shown in Fig.~4), while the volume fractal $D_3$ is generally 
not available (except in numerical simulations), unless one infers the 
corresponding 3D information from stereoscopic triangulation.
A good approximation for the
expected fractal dimension $D_d$ of SOC avalanches is the mean value 
of the smallest
likely fractal dimension $D_{d,min}\approx 1$ and the largest
possible fractal dimension $D_{d,max}=d$. The minimum possible fractal
dimension is near the value of 1 for SOC systems, because the 
next-neighbor interactions
in SOC avalanches require some contiguity between active nodes in a
lattice simulation of a cellular automaton, while smaller fractal
dimensions $D_d < 1$ are too sparse to allow an avalanche to propagate
via next-neighbor interactions. Thus, the mean value of the fractal
dimension of SOC avalanches is expected to be (Aschwanden 2012a),
\begin{equation}
        D_d \approx {D_{d,min} + D_{d,max} \over 2} = {(1+d) \over 2} \ .
\end{equation}
Thus, we expect a mean fractal dimension of $D_3\approx (1+3)/2=2.0$ for
the 3D space, and $D_2\approx (1+2)/2=1.5$ for the 2D space. The example
shown in Fig.~(4) yielded a value of $D_2=1.55\pm0.03$, which is close
to the prediction of Eq.~(8).

Fractals are measurable from the spatial structure
of an avalanche at a given instant of time. Therefore, they enter
the statistics of time-evolving SOC parameters, such as the observed
flux or intensity per time unit, which is proportional to the number of
instantaneously active nodes in a lattice-based SOC avalanche simulation.

\subsection{Spatio-Temporal Evolution and Transport Process}

Let us consider some basic aspects in the time domain of SOC avalanches.
The spatio-temporal evolution of SOC avalanches has been simulated with
cellular automaton simulations (Bak et al.~1987, 1988; Lu and Hamilton
1991; Charbonneau et al.~2001), which produced statistics of the final
avalanche sizes $L$ and durations $T$, but there is virtually no
statistics on the spatio-temporal evolution of the instantaneous
avalanche size or radius $r(t)$ as a function of time $t$, which would
characterize the macroscopic transport process. Statistics on this
spatio-temporal evolution is important to establish spatio-temporal
correlations and scaling laws between $L$ and $T$, which defines the
macroscopic transport process.

\begin{figure}[t]
\centerline{\includegraphics[width=0.8\textwidth]{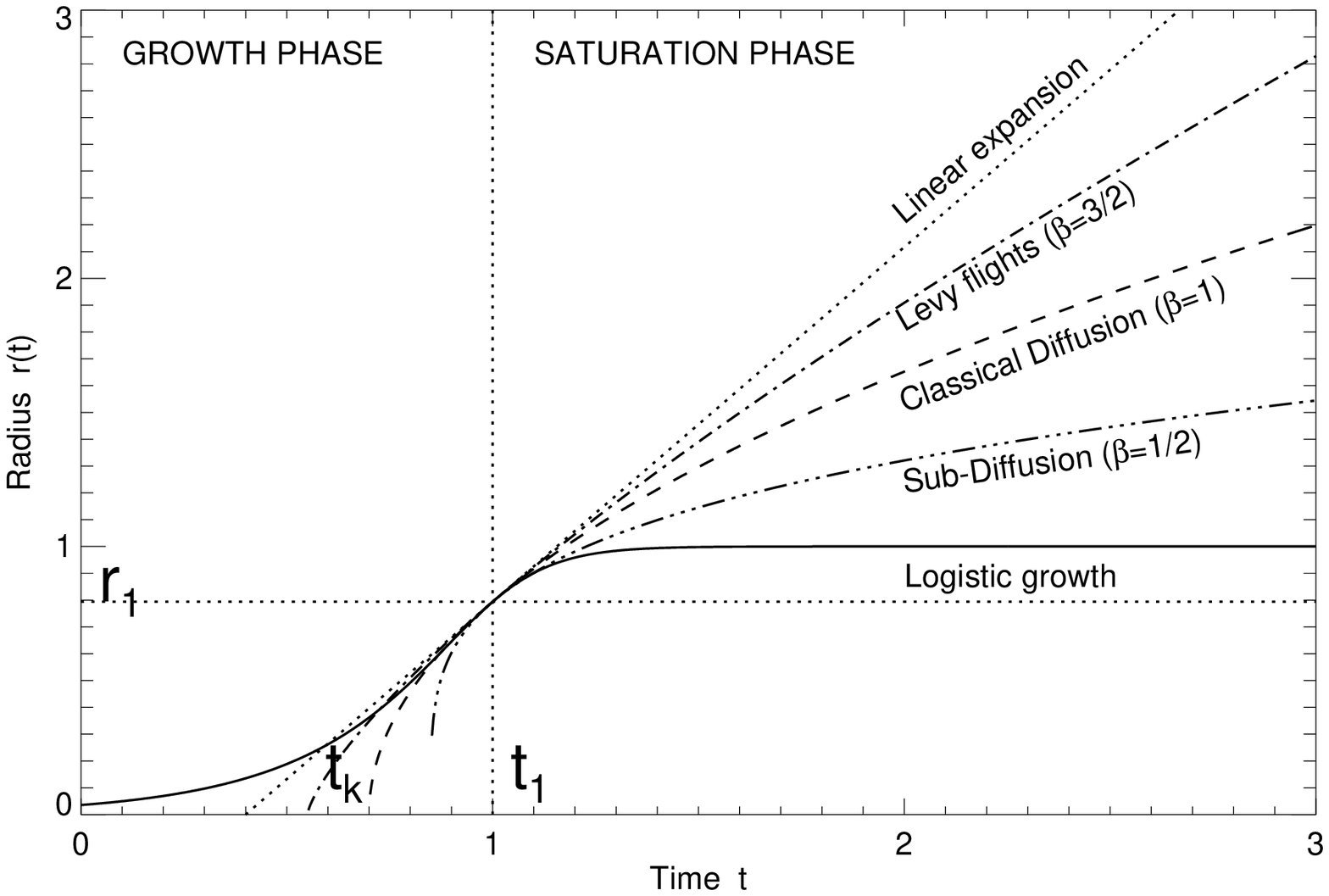}}
\captio{Comparison of spatio-temporal evolution models:
Logistic growth with parameters $t_1=1.0, r_\infty=1.0, \tau_G=0.1$,
sub-diffusion ($\beta=1/2$), classical diffusion ($\beta=1$),
L\'evy flights or hyper-diffusion ($\beta=3/2$), and linear expansion
($r \propto t$).} 
\end{figure}

Ignoring the complexity of the microscopic transport, which is
quantified by an iterative redistribution rule in cellular automaton
simulations, we can measure the radius $r(t)=\sqrt{A(t)/\pi}$ of
a circular 2D area $A(t)$ as a function of time $t$,
which corresponds to the solid (Euclidean) area that is equivalent
to the time-integrated fractal avalanche area.
This has been performed for BTW cellular automaton simulations 
(Aschwanden 2012a), as well as for solar flare data (Aschwanden 2012b; 
Aschwanden and Shimizu 2013; Aschwanden et al.~2013a), and was found 
to fit a diffusion-type relationship,
\begin{equation}
        r(t) = \kappa (t - t_0)^{\beta/2} \ ,
\end{equation}
where $t_0$ is the onset time of the instability, $\kappa$ is the
diffusion coefficient, and $\beta$ is the diffusive spreading exponent:
a value of $\beta \lapprox 1$ corresponds to sub-diffusion, 
$\beta=1$ to classical diffusion, $\beta \gapprox 1$ to 
hyper-diffusion or L\'evy flight, and $\beta=2$ to linear
expansion (Fig.~5). From this macroscopic evolution we expect a
statistical scaling law of the form,
\begin{equation}
        L \propto \kappa\ T^{\beta/2} \ ,
\end{equation}
for the final sizes $L$ and durations $T$ of SOC avalanches.
Substituting this scaling law $L(T)$ into the PFD of length scales
(Eq.~1), we establish a powerlaw distribution function for time scales,
\begin{equation}
        N(T) dT = N(L[T]) {dL\over dT} dT
                = T^{-[1+(d-1)\beta/2]}\ dT = T^{-\alpha_T}\ dT \ .
\end{equation}
with the powerlaw slope of $\alpha_T=1+(d-1)\beta/2$,
which has a value of $\alpha_T =1 + \beta = 2.0$ for 3D-Euclidean space $(d=3)$
and classical diffusion ($\beta=1$). This powerlaw slope for avalanche
time scales is a prediction of universal validity, since it is only
based on the scale-free probability conjecture (Eq.~1), $N(L) \propto L^{-d}$,
and the statistical property of random walk in the transport process.

\subsection{Flux and Energy Scaling}

The original BTW model specified avalanche sizes by the total number of
active nodes, which corresponds to the cluster area of an avalanche in a 
2D lattice. If we want to characterize the area $a(t)$ of an avalanche
as a function of time, which is a highly fluctuating quantity in time, 
we can define also a time-integrated final
area $a(<t)$ that includes all nodes that have been gone unstable at least
once during the course of an avalanche, which is a monotonically increasing
quantity and quantifies the size of an avalanche with a single number
$A=a(t=T)$, which we simply call the time-integrated avalanche area.

In real-world data we observe a signal from a SOC avalanche in form of
an intensity flux $f(t)$ (e.g., seismic waves from earthquakes, hard X-ray
flux from solar flares, or the amount of lost dollars per day in the
stockmarket). Let us assume that this intensity flux is proportional 
to the volume of active nodes in the BTW model, which corresponds to the
instantaneous fractal volume $V_f(t)$ (Eq.~6) in a macroscopic SOC model
(Aschwanden 2012a, 2014), 
\begin{equation}
        f(t) \propto V_f(t) \propto r(t)^{D_d} \ .
\end{equation}
The flux time profile $f(t)$ is expected to fluctuate substantially
in real-world data as well as in lattice simulations, because the 
approximation of the instantaneous volume of a SOC avalanche implies
a highly variable fractal dimension $D_d(t)$, which can vary in 
the range of $D_{d,min} \approx 1$ and $D_{d,max}=d$, with a mean value 
$D_d=(1+d)/2$ (Eq.~8). Occasionally, the instantaneous
fractal dimension may reach its maximum value, i.e., $D_d(t)\lapprox d$,
which defines an expected upper limit $f_{max}(t)$ of
\begin{equation}
        f_{max}(t) \propto V(t) \propto r(t)^d \ .
\end{equation}
This is an important quantity that corresponds to the peak flux 
of an avalanche, which is often measured in astrophysical observations.

Integrating the time-dependent flux $f(t)$ over the time interval
$[0,t]$ yields the time-integrated avalanche volume $e(t)$ up to time $t$,
which is often associated with the total dissipated energy during
an avalanche (tacitly assuming an equivalence between energy and
avalanche volume), using Eq.~(9),
\begin{equation}
        e(t) \propto \int_{t_0}^t V_f(t) dt =
                     \int_{t_0}^t r^{D_d}(t) dt =
                     \int_{t_0}^t \kappa^{D_d} (t-t_0)^{D_d \beta/2} dt =
                     {\kappa^{D_d} \over D_d \beta/2+1}
                        (t-t_0)^{D_d \beta/2 + 1}\ ,
\end{equation}
which is a monotonically increasing quantity with time. We see that
this total dissipated energy depends on the fractal dimension $D_d$
and the diffusion spreading exponent $\beta$, within the framework
of the fractal-diffusive transport model (Eq.~9).

From this time-dependent evolution of a SOC avalanche we can characterize
at the end time $t$ a time duration $T=(t-t_0)$, a spatial scale
$L=r(t=t_0+T)$, an expected flux or energy dissipation rate $F=f(t=t_0+T)$,
an expected peak flux or peak energy dissipation rate $P=f_{max}(t=t_0+T)$,
and a dissipated energy $E=e(t=t_0+T)$, which is identical to the avalanche 
size $S$ in BTW models, i.e., $E \propto S$, for which we expect the 
following scaling laws (using Eqs.~12-14), 
\begin{equation}
        F \propto L^{D_d} \propto T^{D_d \beta/2} ,
\end{equation}
\begin{equation}
        P \propto L^d \propto T^{d \beta/2} \ ,
\end{equation}
\begin{equation}
        E \propto S \propto L^{D_d+2/\beta} \propto  T^{D_d \beta/2+1} \ .
\end{equation}

Finally we want to quantify the occurrence frequency distributions of the
the (smoothed) energy dissipation rate $N(F)$, the peak flux $N(P)$,
and the dissipated energy $N(E)$, which all can readily be obtained by
substituting the scaling laws (Eqs.~15-17) into the fundamental length scale
distribution (Eq.~1), yielding
\begin{equation}
        N(F) dF = N(L[F]) \left| {dL \over dF} \right| dF
        \propto F^{-[1+(d-1)/D_d]} \ dF \ ,
\end{equation}
\begin{equation}
        N(P) dP = N(L[P]) \left| {dL \over dP} \right| dP
        \propto P^{-[2-1/d]}
        \ dP \ ,
\end{equation}
\begin{equation}
        N(E) dE = N(L[E]) \left| {dL \over dE} \right| dE
        \propto E^{-[1+(d-1)/(D_d+2/\beta)]}
        \ dE \ .
\end{equation}
Thus this derivation from first principles predicts powerlaw functions 
for all parameters $L$, $A$, $V$, $T$, $F$, $P$, $E$, and $S$ which are 
the hallmarks of SOC systems.

In summary, if we denote the occurrence frequency distributions
$N(x)$ of a parameter $x$ with a powerlaw distribution with power law index
$\alpha_x$,
\begin{equation}
        N(x) dx \propto x^{-\alpha_x} \ dx \ ,
\end{equation}
we have the following powerlaw coefficients $\alpha_x$ for the parameters
$x=L, A, V, T, F, P, E$, and $S$,
\begin{equation}
        \begin{array}{ll}
        \alpha_L &=  d \\
        \alpha_A &=  1+(d-1)/2 \\
        \alpha_V &=  1+(d-1)/d \\
        \alpha_T &=  1+(d-1)\beta/2 \\
        \alpha_F &=  1+(d-1)/D_d \\
        \alpha_P &=  1+(d-1)/d \\
        \alpha_E &=  \alpha_S = 1+(d-1)/(D_d+2/\beta)\\
        \end{array} \ .
\end{equation}
If we restrict to the case to 3D Euclidean space (d=3), as it is almost
always the case for real world data, the predicted powerlaw indexes are,
\begin{equation}
        \begin{array}{ll}
        \alpha_L &= 3 \\
        \alpha_A &= 2 \\
        \alpha_V &= 5/3 \\
        \alpha_T &= 1+\beta \\
        \alpha_F &= 1+2/D_3 \\
        \alpha_P &= 5/3 \\
        \alpha_E &= \alpha_S = 1+1/(D_3/2+1/\beta)\\
        \end{array} \ .
\end{equation}
Restricting to classical diffusion $(\beta=1)$ and a mean fractal 
dimension of $D_d \approx (1+d)/2$ for $d=3$, we have the 
following absolute predictions of the FD-SOC model,
\begin{equation}
        \begin{array}{ll}
        \alpha_L &= 3   \\
        \alpha_A &= 2   \\
        \alpha_V &= 5/3 \\
        \alpha_T &= 2   \\
        \alpha_F &= 2   \\
        \alpha_P &= 5/3 \\
        \alpha_E &= \alpha_S = 3/2 \\
        \end{array} \ .
\end{equation}
to which we refer to as the {\sl standard FD-SOC model} in this review.
We will see that these powerlaw indices represent a good first estimate
that applies to many astrophysical and other observations interpreted
as SOC phenomena. In some cases, however, the measurements clearly
do not agree with these standard values, which imposes interesting
constraints for modified SOC models.

The scaling laws between SOC parameters $E$, $P$, and $T$ (Eqs.~16-17)
imply the following correlations for standard parameters $d=3$, $D_3=2.0$,
and $\beta=1$,
\begin{equation}
	P \propto T^{3/2} \ , \qquad T \propto P^{2/3} \ ,
\end{equation}
\begin{equation}
	E \propto S \propto T^{2} \ , \qquad T \propto E^{1/2} \propto S^{1/2} \ ,
\end{equation}
\begin{equation}
	E \propto S \propto P^{4/3} \ , \qquad P \propto E^{3/4} \propto S^{3/4} \ ,
\end{equation}
which are sometimes tested in observations and cellular automaton simulations.

\subsection{Coherent and Incoherent Radiation}

Self-organized criticality models can be diagnosed and tested by means
of statistical distributions, e.g., by the omnipresent powerlaw
or powerlaw-like size distributions, and by the underlying scaling laws
that relate the powerlaw slopes of different observables to each other
(see also McAteer \etal~2014 for a description of methods).
The original paradigm of a SOC model, the BTW cellular automaton
simulations (Bak et al.~1987, 1988), produced powerlaw distributions
of two variables, the size $S$, and the time duration $T$. The size $S$
is simply defined by the time-integrated area $A$ of active nodes (pixels) 
in 2D lattice simulations, or by the time-integrated fractal volume $V_f$ 
of active nodes (voxels) in 3D lattice simulations. 

In astrophysical observations, however, the volume of an
avalanche cannot be measured, but rather a flux intensity $F_{\lambda}$ 
in some wavelength regime $\lambda$ is observed, which is not necessarily
proportional to the fractal volume $V_f$, depending on the emission 
mechanism that
is dominant at wavelength $\lambda$. Therefore, for astrophysical
observations in particular, we have to introduce a relationship between 
the observed flux $F_{\lambda}$ and the emitting volume $V_f$ that 
is fractal for a SOC avalanche process. 
For sake of simplicity we characterize this relationship 
with a power exponent $\gamma$ (Aschwanden 2012b,c),
\begin{equation}
	F_{\lambda} \propto V_f^{\gamma} \ .
\end{equation}
This definition allows us to distinguish two categories of physical
processes: {\sl incoherent processes} that have a linear relationship
between the emitting flux and volume ($\gamma=1$), and {\sl coherent
processes} that have a nonlinear relationship,
\begin{equation}
	F_{\lambda} \propto V_f^\gamma \ , \qquad
        \left\{ \begin{array}{ll}
        \gamma > 1 & {(\rm coherent \ process)} \\
        \gamma = 1 & {(\rm incoherent \ process)} \\
        \end{array} \right.
\end{equation}
Incoherent processes are, for instance, free-free emission in
optically thin media, bremsstrahlung, or gyrosynchrotron emission.
Free-free emission is a common emission mechanism in soft X-rays
and EUV, where the total flux scales with the emission measure $EM$
integrated over the entire (fractal) source volume $V_f$.
Coherent processes on the other hand, can occur by wave-particle interactions
in collisionless plasmas, such as loss-cone instabilities, 
electron-beam instabilities, or electron cyclotron maser emission.
The flux level of coherent waves amplifies exponentially or with 
a nonlinear power to the spatial scale of the source, and thus
with a nonlinear power to the source volume. 

What is the resulting modification in the size distribution of
observed fluxes? Incoherent processes are expected to have the
same size distribution as the size distribution of (fractal) avalanche
volumes. For coherent processes, the size distributions that depend on the
flux $F$ will have a modified powerlaw slope, which we can calculate
straightforwardly from the modified scaling laws (Eq.~15-17),
\begin{equation}
        F \propto V_f^\gamma \propto L^{\gamma D_d} 
	  \propto T^{\gamma D_d \beta/2} ,
\end{equation}
\begin{equation}
        P \propto V^\gamma \propto L^{\gamma d} \propto T^{\gamma d \beta/2} \ ,
\end{equation}
\begin{equation}
        E \propto L^{\gamma D_d+2/\beta} \propto  T^{\gamma D_d \beta/2+1} \ .
\end{equation}
resulting into the frequency distributions,
\begin{equation}
        N(F) dF = N(L[F]) \left| {dL \over dF} \right| dF
        \propto F^{-[1+(d-1)/\gamma D_d]} \ dF \ ,
\end{equation}
\begin{equation}
        N(P) dP = N(/[P]) \left| {dL \over dP} \right| dP
        \propto P^{-[1+(d-1)/\gamma d]}
        \ dP \ ,
\end{equation}
\begin{equation}
        N(E) dE = N(L[E]) \left| {dL \over dE} \right| dE
        \propto E^{-[1+(d-1)/(\gamma D_d+2/\beta)]}
        \ dE \ .
\end{equation}
Consequently, the generalized powerlaw coefficients $\alpha_x$ for 
the parameters $x=L, A, V, T, F, P, E$ and $S$ are (Eq.~22), 
\begin{equation}
        \begin{array}{ll}
        \alpha_L &=  d \\
        \alpha_A &=  1+(d-1)/2 \\
        \alpha_V &=  1+(d-1)/d \\
        \alpha_T &=  1+(d-1)\beta/2 \\
        \alpha_F &=  1+(d-1)/(\gamma D_d) \\
        \alpha_P &=  1+(d-1)/(\gamma d) \\
        \alpha_E &=  1+(d-1)/(\gamma D_d+2/\beta)\\
        \alpha_S &=  1+(d-1)/(D_d+2/\beta)\\
        \end{array} \ ,
\end{equation}
where we included also the time-integrated avalanche size $S$
that is generally used in cellular automaton models, which corresponds
in our definition to the time-integrated energy with $\gamma=1$.
The modification with the coherence parameter $\gamma$ predicts 
flatter powerlaw slopes ($\alpha_F,
\alpha_P, \alpha_E$) for flux-related observables ($F, P, E$)
of coherent processes. We will see that coherent emission
processes in radio wavelengths (Section 3.1.4) indeed have been
observed with flatter size distributions than incoherent emission
processes. 

\subsection{Waiting Times and Memory}

Waiting times, also called ``elapsed times'', ``inter-occurrence times'', 
``inter-burst times'', or ``laminar times'', are defined by the time 
interval between two subsequent bursts. The distribution of waiting
times requires to break a continuous time series down into discrete events,
for instance by using a threshold criterion. Consequently, waiting
time statistics requires a separation of time scales, which means
that the burst durations have to be shorter than the waiting times,
otherwise multiple bursts are counted as a single one and the waiting
time between two closely following bursts is missing in the statistics.

\subsubsection{		Stationary Poisson Processes		}

If a process is purely random, also called a ``Poisson process'',
the waiting times $\Delta t=t_{i+1}-t_i$ between subsequent bursts 
at times $t_i$ and $t_{i+1}$ should be uncorrelated
and follow a Poissonian probability distribution function, which 
can be approximated by an exponential function,
\begin{equation}
	P(\Delta t) = \lambda e^{-\lambda \Delta t} \ ,
\end{equation}
where $\lambda$ is the mean burst rate or flare rate. It the flare
rate $\lambda$ is constant, we call this also a ``stationary Poisson
process''. 

A waiting time distribution measured in a global system
loses all timing information from individual local regions, so we
can never conclude from the waiting times of a global
system whether the waiting times in a local region is a random process or not.
However, the opposite is true and can be mathematically proven, i.e., that the
combination of time series with random time intervals produces a combined time
series that has also random time intervals. This property is also called
the {\sl superposition theorem of Palm and Khinchin} (e.g., Cox and Isham 1980; 
Craig and Wheatland 2002) and is analogous to the {\sl central limit theorem}
(Rice 1995). An example that waiting times in local regions can be completely 
different from those of the global system was confirmed in earthquake 
statistics, 
where aftershocks (occurring in the same local region) exhibit an excess of 
short waiting times (Omori's law; Omori 1895), compared with the overall 
statistics of (spatially) independent earthquakes.

\subsubsection{		Non-Stationary Poisson Processes	}

Many SOC processes have variable drivers or spatial subsystems with different
drivers. Consequently the burst rates or flare rates, and thus the waiting
time statistics, may vary in time and/or space. If every spatial system is a 
random system with different flaring rates $\lambda_i$ in individual local 
regions or during individual time epochs, a superposition of many random 
systems is called a ``non-stationary Poisson process'', or
``time-dependent Poisson process''. Let us consider non-stationarity
in the time domain.  A non-stationary Poisson process may be approximated by a 
subdivision into discretized time intervals with piecewise stationary processes 
with occurrence rates $\lambda_1, \lambda_2, ..., \lambda_n$ 
(Wheatland et al.~1998),
\begin{equation}
        P(\Delta t) = \left\{ \begin{array}{ll}
          \lambda_1 e^{-\lambda_1 \Delta t} & {\rm for} \ t_1 \le t \le t_2 \\
          \lambda_2 e^{-\lambda_2 \Delta t} & {\rm for} \ t_2 \le t \le t_3 \\
        ...............                 & .......                       \\
        \lambda_n e^{-\lambda_n \Delta t} & {\rm for} \ t_n \le t \le t_{n+1}
        \end{array} \right.
\end{equation}
where the occurrence rate $\lambda_i$ is stationary during a time interval
$[t_i, t_{i+1}]$, but has different values in subsequent time intervals. The
time intervals $[t_i, t_{i+1}]$ where the occurrence rate is stationary are
called {\sl Bayesian blocks}, a special application of {\sl Bayesian
statistics} (e.g., see Scargle 1998 for astrophysical applications).
If we make a transition to a continuous flaring rate $\lambda(t)$
and use a time-dependent function $f(\lambda)$ to describe the variation of
the flaring rate $\lambda(t)$, we obtain the following waiting time distribution
(Wheatland et al.~1998, 2003),
\begin{equation}
        P(\Delta t) =
        {\int_0^{\infty}
        f(\lambda) \lambda^2 e^{-\lambda \Delta t} d\lambda
        \over \int_0^{\infty} \lambda f(\lambda)\ d\lambda} \ ,
\end{equation}
where the denominator $\lambda_0=\int_0^\infty \lambda f(\lambda) d\lambda$
is the mean rate of flaring. 

\begin{figure}
\centerline{\includegraphics[width=0.9\textwidth]{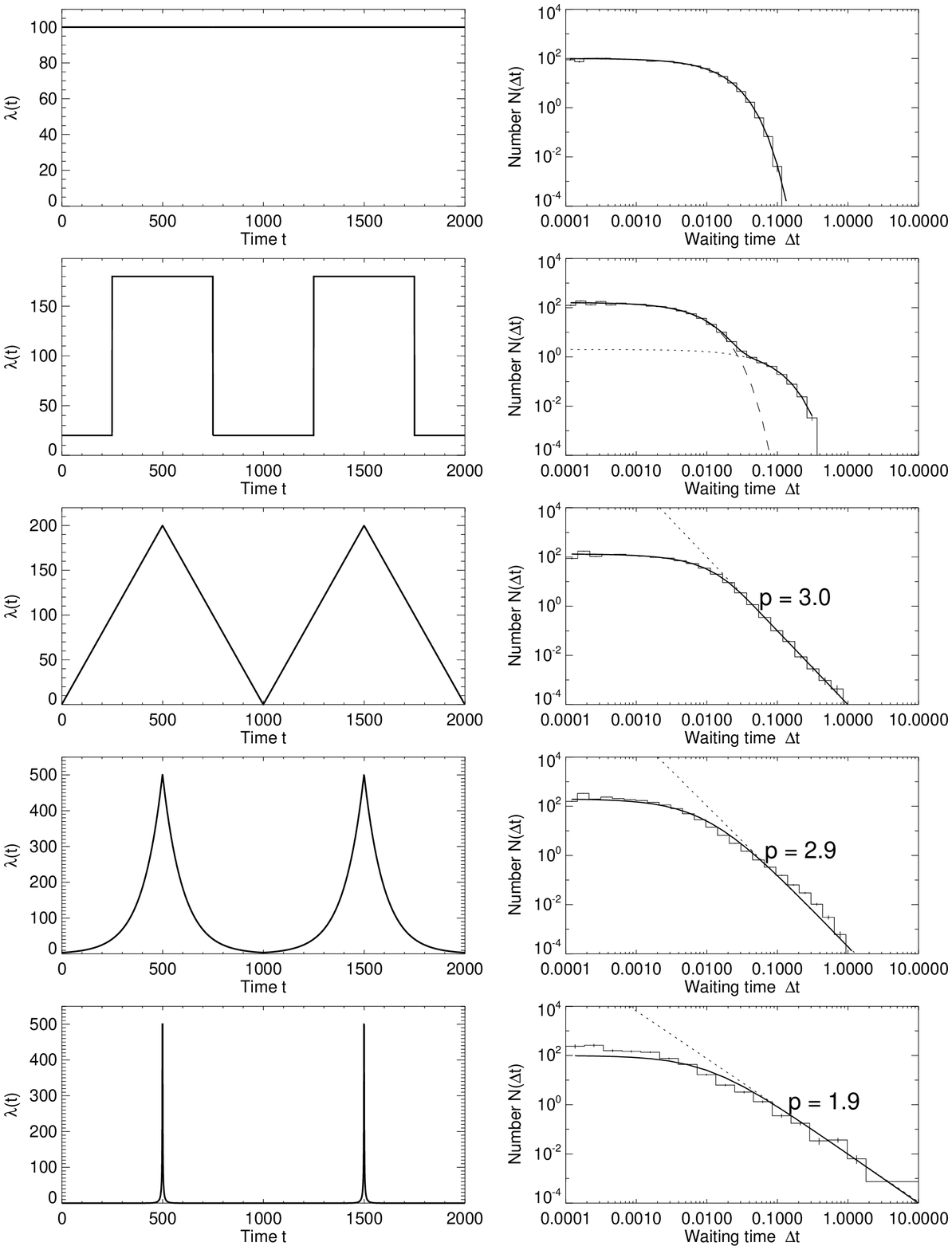}}
\captio{One case of a stationary Poisson process (top) and four cases of
nonstationary Poisson processes with two-step, linear-increasing,
exponentially varying, and $\delta$-function like variations of the
occurrence rate $\lambda(t)$. The time-dependent occurrence rates
$\lambda(t)$ are shown on the left side, while the waiting-time distributions
are shown in the right-hand panels, in the form of histograms
sampled from Monte-Carlo simulations, as well as in the form of the analytical
solutions. Powerlaw fits $N(\Delta t) \propto \Delta t^{-p}$ are indicated 
with a dotted line and labeled with the slope $p$ 
(Aschwanden and McTiernan 2010).}   
\end{figure}

It is instructive to study the functional
shape of waiting time distributions that result from non-stationary
Poisson processes. In Fig.~6 we illustrate five cases, which each can
be derived analytically: (1) a stationary Poisson process with a constant 
rate $\lambda_0$; (2) a two-step process with two different occurrence rates 
$\lambda_1$ and $\lambda_2$;
(3) a nonstationary Poisson process with a linearly increasing occurrence
rate $\lambda(t)=\lambda_0 t / T$, varying like a triangular function for
each cycle, (4) a piecewise constant Poisson process with an
exponentially varying rate distribution,
and (5) a piecewise constant Poisson process with an exponentially
varying rate distribution steepened by a reciprocal factor.
For each case we show the time-dependent occurrence rate $\lambda(t)$ and
the resulting probability distribution $P(\Delta t)$ of
events. We see that a stationary Poisson process produces an exponential
waiting-time distribution, while nonstationary Poisson processes with a
discrete number of occurrence rates $\lambda_i$ produce a superposition of
exponential distributions, and continuous occurrence rate functions
$\lambda(t)$ generate powerlaw-like waiting-time distributions
at the upper end. The analytical derivations of these five cases is
given in Aschwanden (2011a).

Thus we learn from the last four examples
that most continuously changing occurrence rates produce powerlaw-like
waiting-time distributions $P(\Delta t) \propto (\Delta t)^{-p}$
with slopes of $p \lapprox 2, ..., 3$ at large
waiting times, despite the intrinsic exponential distribution that
is characteristic to stationary Poisson processes. If the variability of
the flare rate is gradual (third and fourth case in Fig.~6),
the powerlaw slope of the waiting-time distribution is close to
$p \lapprox 3$. However, if the variability of the flare rate
shows spikes like $\delta$-functions (Fig.~6, bottom), which is highly
intermittent with short clusters of flares, the distribution of waiting
times has a slope closer to $p \approx 2$. This phenomenon is also
called {\sl clusterization} and has analogs in earthquake statistics, where
aftershocks appear in clusters after a main shock (Omori's law; Omori 1895). 
Thus the powerlaw slope of waiting times contains essential information
whether the flare rate is constant, varies gradually, or in form of
intermittent clusters.

Powerlaw-like waiting time distributions can also be produced
by standard BTW sandpile simulations, when correlations exist in the
slowly-driven external driver, producing a ``colored'' power spectrum, 
especially when only avalanches above some threshold are included in 
the waiting-time distribution (Sanchez et al.~2002).

\subsubsection{	Waiting Time Probabilities in the Fractal-Diffusive SOC Model }

The {\sl fractal-diffusive self-organized criticality (FD-SOC)} model
predicts a powerlaw distribution $N(T) \propto T^{-\alpha_T}$ of event 
durations $T$ with a slope of $\alpha_T=[1+(d-1)\beta/2]$ (Eq.~11) that 
derives directly from the scale-free probability conjecture $N(L) \propto
L^{-d}$ (Eq.~1) and the random walk (diffusive) transport 
($L \propto T^{\beta/2}$; Eq.~10). For
classical diffusion ($\beta=1$) and space dimension $d=3$ the predicted
powerlaw is $\alpha_T=2$. From this time scale distribution we can also
predict the waiting time distribution with a simple probability argument.
If we define a waiting time as the time interval between the start time
of two subsequent events, so that no two events overlap with each other
temporally,  the waiting time cannot be shorter than the time
duration of the intervening event, i.e., $\Delta t_i \ge (t_{i+1}-t_i)$.
Let us consider the case of non-intermittent, contiguous flaring, but no
time overlap between subsequent events. In this case the waiting times
are identical with the event durations, and therefore their waiting
time distributions are equal too, reflecting the same statistical 
probabilities,
\begin{equation}
	N(\Delta T) d\Delta t 
	\propto N(T) dT 
	\propto T^{-\alpha_T} \ dT 
	\propto \Delta t^{-\alpha_{\Delta t}} d\Delta t \ ,
\end{equation}
with the powerlaw slope,
\begin{equation}
	\alpha_{\Delta t} = \alpha_T = 1 + (d-1)\beta/2 \ .
\end{equation}
This statistical argument is true regardless what the order of
subsequent event durations is, so it fulfills the Abelian property.
Now we relax the contiguity condition and subdivide the time series into
blocks with contiguous flaring, interrupted by arbitrarily long quiet 
periods when no event happens (Fig.~7). The contributions of waiting times 
from the subset of contiguous time blocks will still be identical
to those of the event durations, while
those time intervals from the intervening quiet periods add a few arbitrarily
longer waiting times, which form an exponential drop-off in the case
of random quiescent time intervals (Fig.~7).
As long as the number of quiet time intervals
is much smaller than the number of detected events, the modified
waiting time distribution will still be similar to the one of
contiguous flaring (Eq.~40), which is $\alpha_{\Delta t}=2.0$ for
classical diffusion $\beta=1$ and space dimension $d=3$. Interestingly,
this predicted slope is identical to that of nonstationary Poisson
processes in the limit of intermittency (Fig.~6 bottom).

\begin{figure}[t]
\centerline{\includegraphics[width=1.0\textwidth]{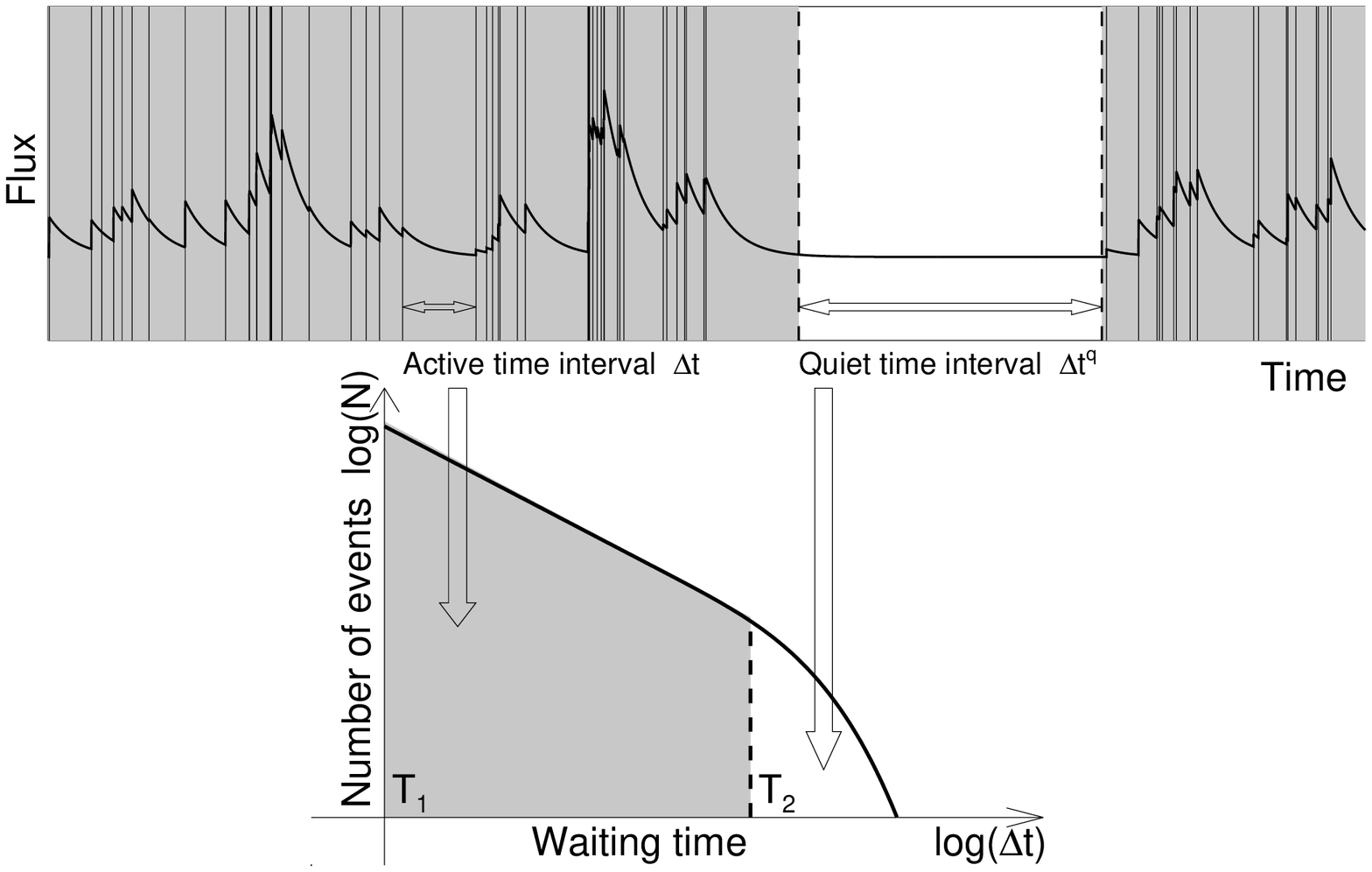}}
\captio{The concept of a dual waiting time distribution is
illustrated, consisting of active time intervals 
$\Delta t \lapprox T_2$ that contribute to a powerlaw
distribution, which is equal to that of time durations,
$N(T)$, and random-like quiescent time intervals
($\Delta t^q$) that contribute to an exponential cutoff.
Vertical lines in the upper panel indicate the start times
of events, between which the waiting times are measured
(Aschwanden 2014).}
\end{figure}

We can define a mean waiting time $\langle \Delta t \rangle$ 
from the total duration of the observing period $T_{obs}$ and 
the number of observed events
$n_{obs}$,
\begin{equation}
	\langle \Delta t \rangle = {T_{obs} \over n_{obs}} \ .
\end{equation}
From the distribution of event durations $T$, we have an inertial 
range of time scales $[T_1, T_2]$, over which we observe a powerlaw 
distribution, $N(T) \propto T^{-\alpha_T}$, with the corresponding 
number of events $[N_1, N_2]$, so that we can define a nominal 
powerlaw slope of $\alpha_T=\log(N_2/N_1)/\log(T_2/T_1)$.
If the mean waiting time of an observed time series becomes shorter 
than the upper limit of time scales, $T_2$, we start to see 
time-overlapping events, a situation we call {\sl ``event pile-up"} 
or {\sl ``pulse pile-up''}. In such a case we expect that the waiting
time distribution starts to be modified, because the time durations
of the long events are underestimated (by some automated detection
algorithm), so that the nominal powerlaw slope that is expected with no
pulse pile-up, $\alpha_{\Delta t}=\log(N_2/N_1)/\log(T_2/T_1)$,
has to be modified by replacing the upper time scale $T_2$ by the
mean waiting time $\langle \Delta t \rangle$,
\begin{equation}
	\alpha_{\Delta t}^{pileup} \ = \ \alpha_{\Delta t}
	\times \left\{ 
	\begin{array}{ll} 
		1 & {\rm for}\ \langle \Delta t \rangle \ > \ T_2 \\
		\log(T_2)/\log{\langle \Delta t \rangle} 
		  & {\rm for}\ \langle \Delta t \rangle \ \le \ T_2 \\
	\end{array}
	\right.
\end{equation}
As a consequence, the measurements of event durations must suffer
the same pile-up effect, and a similar correction is expected
for the time duration distribution $N(T)$,
\begin{equation}
	\alpha_T^{pileup} \ = \ \alpha_T
	\times \left\{ 
	\begin{array}{ll} 
		1 & {\rm for}\ \langle \Delta t \rangle \ > \ T_2 \\
		\log(T_2)/\log{\langle \Delta t \rangle} 
		  & {\rm for}\ \langle \Delta t \rangle \ \le \ T_2 \\
	\end{array}
	\right.
\end{equation}
Thus the predicted waiting time distribution has a slope of 
$\alpha_T=2$ in the slowly-driven limit, but can be
steeper in the strongly-driven limit. We will see below that the
waiting time distributions of solar flares correspond to the
slowly-driven limit during the minima of the solar 11-year cycle,
while their powerlaw slopes indeed steepen during the maxima of the
solar cycle, when the flare density becomes so high that the
slowly-driven limit, and thus the separation of time scales,
is violated. 

\subsubsection{	    Weibull Distribution and Processes with Memory	}

As we stated in a previous section, we can never conclude from the waiting 
times of a global system whether the waiting times in a local region is 
a random process or not. Non-stationary Poisson processes may fit an
observed waiting time distribution perfectly well, with an appropriate
flaring rate function $f(\lambda)$, but the best-fit solution is not unique.
Local regions may have non-random statistics with clustering, memory, and
persistence. Such non-Poissonian processes can, for instance, be
characterized with the more general Weibull distribution, which 
originially has been used to describe particle size distributions
(Weibull 1951). Here we outline the formalism according to an application
to (solar) coronal mass ejections (Telloni et al.~2014).

Generalizing the Poissonian exponential function (Eq.~37) we can define
the waiting time distribution function $P(\Delta t)$
\begin{equation}
	P(\Delta t) = z(\Delta t) \ e^{-\int_0^{\Delta t} z(x) dx} \ ,
\end{equation}
where $z(\Delta t)$ represents the local flaring rate,
\begin{equation}
	z(\Delta t) = { P(\Delta t) \over P(\Delta t \ge \Delta T)} \ ,
\end{equation}
defined by the ratio of the {\sl probability distribution function (PDF)} 
$P(\Delta t)$ and the {\sl Surviving Distribution Function (SDF)} 
$P(\Delta t \ge \Delta T)$. In a memory-less stochastic (Poisson)
process, the probability of occurrence of an event is constant,
e.g., $z(\Delta t)=\lambda$, producing the Poisson distribution (Eq.~37).
If the probability of occurrence changes with time, especially when the
process has memory, $z(\Delta t)$ can be expressed by (Weibull 1951),
\begin{equation}
	z(\Delta t) = \lambda^k k (\Delta t)^{k-1} \ ,
\end{equation}
where $k$ is the key parameter that describes whether the probability of
occurrence decreases or increases with time ($k < 1$ or $k > 1$).
Substituting Eq.~(47) into Eq.~(45) yields than the probability 
density function of a Weibull random variable $\Delta t$ (Weibull 1951),
\begin{equation}
	P(\Delta t) = {k \over \beta} 
	\left({\Delta t \over \beta}\right)^{k-1} \ e^{-(\Delta t / \beta)^k} \ ,
\end{equation}
where $\beta=1/\lambda$ is the reciprocal of the occurrence rate of the
events, $k>0$ is the shape parameter, and $\beta > 0$ is the scale parameter 
of the distribution. 

\begin{figure}[t]
\centerline{\includegraphics[width=1.0\textwidth]{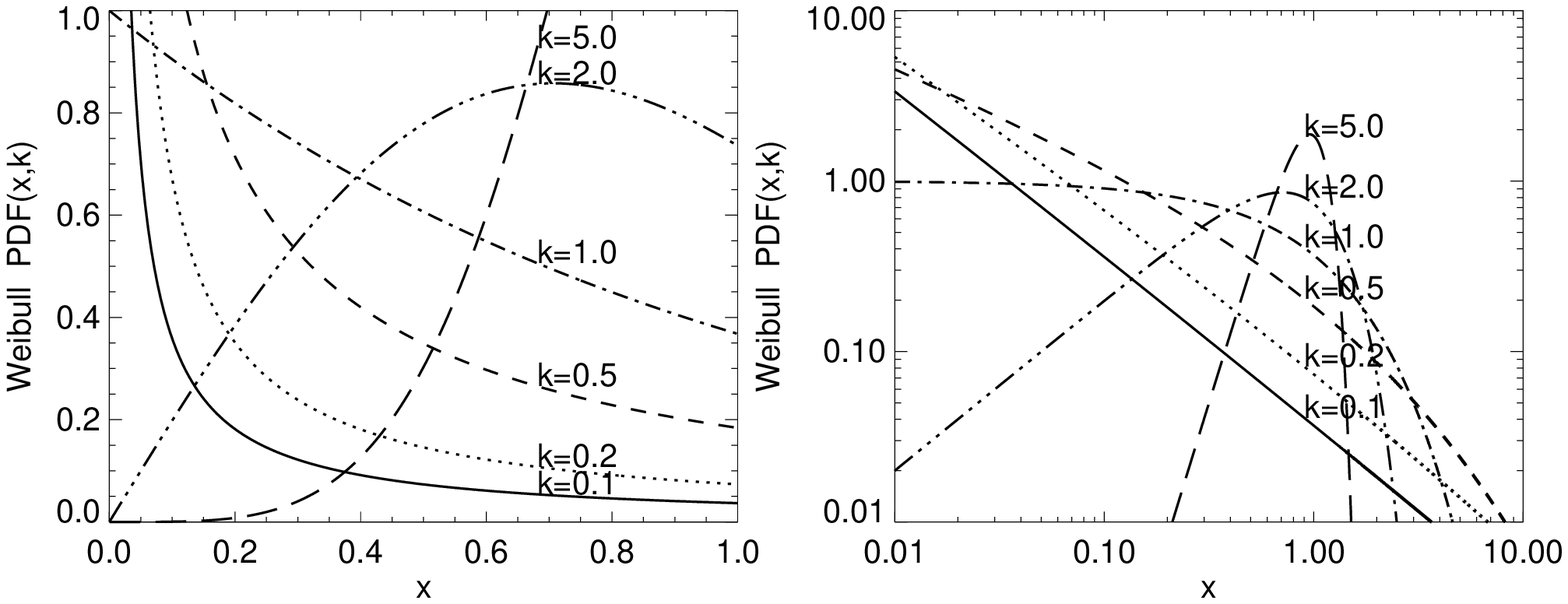}}
\captio{The Weibull probability density function (PDF) $f(x; k, \lambda)$ 
for $k=0.1, 0.2, 0.5, 1.0, 2.0, 5.0$ and $\lambda=1$, in a lin-lin display
(left panel) and in a log-log display (right panel). Notice the asymptotic
limit of a powerlaw function for $k \mapsto 0$.}
\end{figure}

In Fig.~8 we display some forms of the Weibull distribution function for
different shape parameters $k=0.1, ..., 5$. The distribution function
turns into a powerlaw function for $k \mapsto 0$, into an exponential function
for $k=1$, and into a Rayleigh distribution for $k \gg 1$, which is almost 
Gaussian-like. For $k=1$, the process is Poissonian or random 
and has no memory. For $k<1$ the flaring rate decreases over time,
while for $k>1$ the flaring rate is increasing with time, indicating that the
process has some memory and persistence, because a persistent driver with
memory varies the flaring rate with a systematic trend, which causes also
long correlation times among clusters of events. Thus, the Weibull
distribution function allows to model random-like (Poissonian) processes
as well as processes with memory and persistence. 

\subsection{The Separation of Time Scales}

Most of the original numerical simulations of SOC systems were performed
in the slowly-driven limit, which warrants a strict separation of time
scales. In lattice-type cellular automaton simulations, the separation
of time scales is enforced by dropping only one single sand grain 
at a time, or disturbing only one single lattice node at a time.
If nothing happens, the algorithm proceeds with the next input of 
a disturbance. In the alternative case, when a disturbance triggers
an avalanche, the incremental input function is stopped until the
avalanche process ends, after which another input step is continued.
This asymptotic limit of strict time scale separation between the
waiting time scale and durations of subsequent avalanches, is also
called a ``slowly-driven SOC system''. This ideal, but unnatural
condition is, however, not necessarily always enforced in nature.
Especially for SOC systems with time-variable drivers, the trigger rate
can get so high that multiple avalanches are triggered near-simultaneously
and small avalanches occur at various places while a previously
triggered large avalanche is still evolving. If we encounter such
a ``multi-avalanching system'', or multi-avalanching behavior
during some busy periods of time, we may call it a ``fast-driven'' or 
``strongly-driven'' system.  We can adopt the terminology of 
a slow or fast driver being a synonym for the existence or 
non-existence of time scale separation, which can be expressed by 
the ratio of the avalanche duration $T$ to the waiting time $\Delta t$,
\begin{equation}
	\begin{array}{rr}
	{\rm Slow\ driver:} & \mapsto \ \Delta t \gapprox T \\
	{\rm Fast\ driver:} & \mapsto \ \Delta t \lapprox T \\
	\end{array} \ .
\end{equation}
For a fast driver the question arises how this affects the observed 
(powerlaw-like) size distributions that we calculated in the slowly-driven 
limit. The answer depends very much on the event detection method.
Ideally one would use imaging information so that the spatial locations
of two temporally overlapping events can be separately determined and
the time profiles of the two events can be properly disentangled.
In practice, especially in the case of astrophysical observations,
spatial sources of co-temporaneous events cannot be resolved and 
a time series analysis is the only available method. In that case,
superimposed time profiles of different events can still be separated
if they have a characteristic shape, for instance a rapid rise and an
exponential decay, using a deconvolution method. If no proper
deconvolution method is applied, which is unfortunately the case in
almost all published studies with event statistics applied to SOC models,
there will be a systematic bias of underestimating the time duration of
long events, especially when the rule is applied that a previous event
has to end before the next event is detected. This leads predictably
to steeper powerlaw slopes in the time scale distribution $N(T)$.
We will see later on that an increase in the event rate (for instance
the flaring rate during the maximum of the solar cycle) will lead to
substantially steeper powerlaw slopes of the time scale duration
(for solar flare events detected in soft X-rays),
from a value of $\alpha_T\approx 2$ in the slowly-driven regime
(during solar cycle minimum) to a large value of $\alpha_T \lapprox 5$
in the fast-driven regime (during solar cycle maximum, see Fig.~10).
Interestingly, the size distribution of fluxes was not affected
in the strongly-driven regime.
In another study it was demonstrated that low resolution observations 
of a time profile causes an exponential cutoff at large values of
the time scale distribution, which also leads to steeper
powerlaw slopes (Isliker and Benz 2001). Thus, we generally expect
steeper powerlaws or exponential distributions of time scales in the 
limit of strong driving with clustered events that violate the 
separation of time scales, although there are also reports with
flatter powerlaw slopes during episodes of higher event rates
(e.g., Bai 1993; Georgoulis and Vlahos 1996, 1998).

\begin{table}[t]
\begin{center}
\captio{Numerical simulations of SOC cellular automata:
with spatial dimension $d=1,2,3$ and powerlaw exponents of 
avalanche sizes ($\alpha_E = \alpha_S$) and durations ($\alpha_T$),
adapted from Pruessner (2012), and theoretical predictions
of the FD-SOC model (Eq.~22).}
\medskip
\begin{tabular}{llll}
\hline
Reference			&Dimension	&Powerlaw         &Powerlaw         \\
                		&$d$            &slope $\alpha_S$ &slope $\alpha_T$ \\
\hline
\hline
Ruelle \& Sen (1992)		& 1		& 1.0 	          & 1.0             \\
Bak \& Sneppen (1993)		& 1		& $1.0-1.1$       &                 \\
Christensen et al.~(1996)	& 1             & 1.55            & $1.7-1.9$       \\
Aschwanden (2012a)		& 1		& 0.88$\pm$0.09   & 1.17$\pm$0.02   \\
FD-SOC prediction (Aschwanden 2012a) & {\bf 1}  & {\bf 1.00}     & {\bf 1.00}     \\
\hline
Bak et al.~(1987)		& 2             & 0.98            & 0.97            \\
Zhang (1989)			& 2 		& $1.2-1.7$       & 1.5             \\
Dhar (1990) 			& 2		& $1.2-1.3$       & $1.30-1.50$     \\
Manna (1990)			& 2		& 1.22            & 1.38            \\
Manna (1991)			& 2             & $1.25-1.30$     & 1.50            \\
Christensen et al.~(1991)	& 2		& 1.21            & 1.32            \\
Manna (1991), Bonachela (2008)	& 2		& 1.20            & 1.16            \\
Drossel and Schwabl (1992)	& 2             & $1.0-1.2$       & $1.20-1.30$     \\
Olami et al.~(1992)		& 2		& $1.2-1.3$       &	 	    \\
Pietronero et al.~(1994) 	& 2		& 1.25            &                 \\
Priezzhev et al.~(1996)		& 2		& 1.20            &                 \\
L\"ubeck and Usadel (1997)	& 2	        & 1.00, 1.29      & 1.48            \\
Chessa et al.~(1999)		& 2		& 1.27            &                 \\
Lin \& Hu (2002)		& 2		& $1.12-1.37$     &                 \\
Bonachela (2008)		& 2		& 1.30            &                 \\
Charbonneau et al.~(2001)	& 2		& 1.42$\pm$0.01   & 1.71$\pm$0.01   \\
McIntosh et al.~(2002)		& 2		& 1.41$\pm$0.01   &                 \\
Aschwanden (2012a)		& 2		& 1.48$\pm$0.03   & 1.77$\pm$0.18   \\
FD-SOC prediction (Aschwanden 2012a) & {\bf 2} & {\bf 1.29}      & {\bf 1.50}     \\
\hline
Bak et al.~(1987)		& 3             & 1.35            & 1.59            \\
Grassberger \& Manna (1990)	& 3             & 1.33            & 1.63            \\
Christensen et al.~(1991)	& 3             & 1.37$-$1.47     & 1.60            \\
Charbonneau et al.~(2001)	& 3		& 1.47$\pm$0.02   & 1.74$\pm$0.06   \\
McIntosh et al.~(2002)		& 3		& 1.46$\pm$0.01   & 1.71$\pm$0.01   \\
Aschwanden (2012a)		& 3		& 1.50$\pm$0.06   & 1.76$\pm$0.19   \\
FD-SOC prediction (Aschwanden 2012a)& {\bf 3}  &{\bf 1.50}       & {\bf 2.00}     \\
\hline
\end{tabular}
\end{center}
\end{table}

\subsection{	Cellular Automaton Models		}

Since the original BTW model has been a paradigm of SOC models
for 25 years, we should evaluate its predictive potential, since
every theory can only be validated when it is able to make quantitative
predictions for future (or past) measurements. The original BTW model
simulated a complex system by numerical lattice simulations of 
iterating a simple next-neighbor interaction redistribution rule
(generally called a {\sl cellular automaton model},
which produced a distribution with a powerlaw slope of 
$\alpha_E \approx 0.98$ for avalanche sizes in 2D space, or
$\alpha_E \approx 1.35$ for avalanche sizes in 3D space
(Bak et al.~1987). These values are somewhat different from the
predictions of the basic SOC model based on the scale-free probability
conjecture (Sections 2.6 and 2.7), which predicts
$\alpha_E = 9/7 \approx 1.29$ for avalanche sizes in 2D space, and 
$\alpha_E \approx 1.50$ for avalanche sizes in 3D space.
Other extensive BTW simulations with a variety of grid sizes find
$\alpha_E \approx 1.42\pm0.01$ for avalanche sizes in 2D space, and 
$\alpha_E \approx 1.47\pm0.02$ for avalanche sizes in 3D space
(Charbonneau et al.~2001). The latter values are actually almost
consistent with the value $\alpha_E=1.55$ (in 2D space) obtained from
a pre-Bak simulation as a model for propagating brittle failure
in heterogeneous media (Katz 1986). From these few examples it is
already clear that various cellular automaton models produce different
powerlaw slopes, and thus the question arises whether the obtained
powerlaw slopes depend on the numerical details of the setup of 
lattice simulations, or whether they have universal validity that
is independent of numerical redistribution rules and may even apply
to observations in nature.

In order to investigate the universality of cellular automaton models
we compare the obtained powerlaw slope ($\alpha_S = \alpha_E$) of avalanche
sizes (which is the time-integrated volume of all active nodes at
each time step of an avalanche) and the powerlaw slope ($\alpha_T$)
of the avalanche durations $T$. 
An exhaustive collection of cellular automaton models are described
in Pruessner (2012), from which we extract the powerlaw indices of
the mentioned parameters (Table 1). 

Based on the scale-free probability conjecture and the geometric scaling 
laws of the fractal-diffusive SOC model described in Sections 2.6-2.10,
we predict for classical diffusion ($\beta=1$) and a mean fractal
dimension $D_d=(1+d)/2$ the following powerlaw slopes for avalanche
size distributions (Eq.~22): $\alpha_E=1$ for 1D space, 
$\alpha_E=9/7\approx 1.29$ for 2D space, and $\alpha_E=3/2=1.5$ for 3D space, 
which agree with most of the measured slopes of
avalanche sizes in cellular automaton simulations (Table 1).
For event durations we predict: $\alpha_T=1$ for 1D space,
$\alpha_T=3/2=1.5$ for 2D space, and $\alpha_T=2.0$ for 3D space,
which also roughly agrees with the simulations in Table 1. 

Vice versa, the measured values listed in Table 1 can be used to
invert the diffusive spreading exponent $\beta$ and the fractal 
dimension $D_d$ for cellular automata according to Eq.~(22):
\begin{equation}
	\beta = { 2 (\alpha_T - 1) \over (d-1)} \ ,
\end{equation}
\begin{equation}
	D_d = {(d - 1) \over (\alpha_E - 1)} - {2 \over \beta} \ .
\end{equation}

For instance, the 3D cellular automaton simulations listed in Table 1
exhibit a range of $\alpha_T \approx 1.6-1.8$ for the powerlaw slope
of time durations, which is systematically below the prediction of
the standard (FD-SOC) model with $\alpha_T=2.0$. Application of
Eq.~(50) would then imply a diffusive spreading exponent of
$\beta \approx 0.6-0.8$, which is the sub-diffusive regime.
We will see later on that real-world data yield a powerlaw
slope of $\alpha_T \approx 2.0$ (e.q., Table 2), which corresponds
to classical diffusion or random walk ($\beta=1$). This tells us
that the cellular automaton redistribution rules do not necessarily
reflect the behavior of SOC processes found in the real world.

The diffusion or spreading exponent $\beta$ and the
fractal dimension $D_d$ are essentially macroscopic parameters to
describe the average dynamics and inhomogeneous spatial structure
of avalanches, which are microscopically defined in terms of an
iterative mathematical redistribution rule. The diffusion exponent
$\beta$ characterizes the macroscopic transport process (subdiffusive,
classical diffusion, hyper-diffusion), and the fractal dimension
describes the spatial inhomogeneity of an avalanche, in the spirit
of Bak and Chen (1989): {\sl Fractals in nature originate from
self-organized critical dynamical processes}. Cellular 
automata exhibit a range of fractal dimensions and diffusion
exponents, as the values in Table 1 demonstrate, and thus may 
not have universal validity for SOC systems. If we find the same
disparity among astrophysical
observations, as we will survey in the following sections, nature
operates in SOC systems with different spatial inhomogeneities 
and transport processes, which may be related to the 
underlying physical scaling laws in each SOC system. The
cellular automaton world may have (slightly) different SOC 
parameters ($\beta, D_d$) than the astrophysical world, but
we are able to describe the nonlinear dynamics of 
complex systems with the same theoretical framework.

\clearpage

\section{	ASTROPHYSICAL APPLICATIONS	}

We subdivide the astrophysical phenomena that have been associated with
SOC according to solar physics (Sections 3.1, 3.2), the Earth's 
magnetosphere and planets (Section 3.3), and stars and galaxies 
(Section 3.4).  We tabulate the statistics
of SOC parameters mostly in form of measured power law indices. 
In addition, we discuss briefly the theoretical interpretations in each 
case and summarize studies that contain modeling attempts of these
SOC phenomena, often tailored to a specific astrophysical object.

\subsection{Solar Physics: Observations}

The applications of SOC theory to solar data outnumbers all other
astrophysical applications. Therefore, we brake the subject down
into observational statistics from different wavelengths (hard X-rays,
soft X-rays, EUV, radio, etc.) in Section 3.1, and into various aspects 
of theoretical modeling (e.g., cellular automaton simulations,
magnetic fields, magnetic reconnection, plasma magneto-hydrodynamics 
(MHD), coronal heating, particle acceleration, solar wind,
Sun-Earth connection, etc.) in Section 3.2.

\subsubsection{Statistics of Solar Flare Hard X-Rays}

Solar flares provide the energy source for acceleration of nonthermal
particles, which emit bremsstrahlung in hard X-ray wavelengths, once
the non-thermal particles interact with a high-density plamsa via
Coulomb collisions. Most solar flares display an impulsive component
in hard X-rays, produced by accelerated coronal electrons that
precipitate towards the chromosphere and produce intense hard X-ray
emission at the footpoints of flare loops. Therefore, hard X-ray
pulses are a reliable signature of solar flares, often detected
at energies $\gapprox 20$ keV, but for smaller flares down to
$\gapprox 8$ keV. 

Solar flare event catalogs containing the peak rate $(P)$,
fluences $(E)$, and flare durations $(T)$, have therefore been
compiled from a number of spacecraft or balloon-borne hard X-ray
detectors over the last three decades, such as from OSO-7 
(Datlowe et al.~1974), a University of Berkeley balloon flight
(Lin et al.~1984), HXRBS/SMM (Dennis 1985; Schwartz et al.~1992; Crosby 
et al.~1993), BATSE/CGRO (Schwartz et al.~1992; Biesecker et 
al.~1993, 1994; Biesecker 1994), WATCH/GRANAT (Crosby 1996; 
Georgoulis et al.~2001); ISEE-3 (Lu et al.~1993; Lee et al.~1993; 
Bromund et al.~1995); PHEBUS/GRANAT (Perez-Enriquez and 
Miroshnichenko 1999).  RHESSI (Su et al.~2006; Christe et al.~2008; 
Lin et al.~2001), and ULYSSES (Tranquille et al.~2009).
Three examples of hard X-ray peak flux distributions are shown
in Fig.~9. Note that the size distribution of peak counts have
a sharp cutoff at the lower end due to a fixed count rate 
threshold that is generally used in the compilation of hard X-ray
flare catalogs, and thus the powerlaw slope can be determined with
the highest accuracy. Other parameters have generally a gradual
rollover at the low end due to incomplete sampling and finite-resolution
effects, which causes 
truncation effects in the histogram and hampers the accuracy of
the powerlaw fit. The size distribution of solar flare hard X-ray counts,
which has already been pointed out before the SOC concept came along
(Dennis 1985), is still one of the ``cleanest'' powerlaw size distributions
measured in astrophysics (Fig.~9). 

\begin{table}[t]
\begin{center}
\captio{Frequency distributions measured from solar flares in hard X-rays
and $\gamma$-rays. The prediction is based on the FD-SOC model (Aschwanden 2012a).}
\medskip
\begin{tabular}{lllrll}
\hline
Powerlaw        &Powerlaw       &Powerlaw   &Number     &Instrument&References\\
slope of        &slope of       &slope of   &of         &and     &\\
peak flux       &fluence        &durations  &events     &threshold&\\
$\alpha_P$      &$\alpha_E$     &$\alpha_T$ &$n$        &energy  &\\
\hline
\hline
1.8             &               &           &123        &OSO--7($>$20 keV)&Datlowe \etal (1974) \\
2.0             &               &           &25         &UCB($>$20 keV)   &Lin \etal (1984)\\
1.8             &               &           &6775       &HXRBS($>$20 keV) &Dennis (1985)\\
1.73$\pm$0.01   &               &           &12,500     &HXRBS($>$25 keV) &Schwartz \etal (1992)\\
1.73$\pm$0.01   &1.53$\pm$0.02  &2.17$\pm$0.05 &7045    &HXRBS($>$25 keV) &Crosby \etal (1993)\\
1.71$\pm$0.04   &1.51$\pm$0.04  &1.95$\pm$0.09 &1008    &HXRBS($>$25 keV) &Crosby \etal (1993)\\
1.68$\pm$0.07   &1.48$\pm$0.02  &2.22$\pm$0.13 &545     &HXRBS($>$25 keV) &Crosby \etal (1993)\\
1.67$\pm$0.03   &1.53$\pm$0.02  &1.99$\pm$0.06 &3874    &HXRBS($>$25 keV) &Crosby \etal (1993)\\
1.61$\pm$0.03   &               &           &1263       &BATSE($>$25 keV) &Schwartz \etal (1992)\\
1.75$\pm$0.02   &               &           &2156       &BATSE($>$25 keV) &Biesecker \etal (1993)\\
1.79$\pm$0.04   &               &           &1358       &BATSE($>$25 keV) &Biesecker \etal (1994)\\
1.59$\pm$0.02   &               &2.28$\pm$0.08 &1546    &WATCH($>$10 keV) &Crosby (1996)\\
1.86            &1.51           &1.88       &4356       &ISEE--3($>$25 keV) &Lu \etal (1993)\\
1.75            &1.62           &2.73       &4356       &ISEE--3($>$25 keV) &Lee \etal (1993)\\
1.86$\pm$0.01   &1.74$\pm$0.04  &2.40$\pm$0.04 &3468    &ISEE--3($>$25 keV) &Bromund \etal (1995)\\
1.80$\pm$0.01   &1.39$\pm$0.01  &           &110        &PHEBUS($>$100 keV) &Perez-Enriquez \& \\
		&		&	    &	        &                   &Miroshnichenko (1999)\\
1.80$\pm$0.02   &               &2.2$\pm$1.4&2759       &RHESSI($>$12 keV)  &Su \etal (2006)\\
1.58$\pm$0.02   &1.7$\pm$0.1    &2.2$\pm$0.2&4241       &RHESSI($>$12 keV)  &Christe \etal (2008)\\
1.6             &               &           &243        &BATSE($>$8 keV)    &Lin \etal (2001)\\
1.61$\pm$0.04   &               &           &59         &ULYSSES($>$25 keV) &Tranquille \etal (2009)\\
\hline
1.73$\pm$0.07   &1.62$\pm$0.12  &1.99$\pm$0.35 &        &Average            &All HXR observations \\
\hline
{\bf 1.67}	&{\bf 1.50}     &{\bf 2.00} &           &FD-SOC prediction  &Aschwanden (2012a)\\
\hline
\end{tabular}
\end{center}
\end{table}

\begin{figure}[t]
\centerline{\includegraphics[width=0.8\textwidth]{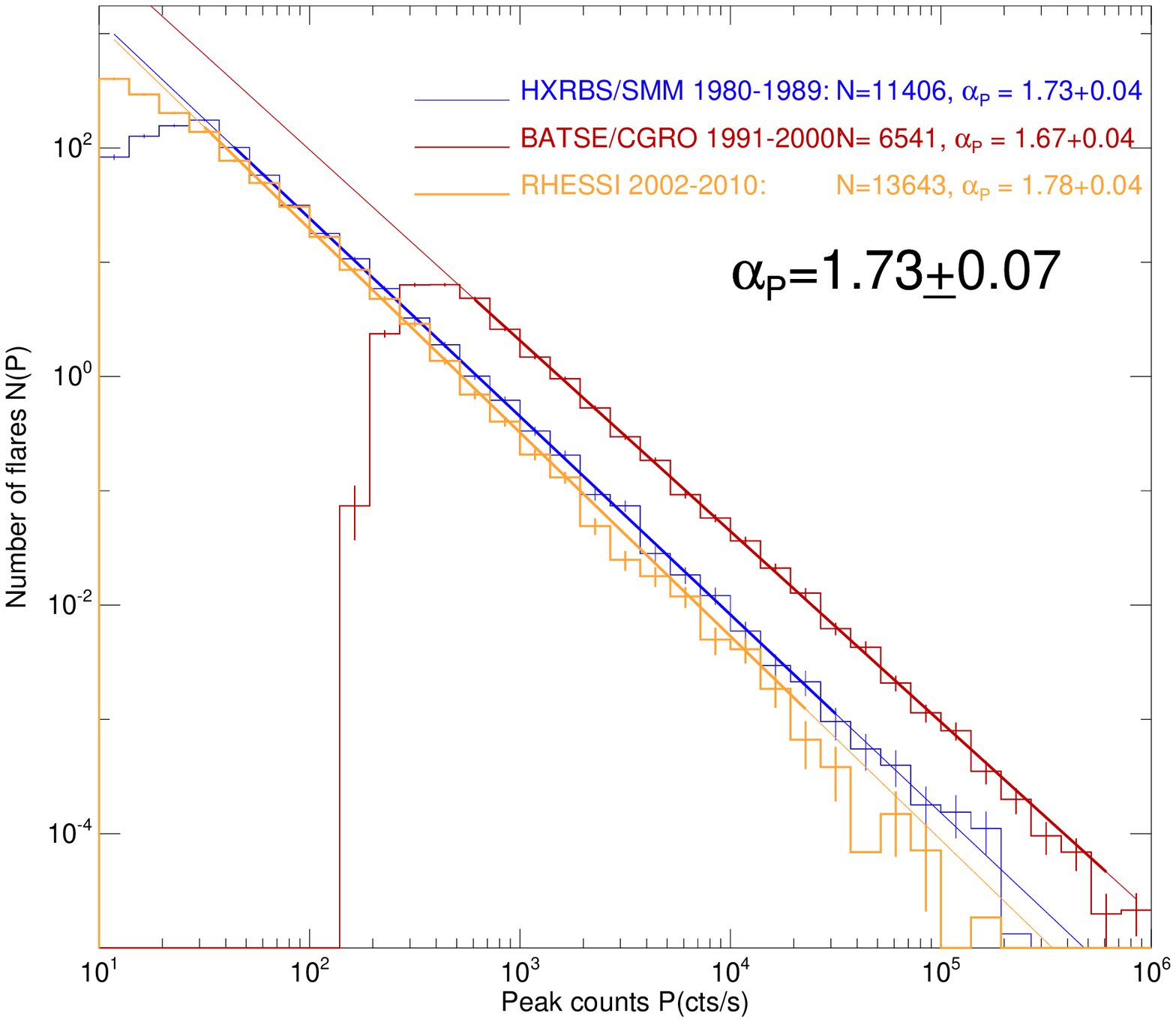}}
\captio{Occurrence frequency distributions of hard X-ray peak count rates
$P$ [cts s$^{-1}$] observed with HXRBS/SMM (1980 -- 1989), BATSE (1991 -- 2000),
and RHESSI (2002 -- 2010), with powerlaw fits. An average pre-flare
background of $40$ [cts s$^{-1}$] was subtracted from the HXRBS
count rates. Note that BATSE/CGRO has larger detector areas, and thus
records higher count rates (Aschwanden 2011b).}
\end{figure}

A compilation of occurrence frequency distribution powerlaw slopes
of solar hard X-ray flare peak fluxes ($\alpha_P$), fluences
or energies ($\alpha_E$), and flare durations ($\alpha_T$) is listed 
in Table 2. In this Table we combined both the powerlaw slopes
$\alpha_E$ from the fluences (which is the time-integrated or
total number of hard X-ray counts per flare) and nonthermal energies
(which are computed from the hard X-ray energy spectrum assuming
a low-energy cutoff at 10 or 25 keV), both representing a
physical quantity in terms of energy. In Table 2 we indicate also
the number of events, which constrains the accuracy of the fitted
powerlaw slopes. Synthesizing the datasets with the largest
statistics (HXRBS/SMM, BATSE/CGRO, RHESSI), the following
means and standard deviations of the powerlaw slopes were found
$\alpha_P=1.73\pm0.07$ for the peak fluxes (Fig.~9), 
$\alpha_E=1.62\pm0.12$ for the fluences or energies, and 
$\alpha_T=1.99\pm0.35$ for the flare durations (Aschwanden 2011b).
The uncertainties of the powerlaw slope quoted in literature
generally include the formal fitting error only,
while the standard deviations given here 
reflect methodical and systematic uncertainties also, since every
dataset has been analyzed from different instruments and with
different analysis methods. One of the largest systematic uncertainties
results from the preflare background subtraction, because the preflare 
flux is often not specified in solar flare catalogs.
Nevertheless, given these systematic
uncertainties, the observed values are consistent with the 
theoretical predictions of the basic fractal-diffusive SOC model, 
based on an Euclidean space dimension of $d=3$, a mean fractal dimension
of $D_3=2$, and classical diffusion $\beta=1$, which yields 
$\alpha_P=1.67$ for peak fluxes,
$\alpha_E=1.50$ for energies, and
$\alpha_T=2.00$ for durations (Eq.~24).
Thus, the basic fractal-diffusive SOC model predicts the correct
powerlaw slopes within the uncertainties of hard X-ray measurements.

Frequency-size distributions of solar flares are generally sampled from
the entire Sun, and thus from multiple active regions that are present
on the visible hemisphere at a given time. This configuration corresponds
to a multi-sandpile situation, and the resulting powerlaw distribution
is composed of different individual active regions, which may have different
physical conditions and sizes. In particular, different sizes may cause
an exponential cut-off at the upper end of the size distribution due to
finite system-size effects. A study of flare statistics on individual
active regions, however, did not reveal significant differences in their
size distributions, and thus the size distributions of individual active
regions seem to follow the universal powerlaw slopes that are invariant,
individually as well as in a superimposed ensemble (Wheatland 2000c),
except for one particular active region (Wheatland 2010). 

Instead of testing powerlaw slopes of size distributions, an equivalent
test is a linear regression fit among SOC parameters. For instance,
statistics of WATCH/GRANAT data exhibited correlations of
$P \propto E^{0.60\pm0.01}$, $T \propto E^{0.53\pm0.02}$, and
$T \propto P^{0.54\pm0.03}$ (Georgoulis et al.~2001), which are
consisent with the predictions of the standard model (Section 2.10),
i.e., $P \propto E^{0.75}$, $T \propto E^{0.50}$, and
$T \propto P^{0.67}$, given the uncertainties of about $\pm0.15$ due
to data truncation effects that are not accounted for in the
linear regression fits. 

Time series analysis of solar hard X-ray bursts has been performed for a
few flares with a variety of methods, such as wavelet analysis
(Aschwanden et al.~1998a), search for quasi-periodic variations
(Jakimiec and Tomczak 2010), search for sub-second time scales
(Cheng et al.~2012), statistics of UV subbursts (used as proxies
for the hard X-ray subbursts) during a flare
that exhibit powerlaw distributions (Nishizuka et al.~2009),
multi-fractal spectral analysis of a hard X-ray time profile 
(McAteer et al.~2007, 2013b), or wavelet and {\sl local intermittency 
measure (LIM)} analysis (Dinkelaker and MacKinnon 2013a,b).
The size distributions $N(t)$ of hard X-ray sub-burst durations during 
a flare were found to be mostly exponential (Aschwanden et al.~1998a), 
probably due to finite system-size effects in each flaring region.
The LIM method can reveal scale-invariant time evolutions, such as
the fragmentation of the energy release cascading from large to
smaller structures (the ``top-down'' scenario), or a small flare event
that is avalanching into a larger structure (the ``bottom-up'' scenario), 
but it was found that neither of the two extremes captures the totality 
of a flare time profile (Dinkelaker and MacKinnon 2013a,b). 

\begin{table}[t]
\begin{center}
\normalsize
\captio{Frequency distributions measured from solar flares in soft X-rays.
Measurements with no preflare background subtraction are marked with parentheses.}
\medskip
\begin{tabular}{llllll}
\hline
Powerlaw        &Powerlaw       &Powerlaw   &log        &Instrument&References\\
slope of        &slope of       &slope of   &range      &        &\\
peak flux       &total fluence  &durations  &           &        &\\
$\alpha_P$      &$\alpha_E$     &$\alpha_T$ &           &        &\\
\hline
\hline
1.8             &               &           &1          &OSO-3   &Hudson \etal (1969)\\
1.75            &1.44           &           &2          &Explorer&Drake (1971)\\
1.64-1.89       &1.5-1.6        &           &2          &Yohkoh  &Shimizu (1995)\\
1.79            &               &           &2          &SMM/BCS &Lee \etal (1995)\\
1.86            &               &           &2          &GOES    &Lee \etal (1995)\\
1.88$\pm$0.21   &               &           &3          &GOES    &Feldman \etal (1997)\\
1.7$\pm$0.4     &               &           &2          &Yohkoh  &Shimojo \& Shibata (1999)\\
1.98            &1.88           &           &3          &GOES    &Veronig \etal (2002a,b)\\
1.98$\pm$0.11   &               &2.02$\pm$0.04    &5    &GOES    &Aschwanden \& Freeland (2012)\\
$(2.11\pm0.13)$ &$(2.03\pm0.09)$ &$(2.93\pm0.12)$ &3    &GOES    &Veronig \etal (2002a)\\
$(2.16\pm0.03)$ &$(2.01\pm0.03)$ &$(2.87\pm0.09)$ &3    &GOES    &Yashiro \etal (2006)\\
\hline
{\bf 1.67}      &{\bf 1.50}     &{\bf 2.00} &           &FD-SOC prediction &Aschwanden (2012a)\\
\hline
\end{tabular}
\end{center}
\end{table}

\subsubsection{Statistics of Solar Flare Soft X-rays}

Solar flares display signatures of thermal emission in soft X-ray 
wavelengths, besides the non-thermal emission detected in hard X-rays.
The emission in both wavelength regimes is produced by the same
flare process, which is called the {\sl chromospheric evaporation
scenario}, but by different physical processes. While hard X-rays
are mostly produced by bremsstrahlung of non-thermal particles
precipitating down into the dense chromosphere, soft X-ray line
and continuum emission is excited by impulsive heating of the 
chromospheric plasma. The precipitating electrons and ions essentially
dictate the heating rate of the chromospheric plasma, while the
energy emitted from the heated thermal plasma (typically to temperatures of
$T_e \approx 10-35$ MK) follows approximately the time integral of
the hard X-ray-driven heating rate, a relationship that has been
dubbed the {\sl Neupert effect}. Because of this intimate relationship
between soft X-rays and hard X-rays in solar flares, similar energy
or size distributions are expected in both wavelength regimes, which
is indeed the case, as the compilations in Table 2 and 3 show. 

Size distributions of soft X-ray peak fluxes, fluences, and durations
were mostly obtained from flare detections with the OSO-3 spacecraft 
(Hudson et al.~1969), the Explorer (Drake 1971), Yohkoh/SXT
(Shimizu 1995; Shimojo and Shibata 1999), the SMM/BCS (Lee et al.~1999),
and the GOES spacecraft (Lee et al.~995; Feldman et al.~1997;
Veronig et al.~2002a,b; Yashiro et al.~2006; Aschwanden and Freeland 2012).
Interestingly, the size distribution of the peak count rates in the
range of $\alpha_P=1.64-1.98$ is similar to the hard X-rays, and thus
implies a proportionality between the hard X-ray counts and the soft X-ray
fluxes, which is different from what is expected from the Neupert
effect. Since the Neupert effect predicts that the time profile of
soft X-rays approximately follows the time integral of the impulsive
hard X-rays, one would expect that the soft X-ray peak flux distribution
should be equal to the hard X-ray fluences, which is however not the case 
(Lee et al.~1995). The different powerlaw slopes indicate a special
scaling law between flare temperatures and densities, i.e., 
$n_e \propto T^{-4/5}$ (Lee et al.~1995), while the Neupert effect
must be considered as an oversimplified rule that neglects any 
temperature dependence.

Some of the size distributions of soft X-ray peak fluxes have been
found to have values steeper than $\alpha_P \ge 2.0$ (Veronig et al.~2002a;
Yashiro et al.~2006), which in hindsight we can understand to be a consequence 
of neglecting the subtraction of the preflare background flux, which makes 
up a substantial amount of the total flux for small flares.

\begin{figure}[t]
\centerline{\includegraphics[width=0.9\textwidth]{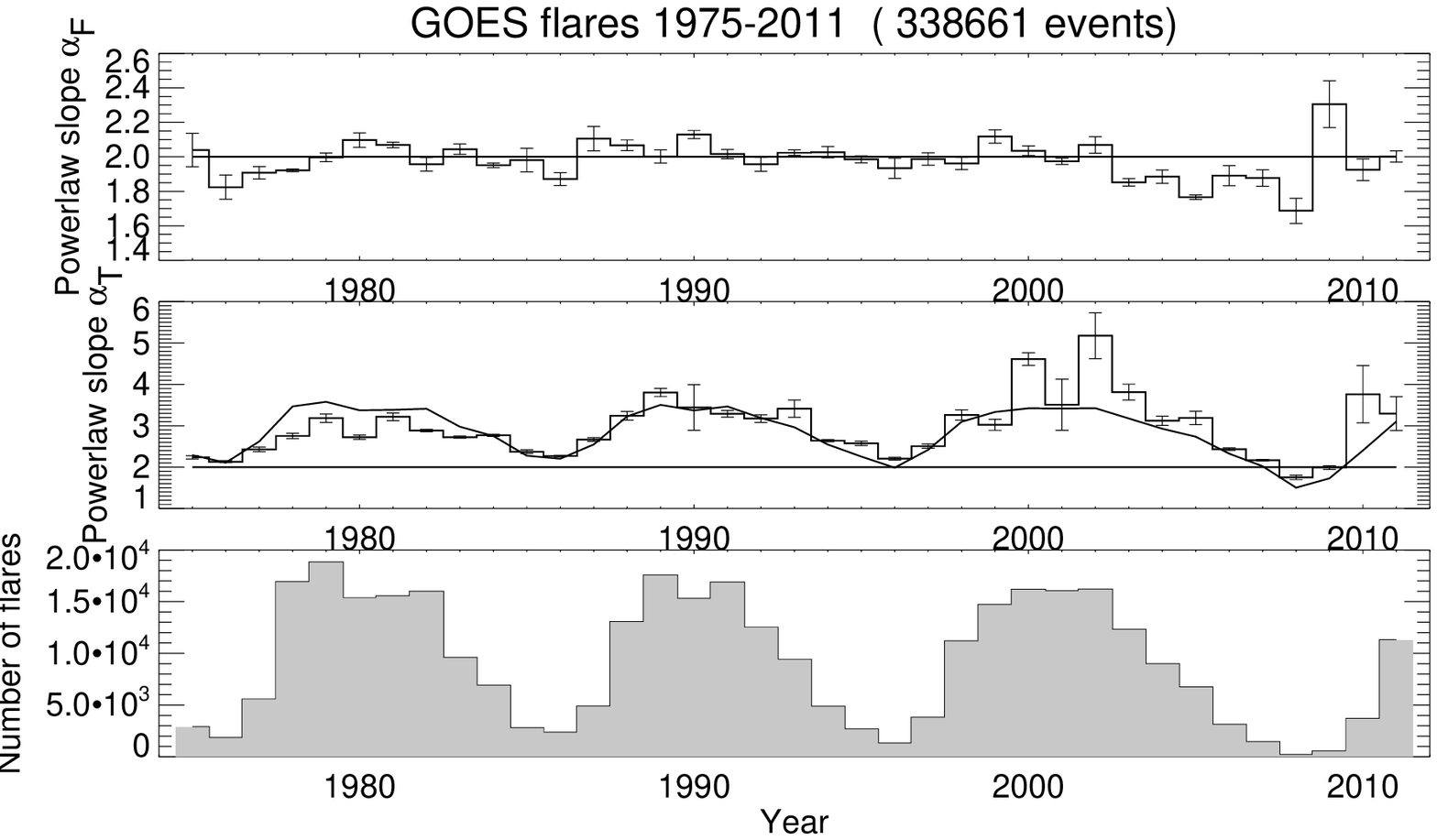}}
\captio{Variation of the power-law slopes $\alpha_P(t)$ of the soft X-ray
1-8 \ang\ peak flux (top panel) and the flare rise time $\alpha_T(t)$,
detected with GOES (middle panel), and the annual variation of the 
sunspot number over 3 solar cycles (bottom panel). The sunspot number
predicts the variation in the powerlaw slope $\alpha_T(t)$ of the flare
time duration (smooth curve in middle panel) as a consequence of the
violation of the separation of time scales (Aschwanden and Freeland 2012).}
\end{figure}

Flare statistics from the GOES satellite could be sampled over a period
of 37 years (1975-2011), which covers about three solar cycles. Since
the soft X-ray flux from the Sun varies by about two orders of magnitude
during each solar cycle, due to the variation of emerging magnetic fields
and the resulting coronal plasma heating rate, which is driven by the 
solar magnetic dynamo, the Sun is an ideal system to study
SOC systems with variable drivers. While the powerlaw of the soft X-ray
peak rate was found to be invariant during different solar cycles, having
a roughly constant value of $\alpha_F=1.98\pm0.11$, the time durations were found
to have a variable slope from $\alpha_T\approx 2.0$ during solar minima
to $\alpha_T\approx 2-5$ during solar maxima (Fig.~10), which was explained 
in terms of a flare pile-up effect (Aschwanden and Freeland 2012). In other
words, the {\sl separation of time scales}, i.e., the waiting times and flare 
durations, is violated during the busy periods of the solar cycle maximum.
In contrast, an opposite trend has been reported for a 158-day
modulation of the flare rate (Bai 1993).

A power spectrum of a time series of the GOES 0.5-4 \ang\ flux during a 
flare-rich episode of two weeks during 2000, containing about 100 
GOES $>$C1.0 flares, has been found to follow a spectral slope of 
$P(\nu) \propto \nu^{-1}$ (Bershadskii and Sreenivasan 2003),
which indeed confirms Bak's original idea that the SOC concept provides 
an explanation for the 1/f-noise (Bak et al.~1987). 

\subsubsection{Statistics of Solar Flare EUV Fluxes}

Large solar flares (with energies of $E \approx 10^{30}-10^{32}$ erg)
exhibit heated plasma with peak temperatures of 
$T_e \approx 10-35$ MK, most conspicously detected in soft X-rays,
which cools down to temperatures of $T_e \approx 1-2$ MK that is readily
detected in the postflare phase in extreme ultra-violet (EUV) wavelengths.
Also small flares, microflares, and nanoflares (with energies of
$E \approx 10^{24}-10^{27}$ erg) radiate mostly in the EUV temperature range. 
Combining these wavelengths, one can obtain statistics of solar flare
energies extending over up to 9 orders of magnitude (Fig.~11), hence the 
term ``nanoflares''. Therefore, gathering flare statistics in EUV is expected
to complement the lower end of the size distribution sampled in the upper
end in soft X-rays and hard X-rays.

\begin{table}[t]
\begin{center}
\normalsize
\captio{Frequency distributions measured in small-scale events in EUV,
UV, and H$\alpha$.}
\medskip
\begin{tabular}{lllll}
\hline
Powerlaw        &Powerlaw       &Powerlaw   &Waveband &References\\
slope of        &slope of       &slope of   &        &\\
peak flux       &total fluence  &durations  &        &\\
                &or energy      &           &        &\\
$\alpha_P$      &$\alpha_E$     &$\alpha_T$ &$\lambda$ (\ang )&\\
\hline
\hline
                &2.3-2.6        &           &171, 195 &Krucker \& Benz (1998)\\
$1.19\pm1.13$   &               &           &195      &Aletti \etal (2000)\\
                &2.0-2.6        &           &171, 195 &Parnell \& Jupp (2000)\\
$1.68-2.35$     &$1.79\pm0.08$  &           &171, 195 &Aschwanden \etal (2000a,b)\\
                &2.31-2.59      &           &171, 195 &Benz \& Krucker (2002)\\
                &2.04-2.52      &           &171, 195 &Benz \& Krucker (2002)\\
$1.71\pm0.10$   &$2.06\pm0.10$  &           &171      &Aschwanden \& Parnell (2002)\\
$1.75\pm0.07$   &$1.70\pm0.17$  &           &195      &Aschwanden \& Parnell (2002)\\
$1.52\pm0.10$   &$1.41\pm0.09$  &           &AlMg     &Aschwanden \& Parnell (2002)\\
                &$1.54\pm0.03$  &           &171+195+AlMg &Aschwanden \& Parnell (2002)\\
$2.12\pm0.05$   &               &           &6563     &Georgoulis \etal (2002)\\
1.5-3.0         &               &           &1-500    &Greenhough \etal (2003)\\
                &               &1.4-2.0    &171,195,284 &McIntosh \& Gurman (2005)\\
   	        &$1.66-1.70$    &$1.96-2.02$&EUV      &Uritsky \etal (2007)\\
$1.86\pm0.05$	&$1.50\pm0.04$  &$2.12\pm0.11$&EUV    &Uritsky \etal (2013)\\
1.5             &               &2.3        &1550     &Nishizuka \etal (2009)\\
2.42$-$2.52     &		&2.02$-$2.66&STEREO 171 &Aschwanden \etal (2013b)\\
2.66$-$2.69	&		&2.50$-$2.52&STEREO 195 &Aschwanden \etal (2013b)\\
2.14$-$2.18	&		&2.15$-$2.24&STEREO 284 &Aschwanden \etal (2013b)\\
2.58$-$2.70	&		&2.61$-$2.74&STEREO 304 &Aschwanden \etal (2013b)\\
\hline
{\bf 1.67}      &{\bf 1.50}     &{\bf 2.00} &FD-SOC prediction &Aschwanden (2012a) \\
\hline
\end{tabular}
\end{center}
\end{table}

\begin{figure}[t]
\centerline{\includegraphics[width=0.9\textwidth]{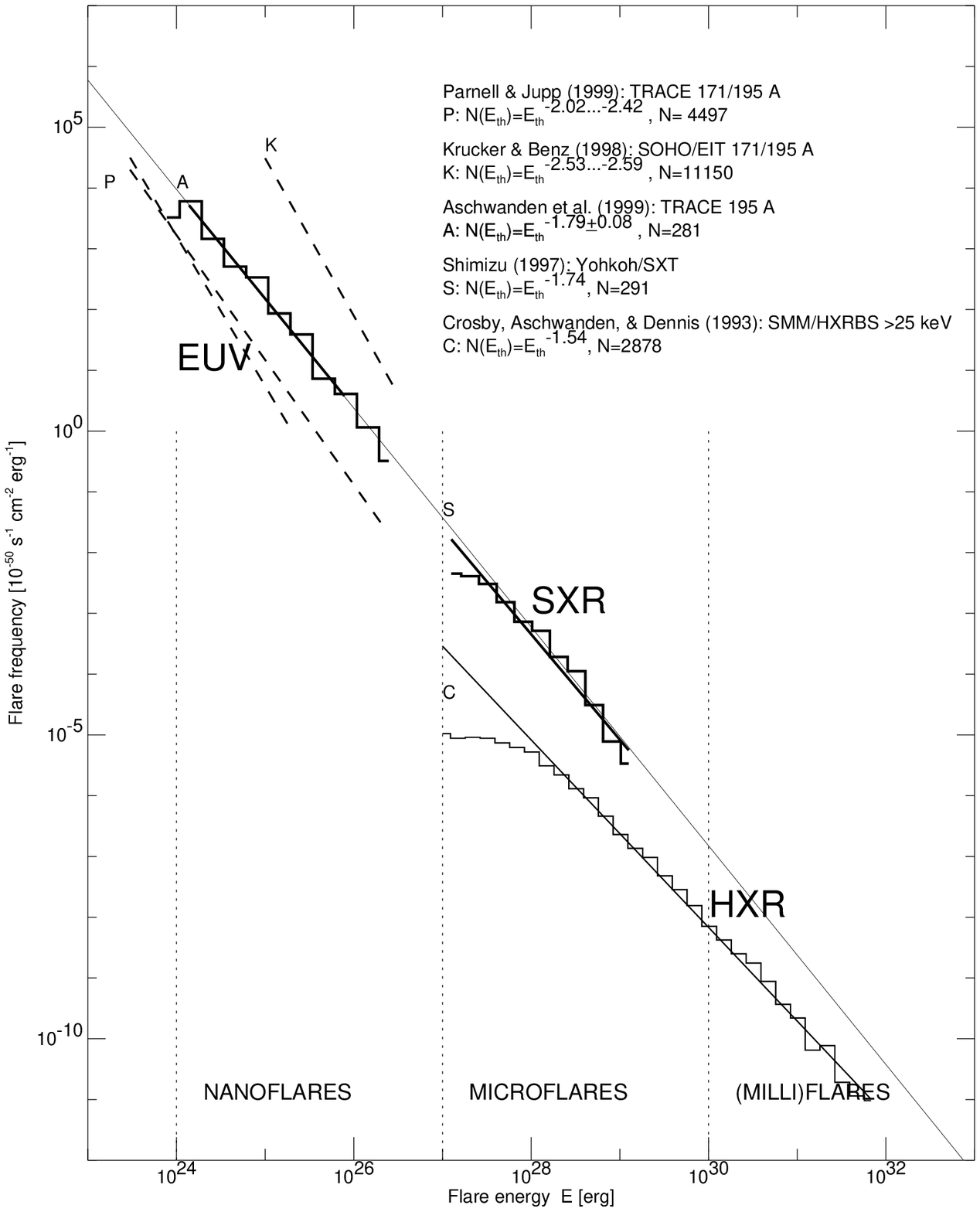}}
\captio{Composite flare frequency distribution in a normalized scale 
in units of $10^{-50}$ flares per time unit ($s^{-1}$), area unit (cm$^{-2}$), 
and energy unit (erg$^{-1}$). The diagram includes EUV flares analyzed in
Aschwanden et al.~(2000b), from Krucker \& Benz (1998), from Parnell 
\& Jupp (2000), transient brightenings in (SXR) (Shimizu 1995), and hard
X-ray flares (HXR) (Crosby et al.~1993). All distributions are specified 
in terms of thermal energy $E_{th}=3 n_e k_B T_e V$, except for the case 
of HXR flares, which is specified in terms of nonthermal energies in $>25$ 
keV electrons. The slope of $-1.8$ is extended over the entire energy 
domain of $10^{24}-10^{32}$ erg (Aschwanden et al.~2000b).}
\end{figure}

A compilation of occurrence frequency distributions of flare samples
observed in EUV is given in Table 4. The range of powerlaw slopes
seems to vary over a much broader range, say within $\alpha_P \approx
1.2-2.1$ for EUV peak fluxes, $\alpha_E \approx 1.3-2.6$ for EUV-inferred
energies, or $\alpha_T \approx 1.4-2.3$ for EUV event durations.
The large scatter, which does not exist in flare statistics in hard X-ray
wavelengths (Table 2), can be attributed to a number of methodical differences.
The most important reason is incomplete temperature coverage when
statistics of nanoflares is obtained in a single (narrowband) EUV filter,
which results into relatively steep powerlaw slopes, while synthesized
energy statistics combined from a broader range of EUV and soft X-ray
filters combined yields the same powerlaw slope of $\alpha_P \approx 1.8$ in
peak fluxes and $\alpha_E \approx 1.5$ in energies as obtained
in soft X-rays and hard X-rays (Fig.~11; Aschwanden and Parnell 2002). 
Equally important is the
scaling law used in the definition of flare energies. The classical
approach is to estimate the thermal flare energy $E_{th}=3 k_B n_e T_e V$
from the peak electron density $n_e$, flare peak temperature $T_e$,
and flare volume $V$. However, since the flare volume $V$ cannot directly
be measured, but only the flare area $A$ instead, the scaling of the
thermal energy depends crucially on the used geometric scaling law.
Some authors used a ``pill-box'' model $V = A h$ with a constant
height $h$, which corresponds to a geometric scaling law $V \propto L^2$,
while a spherical volume scales as $V \propto L^3$. In Section 2.7 we
derived a distribution of $N(A) \propto A^{-2}$ for 2D areas, and a 
distribution of $N(V) \propto V^{-5/3}$ for 3D volumes, which explains
part of the discrepancies among the powerlaw slopes compiled in Table 4.
Other factors that play a role are the geometric scaling of fractal
volumes, the flare selection, the flare detection algorithm, the detection
thresholds, the synchrony in different temperature filters,
the completeness of sampling, truncation effects in small samples, 
the powerlaw fitting method, etc. (e.g., Benz and Krucker 2002). 

Nonetheless, flare statistics from different wavelength 
regimes start to converge, as shown in Fig.~11. What is still needed
is an unified identical detection method that uniformly samples events
from the largest giant flare down to the smallest nanoflare.

\subsubsection{Statistics of Solar Flare Radio Fluxes}

Solar radio bursts are usually subdivided into incoherent 
(gyroemission, gyrosynchrotron emission, free-free emission) and
coherent emission mechanisms (electron beam instability,
loss-cone instability, maser emission). Since incoherent emission 
mechanisms scale with the volume of the emitting source, which could
be a solar flare region, we expect some proportionality between
the flare energy and the radio burst flux, such as for microwave
bursts and type IV bursts (produced by gyrosynchrotron emission).
Consequently we expect powerlaw slopes of their size distributions 
that are similar to other incoherent emission mechanisms of flares
(e.g., bremsstrahlung in hard X-rays or soft X-rays).
On the other hand, since coherent emission mechanisms produce a
highly nonlinear response to some wave-particle instability,
their emitted intensity flux is expected to scale nonlinearly with
the flare volume, and thus may produce quite different size
distributions.

\begin{table}[t]
\begin{center}
\normalsize
\captio{Frequency distributions measured from solar radio bursts,
classified as type I storms (I), type III bursts (III),
decimetric pulsation types (DCIM-P), decimetric millisecond spikes (DCIM-S),
microwave bursts (MW), and microwave spikes (MW-S).}
\medskip
\begin{tabular}{llllll}
\hline
Powerlaw        &Powerlaw   &log        &Waveband  &Radio	&References\\
slope of        &slope of   &range      &frequency &burst	&\\
peak flux       &durations  &           &          &type        &\\
$\alpha_P$      &$\alpha_T$ &           &$f$       &		&\\
\hline
\hline
$1.8$           &           &2          &3 GHz               & MW & Akabane (1956)\\
$1.5$           &           &2          &3, 10 GHz           & MW & Kundu (1965)\\
$1.8$           &           &2          &1, 2, 3.75, 9.4 GHz & MW & Kakinuma \etal (1969)\\
$1.9-2.5$       &           &2          &3.75, 9.4 GHz       & MW & Kakinuma \etal (1969)\\
$1.74-1.87$     &           &2          &1-35 GHz            & MW & Song \etal (2011)\\
$1.65$          &           &2          &2.8 GHz             & MW & Das \etal (1997)\\
$1.71-1.91$     &           &4          &0.100-2 GHz         & MW & Nita \etal (2002)\\
\hline
$1.26-1.69$     &           &3          &110 kHz-4.9 MHz & III    & Fitzenreiter \etal (1976)\\
$1.28$          &           &2          &100 MHz-3 GHz   & III    & Aschwanden \etal (1995)\\
$1.45\pm0.31$   &           &3          &100 MHz-3 GHz   & III    & Aschwanden \etal (1998b)\\
$1.22-1.25$     &           &2.5        &$650-950$ MHz   & III    & Das \etal (1997)\\
$1.33\pm0.11$   &           &3          &100 MHz-3 GHz   & DCIM-P & Aschwanden \etal (1998b)\\
\hline
$2.9-3.6$       &           &1.5        &164, 237 MHz    & I      & Mercier \& Trottet (1997)\\
$4.8\pm0.1$     &           &0.5        &185-198 MHz     & I      & Iwai \etal (2013)\\
$2.99\pm0.63$   &           &3          &100 MHz-3 GHz   & DCIM-S & Aschwanden \etal (1998b)\\
$7.4\pm0.4$     &$5.4\pm0.9$ &0.5       &4.5-7.5 GHz     & MW-S   & Ning \etal (2007)\\
\hline 
{\bf 1.67}      &{\bf 2.00} &           &                & FD-SOC prediction &Aschwanden (2012a) \\
\hline
\end{tabular}
\end{center}
\end{table}

We present a compilation of size distributions gathered from solar
radio bursts in Table 5. Microwave bursts (MW), which are typically
observed in frequencies of $\nu \approx 1-15$ GHz, have been found
to exhibit size distributions with powerlaw slopes within a range
of $\alpha_P \approx 1.7-1.9$ (Akabane 1956; Kundu 1965; 
Kakinuma et al.~1969; Song et al.~2011, 2013; Das et al.~1997;
Nita et al.~2002), similar to the size distributions observed in
solar hard X-ray and soft X-ray bursts, which implies a 
near-proportionality between the flare energy and the radio peak flux. 

Type III bursts, which are
believed to be produced by plasma emission excited by an electron
beam-driven instability, display flatter size distributions in the
order of $\alpha_P \approx 1.2-1.5$ (Fitzenreiter et al.~1976; Das et al.~1997;
Aschwanden et al.~1995, 1998b), which can be explained by a nonlinear
scaling $F \propto E^\gamma$ between radio peak flux $P$ and flare energy $E$.
For the radio peak flux distribution $N(P) \propto P^{-\alpha_P}$, 
and assuming the standard volume scaling $N(V) \propto V^{-5/3}$
(Eq.~5), we expect then, say for a nonlinear exponent $\gamma =2$, 
a powerlaw slope of $\alpha_P=(1+1/2\gamma) \approx 1.25$. 
The fact that relatively flat powerlaw slopes have also been observed 
for other coherent radio bursts, such as $\alpha_P \approx 1.3$ for 
decimetric pulsations (DCIM-P; Aschwanden et al.~1998b), may also 
indicate a nonlinear scaling to the flare volume. 

On the other hand, some very steep size distributions
have been observed, such as $\alpha_P \approx 3-5$ for type I bursts
(Mercier and Trottet 1997; Iwai et al.~2013), or $\alpha_P \approx 3-7$ for 
decimetric and microwave millisecond spike bursts (Aschwanden et al.~1998b; 
Ning et al.~2007), which implies either 
a strong quenching effect that inhibits high levels of radio fluxes, 
or a pulse-pileup problem that violates the separation of time scales 
(i.e., the inter-burst time intervals or waiting times are 
shorter than the burst durations). The latter effect is most likely
to occur in the statistics of fine structure in complex patterns 
of radio dynamic spectra, where a multitude of small pulses occur in 
clusters. Such peculiar types of clustered radio emission are, for instance, 
type I bursts (Mercier and Trottet 1997; Iwai et al.~2013), 
or decimetric millisecond spikes (Aschwanden et al.~1998b). 
Low-resolution radio observations tend to cause an exponential cutoff 
at large radio flux values, even when the actual distribution has a 
powerlaw shape (Isliker and Benz 2001), and thus explains the trend
of steeper powerlaw slopes. Also stochastic models
of clustered solar type III bursts produce powerlaw-like size
distributions with an exponential cutoff (Isliker et al. 1998b).

From the statistics of solar radio bursts we learn that we can
discriminate between three diagnostic regimes (as grouped in Table 5): 
(1) the incoherent regime where the radio burst flux is essentially 
proportional to the flare volume ($\alpha_P \approx 1.7-1.9$); 
(2) the coherent regime 
that implies a nonlinear scaling between the radio peak flux and the
flare volume $P \propto V^\gamma$ with $\gamma \approx 2$ and
$\alpha_P \approx 1.2-1.5$; and (3) the exponential regime with
clustered bursts that violate the {\sl separation of time scales}
with steep slopes $\alpha_P \approx 2-7$ and have an exponential 
cutoff.  Thus, the powerlaw slopes offer a useful diagnostic 
to quantify scaling laws between the radio flux (emissivity)
and the flare volume.

\begin{table}
\begin{center}
\normalsize
\captio{Frequency distributions of solar energetic particle (SEP) events.}
\medskip
\begin{tabular}{llllll}
\hline
Powerlaw        &Powerlaw       &Spacecraft &Energy  		&Reference\\
slope of        &slope of       &           &range   		&\\
peak flux       &total flux or  &           &                	&\\
                &total energy   &           &                	&\\
$\alpha_P$      &$\alpha_E$     &           &$E_{min}$         	&\\
\hline
\hline
$1.10\pm0.05$   &               &IMP4-5     &20-80 MeV protons & Van Hollebeke et al.~(1975)\\
$1.40\pm0.15$   &               &           &$>10$ MeV protons & Belovsky and Ochelkov (1979)\\
$1.13\pm0.04$   &               &IMP8       &24-43 MeV protons & Cliver et al.~(1991)\\
$1.30\pm0.07$   &               &IMP8       &3.6-18 MeV electrons & Cliver et al.~(1991)\\
                &$1.32\pm0.05$  &IMP,OGO    &$>10$ MeV protons & Gabriel \& Feynman (1996)\\
                &$1.27\pm0.06$  &IMP,OGO    &$>30$ MeV protons & Gabriel \& Feynman (1996)\\
                &$1.32\pm0.07$  &IMP,OGO    &$>60$ MeV protons & Gabriel \& Feynman (1996)\\
$1.47-2.42$     &               &           &$>10$ MeV protons & Smart \& Shea (1997)\\
$1.27-1.38$     &               &           &$>10$ MeV protons & Mendoza et al.~(1997)\\
$1.00-2.12$     &               &IMP        &$>10$ MeV protons & Miroshnichenko et al.~(2001)\\
1.35            &               &           &$>10$ MeV protons & Gerontidou et al.~(2002)\\
                &1.34$\pm$0.02  &           &$>10$ MeV protons & Belov et al.~(2007)\\
                &1.46$\pm$0.03  &           &$>100$ MeV protons& Belov et al.~(2007)\\
                &1.22$\pm$0.05  &           &$>10$ MeV protons & Belov et al.~(2007)\\
                &1.26$\pm$0.03  &           &$>100$ MeV protons& Belov et al.~(2007)\\
$1.56\pm0.02$   &               &           &$>10$ MeV protons & Crosby (2009)\\
\hline
{\bf 1.67}	&{\bf 1.50}     &           & FD-SOC prediction& Aschwanden (2012a)\\
\hline
\end{tabular}
\end{center}
\end{table}

\begin{figure}[t]
\centerline{\includegraphics[width=0.7\textwidth]{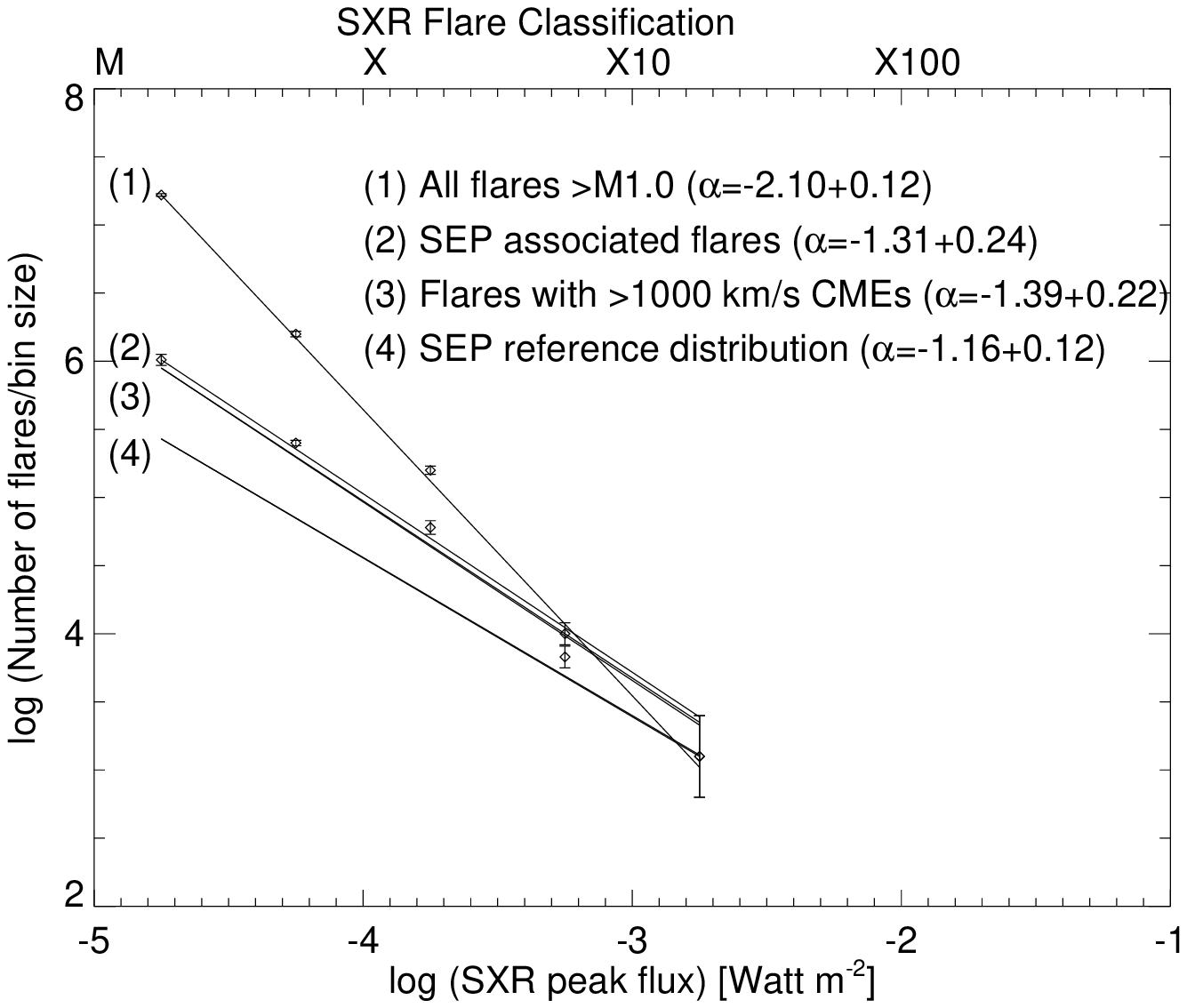}}
\captio{Size distributions for (1) peak 1-8 \ang\ fluxes of all
$\ge M1.0$ soft X-ray flares; (2) subset of flares
associated with $>10$ MeV proton events; (3) subset
of flares with $\ge 1000$ km s$^{-1}$ CMEs; and
(4) peak proton fluxes of $>10$ MeV SEP events.
All SEP events were observed during 1996-2005 (adapted from 
Cliver et al.~2012).}
\end{figure}

\subsubsection{Statistics of Solar Energetic Particle (SEP) Events }

Solar energetic particle (SEP) events represent a subset of large
solar flares that produce protons, electrons, and helium ions, with
energies of $\gapprox 25$ keV to 1 GeV. It was noted early on
that the size distribution of peak counts of SEP events is flatter 
($\alpha_E \approx 1.2-1.4$) than those of flare electromagnetic 
emission (Hudson 1978). Size distributions of the fluences of SEP 
events were gathered in a typical range of $\alpha_E \approx 1.2-1.4$  
(Van Hollebeke et al.~1975; Belovsky and Ochelkov 1979;
Cliver et al.~1991; Gabriel and Feynman 1996; Smart \& Shea 1997;
Perez Enriquez and Miroshnichenko 1999; Miroshnichenko et al.~2001;
Gerontidou et al.~2002; Belov et al.~2007).
The powerlaw slopes of the size distributions of peak fluxes and fluences
are listed in Table 6, which clearly exhibit a much flatter range
($\alpha_P \approx \alpha_E \approx 1.2-1.4$) than those measured
in hard (Table 2) and soft X-rays (Table 3), as noted earlier
(Hudson 1978). Cliver et al.~(2012) interpreted this discrepancy
as a selection effect of SEP events being preferentially associated
with larger flares, and thus the SEP events are drawn from a subset
of flares that do not form a statistically representative sample. 
This interpretation has been demonstrated by sampling subsets
of flares that are associated with $>10$ MeV proton events,
or with $\ge 1000$ km s$^{-1}$ CMEs, which exhibited a similar
flat powerlaw slope as the SEP events themselves (Fig.~12;
Cliver et al.~2012). Note that a powerlaw function fits the size
distribution of SEP fluences only in the low-fluence part, while
the high-fluence part is better fitted by an exponential cutoff
function, i.e., $N(E) \propto E^{-\alpha_E}
\times \exp{(E/E_0)}$, based on SEP data from 41 solar cycles
from 1561 to today (Miroshnichenko and Nymmik 2014).

Alternatively, Kahler (2013) challenges the interpretation of a
selection bias and suggests that the difference can be explained
by the dimensionality of the SOC system. If we take the general
expression of the powerlaw slope $\alpha_E$ for energy or fluences
(Eq.~36), we expect in the standard model, for classical 
diffusion $(\beta=1)$ and incoherent processes $(\gamma = 1)$,
and inserting the mean value of the fractal dimension
$D_d=(1+d)/2$, a powerlaw slope of
\begin{equation}
        \alpha_E = 1+{2 (d-1) \over (d+5)} \ ,
\end{equation}
which yields $\alpha_E=3/2=1.5$ in 3D space, but a flatter slope
of $\alpha_E=9/7 \approx 1.3$ in 2D space (or even a limit of
$\alpha_E=1$ for d=1). This is conceivable
if the avalanche spreads over a 2D surface only, such as a
reconnection current sheet, or the surface of a shock wave. 
This would invalidate a close physical connection between flares and
SEP events, and provide this way a diagnostics of the dimensionality of
the particle acceleration process in flares and SEP events 
(Kahler 2013).

Because of the high energies of SEP events, which can harm 
astronauts or electronic equipment in space, statistical 
information that improves their predictability is highly desirable 
(Gabriel and Patrick 2003), but statistical studies demonstrate 
that it is not possible to predict the time of occurrence of
SEP events within narrow limits (Xapsos et al.~2006).

\begin{table}[t]
\begin{center}
\normalsize
\captio{Waiting time distributions measured from solar flares 
hard X-ray events, soft X-ray events, coronal mass ejections, 
and radio bursts. The waiting time distribution (WTD) functions 
are abbreviated as: PL=powerlaw, E=exponential, PE=powerlaw with 
exponential rollover, DE=double exponential.}
\medskip
\begin{tabular}{lrllll}
\hline
Observations:   &Number         &Time           &WTD    &Powerlaw   	&References\\
Spacecraft or   &of events      &range      &   &       &         \\
instrument      &               &$\Delta t$ &           &$\alpha_{\Delta t}$ &  \\
\hline
\hline
HXRBS/SMM	& 8319		& $1-100$ min	& PL	& $0.75\pm0.1$	& Pearce \etal (1993)\\
BATSE/CGRO	& 6596          & $2-400$ min	& E	&		& Biesecker (1994)\\
WATCH/GRANAT	& 182           & $10-300$ min  & PE    & $0.78\pm0.13$ & Crosby (1996)\\
ICE/ISEE-3	& 6916          & $0.01-20$ hrs & DE    &               & Wheatland \etal (1998)\\
SMM/HXRBS	& 12,772	& $0.01-500$ hrs& PL    & $2.0$		& Aschwanden \& McTiernan (2010)\\
BATSE/CGRO	& 4113		& $0.01-200$ hrs& PL    & $2.0$		& Aschwanden \& McTiernan (2010) \\
BATSE/CGRO	& 7212		& $1-5000$ hrs  & PL    & $2.14\pm0.01$	& Grigolini \etal (2002)\\
RHESSI		& 11,594	& $2-1000$ hrs  & PL	& $2.0$		& Aschwanden \& McTiernan (2010)\\
\hline
GOES 1-8 A	& 32,563	& $1-1000$ hrs  & PL    & $2.16\pm0.05$ & Wheatland (2000a)\\ 
GOES 1-8 A	& 32,563	& $1-1000$ hrs  & PL    & $2.4\pm0.1$   & Boffetta \etal (1999)\\ 
                &               &               &       &               & Lepreti \etal (2001)\\     
GOES 1-8 A      & 4645          & $1-1000$ hrs  & PL    & $2.26\pm0.11$ & Wheatland (2003)\\
GOES 1-8 A      & (sol min)     & $1-1000$ hrs  & PL    & $1.75\pm0.08$ & Wheatland (2003)\\
GOES 1-8 A      & (sol max)     & $1-1000$ hrs  & PL    & $3.04\pm0.19$ & Wheatland (2003)\\
\hline
SOHO/LASCO      & 4645          & $1-1000$ hrs  & PL    & $2.36\pm0.11$ & Wheatland (2003)\\
SOHO/LASCO      & (sol min)     & $1-1000$ hrs  & PL    & $1.86\pm0.14$ & Wheatland (2003)\\
SOHO/LASCO      & (sol max)     & $1-1000$ hrs  & PL    & $2.98\pm0.20$ & Wheatland (2003)\\
\hline
FD-SOC prediction & (sol min)   &               & PL    & {\bf 2.00}    & Aschwanden (2012a)\\
\hline
\end{tabular}
\end{center}
\end{table}

\subsubsection{Statistics of Solar Flare Waiting Times}

In the simplest scenario we could envision that solar flares occur randomly 
in space and time. However, there are subsets of flares that occur 
simultaneously at different locations (called {\sl ''sympathetic flares''}; 
Fritzova-Svestkova et al.~1976; Wheatland 2006; Moon et al.~2003), 
as well as flare events that subsequently occur at the same location 
(called ``homologous flares''; Fokker 1967), which indicates a
spatial or temporal clustering that is not random. We outlined the concept
of random processes in Section 2.12, which can produce an exponential
waiting time distribution (for stationary Poisson processes), as well as
powerlaw-like distributions of the waiting time (for non-stationary
Poisson processes; Fig.~6). Moreover, both exponential and powerlaw
distributions can be generated with the Weibull distribution (Section 2.12.3).
The functional shape of the waiting time distribution depends moreover
on the definition of events in a time series, where powerlaws are found
more likely to occur when a threshold is used (Buchlin 2005). 
Allowing an overlap of time scales between burst durations and quiet
times, agreement was found between the waiting time distributions 
sampled with different thresholds (Paczuski et al.~2005; Baiesi et al.~2006).
In summary, the finding of powerlaw-like waiting time distributions has no
unique interpretation, because it can be consistent with both a
random process without memory (in the case of a non-stationary Poisson 
process) or with a non-random process with memory (in the case of a
Weibull distribution with $k \neq 1$). This dichotomy of stochasticity
versus persistence or clustering has been noted in SOC processes before,
for earthquakes that have aftershocks with an excess of short waiting times 
(Omori's law; Omori 1895). 

In Table 7 we compile studies on waiting times of solar flare phenomena,
grouped into hard X-ray events, soft X-ray events, coronal mass ejections, 
radio bursts, and solar wind fluctuations. Statistics in hard X-rays were
obtained from HXRBS/SMM (Pearce et al.~1993; Aschwanden and McTiernan 2010),
BATSE/CGRO (Biesecker 1994; Grigolini et al.~2002; Aschwanden and McTiernan 2010),
WATCH/GRANAT (Crosby 1996); ICE/ISEE-3 (Wheatland et al.~1998; Wheatland 
and Eddey 1998), and RHESSI (Aschwanden and McTiernan 2010).

All waiting time distributions observed for hard X-ray bursts 
have been reconciled with a single common model that represents 
a limit of intermittency,
\begin{equation}
	P(\Delta t) = \lambda_0 (1 + \lambda_0 \Delta t)^{-2} \ ,
\end{equation}
which has a powerlaw slope of $\alpha_{\Delta t}=2.0$ for large
waiting times ($\Delta t \approx 1-1000$ hrs) and flattens
out for short waiting times $\Delta t \lapprox 1/\lambda_0$, which
is consistent with a highly intermittent flare productivity in short 
clusters with high rates, as it can be analytically derived 
(Aschwanden and McTiernan 2010), depicted in Fig.~6 (bottom panel).
A similar functional form of the waiting time distribution is obtained 
with the diffusion entropy method (Grigolini et al.~2002).

In addition, the powerlaw slope of $\alpha_{\Delta t}=2$ is also
predicted by the fractal-diffusive model (Section 2.12.3) in the
slowly-driven limit, while steeper observed slopes are consistent 
with the predicted modification for strongly-driven systems (Eq.~43).
The results compiled in Table 7 indeed yield higher values of
$\alpha_{\Delta t}\approx 3$ during periods of high flare activity,
as it occurs during the solar cycle maximum. 

In soft X-rays, a similar powerlaw slope of $\alpha_{\Delta t} \approx
2.1-2.4$ was found, which could be fitted with a non-stationary Poisson
process (Wheatland 2000a; Moon et al.~2001), with a shell model 
of turbulence (Boffetta et al.~1999), or with a Levy function 
(Lepreti et al.~2001). These
different interpretations underscore the ambiguity of powerlaw 
distributions, which do not allow to discriminate between SOC and
turbulence processes. Moreover, the powerlaw slope of waiting time
distributions varies during the solar cycle, which implies a 
time-variable SOC driver (Wheatland and Litvinenko 2002). 
The flaring rate was found to vary among different active regions,
as well as during the disk transit time of a single active region
(Wheatland 2001).
The variability in the flare rate was found to correlate with the
sunspot number, however with a time lag of about 9 months, which reflects
the hysteresis of the coronal response to the solar dynamo
(Wheatland and Litvinenko 2001), a result that can be used for
statistical flare forecasting (Wheatland 2004; Wheatland and Craig 2006).
Additional tests whether the waiting time of solar flares is
random (multi-Poissonian) or clumped in persistent clusters
(with some memory) have been carried out with a Hurst analysis,
finding a Hurst exponent of $H=0.74\pm0.02$ (compared with
$H=0.5$ for a pure stochastic process) (Lepreti et al.~2000),
or by fitting a Weibull distribution (Section 2.12.3), 
finding two statistical components for coronal mass ejections,
a continuous random process during solar minima, and another 
component with temporary persistence and memory during solar
maxima (Telloni et al.~2014), similar to the FD-SOC scenario 
(Fig.~7 and section 2.12.3), or the aftershocks 
in earthquake statistics (Omori's law).

Does the waiting time give us some information about the energy build-up
in solar flares? Early studies suspected that the waiting time is the longer
the more energy is built up, which predicts a correlation between the
waiting time and the energy of the flare (Rosner and Vaiana 1978).   
However, several observational studies have shown that no such
correlation exists (e.g., Lu 1995b;  Crosby 1996;  Wheatland 2000b;   
Georgoulis et al.~2001; Moon et al.~2001), not even between subsequent 
flares of the same active region (Crosby 1996; Wheatland 2000b).  
The original SOC model of BTW assumes that avalanches occur randomly
in time and space without any correlation, and thus a waiting-time
interval between two subsequent avalanches refers to two different
independent locations (except for sympathetic flares), and thus bears
no information on the amount of energy that is released in each
spatially separated avalanche. In solar applications, flare events
seem to deplete only a small amount of the available free energy,
and thus no correlation between waiting times and flare magnitudes
are expected to first order. In contrast, however, recent studies
that analyze the probability differences of subsequent events from
the GOES flare catalog, and compare them with randomly re-shuffled data,
find non-trivial correlations between waiting times and dissipated
energies. Flares that are close in time tend to have a second event
with large energy. Moreover, the flaring rate as well as the probability
of other large flares tends to increase after large flares 
(Lippiello et al.~2010), similar to the clustering of coronal 
mass ejections (CMEs) (Telloni et al.~2014), and aftershocks of
earthquakes (Omori 1895).

\begin{table}[t]
\begin{center}
\normalsize
\captio{Area fractal dimension $D_2$ of scaling between length scale $L$
and fractal area $A(L) \propto L^{D_2}$ measured from various solar 
phenomena observed in different wavelength regimes: WL =  white light,
H-$\alpha$ = visible spectral line in the Balmer series produced in
hydrogen at 6562.8 \ang , MG = magnetogram measured with Zeeman effect,
e.g., from Fe XIV 5303 \ang\ line;  EUV = extreme ultra-violet, 
SXR = soft X-rays. The methods are: PA = perimeter vs. area, 
LA = linear size vs. area, and BC = box-counting.} 
\medskip
\begin{tabular}{lllll}
\hline
Phenomenon  & Wavelength&Method & Area fractal  & Reference: \\
	    & regime    &       & dimension $D_2$ & \\
\hline
\hline
Granules	& WL 	& PA  	& $1.25$ 	& Roudier \& Muller (1987)\\  
Granules	& WL	& PA  	& $1.30$        & Hirzberger \etal (1997)\\   
Granular cells	& WL	& PA  	& $1.16$        & Hirzberger \etal (1997)\\   
Granules 	& WL    & PA  	& $1.09$        & Bovelet \& Wiehr (2001)\\ 
Super-granulation& MG	& PA  	& $1.25$        & Paniveni \etal (2005)\\ 
Super-granulation& MG	& PA  	& $1.2, 1.25$   & Paniveni \etal (2010)\\ 
Small scales	& MG	& PA  	& $1.41\pm0.05$ & Janssen \etal (2003)\\   
Active regions	& MG    & LA  	& $1.56\pm0.08$ & Lawrence (1991)\\   
		&	&	&		& Lawrence and Schrijver (1993)\\  
Plages 		& MG 	& LA  	& $1.54\pm0.05$ & Balke \etal (1993)\\  
Active regions	& MG	& LA  	& $1.78-1.94$   & Meunier (1999)\\  
                & MG    & PA  	& $1.48-1.68$   & Meunier (1999)\\   
Active regions  & MG    &       & $1.71-1.89$   & Meunier (2004)\\
\ $-$ Cycle minimum& MG &       & $1.09-1.53$   & Meunier (2004)\\
\ $-$ Cycle rise&    MG &       & $1.64-1.97$   & Meunier (2004)\\
\ $-$ Cycle maximum& MG &       & $1.73-1.80$   & Meunier (2004)\\
Quiet Sun	& MG    &       & multifractal  & Lawrence \etal (1993)\\ 
Active regions 	& MG    &       & multifractal  & Lawrence \etal (1993)\\ 
Active regions  & MG 	& BC    & multifractal  & Cadavid \etal (1994)\\ 
Active regions  & MG 	& BC    & multifractal  & Lawrence \etal (1996)\\ 
Active regions  & MG    & BC    & $1.25-1.45$   & McAteer \etal (2005)\\  
Active regions  & MG    &       & multifractal  & Conlon \etal (2008)\\
Active regions  & MG    &       & multifractal  & Hewett \etal (2008)\\
Active regions  & MG    &       & multifractal  & Conlon \etal (2010)\\
Quiet Sun network& EUV  & BC  	& $1.30-1.70$   & Gallagher \etal (1998)\\ 
Ellerman bombs	& H$\alpha$ &BC & $1.4$         & Georgoulis \etal (2002)\\  
Nanoflares	&EUV 171 \ang\ &BC&$1.49\pm0.06$& Aschwanden \& Parnell (2002)\\ 
Nanoflares      &EUV 195 \ang\ &BC&$1.54\pm0.05$& Aschwanden \& Parnell (2002)\\ 
Nanoflares	& SXR   & BC    & $1.65$        & Aschwanden \& Parnell (2002)\\ 
Flare 2000-Jul-14 &EUV 171 \ang\ &BC& $1.57-1.93$&Aschwanden \& Aschwanden (2008a)\\ 
Flares          & EUV   & BC    & $1.55\pm0.11$ & Aschwanden \etal (2013a)\\
\hline
FD-SOC prediction &     & $d=2$ & {\bf 1.50}    & Aschwanden (2012a)\\
\hline
\end{tabular}
\end{center}
\end{table}

\subsubsection{		Solar Fractal Measurements		}

{\sl ``Fractals in nature originate from self-organized critical dynamical 
processes} (Bak and Chen 1989). In principle, SOC avalanches could be
non-fractal and encompass space-filling solid volumes, as the sandpile
analogy suggests. However, using the BTW model as a paradigm for SOC
avalanches, it is quite clear from inspecting numerical simulations
that the next-neighbor interactions propagate in ``tree-like'' patterns
that can indeed be quantified with a fracal dimension (e.g., Aschwanden 2012a).
Also the EUV images of solar flares show highly fragmented postflare loops
that can be characterized with a fractal dimension (Aschwanden and 
Aschwanden 2008a). 

Interestingly, measurements of the area 
fractal dimension $D_2$ in solar data have been published over the same 
25-year era as SOC publications exist (Table 1). Reviews on fractal
analysis of solar flare data can be found in Aschwanden (2011a, chapter 8)
and in McAteer (2013a). While cellular automaton
simulations allow for various Euclidean space dimensions ($d=1,2,3$),
solar observations are restricted to the 2D-case $(d=2)$, for which the 
standard model predicts an mean area dimension of $D_2=(1+d)/2=1.5$, with a
lower limit of $D_{2,min} \gapprox 1.0$ and an upper limit of $D_{2,max}=2.0$.
The observed fractal dimensions listed in Table 8 indeed cover the full range
of $D_2=[1.09,1.97]$ and have a median value of $D_{2,med}=1.54$, or a mean
and standard deviation of $D_2=1.54\pm0.25$. These fractal dimensions have
been measured in a variety of solar phenomena: from granulation (Roudier 
and Muller 1987; Hirzberger \etal 1997; Bovelet and Wiehr 2001), 
super-granulation (Paniveni et al.~2005, 2010), active regions (Lawrence 1991;
Cadavid \etal 1994; Lawrence \etal 1996; McAteer \etal 2005; 
Lawrence and Schrijver 1993; Meunier 1999, 2004), 
plages (Balke \etal 1993), quiet-Sun network (Lawrence \etal 1993;
Gallagher \etal 1998), Ellerman bombs (Georgoulis \etal 2002),
nanoflares (Aschwanden and Parnell 2002), to large flares 
(Aschwanden and Aschwanden 2008a; Aschwanden \etal 2013a). 
The lowest fractal dimensions $D_2\approx 1.1-1.3$ are 
measured in granules and in the quiet-Sun netowrk, which consist of
elongated curvi-linear structures, while active region and flare areas
have a higher fractal dimension of $D_2\approx 1.4-1.8$, consisting of
chains of coherent patchy areas, as expected for SOC avalanches with
isotropic next-neighbor interactions. A lower threshold fractal dimension
of $D_2 \gapprox 1.2$ (and 1.25) was found as a necessary condition for
an active region to produce M-class (and X-class) flares (McAteer \etal 2005).
In the overall, we can say that most solar observations are consistent with 
a predicted area fractal dimension of $D_2=(1+d)/2=1.5$. 

For the application of SOC models to solar flares, which have a 3D geometry,
we cannot measure the volume fractal dimension $D_3$ directly. If we rely
on the simple mean-value theorem, $D_d=(1+d)/2$ (Eq.~8), we expect a 
volume fractal dimension of $D_3=2.0$. 
Attempts have been made to determine
the 3D volume fractal dimension $D_3$ from observations of 20 large-scale
solar flares, using a fractal loop arcade model, which yielded a mean
value of $D_3=2.06\pm0.48$ (calculated from Table 1 in Aschwanden and 
Aschwanden 2008b). 
Thus, we can conclude that the solar flare observations 
are consistent with the volume fractal dimension predicted by the standard 
SOC model, i.e., $D_3=(1+d)/2=2.0$ (for $d=3$). 

\begin{figure}[t]
\includegraphics[width=1.0\textwidth]{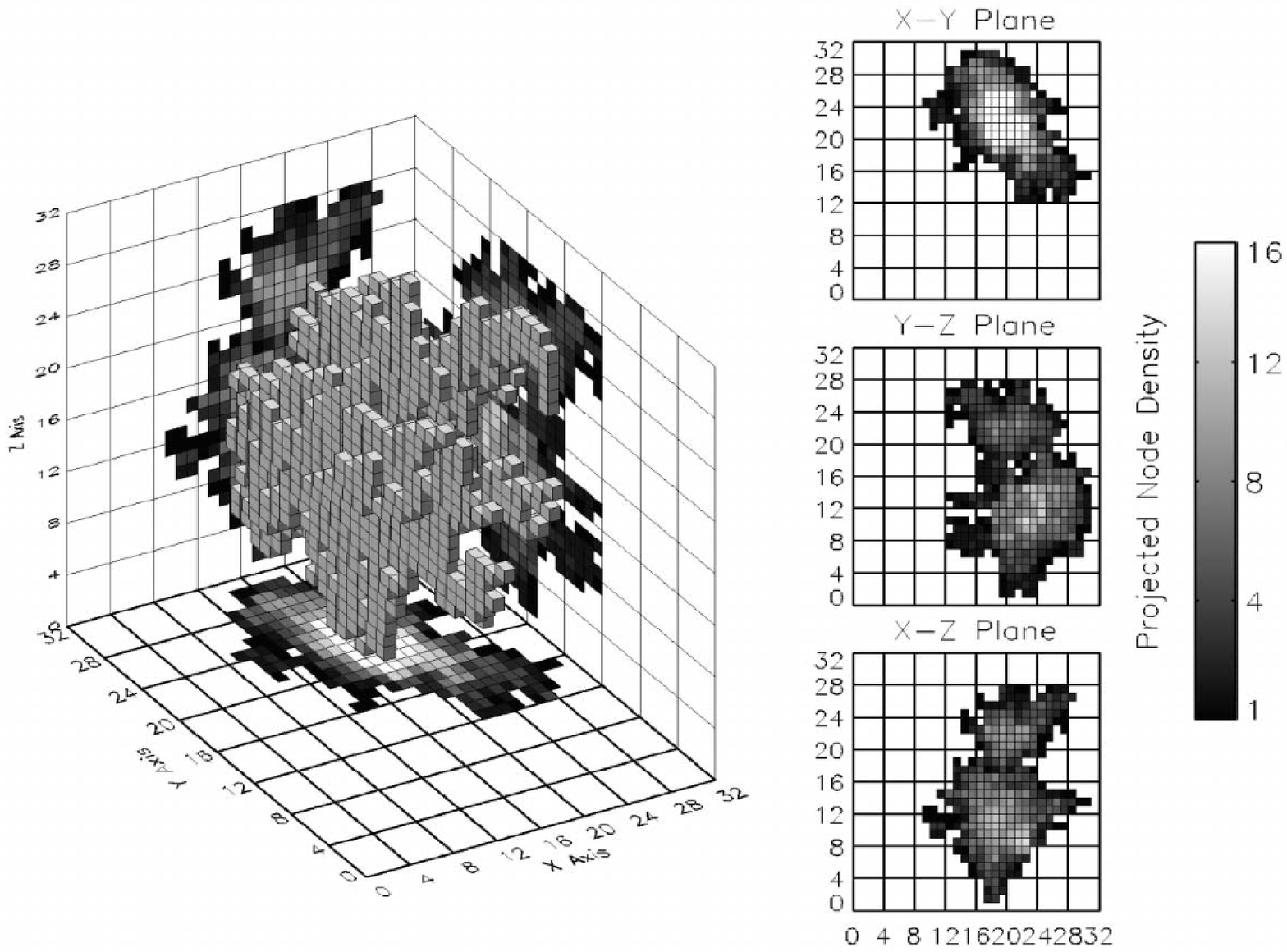}
\captio{Relationship between the 3D fractal volume of a SOC avalanche 
(left panel) and the three 2D projections that show the 2D fractal areas
(right panels), simulated in a $32^3$ lattice (McIntosh and Charbonneau 2001).}
\end{figure}

This mean-value theorem predicts
a scaling law between the fractal avalanche area $A_f \propto L^{D_2}$ and 
the fractal avalanche volume $V_f \propto L^{D_3}$,
\begin{equation}
	V_f \propto A_f^{\delta} \ , \qquad \delta = {D_3 \over D_2} =
	{1+3 \over 1+2} = {4 \over 3} \approx 1.33 \ ,
\end{equation}
which is lower than the Euclidean scaling law, $V \propto A^{3/2} = A^{1.5}$.
Cellular automaton simulations yield an intermediate value for the 
exponent, $\delta = 1.41 \pm 0.04$ (Fig.~13; McIntosh and Charbonneau 2001).

The volume fractal dimension is important to derive the correct scaling law 
between the length scale $r(t)$ of a SOC avalanche at a given time $t$ and 
the instantaneous fractal avalanche volume $V_f(t)$ (Eq.~12), being 
proportional to the observed flux $f(t)$, as well as for the total 
time-integrated energy $e(t)$ (Eq.~14). It affects the powerlaw 
slopes of the size distributions of avalanche areas ($\alpha_A$),
avalanche volumes ($\alpha_V$), flux ($\alpha_F$), and total
energy ($\alpha_E$) (see Eqs.~22 and 36). 

The fractal dimension measured in magnetograms can diagnose both 
SOC behavior or turbulence, but cannot discriminate between the two 
interpretations because both processes have fractal-like structures. 
An alternative method to measure the fractal structure or 
intermittency of a turbulent magnetic field is the structure function 
(Frisch 1995; Abramenko et al.~2003),
\begin{equation}
	S_q({\bf r} \langle |{\bf B}_z ({\bf x} +{\bf r})
	- {\bf B}_z({\bf x}) | {\rangle}^q \propto ({\bf r})^{\zeta(q)} \ ,
\end{equation}
where $q$ is an order of a statistical moment, ${\bf r}$ is a
separation vector, ${\bf x} = (x,y)$ is a spatial location in a
magnetogram, and ${\bf B}_z$ is the observed line-of-sight 
longitudinal magnetic field. This structure function essentially
measures the level of long-range correlations, which are an intrinsic
property of SOC avalanches. Analysis of magnetograms before solar
flares have shown an increase in the degree of intermittency and
in the maximum of the correlation length, which was interpreted in
terms of enhanced turbulence as a precursor to a SOC avalanche
of a solar flare (Abramenko \etal 2002, 2003; Abramenko and
Yurchyshyn 2010). The presence of a topologically complex,
asymmetrically fragmented magnetic network can trigger a 
magnetic instability acting as an energy source for a coronal
dissipation event (Uritsky and Davila 2012). 

\begin{figure}[t]
\centerline{\includegraphics[width=1.\textwidth,clip]{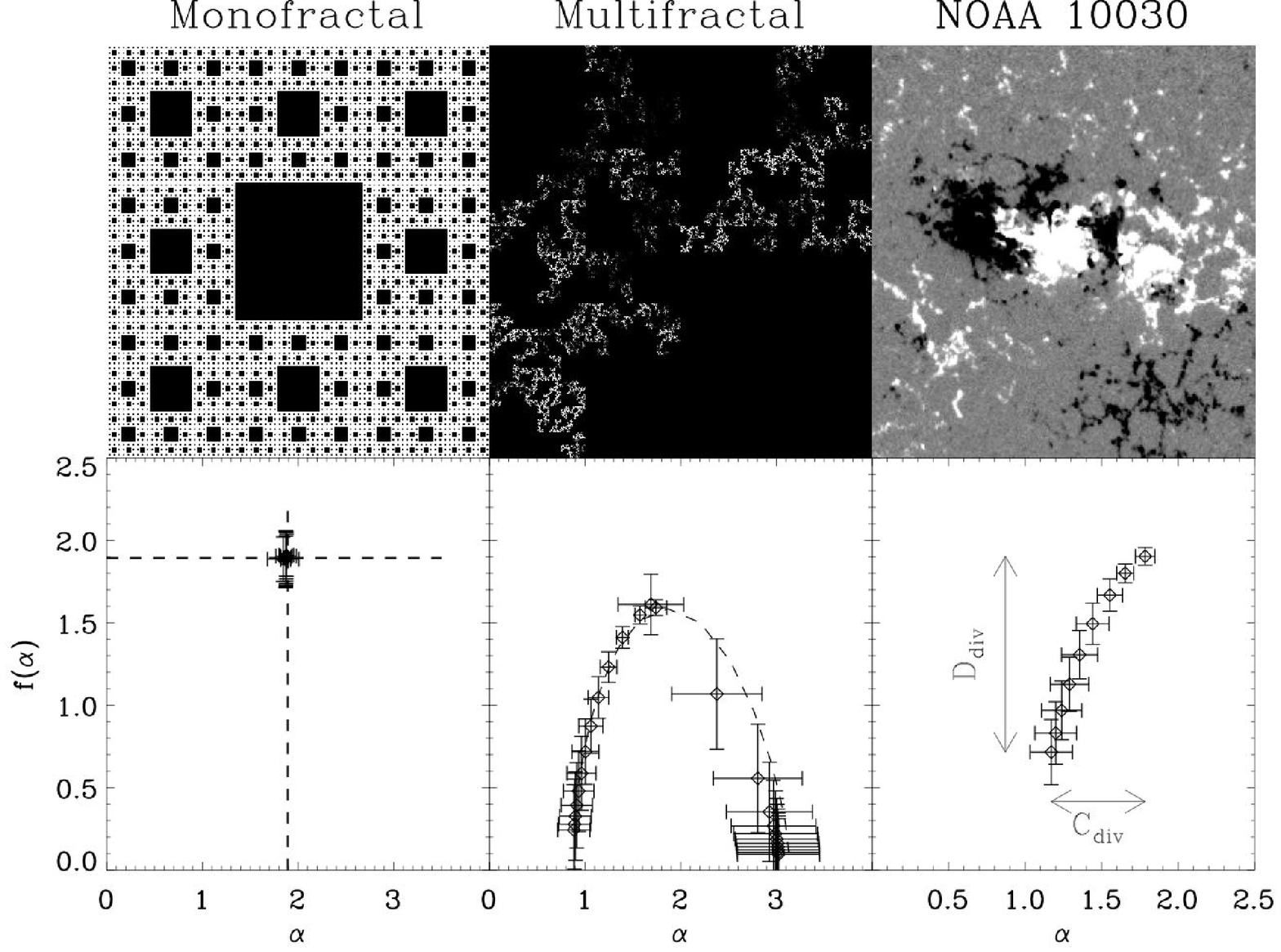}}
\captio{A monofractal image of the Sierpinski carpet (left), a theoretical
multifractal image (middle), and an observed multifractal solar magnetogram
of active region NOAA 10030 (right), along with the singularity spectra
$f(\alpha)$ (bottom panels) determined for these structures
(Conlon et al.~2008).}   
\end{figure}

A generalization to fractal areas, $A_f \propto L^D$, is
a multi-fractal system, which has a spectrum $f(D)$ of
fractal dimensions $D$,
\begin{equation}
	A_f \propto L^{f(D)} \ ,
\end{equation}
where $D$ is called the strength or significance, and $f(D)$
the singularity spectrum. This singularity spectrum has a peak
at $f(D)_{max}$ and a minimum of $f(D)_{min}$ (Fig.~14), which is 
characterized by the terms {\sl contribution diversity}
$C_{div}=D_{max}-D_{min}$ and {\sl dimensional diversity}
$D_{div}=f(D)_{max}-f(D)_{min}$, both being measures of the
geometric complexity and richness of a fractal structure.
Multi-fractal analysis of solar data has been carried out
on magnetograms (Abramenko 2005; Conlon \etal 2008; 
Iospha \etal 2008; Hewett \etal 2008; Dimitropoulou \etal 2009).
Flare-quiet active regions were found to have a lower degree
of multi-fractality than flaring active regions (Abramenko 2005).
Multi-fractality was also found to increase during magnetic flux
emergence in active regions, while a decrease occurred when the
active regions evolved to large-scale, coherent structures
(Conlon \etal~2008; McAteer \etal~2010). 
While the exact energy scaling is not
known, spatial complexity and flare productivity seem to be related,
even when they have different fractal dimensions (Hewett \etal 2008).
The 2D photospheric magnetic field contains the footprints of 
3D magnetic structures in the solar corona, which explains
numerous correlations between photospheric and coronal phenomena
(e.g., Dimitropoulou \etal 2009; Uritsky \etal 2013). 
On the other side, solar active
regions with major flares were not found to exhibit a higher
level of fractality, multi-fractality, or non-Kolmogorov
turbulence than non-flaring regions (Georgoulis \etal~2012).

While the previous discussion applies to fractal geometries
in 2D space with two spatial dimensions, the concept of fractals
has also been applied to a time series $f(t)$, where a fractal 
dimension is measured in the 2D space of $f$ versus $t$. 
We can easily imagine that a constant function $f(t)=const$
represents a straight line in a 2D box $[t,f]$, and thus has
the fractal dimension of $D_2=1$, while an erratically fluctuating
noise time series renders a plotted box $[t,f]$ almost black,
and thus has an almost space-filling Euclidean dimension
$D_2=2$. So, a fractal (or multi-fractal) dimension of a 
time series is essentially a measure of the time variability, 
and has been applied to solar radio burst data (Higuchi 1988; 
Watari 1996), or daily flare indices (Watari 1995; Sen 2007).
A multi-fractal spectrum of the hard X-ray time profile of a
solar flare was used to discriminate thermal and non-thermal
emission based on their different temporal signatures (McAteer et al.~2013b).
In principle, such a dimensional time variability analysis could also
be applied to SOC simulations, and this way could characterize the
predicted waiting time distribution, but we are not aware of 
such studies.

\begin{table}[t]
\begin{center}
\normalsize
\captio{Measurements of powerlaw slopes of solar flare size distributions 
of geometric parameters: length scales ($\alpha_L$), flare areas ($\alpha_A$), 
and flare volumes ($\alpha_V$). The references are: B1998 = Berghmans et al.~(1998),
A2000 = Aletti et al.~(2000); 
A2000b = Aschwanden et al.~(2000b); AP02 = Aschwanden \& Parnell (2002);
A2012b = Aschwanden (2012b), A2013 = Aschwanden et al.~(2013a); and
A2012a = Aschwanden et al.~ (2012a), L2010 = Li et al.~2012.}
\medskip
\begin{tabular}{lrrllll}
\hline
Instrument   & Wavelength & Number    & Length    & Area          & Volume        & References \\
	     & or energy  & of events & exponent  & exponent      & exponent      & \\
	     & $\lambda, \epsilon$&$N$& $\alpha_L$& $\alpha_A$    & $\alpha_V$    & \\
\hline
\hline
SOHO/EIT     & 304 \ang   &13,067 &               & 2.7           &               & B1998 \\
SOHO/EIT     & 195 \ang   &13,607 &               & 2.0           &               & B1998 \\
SOHO/EIT     & 195 \ang   &       &               & 1.26$\pm$0.04 &               & A2000 \\
SOHO/EIT     & 195 \ang   &       &               & 1.36$\pm$0.05 &               & A2000 \\
TRACE        &171-195 \ang&   281 & 2.10$\pm$0.11 & 2.56$\pm$0.23 & 1.94$\pm$0.09 & A2000b\\
TRACE/C      &171-195 \ang&       & 3.24$\pm$0.16 & 2.43$\pm$0.10 & 2.08$\pm$0.07 & AP2002\\ 
TRACE/A      & 171 \ang   &   436 & 2.87$\pm$0.24 & 2.45$\pm$0.09 & 1.65$\pm$0.09 & AP2002\\
TRACE/B      & 171 \ang   &   436 & 2.77$\pm$0.17 & 2.34$\pm$0.10 & 1.75$\pm$0.13 & AP2002\\
TRACE/A      & 195 \ang   &   380 & 2.59$\pm$0.19 & 2.16$\pm$0.18 & 1.69$\pm$0.05 & AP2002\\
TRACE/B      & 195 \ang   &   380 & 2.56$\pm$0.17 & 2.24$\pm$0.04 & 1.63$\pm$0.04 & AP2002\\
Yohkoh/SXT   & AlMg       &   103 & 2.34$\pm$0.27 & 1.86$\pm$0.13 & 1.44$\pm$0.07 & AP2002\\
TRACE+SXT &171,195,AlMg   &   919 & 2.41$\pm$0.09 & 1.94$\pm$0.03 & 1.55$\pm$0.03 & AP2002\\
AIA/SDO      & 335 \ang   &   155 & 1.96          &               &               & A2012b\\
AIA/SDO      &  94 \ang   &   155 & 3.1$\pm$0.6   & 2.0$\pm$0.1   & 1.5$\pm$0.1   & A2013 \\
AIA/SDO      & 131 \ang   &   155 & 3.5$\pm$0.5   & 2.2$\pm$0.2   & 1.7$\pm$0.2   & A2013 \\
AIA/SDO      & 171 \ang   &   155 & 3.5$\pm$1.2   & 2.1$\pm$0.5   & 1.7$\pm$0.2   & A2013 \\
AIA/SDO      & 193 \ang   &   155 & 3.5$\pm$0.9   & 2.0$\pm$0.3   & 1.7$\pm$0.2   & A2013 \\
AIA/SDO      & 211 \ang   &   155 & 2.7$\pm$0.6   & 2.1$\pm$0.3   & 1.6$\pm$0.2   & A2013 \\
AIA/SDO      & 304 \ang   &   155 & 2.9$\pm$0.6   & 2.1$\pm$0.2   & 1.7$\pm$0.1   & A2013 \\
AIA/SDO      & 335 \ang   &   155 & 3.1$\pm$0.4   & 1.9$\pm$0.2   & 1.6$\pm$0.1   & A2013 \\
AIA/SDO      & 94-335 \ang&   155 & 3.2$\pm$0.7   & 2.1$\pm$0.3   & 1.6$\pm$0.2   & A2013 \\
RHESSI       & 6-12 keV   &  1843 &               & 2.65$\pm$0.08 &               & L2012 \\
\hline
FD-SOC prediction &       &       & {\bf 3.00}    & {\bf 2.00}    & {\bf 1.67}    & A2012a\\
\hline
\end{tabular}
\end{center}
\end{table}

\subsubsection{		Flare Geometry Measurements  			} 		 

The most fundamental assumption in the SOC standard model is the
scale-free probability conjecture, i.e., $N(L) \propto L^{-d}$ (Eq.~1),
which should be easy to test with imaging solar observations, but there
is surprisingly little statistics available. For solar flare observations
we expect that SOC systems have an Euclidean dimension of $d=3$, and thus 
the prediction for the size distribution of flare length scales is
$N(L) \propto L^{-3}$. The directly measured quantity in solar flares
is usually the Euclidean flare area $A$, which relates to the length 
scale by $L \propto A^{1/2}$ (Eq.~2), and thus a size distribution 
of $N(A) \propto A^{-2}$ is expected.

In Table 9 we compile measurements of the flare geometry in terms
of length scales $L$, flare areas $A$, or flare volumes $V$, which
obey the expected geometric scaling lasw of the SOC standard model
within the uncertainties: $\alpha_L=3$, $\alpha_A=2$, $\alpha_V=5/3$.
The recent measurements in 7 different wavelengths using AIA/SDO have
been derived from the time-integrated areas $A(t<T)$ at the end time
of the flares (at $t=T$), measured with five different thresholds
and normalized (Aschwanden et al.~2013a). Note that the 
instantaneous flare areas $a(t)$ at some given time $0 < t <T$ 
are fractal, approximately following the scaling $a(t) \propto
r(t)^{D_2}$ as a function of the instantaneous radius $r(t)$ of the
Euclidean flare area, while the time-integrated flare areas are 
nearly solidly filled and outline essentially the Euclidean area
$A \propto L^2$ at the end time $(t=T)$ of the flare when
$r(t=T)=L$. These statistical measurements represent the most direct
observational test of the scale-free probability conjecture,
$N(L) \propto L^{-3}$, and this way corroborate the standard SOC model. 

Area measurements have also been carried out for supra-arcade
downflows during flares, which were found not to be compatible with
a powerlaw distribution (McKenzie and Savage 2001), a result that
is not surprising given the small range of measured areas (covering 
about a half decade). 

The geometric measurements are also of fundamental importance for
deriving and testing physical scaling laws, which are generally
expressed by a length scale (i.e., a coronal loop length), or by
a volumetric emission measure (which is proportional to the total 
flare volume), or thermal energy (which is also proportional to the
total flare volume). We will discuss such theoretical scaling laws
in Section 3.2.7.

\subsubsection{		Solar Wind Measurements 		}

The solar wind is a turbulent magneto-fluid, consisting of charged 
particles (electrons, protons, alpha particles, heavy ions) 
with typical energies of $1-10$ keV, which escape the Sun's gravity 
field because of their high kinetic (supra-thermal) energy and the 
high temperature of the solar corona. The solar wind has two different
regimes, depending on its origin, namely a fast solar wind 
with a speed of $v \lapprox 800$ km s$^{-1}$ originating from
open-field regions in coronal holes, and a slow solar wind 
with a speed of $v \lapprox 400$ km s$^{-1}$ originating from low
latitudes in the surroundings of coronal streamers. The dynamics
of the solar wind was originally explained by Parker (1958) as a
supersonic outflow that can be derived from a steady-state solution 
of the hydrodynamic momentum equation. Later refinements take the
super-radial expansion of the coronal magnetic field, the 
average macro-scale and fluctuating meso-scale electromagnetic
field in interplanetary space, and the manifold micro-scale kinetic
processes (such as Coulomb collisions and collective wave-particle
interactions) into account. The properties of the solar wind 
that can be measured from the solar corona throughout the heliosphere
are plasma flow speeds, densities, temperatures, magnetic fields, 
wave spectra, and particle composition, which all exhibit complex 
spatio-temporal fluctuations.  Most of the observations of the solar 
wind were made in-situ (with the Mariner, Pioneer, Helios, ISEE-3, 
IMP, Voyager, ACE, WIND, Cluster, Ulysses, or STEREO spacecraft), 
complemented by remote-sensing imaging (with STEREO) and radio 
scintillation measurements. 

\begin{figure}[t]
\centerline{\includegraphics[width=0.7\textwidth]{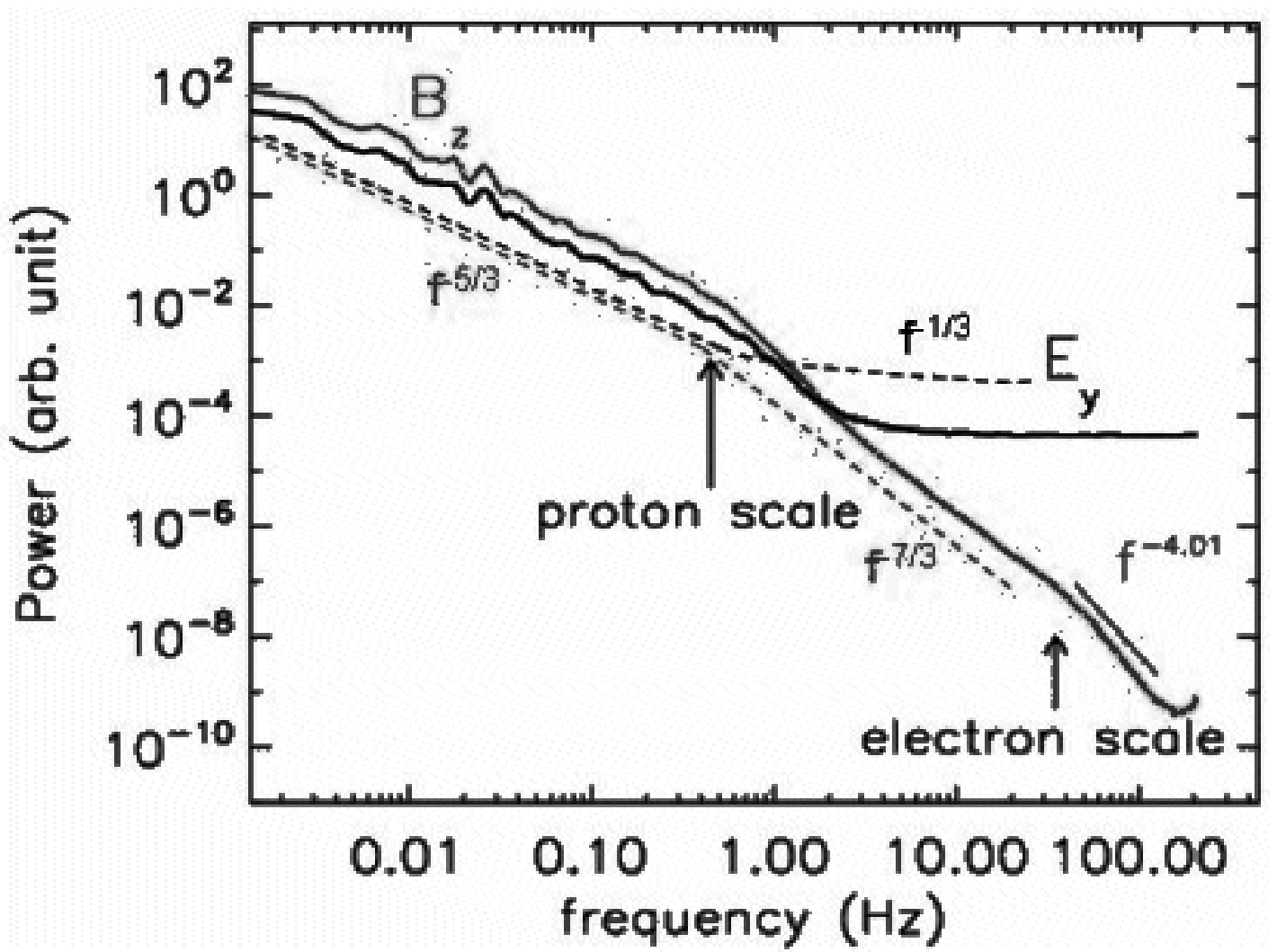}}
\captio{A spectrum of the solar wind is shown, based on 
{\sl CLUSTER} observations
from large scales ($\approx 105$ km) down to small scales ($\approx
3$ km) is shown, with the proton and electron gyroradius scale indicated.
The solar wind spectrum is interpreted in terms of a turbulent MHD cascade,
with the theoretically predicted slopes of $f^{-5/3}$ and $f^{-7/3}$ from
gyro-kinetic theory. The plot proves that the energy continues cascading
below the proton scale down to the electron scale, where it is converted
to heat (via electron Landau damping resonance) causing the steepening of
the $B_z$ spectrum to $f^{-4}$ (Howes et al.~2008;  Sahraoui et al.~2009; 
credit: ESA, CLUSTER).}
\end{figure}

The dynamics of the solar wind is often characterized by the 
MHD turbulent cascade model. 
The solar wind power spectrum exhibits fully developed turbulence 
of the Kolmogorov type, $P(\nu) \propto \nu^{-5/3}$,
in interplanetary space and near Earth (Fig.~15), 
while the input spectrum in the lower corona is of the 1/f-noise type, 
$P(\nu) \propto \nu^{-1}$ (Matthaeus and Goldstein 1986; 
Nicol et al.~2009). 
The MHD turbulent cascade starts at the largest scales fed by
MHD waves with a $1/f$-noise spectrum in the lower corona, while 
turbulent interactions produce a cascade of energy through vortices 
and eddies to progressively smaller sizes with a spectrum of
$\nu^{-5/3}$, and final energy dissipation at the smallest scales by
heating of electrons, with a spectrum of $\nu^{-11/3}$ (Meyrand
and Galtier 2010). 
The analysis of MHD turbulence in solar wind data includes determining 
power spectra and structure functions, waiting time distributions
of solar-wind bursts,
identifying the phenomenology or MHD turbulence (Kolmogorov 1941; 
Kraichnan 1974), characterizing self-similarity and intermittency, 
and identifying the most intermittent structures, such as shock waves, 
small random events, current cores, and 1D current sheets (e.g., 
Horbury and Balogh 1997; Veltri 1999).    

\begin{table}[t]
\begin{center}
\normalsize
\captio{Powerlaw slopes measured in the size distributions of magnetic 
field fluctuations in the solar wind. The burst energy $E$ is defined 
as the area-integrated and time-integrated Poynting flux, derived from 
the Akasofu parameter.} 
\medskip
\begin{tabular}{llllll}
\hline
Instrument  & Powerlaw  & Powerlaw  & Powerlaw & Powerlaw 	& Reference: \\
	    & slope of  & slope of  & slope of & slope of   & \\
	    & area      & energy    & duration & waiting time & \\
	    & $\alpha_A$& $\alpha_E$& $\alpha_T$&$\alpha_{\Delta t}$ &\\
\hline
\hline
WIND 	    &           &$\approx 1.8$&$\approx 2.2$&1.67   & Freeman et al.~(2000a)\\  

ACE 	    &           & 1.5 	    & 2.46     & 1.6 	& Moloney \& Davidsen (2011)\\  
\hline
FD-SOC prediction & {\bf 2.0}  & {\bf 1.5} & {\bf 2.0}& {\bf 2.0} & Aschwanden (2012a)     \\
\hline
\end{tabular}
\end{center}
\end{table}

Recent interpretations of the dynamics of the solar wind include
self-organization and SOC systems. The strongest argument for a SOC 
interpretation is the fact that powerlaw size distributions were found 
for energy fluctuations ($E_B \propto B^2$), durations ($T$), 
and waiting times $(\Delta t)$ in the solar wind (Freeman 
et al.~2000a; Moloney and Davidsen 2011). Although the powerlaw shape
of the waiting time distribution of solar wind bursts is not 
exponential, hence inconsistent with the original BTW model  
(Boffetta et al.~1999; Freeman et al.~2000a),
it can be reproduced with a non-stationary Poisson process (Fig.~6;
Wheatland et al.~1998; Aschwanden and McTiernan 2010). This was also 
demonstrated with MHD simulations (Watkins et al.~2001; Greco et al.~2009a,b),
and by cellular automaton simulations with correlations in the driver
that produces a ``colored'' power spectrum (Sanchez et al.~2002).

SOC systems produce fractal spatio-temporal structures.
The fractal nature of magnetic energy density fluctuations in the
solar wind has been verified observationally (Hnat et al.~2007; 
Rypdal and Rypdal 2011a,b). Moreover, solar wind turbulence is
found to be multi-fractal, requiring a generalized model with multiple 
scaling parameters to analyze intermittent turbulence
(Maczek and Szczepaniak 2008; Macek and Wawrzasek 2009; Macek 2010),
although a single generalized scaling function is sometimes sufficient 
too (Chapman and Nicol 2009; Rypdal and Rypdal 2011). However, 
the fractal geometry of solar wind bursts seems not to be self-similar, 
since the ratio of kinetic ($E_k$) to magnetic energy ($E_B\propto B^2$)
is frequency-dependent, with a magnetic energy spectrum of 
$\propto E_B^{-5/3}$ and a kinetic energy spectrum of $\propto E_k^{3/2}$
(Podesta et al. 2006a,b, 2007). 
It was suggested that the interplanetary magnetic field (IMF) is 
clustered (self-organized) by low-frequency magnetosonic waves, leading 
to a fractal structure with a Hausdorff dimension of $D=4/3$ and a 
turbulent power spectrum with $\nu^{-5/3}$ (Milovanov and Zelenyi 1999).

In the end, can we claim that the dynamics of the solar wind is
consistent with a SOC system? Observationally we find that 
magnetic field and kinetic energy fluctuations measured in the solar
wind exhibit powerlaw distributions, which is consistent with a SOC
system. One argument against the SOC interpretation is the observed
powerlaw distribution of waiting times (Boffetta et al.~1999), but
this argument applies only with respect to the original BTW model, 
while it presents no obstacle for nonstationary Poisson processes. 
Another (Occam's razor) argument was that a SOC interpretation is
not needed when turbulence can already explain solar wind spectra
(Watkins et al.~2001). Considering the spatial structure of the
solar wind, a fractal (or multi-fractal) property was identified,
another hallmark of SOC models. What about the driver, instability, 
and avalanches expected in a SOC system? The driver mechanism 
is the acceleration of the solar wind in the solar corona itself,
a process that basically follows the hydrodynamic model of Parker (1958),
and may be additionally complicated by the presence of nonlinear
wave-particle interactions, such as ion-cyclotron resonance
(e.g., for a recent review see Ofman 2010). Then, the instability 
threshold, triggering extreme bursts of magnetic field fluctuations, 
the avalanches of solar wind SOC events, can be caused by 
dissipation of Alv\'en waves, onset of turbulence, or by the 
ion-cyclotron instability. Thus, in principle the generalized SOC 
concept can be applied to the solar wind, if there is a system-wide 
threshold for an instability that causes extreme magnetic field 
fluctuations. On the other side, the MHD turbulent cascade model explains
naturally two particular spatial scales with enhanced energy
dissipation (i.e., the proton the electron gyroradii), which
is in contrast with the scale-freeness of energy dissipation in 
classical SOC models. Nevertheless, the solar wind dynamics can be 
described by multiple models that do not exclude each other: 
(1) the MHD turbulent cascade model describes the power spectrum 
of the solar wind, (2) kinetic theory captures the microscopic 
physics of wave-particle interactions and the evolution of particle
velocity distributions in the solar wind, and (3) SOC models quantify 
the statistics and macroscopic size distributions of extreme events 
in the solar wind.

\subsubsection{		Solar-Terrestrial Effects	}

A solar-terrestrial effect that has been modeled in terms of
SOC models is the connection between solar flare occurrence
and temperature anomalies on Earth. The scaling of
the Earth's short-term temperature fluctuations and solar flare
intermittency was analyzed in terms of the spreading exponent
and the entropy of diffusion, finding that both have a L\'evy flight 
statistics with the same exponent $\alpha_{\Delta t}=2.1$ in
the waiting-time distribution (Scafetta and West 2003).
The same data were re-analyzed by Rypdal and Rypdal (2010a),
who found that only the integrated solar flare index is consistent
with Lev\'y flight, while the global temperature anomaly follows
a persistent fractional Brownian motion. The persistence 
(long-range memory) of solar activity was investigated further
and it was found that the sunspot number and the total solar
irradiance are long-range persistent, while the solar flare index
is very weakly persistent, with a Hurst exponent of $H < 0.6$
(Rypdal and Rypdal 2012). A stochastic theory to model the
temporal fluctuations in avalanching SOC systems has been 
developed to understand these solar-terrestrial observations 
(Rypdal and Rypdal 2008a,b). Three other Earth climate factors
(average daily temperature, vapor pressure, and relative humidity)
were analyzed and found to exhibit power-law distributions 
and thus believed to constitute a SOC system (Liu et al.~2013).

The prediction of solar-terrestrial effects, such as 
geoeffective solar eruptions and SEP events, 
resulting in space-weather storms and magnetospheric disturbances,
are of course of highest interest for our society. Statistics
of the most extreme events need to be derived from the 
rarest events at the upper end of the size distributions,
where a powerlaw extrapolation is often questionable, and thus
has been modeled with different cutoff functions, often
associated with finite-system size effects. The best relevant data 
we have at hand is the solar flare statistics from the last 40 years, 
while geological tracers (nitrate concentrations in polar ice
cores or select radionuclides) extend over millenia, but are
not reliable proxy records of solar flares or SEP events
(Schrijver et al.~2012), because nitrate spikes in ice cores
can also be caused by biomass burning plumes (Wolff et al.~2012).
Theoretical studies focus on extreme
value and record statistics in heavy-tailed processes with
long-range memory (Schumann et al.~2012). The inclusion of
memory and persistence is obviously very important, because
the predicted number of extreme events during a clustered
time interval can be much larger than predicted in a purely
stochastic SOC model, such as in the original BTW model
(Strugarek and Charbonneau 2014).

\clearpage %%%%%%%%%%%%%%%%%%%%%%%%%%%%%%%%%%%%%%%%%%%%%%%%%%%%%%%%%%%%%%%%%%

\subsection{		Solar Physics: Theoretical Models		}

\subsubsection{		Solar Cellular Automaton Models 		}

One of the most ingenious simplications of reducing complexity in nature
to simplicity in theoretical modeling is the approach of numerical
lattice simulations, also called cellular automaton models.
In the case of simulating avalanches in a SOC system, Bak et al.~(1987,
1988), slightly preceded by Katz (1986), just defined a discretized
step in the evolution of an avalanche by a simple next-neighbor 
redistribution rule in a 3D lattice grid, the so-called BTW model,
\begin{equation}
        \begin{array}{ll}
        z(i,j,k)=z(i,j,k)+1 & {\rm initial\ input} \\
        z(i,j,k)=z(i,j,k)-6 & {\rm if}\ z(i,j,k) \ge 6 , \\
        z(i\pm 1,j\pm 1,k\pm 1)=z(i\pm 1,j\pm 1,k\pm 1)+1 &
        \end{array}
\end{equation}
which of coarse can also be generalized to a 2D avalanche with
4 neighbor nodes, or a 1D avalanche with 2 neighbor nodes. The crucial 
part is a critical threshold that decides whether an avalanche
starts/continues, or stops, which is $z_{crit}=6$ in 3D,
$z_{crit}=4$ in 2D, or $z_{crit}=2$ in 1D. Such cellular automaton
simulations require millions of time steps until the system becomes
critical, and another few millions to produce sufficient statistics
of avalanche sizes. Then, the direct output of such numerical simulations
are statistical distributions of avalanche sizes and durations.
The size of an avalanche is generally defined as the time-integrated 
sum of all nodes that were active during any time step of an avalanche.
Summarizing similar BTW-type lattice simulations, we compiled in Table 1
a list of powerlaw slopes that resulted from the avalanche
sizes in 1D, 2D, and 3D lattice grids, which exhibit a
dependence on the dimensionality of the system, as well as some
scatter among the results from identical dimensions, due to slightly
different definitions of the redistribution rules and different system 
sizes. 

The first applications of BTW cellular automaton simulations to 
solar flares were made by Lu and Hamilton (1991), who interpreted
the avalanches in terms of small magnetic reconnection events, where
unstable magnetic energy is dissipated, and demonstrated that the 
powerlaw slopes of numerically simulated avalanche sizes, durations, 
and instantaneous peak sizes match the observed frequency distributions 
of hard X-ray fluences $E$, flare durations $T$, and peak fluxes $P$. 
The powerlaw slopes were found to be essentially invariant when the size
of the system (i.e., the cartesian lattice grid) was changed
(Lu et al.~1993). 

While the BTW model arranges an isotropic
redistribution (in all next-neighbor directions), with the magnetic
field strength $B$ being the redistributed quantity, in the application
to solar flares (Lu and Hamilton 1991), an anisotropic cellular 
automaton model with a one-directional redistribution along
the direction with the largest magnetic field gradient was
proposed by Vlahos et al.~(1995), in order to mimic the inhomogeneity
of active regions in general, and the directivity of the dominant
magnetic field in the solar corona in particular. The anisotropic
BTW model produced steeper powerlaw distributions than the isotropic
standard model, a property that was utilized to construct a hybrid
model with a steep powerlaw slope for nanoflares and a flatter slope
for large flares (Vlahos et al.~1995; Georgoulis et al.~1995, 1998,
Georgoulis and Vlahos 1996, 1998), which was 
believed to match the observations (Fig.~11). However, 
the anomalously steeper powerlaw slopes reported for nanoflares 
early on (Benz and Krucker 1998; Parnell and Jupp 2000), have been 
downward-corrected later on due to inadequate modeling effects
(McIntosh and Charbonneau 2001; Benz and Krucker 2002; 
Aschwanden and Parnell 2002), and are now
more consistent with the size distribution of larger flares (Fig.~11).
Moreover, an anomalously steep powerlaw slope $\alpha_P > 2$ for the
energy cannot be reconciled with the standard SOC model based on 
the scale-free probability conjecture and diffusive transport 
(Sections 2.6-2.11).

The Sun often displays multiple sunspot groups or active regions 
at the same time, at least during the solar maximum. This implies, in
Bak's sandpile analogy, that solar flare statistics originates from
multiple simultaneous sandpiles. Consequently the size distributions
of active regions has to be folded into the event distributions, an
effect that still produced size distributions close to a single powerlaw 
(Wheatland and Sturrock 1996; Wheatland 2000c).

\begin{figure}[t]
\includegraphics[width=0.49\textwidth]{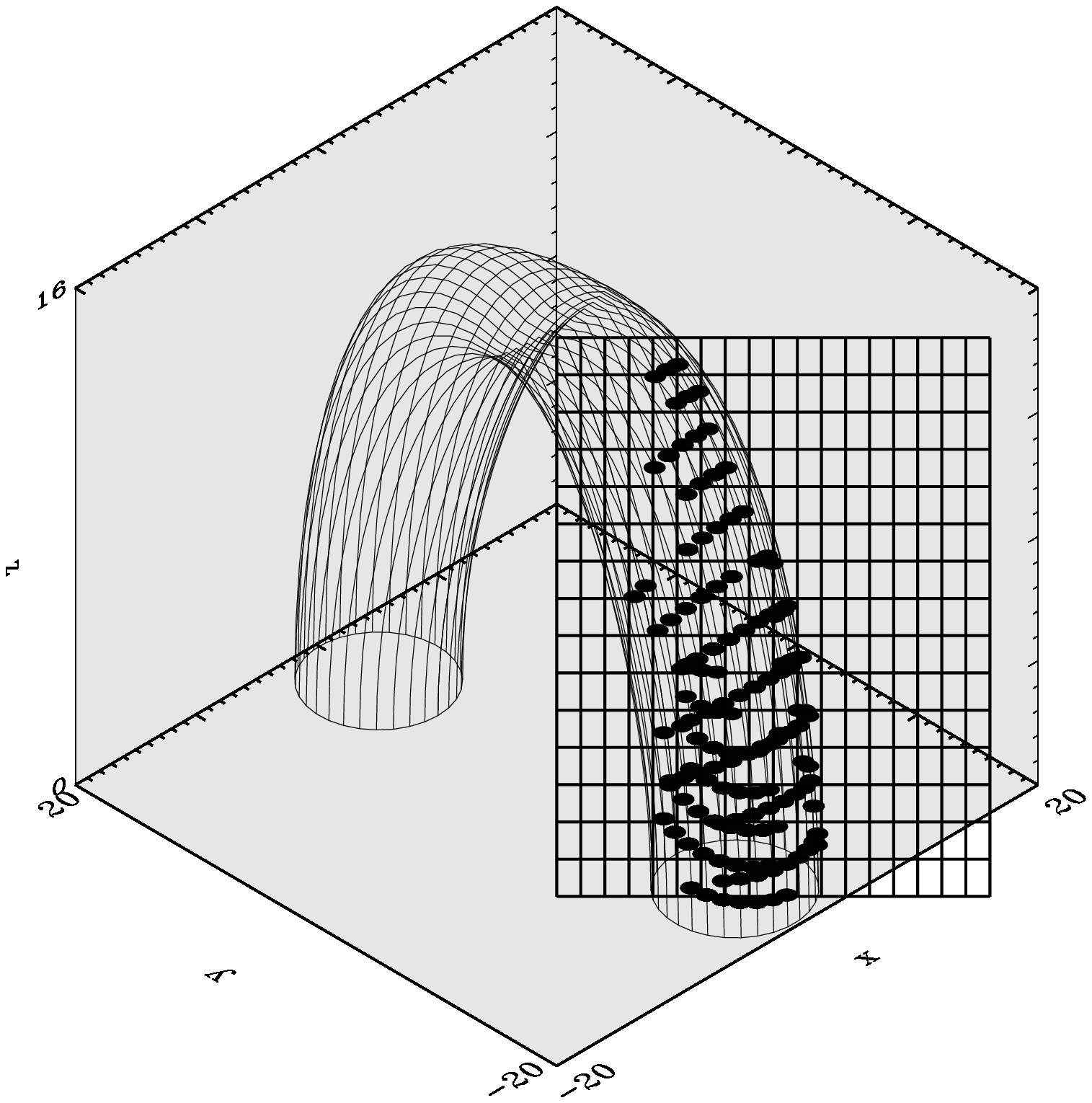}
\includegraphics[width=0.49\textwidth]{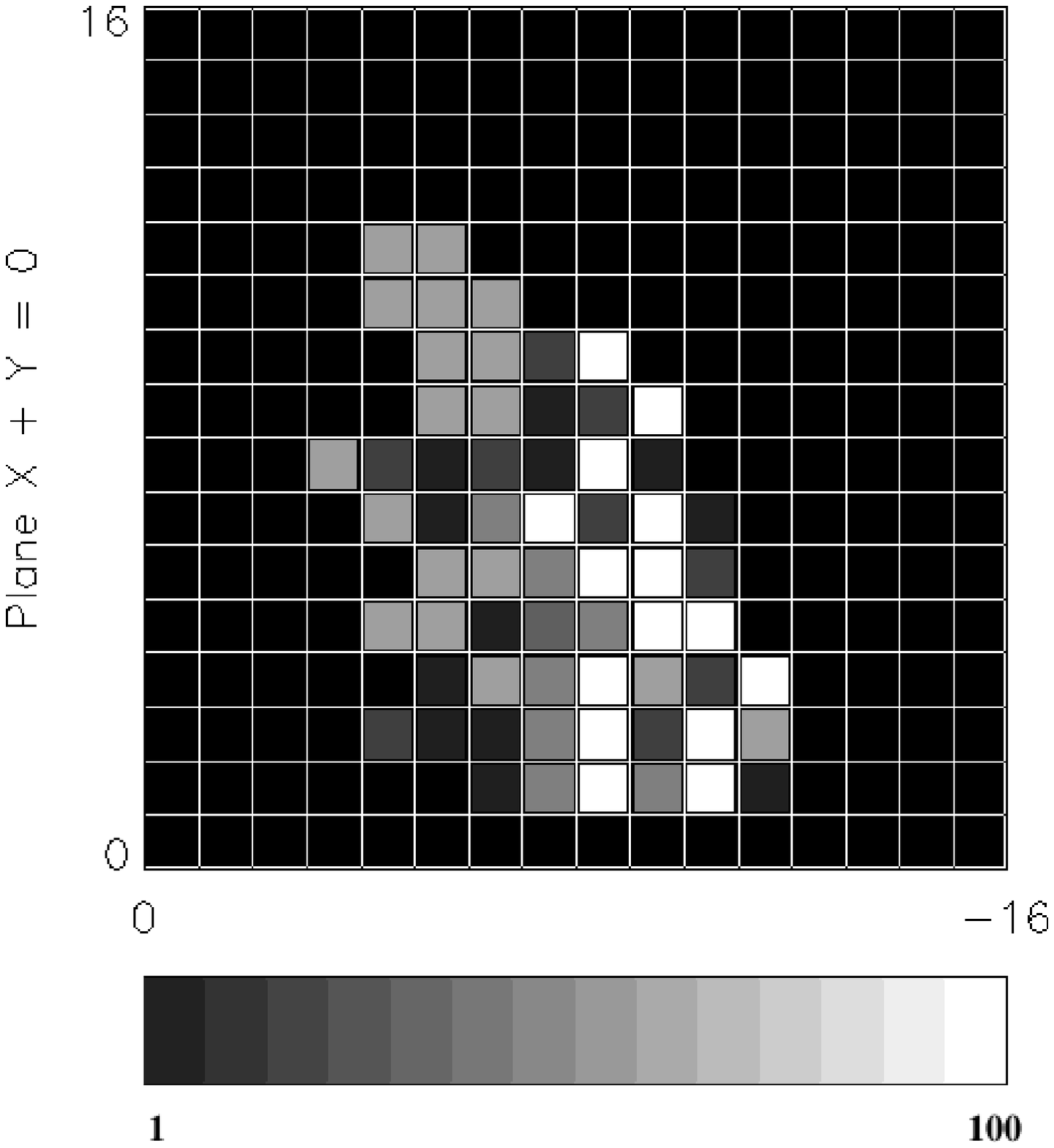}
\captio{Transformation of the flat 2-D lattice geometry of the
divergence-free braiding BTW model onto a pseudo 3-D loop envelope 
and plane-of-sky for an arbitrary observer's line-of-sight direction 
(left panel). The fractal area of the avalanche is projected
into the observer's plane (right panel) (Morales and Charbonneau 2009).}
\end{figure}

Variants or alternatives to the BTW model that have been applied to simulate 
the size distributions of solar flares, to name a few, include 
3D vector fields with periodic, constant, and symmetric boundaries (Galsgaard 1996),
a BTW model with additional nonlocal (remote) triggering that accomodates 
sympathetic flaring (MacKinnon and Macpherson 1997; MacPherson and MacKinnon 1999), 
lattice models with non-stationary driving that reproduce the observed
waiting time distributions (Norman et al.~2001), 
emergence of magnetic flux in evolving active regions (Vlahos et al.~2002),
lattice models with deterministic drivers based on the BTW model
(Strugarek et al.~2014) or on 
a finite driving rate version of the Olami-Feder-Christensen (OFC, 
Olami et al.~1992) model (Hamon et al.~2002), which usually is not in 
a SOC state but rather ``on the edge of SOC'', 
prediction of solar flares by data assimilation to the BTW model 
(Belanger et al.~2007), 
a divergence-free field braiding
cellular automaton model (Fig.~16; Morales and Charbonneau 2008a,b, 2009),
or drivers with diffusive characteristics that mimic a turbulent substrate
(Baiesi et al.~2008).
The 3D vector field simulations revealed two necessary criteria for the
generation of powerlaws: a contiuous driver that produces large scale regions
with coherent tension, and a partial (rather than a complete) release of 
the triggering quantity (Galsgaard 1996).

\subsubsection{		Analytical Microscopic Solar SOC Models  	}

While cellular automaton models are most powerful in simulating SOC processes,
the iterative numerical scheme is generally non-detererministic and unpredictable 
(with some exceptions, e.g., see Strugarek and Charbonneau 2014 for a
discussion on prediction from numerical SOC models), 
while analytical models are deterministic and give us
direct physical insights into the dynamics of a SOC system. Let us review
a few of the analytical approaches that have been employed to model solar
SOC processes.

\begin{figure}[t]
\centerline{\includegraphics[width=0.5\textwidth]{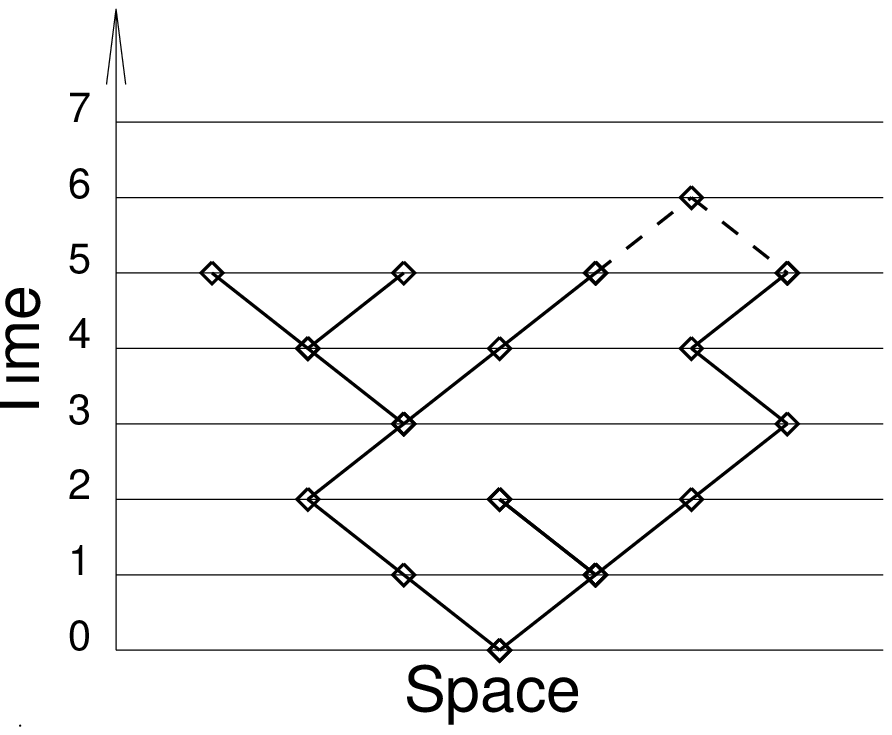}}
\captio{An avalanche represented by the tree generation of a branching
process. Propagation to a site that is already activated by more than
one neighbor (dashed lines) is ignored (Litvinenko 1998).}
\end{figure}

A 1D cellular automaton model was constructed in terms of a branching 
probability $p$, which yields a probabilty of $N(s) =s\ p^{s-1} (1-p)^2$ 
after $s$ time steps, and this way leads to a size distribution of
$N(s) \propto s^{-2}$ (MacKinnon et al.~1996). This branching-type
model was extended to a 2D version by introducing some {\sl ad hoc} 
functions that could produce powerlaw-like size distributions 
(Macpherson and MacKinnon 1999). This exercise demonstrated that the 
statistical redistribution rule of a cellular automaton model with
higher dimensions ($d \ge 2$) cannot easily be formulated in terms of analytical
branching probabilities. In an attempt to generalize this 1-D branching 
process to higher dimensions, Litvinenko (1998) points out a result from 
a tree branching process (Fig.~17), for which an asymptotic limit was found
with the following analytical expression (Otter 1949),  
\begin{equation}
        <N(s)> \ \propto s^{-3/2} \exp{\left(-{s \over s_0}\right)} \ ,
\end{equation}
that is close to observed frequency distributions of flare energies,
if we identify the size $s$ with the energy $E$ of flares. A similar
description of activation in a forest-fire model was adopted by
Christensen et al.~(1993). 

Size distributions of physical parameters and scaling laws can be derived 
from energy balance equations. Such an approach has been pursued by
Wheatland and Glukhov (1998) using a probability balance equation or
``master equation'' (Gardiner 1983; Van Kampen 1992),
\begin{equation}
	{d \over dE} (e P) + P \int_0^E \alpha(E,E') dE' -
	\int_E^\infty P(E') \alpha(E', E) dE' = 0 \ ,
\end{equation}
where E is the flare energy, $P(E)$ the probability distribution,
$\alpha(E,E')=\alpha(E-E')$ the probability for a transition from
energy state $E$ to $E'$, and $e(E)$ is an arbitrary energy increase
function of $E$ in the active region. A general solution of this
master equation was not found, but powerlaw distributions for the
flare energy distribution $N(E)$ were found for the special case
when the energy supply rate $e(E)$ does not depend on the free
energy of the system (Wheatland and Glukhov 1998). The same
energy balanche equation was applied to quantify how the energy
supply in the solar corona relates the flaring rate and free
energy of the system, leading to a hysteresis of about 9 months
during the 11-year solar cycle (Wheatland and Litvinenko 2001).
Further semi-analytical work and Monte-Carlo simulations of this 
{\sl jump-transition model} with a time-dependent driver (of the
energy input rate) illustrated how the SOC system responds in
form of modified flare energy and waiting time distributions
(Wheatland 2008, 2009; Kanazir and Wheatland 2010).

The spatio-temporal transport process of a SOC avalanche can 
macroscopically be approximated by a fractal-diffusive relationship,
$r(t) = \kappa (t-t_0)^{\beta/2}$ (Eq.~9), where $\beta=1$
corresponds to classical diffusion or random walk. The process
of a random walk of particles through a fractal environment in
3D space was analytically described in Isliker and Vlahos (2003).
Particles propagate freely in space not occupied by the fractal,
but are scattered off into random directions when they hit a
boundary of a fractal structure. This spatio-temporal transport 
process turns into a classical random walk in the limit of very
sparse fractals, but produces enhanced diffusion (hyper-diffusion)
with $\beta > 1$ for fractal dimensions $D_d > 2$. Since the
diffusive spreading exponent $\beta$ is a free parameter in the 
standard SOC model (Section 2.9), the analytical derivations of
particle transport in fractal structures can give us physical
insight into the nature of the diffusion process and the values 
of the spreading exponent $\beta$. 

The redistribution rule that involves the next neighbors in a 
lattice grid during one time step of a SOC avalanche, has an
extremely simple discretized form (Eq.~57), but is hard to
capture in the continuum limit, suitable for analytical models.
Lu (1995c) envisions avalanches in a continuum-driven dissipative system,
which is characterized by a coupled equation system of a one-dimensional
diffusion process,
\begin{equation}
        {\partial B(x,t) \over \partial t}
        = {\partial \over \partial x}
        \left[ D(x,t) {\partial B \over \partial x} \right] + S(x,t) \ ,
\end{equation}
\begin{equation}
        {\partial D(x,t) \over \partial t}
        = {Q(|\partial B/\partial x]) \over \tau}
        - {D(x,t) \over \tau} \ ,
\end{equation}
where $B(x,t)$ is a scalar field, $D(x,t)$ is a spatially and temporally
varying diffusion term, $S(x,t)$ is a source term, $Q(|\partial B/\partial x|)$
is a double-valued Heavyside function that has a low or high state and 
depends on the time history and an instability threshold.
Isliker et al.~(1998a) discretize the 3-D cellular automaton redistribution
rule into a differential equation that represents a diffusion process,
\begin{equation}
        {\partial {\bf B}({\bf x},t) \over \partial t}
        = \eta \nabla^2 {\bf B}({\bf x},t) + {\bf S}({\bf x},t) \ ,
\end{equation}
with a source term ${\bf S}({\bf x},t)$ and a diffusion coefficient
$\eta = {1/7} (\Delta h^2/\Delta t)$.
This differential equation contains a continuous function
${\bf B}({\bf x},t)$ that behaves the same way as the nearest neighbors
during one redistribution step, but a singularity occurs at the center
location at $\Delta h \mapsto 0$, which requires a modification of the
cellular automaton rule. Liu et al.~(2002) and Charbonneau et al.~(2001)    
transform the cellular automaton rule into a finite difference equation,
\begin{equation}
        {\partial B \over \partial t} = - {\partial^2 \over \partial x^2}
        \kappa (B_{xx}^2) {\partial^2 B \over \partial x^2} \ ,
\end{equation}
where $\kappa (B_{xx}^2)$ is a diffusion coefficient that depends on the local
curvature $B_{xx}^2$. This is a fourth-order nonlinear hyperdiffusion
equation, which is interpreted as continuum limit of the cellular automaton
rule, compatible with MHD in the regime of strong magnetic field and
strong MHD turbulence (with high effective magnetic diffusity).

\begin{figure}[t]
\centerline{\includegraphics[width=0.8\textwidth]{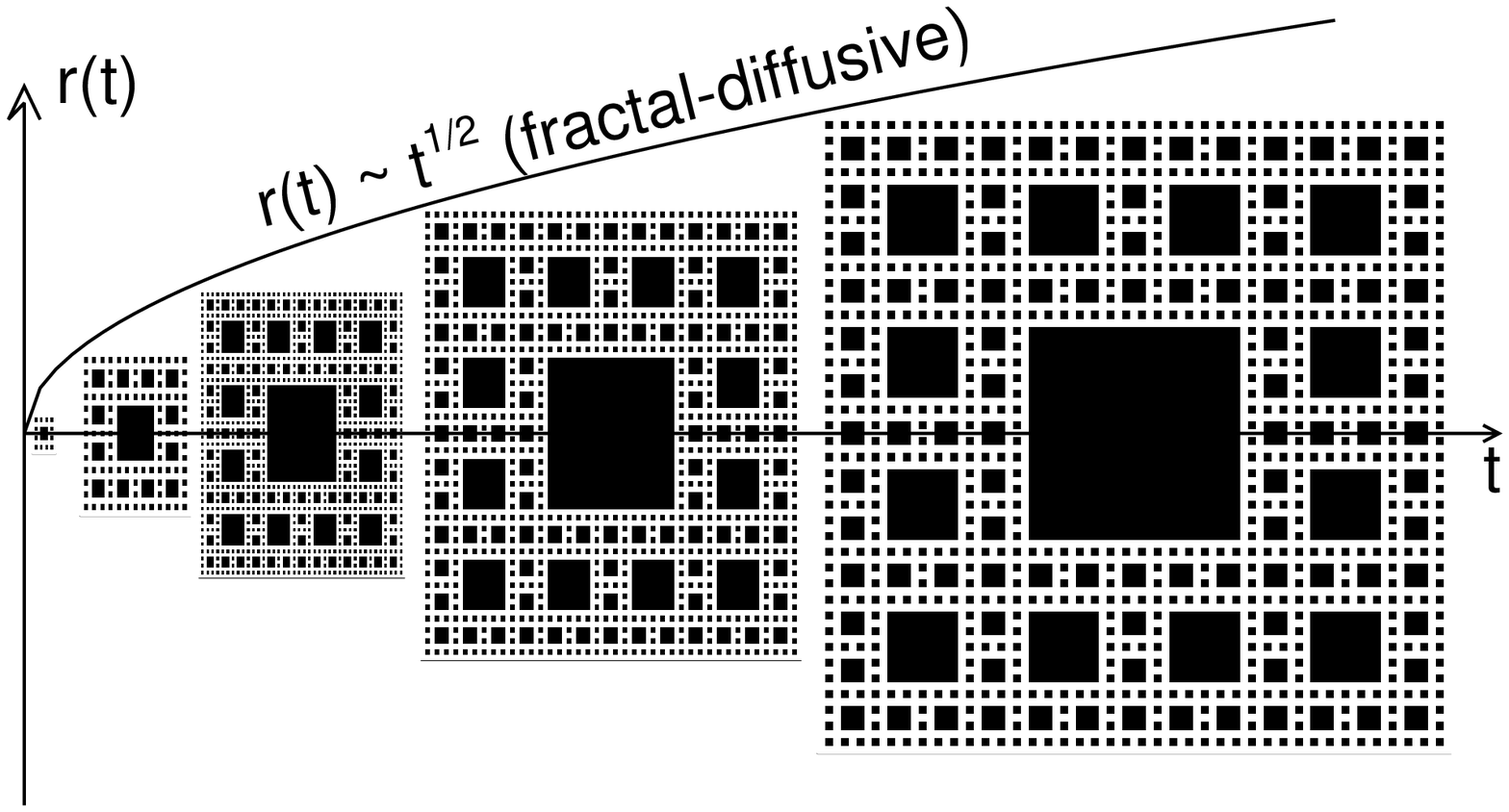}}
\captio{A cartoon that illustrates the concept of fractal-diffusive
avalanche evolution. The Euclidean radius $r(t)$ evolves like a
diffusive random walk, such as $r(t) \propto t^{1/2}$ for classical
diffusion, while the avalanche area is fractal (black substructures).
The instantaneous fractal area $A_f(t) \propto r(t)^{D_t}$ consists
of the number active nodes and is proportional to the energy dissipation
rate $dE(t)/dt$ or flux $F(t)$ at a given time $t$ (Aschwanden 2014).}
\end{figure}

\subsubsection{		Analytical Macroscopic Solar SOC Models  	}

Analytical descriptions of the macroscopic evolution of solar flares
go back to Rosner and Vaiana (1978), who demonstrated that the two assumptions
of (1) an exponentially growing energy storage, $W(t) \propto \exp(\tau/\tau_G)$,
with growth time $\tau_G$, and (2) a random-like interuption with an
exponential distribution of saturation times, $N(\tau) \propto exp(-\tau/t_s)$,
leads directly to a powerlaw distribution of flare energies,
$N(E) \propto E^{-\alpha_E}$, with a powerlaw index,
\begin{equation}
	\alpha_E = (1 + {\tau_G \over t_s}) \ .
\end{equation}
However, observational data analysis did not confirm any correlation
between waiting times (called energy storage times in Rosner and Vaiana 1978)
and flare energies (Lu 1995b; Crosby 1996; Wheatland 2000b; 
Georgoulis et al.~2001; Moon et al.~2001). Also, flare time profiles
show very rarely a simple exponential increase with an abrupt drop. 
Moreover, this model assumes an exponential distribution of waiting times,
while observations exhibit powerlaw distributions with slopes of
$\alpha_{\Delta t} \approx 2-3$ (Fig.~6). 

Variations of this original model in terms of powerlaw-like growth (rather 
than exponential), or logistic growth (Aschwanden et al.~1998b), predict 
strong deviations from powerlaw distributions of flare energies and durations 
(Aschwanden 2011a, chapter 3). 

A better matching macroscopic description of SOC systems was developed in 
terms of a fractal-diffusive transport process (Aschwanden 2012a,b), which
can accomodate a fluctuating time profile of the energy dissipation rate
or observed flux, a fractal spatial structure, diffusive 
transport (Fig.~9), and spatio-temporal scaling laws that predict powerlaw
functions of all physical SOC variables. These scaling laws are in agreement with
virtually all measurements made in solar flares. The time evolution of
the avalanche radius $r(t)$, the fractal dimension $D_d(t)$ (in
Euclidean space with dimension $d$), the average 
energy dissipation rate or flux $f(t)$, the peak energy dissipation rate
or peak flux $p(t)$, and time-integrated energy or fluence $e(t)$ are,
\begin{equation}
        r(t) = \kappa (t - t_0)^{\beta/2} ,
\end{equation}
\begin{equation}
        D_d(t) = 1 + (d-1) \rho(t) \ ,
\end{equation}
\begin{equation}
        f(t) = f_0 t^{D_d(t) \beta/2}  \ .
\end{equation}
\begin{equation}
        p(t) = p_0 t^{d \beta/2} \ ,
\end{equation}
\begin{equation}
        e(t) = \int_{t_0}^t f(t) dt \approx f_0 t^{<D_d> \beta/2}  \ .
\end{equation}
where $\rho(t)$ is a random function varying in the range of $[0,1]$.
The fractal dimension $D_d(t)$ and the flux time profile $f(t)$ can fluctuate 
randomly, but the Euclidean radius $r(t)$ and the total (time-integrated)
energy $e(t)$ are monotonically increasing quantities during an avalanche.
The scaling laws between these parameters and the resulting size distributions
are defined in terms of the fractal-diffusive transport and the scale-free
probability conjecture (Sections 2.6-2.11), and thus all powerlaw indices
are predicted from first principles in this model. The same SOC model 
has been applied to a host of astrophysical phenomena (Aschwanden 2014). 

\subsubsection{		Solar Magnetic Field Models and SOC		}

Since magnetic energy is believed to be the ultimate source of many phenomena 
in the solar corona, from sunspots, coronal loops, nanoflares, microflares, 
to large flares, eruptive filaments, and coronal mass ejections, magnetic 
processes clearly play a paramount role in solar SOC models. In cellular
automaton models, the size of avalanches is measured by the number of all
active nodes. In the original solar cellular automaton model, a magnetic field
variable $B_{i,j}$ is assigned to each node $(x_i,y_j)$ in a 2D lattice grid,
and an isotropic gradient $\Delta B$, also called ``field curvature'',
is defined (Lu and Hamilton 1991; Charbonneau et al.~2001),
\begin{equation}
	\Delta B = B_{i,j} - {1 \over 2d} \sum_{nn=1}^{2d} B_{nn} \ ,
	\qquad |\Delta B| > B_c \ ,
\end{equation}
where $nn$ symbolizes the indices $(i\pm1, j\pm1)$ of all next neighbors,
while $B_c$ is the critical threshold, and $d$ the Euclidean space dimension. 
The next-neighbor redistribution rule is then (Charbonneau et al.~2001),
\begin{equation}
	B_{i,j} \mapsto B_{i,j} - {2d \over 2d+1} B_c \ ,
\end{equation}
\begin{equation}
	B_{nn} \mapsto B_{i,j} + {1 \over 2d+1} B_c \ .
\end{equation}
Although the redistribution  rule is locally conservative in the magnetic
field strength (i.e., $B_{i,j}+\sum B_{nn}$ is constant), the magnetic
energy is not conserved due to its quadratic dependence on the field
strength (i.e., $E_B = B^2/8\pi$). For every redistribution step, the
lattice energy decreases by an amount,
\begin{equation}
	\Delta E = {2d \over 2d + 1} \left( 2 {|\Delta B| \over B_c} - 1 
		\right) B_c^2 \ ,
\end{equation}
which is slightly larger than the minimum ``quantum'' that can be released
by the lattice, 
\begin{equation}
	\Delta E_{min} = {2d \over 2d + 1} B_c^2 \ ,
\end{equation}
Although the amount of dissipated energy per node is not exactly
constant, because $\Delta E \gapprox \Delta E_{min}$, it is in the
spatial and temporal average sufficiently close to the constant
$\Delta E_{min}$ so that we can assume an approximate proportionality 
between the time-integrated avalanche volume $V$ and the time-integrated 
energy $E$, and thus can apply the flux and energy scaling for
incoherent processes (Section 2.10-11).

\begin{figure}[t]
\centerline{\includegraphics[width=0.4\textwidth]{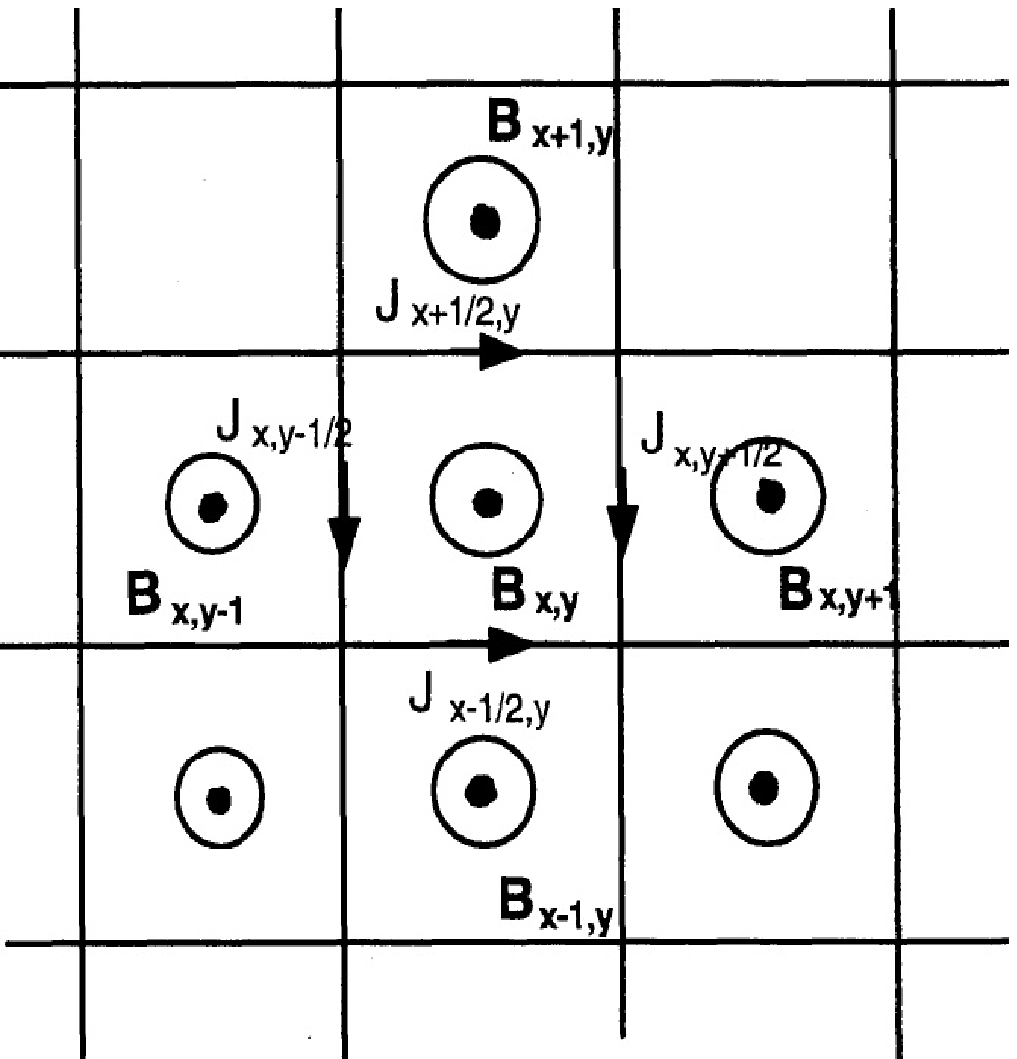}}
\captio{Cellular automaton model containing magnetic fluxtubes,
each one characterized by a magnetic field $B_z(x\pm1,y\pm1)$ and by
four segments of currents $J(x,y\pm{1\over 2})$ and $J(x\pm{1\over 2},y)$
at the cell boundaries (Takalo et al.~1999a).}  
\end{figure}

While this concept of relating the magnetic energy to the avalanche
volume provides a physical unit to a SOC avalanche,
the immediate question arises whether such a SOC system is consistent 
with the physics of magnetohydrodynamics (MHD). Maxwell's equations
applied to a coronal plasma define an electric current density {\bf j},
\begin{equation}
        {\bf j} = {c \over 4 \pi} (\nabla \times {\bf B}) \ ,
\end{equation}
yielding together with Ohm's law (with electric conductivity $\sigma$)
the so-called induction equation,
\begin{equation}
        {\partial {\bf B} \over \partial t} 
	= \nabla \times ({\bf v} \times {\bf B})
        + \eta \nabla^2 {\bf B} \ ,
\end{equation}
which contains a convective and a magnetic diffusion term (with 
magnetic diffusity $\eta = c^2/4\pi \sigma$), and fulfill the divergence-free
condition for the magnetic field,
\begin{equation}
        \nabla \cdot {\bf B} = 0 \ .
\end{equation}
Since the divergence-free condition is linear, it can easily be satisfied
by a suitable choice of a redistribution rule, at least locally during
each redistribution step, and globally within the threshold limit $B < B_c$.

The transformation of a cellular automaton redistribution rule into
a discretized MHD differencing scheme started with Takalo et al.~(1999a)
for an application to a magnetotail field model, and with Vassiliadis
et al.~(1998) for an application to solar flares. The curl of the
current ${\bf j}$ at the cell boundaries is defined in terms of the
magnetic field vectors in each neighbor cell, as shown in Fig.~19 and
defined by Amp\`ere's and Ohm's law (Eqs.~70 and 71). This way,
a resistivity can be defined as a function of the current at the flux
tube boundary, as expected from a current-driven instability.
Anisotropic cellular automata correspond to a nonlinear resistivity,
while isotropic ones can be associated with hyper-resistivity
(Vassiliadis et al.~1998). In the continuum limit, however,
singularities can arise (Isliker et al.~1998a), which largely
disappear in 3D models (Isliker et al.~2000). For solar flare applications,
a threshold of a critical current $j_c$ was found to be physically
more appropriate (Isliker et al.~2001), than a threshold of a critical
magnetic field $B_c$ as used in the original SOC models (Lu and Hamilton 1991).

While the original SOC models have random drivers that incur little
disturbances at random places, solar SOC models became more realistic
by prescribing drivers that mimic the photospheric magneto-convection
(at the lower boundary of the computation box) and drive MHD turbulence 
in 2D (Georgoulis et al.~1998),
drivers that lead to collision of large-amplitude torsional Alfv\'en wave
packets (Wheatland and Uchida 1999), 
drivers that conserve helicity (Chou 1999; 2001),
by calculating linear force-free fields (Vlahos and Georgoulis 2004), 
by calculating an initial nonlinear force-free field from an observed
magnetogram (Dimitropoulou et al.~2011), 
by using a sequence of observed vector magnetograms as an initial
condition (Dimitropoulou et al.~2013), or 
by designing divergence-free ($\nabla \cdot {\bf B} = 0$) redistribution
rules (Fig.~16; Morales and Charbonnau 2008a,b; 2009).
Several of these SOC simulations were designed to mimic coronal heating
according to the field line braiding scenario postulated by Parker (1988),
where the SOC driver is represented by the photospheric convection-driven
random motion of coronal loop footpoints, while SOC avalanches are
triggered by magnetic reconnection above some critical threshold angle 
of magnetic field misalignments (Krasnoselskikh et al.~2002;
Morales and Charbonnau 2008a,b; 2009; Uritsky etal.~2013). 
In one recent study, the photospheric statistics of avalanches 
(measured from magnetograms) and coronal statistics (measured from
extreme-ultraviolet images) was performed simultaneously and  
scaling relationships were found between these two type of events, i.e., 
$L_{cor} \propto L_{phot}^{1.39}$ and $T_{cor} \propto T_{phot}^{0.87}$,
a correlation that implies a stochastic coupling between photospheric 
magnetic energy injection (into the corona) and coronal heating events 
(Uritsky et al.~2013). This stochasticity corroborates the findings of
Dimitropoulou \etal~(2009) on the lack of correlations between fractal
properties of the photosphere and corona.

All these recent studies clearly demonstrate an advancement from the 
simple original cellular automaton algorithms to more sophisticated 
data-driven physical models. These physical models often are able
to reproduce the standard size distributions and waiting time
distributions that are predicted from the standard SOC model
(Sections 2.6-2.12). For instance, the dynamic
data-driven integrated flare SOC model of Dimitropoulou et al.~(2013)
obtains the following powerlaw slopes: $\alpha_P=1.65\pm0.11$ for peak
energies, $\alpha_E=1.47\pm 0.13$ for energies, and $\alpha_T=
2.15\pm0.15$ for the duration of large flares, which agrees
well with the standard model ($\alpha_P=1.67$, $\alpha_E=1.5$,
$\alpha_T=2.0$; Eq.~24). It proves the robustness of
the generic standard SOC model, regardless of the specific physics 
that is involved in a particular phenomenon. Vice versa, deviations from
the predicted powerlaw size distributions of the standard model 
can reveal crucial hints which assumptions of the standard 
SOC model are violated, implying possible refinements to the model.

\subsubsection{	    Magnetic Reconnection in Solar Flares and SOC	}

Once the interpretation of solar flares in terms of a SOC system was
introduced (Lu and Hamilton 1991), physical scaling laws were envisioned 
that could explain the observed size distributions of time scales $T$,
peak fluxes $P$, and (time-integrated) fluences or total energies $E$.
A minimal magnetic reconnection model was formulated in terms of
Alfv\'enic time scales $\Delta T$, dissipated magnetic energy $\Delta E$,
and energy release rate $\Delta P=\Delta E/\Delta T$ (Lu et al.~1993;
Nishizuka et al.~2009),
\begin{equation}
	\Delta E = (\Delta L)^3 \langle {B_{\perp}^2 \over 8\pi} \rangle \ ,
\end{equation}
\begin{equation}
	\Delta T = {\Delta L \over v_A} \zeta \ ,
\end{equation}
\begin{equation}
	\Delta P = {\Delta E \over \Delta T} = 
	(\Delta L)^2 \langle {B_{\perp}^2 \over 8\pi} \rangle 
	{v_A \over \zeta} \ ,
\end{equation}
where $v_A=B/(4\pi \rho)^{1/2}$ is the Alfv\'en velocity, $\rho$ the
plasma density, and $\zeta \approx 10^1,...,10^2$ is a constant estimated 
from the reconnection scenario of Parker (1979). This equation system
can be fitted to the observed correlations, i.e., $E \propto P^{1.82}$,
$E \propto T^{1.77}$ and $P \propto T^{0.90}$ (Lu et al.~1993), 
using suitable scaling laws 
for the free variables $B$, $\rho$, and $\zeta$. On the other side, the 
standard model (Eqs.~15-17) predicts from first principles 
the relationships $E \propto P^{(D_d+2/\beta)/d} \approx P^{1.33}$,
$E \propto T^{1+D_d\beta/2} \approx T^{2.00}$, and $P \propto T^{d\beta/2} 
\propto T^{1.50}$ (with $d=3$, $D_d=(1+d)/2$, and $\beta=1$),
which is not too far off from the measurements, given the large
scatter in the correlations. Thus, an interpretation 
in terms of a magnetic reconnection model does not provide a unique fit 
to the observed size distributions or correlations of SOC parameters
in solar flares, but 
it allows to test consistency between model and observations, and places 
physical units on the SOC parameters. Lu et al.~(1993) predicted
powerlaw behavior down to nanoflare events with energies of 
$\Delta E \approx 3 \times 10^{25}$ ers, durations of $\Delta T \approx
0.3$ s, and length scales of $\Delta L \approx 400$ km.

An alternative SOC reconnection model applied to solar flares is the
separator reconnection scenario (Longcope and Noonan 2000), where
currents flowing along the network of magnetic field separators are
sporadically dissipated. Scaling this system to solar length scales and
inductances yields typical energies of $E \approx 4 \times 10^{28}$ ergs,
waiting times of $\Delta t \approx 300$ s, and a heat flux of
$F \approx 2 \times 10^6$ ergs s$^{-1}$ cm$^{-2}$. The observed flare 
energy distribution $N(E) \propto E^{-3/2}$ requires a probability
of $P(L) \propto L^{-1}$ for separator length scales $L$ (Wheatland 2002; 
Wheatland and Craig 2003), which corresponds to a size distribution of 
$N(L) dL \propto L^{-2} dL$. Generalizing the flare geometry 
to $d=1,...,3$ dimension depending on the reconnection topology 
($E \propto L^d$), size distributions of $4/3 \le \alpha_E \le 2$
were predicted (Craig 2001).

Solar flares produced by cascades of reconnecting magnetic loops were
simulated in form of a SOC model by Hughes et al.~(2003). This model
produces a powerlaw distribution of flare energies with a 
slope of $\alpha_E=3.0\pm0.2$. This prediction disagrees with most
flare observations, which find $\alpha_E\approx 1.5$, but it 
corroborates anisotropic SOC models. Despite discrepancies, the model
still gives us some insight into the topology of energy dissipation regions.
The standard model predicts a probability distribution of 
$N(L) \propto L^{-3}$ for length scales (Eq.~1), and thus the
model of Hughes et al.~(2003) can be reconciled with the standard
SOC model if the dissipated energy volume is proportional to the length scale,
i.e., $E \propto L$, which requires a 1D geometry of the dissipation
region, such as separators of magnetic domains.

\subsubsection{	   Particle Acceleration in Solar Flares and SOC	}

We can consider a hierarchy of SOC systems in our universe: our universe
may be just one particular event in a multi-verse; galaxies are singular
events in our universe; stars are singular events in a galactic system;
planets are singular events in a solar system; solar flares are individual 
events in the solar corona; and accelerated particles are singular events
of a solar flare hard X-ray burst. In the latter example we would
consider the energy spectrum of accelerated particles as the energy
distribution in a SOC system, while the acceleration process of each
particle is an avalanche, driven by some electro-magnetic field in
a magnetic reconnection region or shock structure. The threshold for
particle acceleration is the {\sl run-away regime} in a thermal
plasma, which requires a velocity of a few times the thermal speed.
Once the particle gets accelerated out of the thermal bulk distribution, 
either by a DC electric field,
by wave-particle interactions, or by a quasi-parallel shock structure,
it ends up with a final energy $E \gg E_{th}$ when it leaves the
acceleration region, and the ensemble of all
accelerated particles in a solar flare produce an energy spectrum that
is often close to a powerlaw, $N(E) \propto E^{-\epsilon_E}$. What
powerlaw slope does the standard SOC model predict? The scale-free probability
conjecture, $N(L) \propto L^{-d}$ (Eq.~1), would still be applicable, 
since the probability to accelerate a particle in a subvolume with
length scale $L$ is reciprocal to the volume size. Also the fractal-diffusive
transport process, $L \propto T^{\beta/2}$ (Eq.~9), could still yield
an appropriate model for any stochastic and diffusive (wave-particle or
shock) acceleration process. However, the fractal dimension could
vary from a straight trajectory with $D_d \gapprox 1$ and $\beta \approx 2$
to a random path with $D_d \approx (1+d)/2$ and $\beta \approx 1$. 
Consequently, we predict powerlaw slopes for the energy spectrum in the range of 
$\epsilon_E=1 + 1/(\gamma D_3/2+1/\beta) \approx 1.5,...,1.67$ (Eq.~36), 
either for $D_3=1,...,2$ or $\beta=1,...,2$. This is a relatively narrow 
range that should be testable. 
However, finite system-size effects are expected in relatively small magnetic
reconnection regions, which will lead to a gradual cutoff at the upper
end of the energy spectrum, with a steeper powerlaw slope if the 
energy spectrum is fitted with a double powerlaw function.
Nevertheless, the standard SOC system predicts a lower limit of 
$\alpha_E \ge 1.5$ for all particle spectra.

Particle energy spectra in anomalous cosmic rays (Stone et al.~2008; Decker
et al.~2010), super-Alfv\'enic ions in the solar wind (Fisk and Gloeckler
2006), and the hardest energetic electron spectra in solar flares 
(Holman et al.~2003) exhibit all powerlaws of approximately 
$N(E) \propto E^{-1.5}$.
A model of energetic particles accelerated during multi-island magnetic
reconnection that reproduces this energy spectrum was derived by
Drake et al.~(2013). The omni-directional particle distribution $f(v,t)$
was derived by including pitch-angle scattering, which yields a velocity
dependence of $v^{-5}$ and corresponds to an energy flux of $E^{-1.5}$
(Drake et al.~2013).
Numerical simulations of electron acceleration by random DC electric 
fields constituting a SOC system were performed by Anastasiadis et al.~(1997),
yielding energy spectra with powerlaw slopes of $\alpha_E \approx 1.58-1.64$.
Similar flat powerlaw spectra were simulated by Dauphin et al.~(2007),
although it was recognized that most observed X-ray spectra are steeper
(probably due to finite system-size effects). 
Fermi acceleration in plasmoids interacting with fast shocks 
via fractal reconnection produces also similar energy spectra,
which can be derived from the first-order Fermi process,
\begin{equation}
	N(E) \propto E^{-3/2} \exp{\left( \sqrt{C} \over
	2 \pi \tau \sqrt{E} \right)} \approx E^{-3/2} \ ,
\end{equation}
where $C$ is a constant of the scaling $E(t) \propto C/L^2=C/(L_0-2ut)^2$,
$u$ is the shock velocity, and $\tau$ is the escape time scale.
The exponential cutoff in the energy spectrum (Eq.~81) represents
a deviation from an ideal powerlaw, but the superpositions of many
such spectra produced in a fractal reconnection avalanche can naturally
produce an ideal powerlaw spectrum (Nishizuka and Shibata 2013). 
Therefore, the observed powerlaw spectra can be produced by both,
either by a first-order Fermi process, or by a fractal SOC model.
	
\subsubsection{		Hydrodynamic Flare Models and SOC 	}

Measuring powerlaw slopes of different physical parameters in SOC systems
provides a direct diagnostics or test of physical scaling laws. 
Hydrodynamic simulations or scaling laws were employed in a few studies
in the context of SOC systems. 

A shell model of MHD turbulence was used to demonstrate that chaotic
dynamics with destabilization of the laminar phases and subsequent
restabilization due to nonlinear dynamics can reproduce the observed
waiting time distribution of $N(\Delta t) \propto (\Delta t)^{-2.4}$,
implying long correlation times, in contrast to classical SOC models
that predict Poisson statistics of uncorrelated random events 
(Boffetta et al.~1999). A numerical simulation of a 1D MHD model of
coronal loops was able to produce a similar waiting time distribution,
$N(\Delta t) \propto (\Delta t)^{-2.3}$, a result that was also used
to underscore the existence of sympathetic flaring (Galtier 2001).

A frequently used hydrodynamic scaling law used in the study of the
solar corona is the RTV law (Rosner, Tucker, and Vaiana 1978), which
can be derived from the energy balance between a constant heating rate
and the conductive and radiative losses of a 1D coronal loop.
This scaling law can be expressed by two equations,
\begin{equation}
	T_e \propto (p L)^{1/3} \propto n_e^{1/2} L^{1/2} \,
\end{equation}
\begin{equation}
	H \propto T_e^{7/2} L^{-2} \ ,
\end{equation}
where $T_e$ is the maximum electron temperature at the loop apex, 
$p = 2 n_e k_B T_e$ is the total (electron and ion) pressure, 
$n_e$ is the electron density, $L$ is the loop (half) length,
and $H$ is the constant heating rate. A consequence of the RTV scaling law
is the scaling of the emission measure $EM$ and thermal energy $E_{th}$
for an ensemble of loops filling a volume $V \propto L^3$,
\begin{equation}
	EM \propto n_e^2 V \propto n_e^2 L^3 \ ,	
\end{equation}
\begin{equation}
	E_{th} = 3 n_e k_B T_e V \propto T_e^3 L^2 \ .
\end{equation}
Conveniently, all these scaling laws can be expressed in terms of
powerlaw functions, which makes it analytically straightforward
to calculate the slopes of powerlaw distribution functions.
Since all these scaling laws represent relationships between three
physical parameters, two distribution functions need to be known
to predict the distribution function of the third parameter.
For instance, if we use the scale-free probability conjecture,
$N(L) \propto L^{-3}$ in 3D space (Eq.~1), and a heating rate
distribution,
\begin{equation}
	N(H) \propto H^{-\alpha_H} \ , 
\end{equation}
the distribution functions of all physical parameters 
($n_e, T_e, EM, E_{th}$) can be derived as a function of the
variable $\alpha_H$. In practice, there are additional corrections
due to truncation effects, say a flux or emission measure threshold
($EM > EM_{thres}$) due to the instrumental sensitivity limit in 
sampling of solar events. These truncation effects, however, can
be either quantified by Monte-Carlo simulations or by analytical
calculations (see Appendix A in Aschwanden and Shimizu 2013).

\begin{figure}[tpbh]
\centerline{\includegraphics[width=0.93\textwidth]{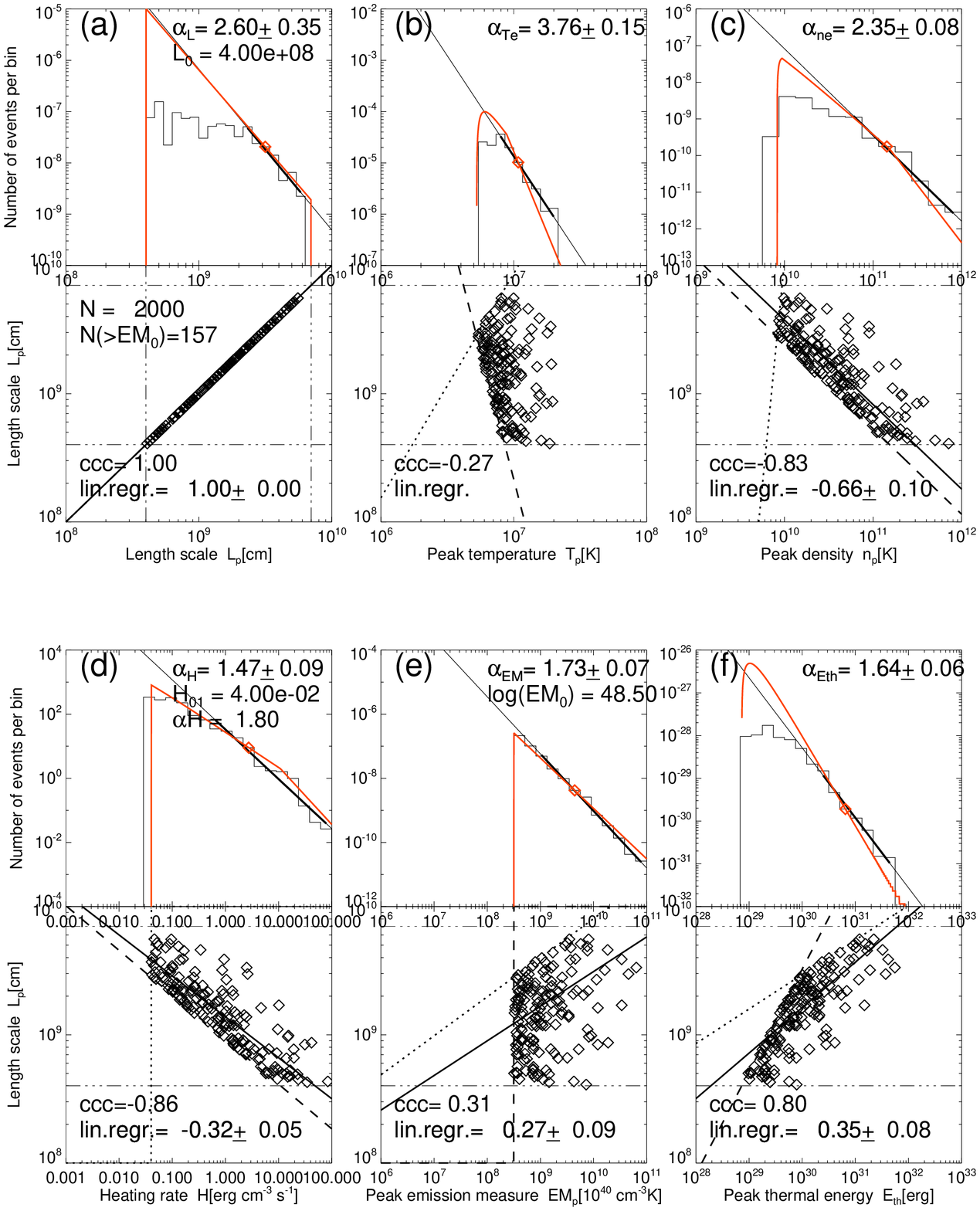}}
\captio{Monte-Carlo simulations of size distributions (red curves)
of flare parameters (diamonds) using the RTV relationships
and a heating rate distribution $N(H) \propto H^{-\alpha_H}$
with a minimum value $H_0=0.4$ erg cm$^{-3}$ s$^{-1}$ and powerlaw slope
$\alpha_H=1.8$, an emission measure threshold of
$EM_0 \ge 10^{48.5}$ cm$^{-3}$. The size distributions derived from
analytical calculations are overlaid with red curves.
(Aschwanden and Shimizu 2013).}
\end{figure}

\begin{figure}[tpbh]
\centerline{\includegraphics[width=0.9\textwidth]{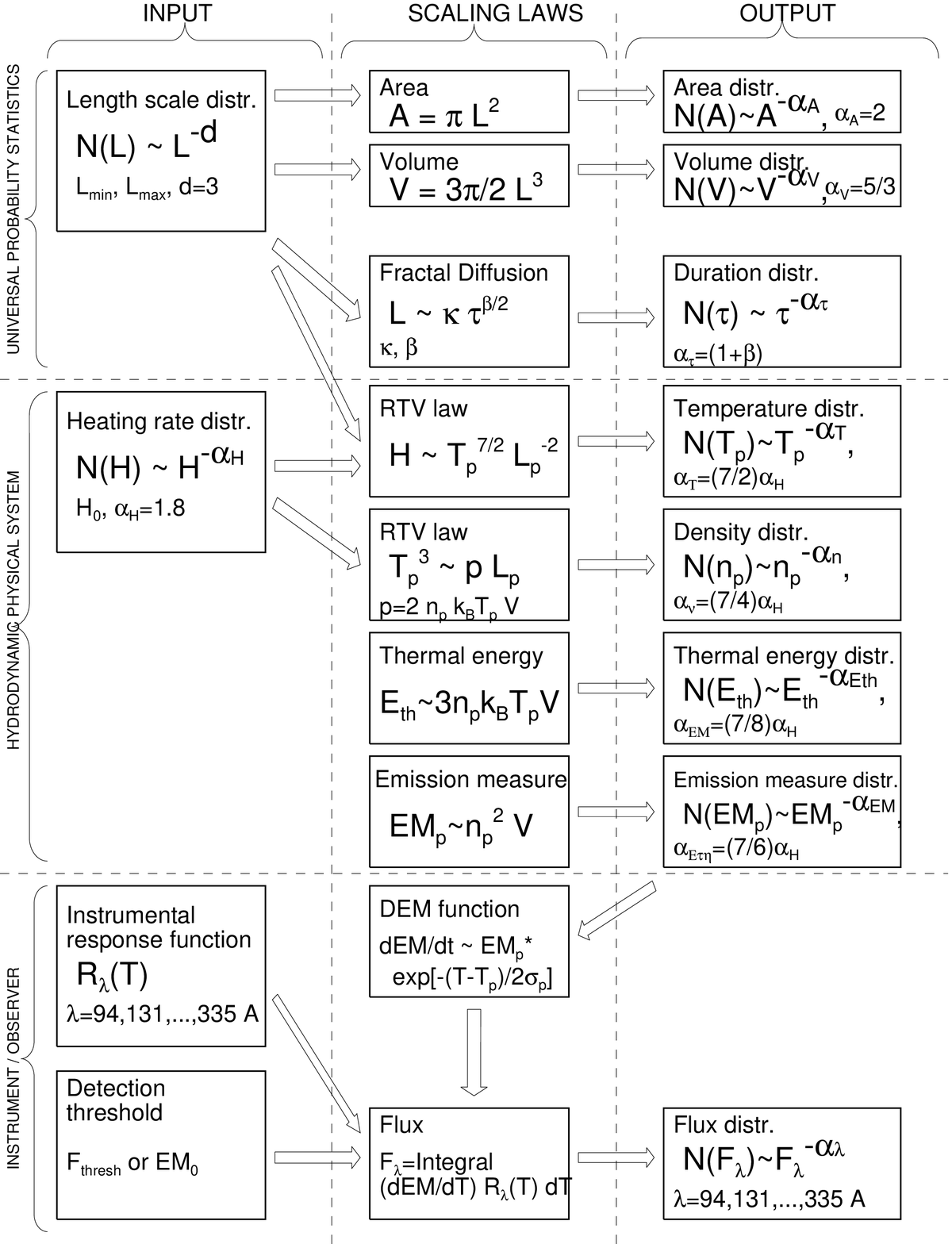}}
\captio{Flow chart of input parameters (left), scaling laws (middle),
and output distribution functions (right) of the fractal-diffusive
SOC model applied to solar flares. The spatio-temporal parameters
($L, A, V, \tau$) follow from universal probability statistics
(top part of diagram), while the physical parameters and their
scaling laws are specific to the hydrodynamics of solar flares
(middle part of diagram), and the instrumental response functions
as a function of temperature and wavelengths are specific to the
observer (bottom part of diagram). The given powerlaw indices $\alpha_x$
are approximative values for dimensionality $d=3$
(Aschwanden and Shimizu 2013).}
\end{figure}

While the original RTV law was applied to a single coronal loop,
supposedly to be in approximate energy balance between the heating
rate ($H$) and the conductive losss rate ($-E_{cond}$) and the
radiative loss rate ($-E_{rad}$), 
\begin{equation}
	H - E_{cond} - E_{rad} = 0 \ ,
\end{equation}
the same scaling law can also be applied to the peak time $t_{peak}$ 
of a flare, just at the turnover time between dominant heating and 
dominant cooling when energy balance occurs for a brief moment 
(Aschwanden and Shimizu 2013),
\begin{equation}
	\begin{array}{ll}
	H(t) - E_{cond}(t) - E_{rad}(t) \ge 0 & {\rm for}\ t < t_{peak} \\
	H(t) - E_{cond}(t) - E_{rad}(t) =   0 & {\rm for}\ t = t_{peak} \\
	H(t) - E_{cond}(t) - E_{rad}(t) \le 0 & {\rm for}\ t > t_{peak} \\
	\end{array}
\end{equation}
Since the emission measure $EM_p$, peak temperature
$T_p$, and length scale $L_p$ can be directly measured with multi-wavelength
imaging observations in solar flares, such as with AIA/SDO, the RTV law
can then be tested and the resulting size distributions of the other 
physical parameters ($n_p, E_{th}, H$) can be predicted. An example
is shown in Fig.~20, where the size distributions of the parameters
$L, T_p, n_p, H, EM_p, E_{th}$ are shown, as log-log histograms of the
observed and derived variables, and in form of scatterplots as a function
of the length scale $L$ to visualize the truncation effects caused by 
the flux threshold detection limit ($EM \ge EM_{thresh}$), and then 
compared with Monte-Carlo simulations of the size distributions 
(red curves in Fig.~20). A powerlaw index of $\alpha_H \approx 1.8$
is inferred for the unknown size distribution of heating rates $H$.

The most interesting size distribution or scaling relationship
concerns the heating rate $H$, which holds the secret of the coronal
heating process and/or energy dissipation process of flares.
From full-Sun simulations of the corona composed as an ensemble of
myriads of individual 1D loops, a scaling law of the heating flux
$F_H \propto B L^{-1}$ was found (Schrijver et al.~2004), which
corresponds to a volumetric heating rate of,
\begin{equation}
	H \propto {F_H \over L} \approx B L^{-2} \ .
\end{equation}
A different scaling is obtained from a magnetic reconnection scenario
in Petschek's theory, by assuming that the loop apex temperature is
balanced by conductive cooling, $T_e \propto (2 H L^2)^{2/7}$
(Shibata and Yokoyama 1999, 2002),
\begin{equation}
	H \approx \left( {B^2 \over 4 \pi} \right) {v_A \over L} \ .
\end{equation}
A similar scaling law was derived for magnetic reconnection
processes with the Sweet-Parker reconnection scenario
(Sweet 1958; Parker 1957; Cassak et al.~2008).
Obviously, staticstics on the size distributions $N(B)$ of the 
magnetic field are required in order to infer the heating rate distribution
$N(H)$. The flow chart in Fig.~21 summarizes how the observed
distributions are related to the model assumptions and physical
scaling laws for solar flare events or coronal heating events.

\subsubsection{		The Role of Nanoflares 			}
 
It was pointed out early on that powerlaw distributions
$N(E) \propto E^{-\alpha}$ of energies with a slope flatter than the 
critical value of
$\alpha_E=2$ imply that the energy integral diverges at the upper end,
and thus the total energy of the distribution is dominated by the
largest events (Hudson 1991), 
\begin{equation}
        E_{tot} = \int_{E_{min}}^{E_{max}} E \ N(E) dE
          = \int_{E_{min}}^{E_{max}} (\alpha -1) E^{1-\alpha_E} dE
          = \left( {\alpha - 1 \over 2-\alpha} \right)
            \left[ E_{max}^{2-\alpha} - E_{min}^{2-\alpha_e} \right] \ .
\end{equation}
Therefore, in the opposite case, when 
the powerlaw distribution is steeper than the critical value, it will
diverge at the lower end, and thus the total energy budget will
be dominated by the smallest detected events, an argument that was
used for dominant nanoflare heating in some cases with insufficient
wavelength coverage of solar nanoflare statistics (e.g.,
Krucker and Benz 2000). The powerlaw slope $\alpha_E$ for energies
depends sensitively on its definition (e.g., Benz and Krucker 2002),
in particular on the assumptions
of the flare volume scaling $V(A)$ that has to be inferred from
measured flare areas $A$ in the case of thermal energies,
$E_{th}=3 n_e k_B T_e V$. Large flares (of M and X GOES class)
were found to exhibit a powerlaw slope of $\alpha_{Eth}=1.66\pm0.13$ 
for the thermal energies $E_{th}$ (Fig.~20), which closely matches 
the powerlaw distributions of non-thermal energies determined from 
hard X-ray producing electrons, e.g., $\alpha_{nth}=1.53\pm0.02$ 
for a much larger sample including smaller flares (Crosby et al.~1993). 
Thus, based on the statistics of large flares we do not see any 
evidence that would support nanoflare heating, at least not for 
flares with energies $\gapprox 10^{29}$ erg.  In contrast, recent 
flare area measurements based on RHESSI hard 
X-ray images yield a steeper distribution of flare areas,
with $\alpha_A \approx 2.7$, and thus also a steeper distribution
of total flare energies, with $\alpha_E \approx 2.3$ 
(Li et al.~2012). At this point it is not clear how the flare area 
statistics from high-resolution imaging in EUV compares with the coarse
Fourier imaging in hard X-rays. We have also to be aware that
synthesized flare energy statistics combined from all scales 
(Figs.~11) are composed of measurements with different event
selection criteria, different detection methods, different
energy definitions, and different activity levels of the solar cycle. 
What is needed in future studies is a homogeneous flare statistics 
from the largest to the smallest flare events, using the same method
and identical time intervals (since the flaring rate varies 
orders of magnitude during the solar cycle) in order to obtain 
a self-consistent flare energy distribution on all scales.

\clearpage
\subsection{		PLANETS						}

Now we start our journey to review SOC interpretations in planetary
atmospheres and solar system bodies, starting with the Earth's 
magnetosphere (Section 3.3.1) and atmosphere (Section 3.3.2), and then 
continuing to lunar craters (Section 3.3.3), the asteroid belt (Section 3.3.4), 
Mars (Section 3.3.5), Saturn's ring system (Section 3.3.6), Jovian and
Neptunian Trojans (Section 3.3.7), Kuijper belt objects (Section 3.3.8),
and extrasolar planets (Section 3.3.9).

\subsubsection{   	The Earth's Magnetosphere		  }

In the Earth's magnetosphere, a number of phenomena have been interpreted 
as features of a SOC system, such as geomagnetic substorms, current disruptions, 
magnetotail current disruptions and associated magnetic field fluctuations, 
bursty bulk flow events, and auroras seen in UV and optical wavelengths.
Some of these are discussed briefly in the following, while a more detailed 
treatment is given in the review by Sharma \etal~(2014). Magnetospheric 
SOC phenomena have also been reviewed previously (Aschwanden 2011a: 
chapters 1.6, 5.5, 7.2, 9.4, 10.5).

Most magnetospheric phenomena result from the interaction of the Earth's
(or some other planet's) magnetic field with the ambient solar wind in the
heliosphere, at the  magnetopause, in the magneto-tail,
and in the polar regions of the planet. The solar wind brings to Earth  
disturbances associated with solar flares, coronal mass ejections, shock 
waves and solar energetic particles, causing magnetospheric storms, substorms, 
and auroral activities. The solar wind is thus the driver of the processes 
in space weather. Substorms involve many processes, including magnetic 
reconnection, ballooning-mirror modes, current disruption, etc., which cause 
a fast unloading of the highly stressed geotail system 
(Papadopoulos et al.~1993; Baker et al.~1996, Horton and Doxas 1996). 
In addition, multi-scale intermittent turbulence 
of overlapping plasma resonances play important role in substorms (Chang 1999a). 
Brief magnetospheric disturbances occur when the interplanetary magnetic
field (IMF) flips southward, which triggers magnetic reconnection at the
dayside magnetopause and transfers momentum and energy from the solar wind to the
magnetosphere. Part of the transferred energy is stored in the magnetotail,
where also magnetic reconnection and field relaxation events can occur during 
{\sl magnetospheric substorms}. A magnetospheric substorm
has three phases (Fig.~22): (1) the growth phase (when energy from the
solar wind is transferred to the dayside magnetosphere), (2) the substorm
expansion phase (when the energy stored in the magnetotail is released, 
the magnetosphere relaxes from the stretched tail, and the tail snaps into
a more dipolar configuration and energizes particles in the plasma sheet),
and (3) the recovery phase (during which the magnetosphere returns to its
quiet state). The whole process causes changes in the auroral morphology
(Fig.~22) and induces currents in the polar ionosphere, with the resultant 
heating leading to the auroral displays. The frequency of substorms is about 
6 per day on average, but larger during geomagnetic storms.

\begin{figure}[t]
 \centerline{\includegraphics[width=0.79\textwidth]{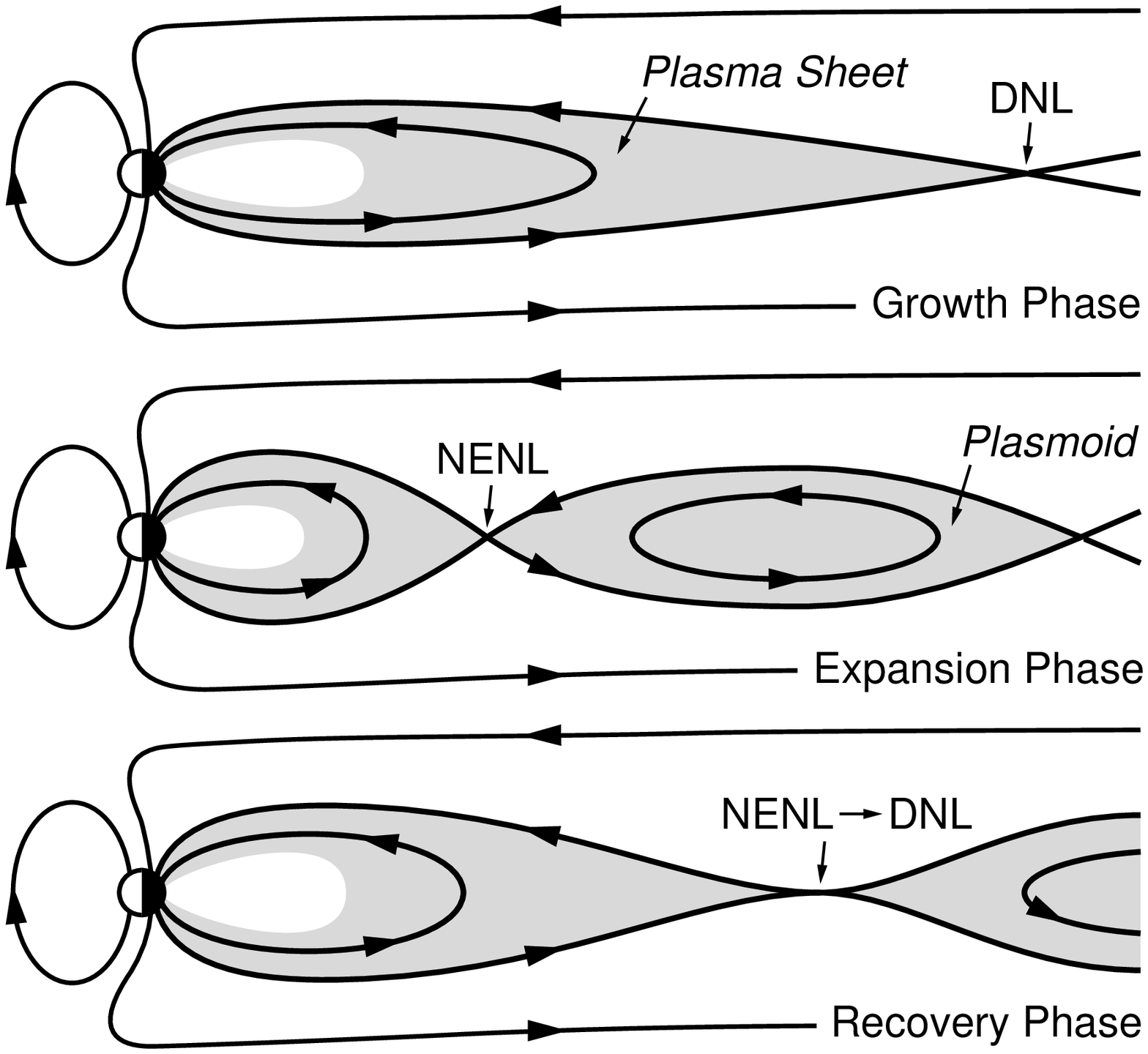}
             \includegraphics[width=0.21\textwidth]{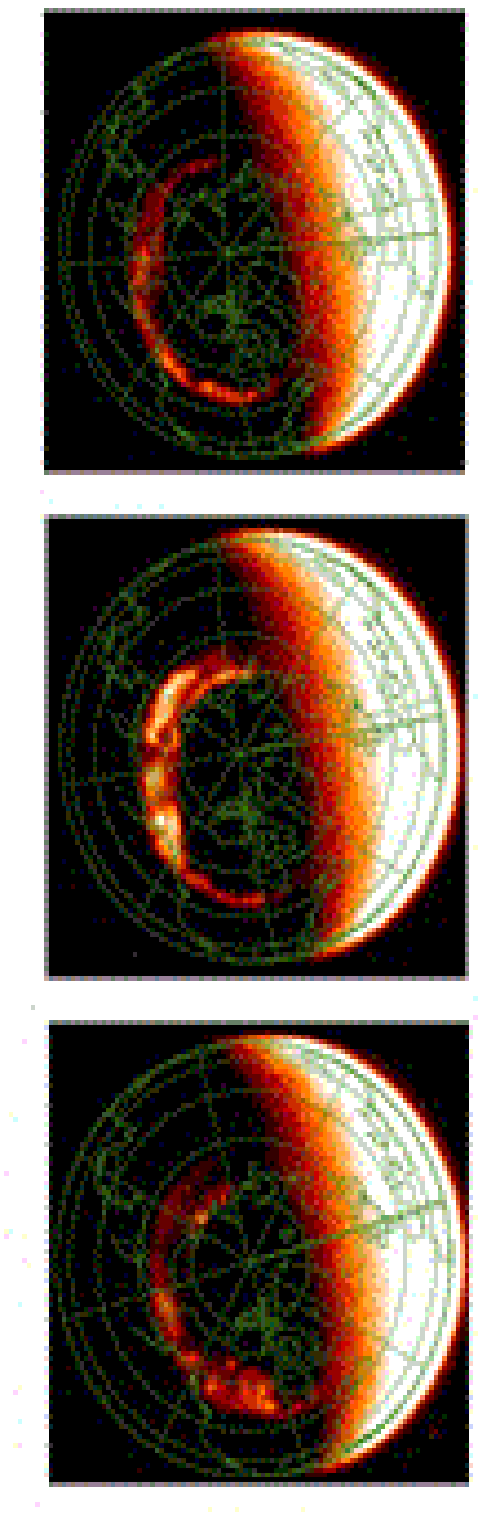}}
\captio{The three phases of a geomagnetic substorm are shown: the growth
phase (top), the expansion phase (middle), and the recovery phase (bottom)
(Baumjohann and Treuman 1996). The accompanying three auroral images
were obtained with the IMAGE WIC instrument (credit: NASA).}
\end{figure}

\begin{table}[t]
\begin{center}
\captio{Frequency distributions measured from magnetospheric phenomena.
Values determined with non-standard methods are marked with parentheses.
The data sets of Uritsky et al.~(2002) refer to different observing periods,
the data sets of Kozelov et al.~(2004) to different luminosity threshold levels,
and the data sets of Uritsky et al.~(2008) to different latitude zones (HL = 
high latitude events, LLs = low-latitude small-scale events, and LLl =
low-latitude large-scale events. The predictions (marked in boldface) are 
based on the FD-SOC model (Aschwanden 2012a).}
\medskip
\begin{tabular}{llllll}
\hline
Phenomenon	&Powerlaw       &Powerlaw        &Powerlaw   	&Powerlaw   	&References:\\
                &slope of       &slope of        &slope of   	&slope of   	&\\
                &area           &peak flux       &fluence    	&durations  	&\\
                &               &                &           	&           	&\\
                &$\alpha_A$     &$\alpha_P$      &$\alpha_E$ 	&$\alpha_T$ 	&\\
\hline
\hline
geotail flow bursts &           &                &           	&1.59$\pm$0.07  &Angelopoulos et al.~(1999)\\
AE index        &               &                &           	&1.24           &Takalo (1993, 1999a)\\
AU index        &               &                &           	&1.3            &Freeman et al.~(2000b)\\
aurora UV (substorms)&(1.21$\pm0.08)$ &(1.05$\pm$0.08)&      	&           	&Lui et al.~(2000)\\
aurora UV (quiet)&(1.16$\pm0.03)$ &(1.00$\pm$0.02) &         	&               &Lui et al.~(2000)\\
aurora UV Jan 1997 &1.73$\pm$0.03  &1.66$\pm$0.03&1.46$\pm$0.04	&2.08$\pm$0.12  &Uritsky et al.~(2002)\\
aurora UV Feb 1997 &1.74$\pm$0.03  &1.68$\pm$0.03&1.39$\pm$0.02	&2.21$\pm$0.11  &Uritsky et al.~(2002)\\
aurora UV Jan 1998 &1.81$\pm$0.04  &1.73$\pm$0.02&1.62$\pm$0.03	&2.24$\pm$0.11  &Uritsky et al.~(2002)\\
aurora UV Feb 1998 &1.92$\pm$0.04  &1.82$\pm$0.03&1.61$\pm$0.04	&2.39$\pm$0.11  &Uritsky et al.~(2002)\\
aurora UV       &1.85$\pm$0.03  &1.71$\pm$0.02   &1.50$\pm$0.02	&2.25$\pm$0.06  &Kozelov et al.~(2004)\\
aurora TV 2.0 kR&1.98$\pm$0.04  &2.02$\pm$0.02   &1.74$\pm$0.03	&2.53$\pm$0.07  &Kozelov et al.~(2004)\\
aurora TV 2.5 kR&1.85$\pm$0.04  &1.92$\pm$0.02   &1.66$\pm$0.04	&2.38$\pm$0.05  &Kozelov et al.~(2004)\\
aurora TV 2.R kR&1.86$\pm$0.05  &1.84$\pm$0.03   &1.60$\pm$0.02	&2.33$\pm$0.06  &Kozelov et al.~(2004)\\
aurora UV HL    &1.87$\pm$0.05  &1.81$\pm$0.02   &1.57$\pm$0.02 &2.30$\pm$0.11  &Uritsky et al.~(2008)\\ 
aurora UV LLs   &2.11$\pm$0.16  &2.16$\pm$0.09   &1.83$\pm$0.04 &3.21$\pm$0.33  &Uritsky et al.~(2008)\\
aurora UV LLl   &1.09$\pm$0.14  &1.32$\pm$0.14   &1.04$\pm$0.12 &1.26$\pm$0.44  &Uritsky et al.~(2008)\\
Outer radiation belt&           &1.5$-$2.1       &1.5$-$2.7   	&               &Crosby et al.~(2005)\\
Ionospheric disturbances &      &                &           	&1.8-2.5        &Bristow (2008)\\
\hline
FD-SOC prediction: &{\bf 2.00}  &{\bf 1.67}	 &{\bf 1.50}    &{\bf 2.00}     &Aschwanden (2012a)\\
\hline
\end{tabular}
\end{center}
\end{table}

How did the SOC concept came into play for magnetospheric processes? 
The bursty nature of magnetospheric phenomena, such as localized current 
disruptions in auroral blobs (Lui et al.~1988), bursty bulk flow events 
in the geotail (Angelopoulos et al.~1996, 1999), and the powerlaw magnetic 
field spectra in the magnetotail (Hoshino et al.~1994), have been interpreted 
in terms of an open, dissipative nonlinear system near a forced or 
self-organized critical state (Chang 1992, 1998a,b; 1999a,b;  
Klimas et al.~2000; Chang et al.~2003; Chapman and Watkins 2001;  
Consolini and Chang 2001; Consolini 2002). Probability or size distributions 
with a powerlaw shape (Table 11), the hallmark of SOC systems, have been 
measured from auroral blobs in UV (Lui et al.~2000; Uritsky et al.~2002, 2003, 2006) 
and optical light (Kozelov et al.~2004), from the {\sl auroral electron jet index (AE)} 
(Takalo 1993; Consolini 1997, 2002), from magnetospheric 
substorm-related tail current disruptions (Consolini and Lui 1999),
from geotail flow bursts (Angleopoulos et al.~1999), from ionospheric velocity 
fluctuations driven by the interplanetary magnetic field (Bristow 2008),
and from electron bursts in the outer radiation belt (Crosby et al.~2005). 
A powerlaw slope of $\alpha_{\Delta t}\approx 1.3$ was also determined
for waiting times in an AE index time series (Lepreti et al.~2004).
Critical finite-size scaling and a fractal dimension of $D_2=1.54\pm0.02$
was found for auroral blobs (Uritsky et al.~2006), which agrees with
the mean-value estimate $D_2=(1+d)/2=1.5$ (Eq.~8) in the FD-SOC model.
The powerlaw behavior, as observed in many magnetospheric phenomena (Table 11),
provides the main basis for interpretation as SOC processes. 
The measurements listed in Table 11
were obtained from UVI onboard the POLAR spacecraft (Lui et al.~2000; Uritsky
et al.~2002), from all-sky TV cameras at the Barentsburg Observatory 
(Kozelov et al.~2004), the GEOTAIL spacecraft (Angelopoulos et al.~1999), 
from the WIND spacecraft (Freeman et al.~2000b), with the SuperDARN radar network 
(Bristow 2008), and the STRV microsatellites (Crosby et al.~2005).

\begin{figure}[tpbh]
\centerline{\includegraphics[width=0.9\textwidth]{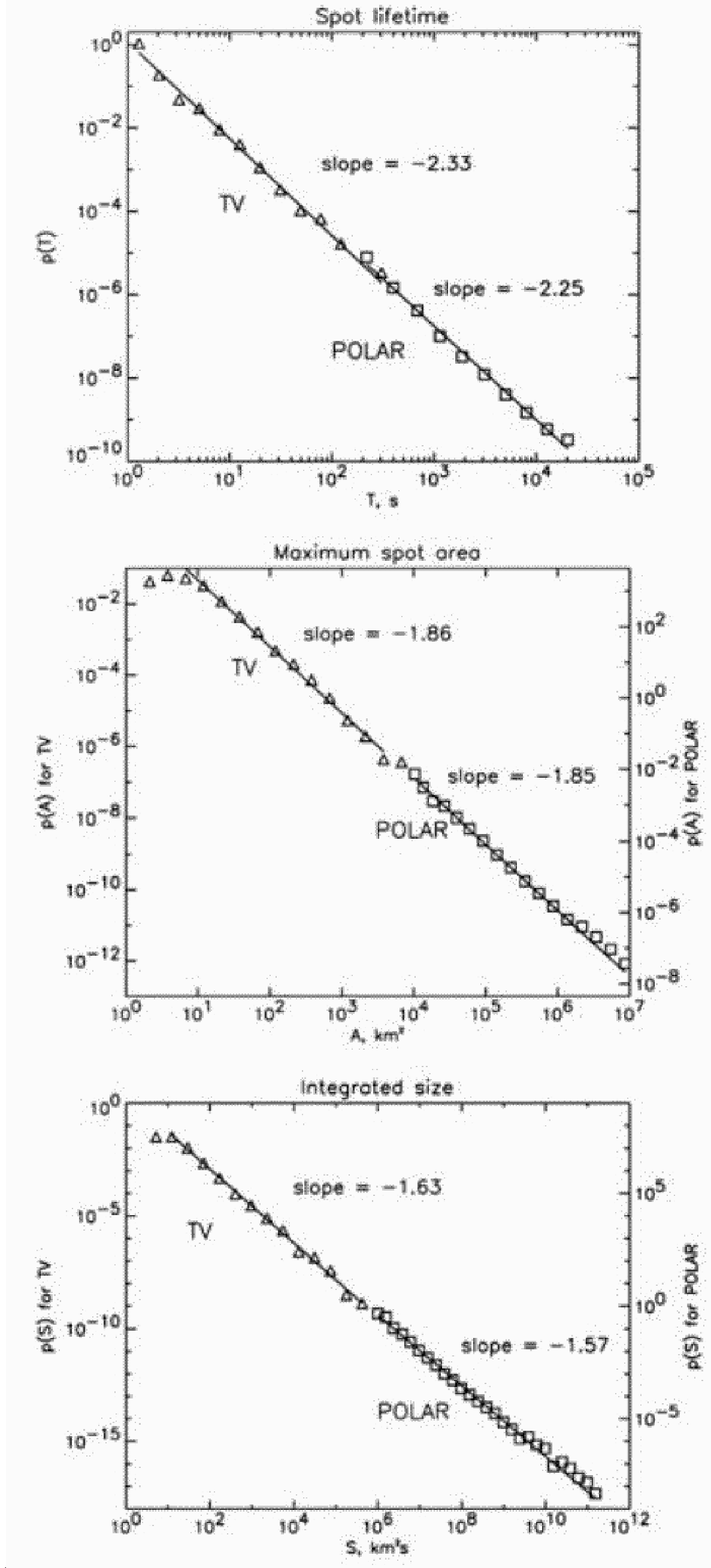}}
\captio{Combined probability distributions of auroral blob parameters
(top panel: event duration $T$; middle panel: maximum area $A$;
bottom panel: time-integrated size $S$, approximately
proportional to the or total energy $E$) 
measured with ground-based TV cameras (Kozelov et al.~2004)
and with the UVI/POLAR spacecraft (Uritsky et al.~2002).}
\end{figure}

The agreement between the statistics of large auroral events observed with 
POLAR/UVI (Uritsky et al.~2002) and small auroral events observed 
with a TV camera (Kozelov et al.~2004) is excellent (Fig.~23) and covers
a combined (but not overlapping) range of 10 orders of magnitude in energy. 
However, we can see in Table 11 a glaring discrepancy between
the measurements made by Lui et al.~(2000) and by Uritsky et al.~(2002),
while the latter agree surprisingly well with 
the predictions of the standard fractal-diffusive SOC model (Aschwanden 2012a). 
The measurements by Lui et al.~(2000) yield much flatter powerlaw 
slopes, close to unity. This discrepancy has been convincingly explained in terms 
of a different methodology and event definition: The statistics carried out by
Lui et al.~(2000) refers to equidistant time snapshots that count large avalanche
events multiple times, while the analysis of Uritsky et al.~(2002) 
determines the time-integrated avalanche areas and energies, consistent with
standard definitions of SOC parameters (also used in the FD-SOC model: see
section 2.10 and Eqs.~14 and 20). Another anomaly that was found
is the latitude dependence of the size distribution of auroral events 
(see Uritsky et al.~2008 in Table 11), which indicates substantially 
different scaling regimes
of bursty energy dissipation in the inner and outer portion of the
geotail plasma sheet (Uritsky et al.~2008, 2009).

The SOC interpretation of magnetospheric phenomena has also stimulated
cellular automaton simulations and alternative aspects of SOC modeling,
such as finite system-size effects (Chapman et al.~1998, 1999; 
Chapman et al.~2001), powerlaw robustness under varying loading 
(Watkins et al.~1999), the discretization in terms of MHD equations 
(Takalo et al.~1999a,b), renormalization group analysis (Tam et al.~2000;
Chang et al.~2004),
the scaling of the critical spreading exponents (Uritsky et al.~2001),
phase transition-like behavior (Sitnov et al.~2000; Sharma et al.~2001), 
aspects of percolation and branching theory (Milovanov et al.~2001;
Zelenyi and Milovanov 2004),
chaotic turbulence models (Kovacs et al.~2001), forced SOC models
(Consolini 2001; Chang et al.~2003), modeling of energetic particle 
spectra in magnetotail (Milovanov and Zelenyi 2002), MHD modeling
of the plasma sheet dynamics near a SOC state (Klimas et al.~2004),
aspects of complexity systems (Dendy et al.~2007),
the framework of thermodynamics of rare events (Consolini and Kretzschmar 2007),
kinetic theory of linear fractional stable motion (Watkins et al.~2009b),
avalanching with an intermediate driving rate (Chapman and Watkins 2009;
Chapman et al.~2009), and multi-fractal and fractional L\'evy flight models
(Zaslavskii et al.~2007, 2008; Rypdal and Rypdal 2010b). 

The Earth's magnetosphere is a large-scale natural system driven by the 
turbulent solar wind and exhibits non-equilibrium phenomena (Sharma and 
Kaw 2005), including SOC discussed here. In general the properties of such 
systems are characterized as a combination of global and multiscale features, 
and have been studied extensively using the techniques of nonlinear dynamics 
and complexity science (Sharma 1995; Klimas et al.~1996; Vassiliadis 2006). 
The first evidence of global coherence of the magnetosphere was obtained 
from time series data of AE index in the form of low-dimensional dynamics 
(Vassiliadis et al.~1990; Sharma et al.~1993).  This result is consistent 
with the morphology of the magnetosphere derived from observations and 
theoretical understanding (Siscoe 1991), and simulations using global MHD models
(Lyon 2000; Shao et al.~2003). The recognition of the low dimensional 
dynamics of the magnetosphere has stimulated a new direction in the studies 
of the solar wind-magnetosphere coupling and such systems in nature.
Among these is the forecasting of the global conditions of space weather, 
viz. the AL and AE indices for substorms (Vassiliadis et al.~1995; 
Ukhorskiy et al.~2002, 2003, 2004; Chen and Sharma, 2006) and the 
disturbance time index Dst for magnetic storms (Valdivia et al.~1996; 
Boynton et al.~2011). The forecasting of regional space weather
requires data from the spatially distributed stations around the globe  
(Valdivia et al.~1999a,b; Chen et al.~ 2008) and the predictability is 
largely determined by the availability of long time series data from 
the network of observing stations. The spatio-temporal dynamics of many 
systems are studied using such data, including the images obtained from 
satellite-borne imagers, by defining new variables computed from the data. 
For example, the fragmentation parameter (Rosa et al.~1998, 1999) represent 
the complexity of the spatial structure and has been used to model
the dynamics of the solar atmosphere using the hard X-ray images from 
SOHO spacecraft. Further, the low-dimensionality of the magnetosphere 
has stimulated the development of models with a small number of equations 
(Vassiliadis et al.\~1993, Horton and Doxas 1996).

The multiscale nature of the magnetosphere, expressed in many ways 
including the power law dependence of the scales, is a reflection of 
turbulence and plays an essential role in the accuracy of the forecasts. 
An early recognition of this was in the analogy of the dynamics of the 
magnetosphere to turbulence generated by a fluid flow past an obstacle
(Rostoker 1984). The power law dependence of the AE index and of the 
solar wind provided quantitative measures of the power law indices and 
also the differences (Tsurutani et al.~1990). The scaling laws, which 
have been studies in detail using techniques such as the structure 
functions (Takalo et al.~1993), have many implications. The first is the
characterization in terms of SOC, as  discussed earlier in this section. 
The second is that the predictability of a multiscale system could not 
quantified readily in terms of the characteristic quantities such as 
the Lyapunov exponents (Vassiliadis et al.~1991) in a low-dimensional 
dynamical system. The presence of many scales as well as the non-equilibrium
nature imply that predictions should be based on  the statistical properties 
of the dynamical trajectories, e. g., using a mean-field approach 
(Ukhorskiy et a.~2004).  Further, this approach is suitable for analyzing 
the predictability of extreme events (Sharma et al.~2012).

In summary, let us ask: What is the merit of the SOC concept in the
context of magnetospheric phenomena? The standard fractal-diffusive 
SOC model (Sections 2.6-2.11) predicts the probability distribution 
functions for each parameter as a function of the dimensionality ($d$),
diffusive spreading exponent ($\beta$), fractal dimension ($D_d$),
and type of (coherent/incoherent) radiation process ($\gamma$). The
waiting time distributions are predicted by the FD-SOC model to follow 
a powerlaw with a slope of $\alpha_{\Delta t} \approx 2$ during 
active and contiguously flaring episodes, while an exponential cutoff
is predicted for the time intervals of quiescent periods. This dual
regimes of the waiting time distribution predict both persistence and
memory during the active periods, and stochasticity during the quiescent
periods. All these predictions of the FD-SOC model provide useful
constraints of the physical parameters and underlying scaling laws.
Significant deviations from the size distributions predicted by the FD-SOC model
could imply problems with the measurements or data analysis, such as 
indicated by the contradicting results of Lui et al.~(2000) and Uritsky et 
al.~(2002) in the case of auroral size distributions.

Let us emphasize again that the generic FD-SOC model is considered
to have universal
validity and explains the statistics and scaling between SOC parameters,
but does not depend on the detailed physical mechanism that governs
the instabilities and energy dissipation in a particular SOC process. 
The physical process may be well described by a number of established 
models, such as turbulence theory, kinetic theory, wave-particle 
interactions, and other branches of plasma physics.
There was also a debate whether magnetospheric substorms are SOC or
forced-SOC (FSOC) (e.g., Chang et al.~2003), an issue that largely 
disappears in our generalized 
FD-SOC concept, where a slow driver is required to bring the SOC system 
continuously near to the instability limit, but is does not matter
whether the driver is internally, externally, or is globally organized.
 
\subsubsection{		Terrestrial Gamma-Ray Flashes 			}

{\sl Terrestrial gamma-ray flashes (TGF)} are gamma-ray bursts of
terrestrial origin that have been discovered with the {\sl Burst and
Transient Experiment (BATSE)} onboard the {\sl Compton Gamma Ray 
Observatory (CGRO)} and have been studied with RHESSI,
Fermi, and AGILE since. These TGF bursts are produced by high-energy 
photons of energy $>100$ keV and last up to a few milliseconds.
They have been associated with strong thunderstorms mostly
concentrated in the Earth's equatorial and tropical regions, 
at a typical height of 15-20 km (Fishman et al.~1994;
Dwyer and Smith 2005; Smith et al.~2005). The physical 
interpretation is that the TGF bursts are produced by
bremsstrahlung of high-energetic electrons that were accelerated
in large electric potential drops within thunderstorms. 
However the gamma-rays produced in thunderstorms (at ~5 km) can not 
readily propagate to higher altitudes due to atmospheric
absorption. A mechanism for the generation of gamma rays that can reach 
the satellite-borne instruments is through the excitation of whistler 
waves by the relativistic elctrons generated in the thunderstorms 
(Kaw et al.~2001; Milikh et al.~2005). The whistler waves form a 
channel by nonlinear self-focusing and the relativistic electrons 
propagate in this channel to higher altitudes (~30 km). The 
gamma-ray generated at this altitude can escape the atmosphere
and thus account for the BATSE/CGRO results.  

A size distribution of the gamma-ray emission from TGF events needs
to be corrected for the distance from the TGF-producing thunderstorm 
to the detecting spacecraft (in Earth orbit). In a combined analysis
of TGF data from the RHESSI and Fermi satellites, corrected for 
their different orbits, different detection rates, and relative 
sensitivies, a true fluence distribution was derived,
which was found to have a powerlaw shape of
$\alpha_E=2.3\pm0.2$ if a sharp cutoff was assumed, or a
slope of $\alpha_E \le 1.3-1.7$ when a more realistic roll-over
of the RHESSI lower detection threshold is assumed 
(Ostgaard et al.~2012). 

\begin{table}[t]
\begin{center}
\captio{Frequency distributions measured from planetary phenomena.}
\medskip
\begin{tabular}{lllll}
\hline
Phenomenon      		&Instrument	&Powerlaw	&Powerlaw       &References:\\
                		&		&slope of	&slope of       &\\
                		&		&length		&fluence        &\\
                		&		&$\alpha_L$     &$\alpha_E$     &\\
\hline
\hline
Terrestrial $\gamma$-ray flashes&               &               &1.3$-$1.7      &Ostgaard \etal (2012)\\
Lunar craters                   &Ranger 7,8,9   &3.0		&		&Cross (1966)\\
Meteorites, space debris	&		&2.75		&		&Sornette (2004)\\
Asteroid belt	   	   & Spacewatch Surveys &2.8		&		&Jedicke \& Metcalfe (1998)\\
Asteroid belt ($<5$ km) 	& Sloan Survey  &2.3		&		&Ivezic \etal (2001)\\
Asteroid belt ($>5$ km)		& Sloan Survey  &4.0		&		&Ivezic \etal (2001)\\
Asteroid belt			& Subaru Survey &2.3		&		&Yoshida \etal (2003)\\
				& 		&		&		&Yoshida \& Nakamura (2007)\\
Jovian Troyans ($<40$ km)       & Hawaii 2.2 m  &3.0$\pm0.3$    &               &Jewitt \& Trujillo (2000)\\
Jovian Troyans ($>40$ km)       & Hawaii 2.2 m  &5.5$\pm0.9$    &               &Jewitt \& Trujillo (2000)\\
Neptune Trojans			& Subaru Survey &5$\pm$1        &               &Sheppard \& Trujillo (2010)\\
Kuiper belt objects             &               &4.3            &               &Fraser et al.~(2008)\\
Saturn ring			& Voyager 1     &2.74$-$3.11	&		&Zebker \etal (1985)\\
Saturn ring			& Voyager 1     &2.74$-$3.11	&		&French \& Nicholson ((2000)\\
Extrasolar planets               & Kepler        &2.48           &               &Catanzarite and Shao (2011)\\
\hline
FD-SOC prediction:              &		&{\bf 3.00}	&{\bf 1.50}     &Aschwanden (2012a)\\
\hline
\end{tabular}
\end{center}
\end{table}

We can consider a part of the Earth's atmosphere that contains
a thunderstorm as a SOC system of finite size, where the
electrostatic charging process represents the driver, 
the critical condition for electric discharging is given
by an electric conductivity threshold, and the spontaneously
triggered gamma-ray flashes or lightenings represent the avalanches. 
The PDF is then given
by the scale-free probability conjecture (Eq.~1), which
together with the fractal-diffusive transport predicts an
energy or fluence distribution with a powerlaw slope of 
$\alpha_E=1.5$ in 3D space, which matches the observed
and corrected fluence distribution with a slope of 
$\alpha_E \approx 1.3-1.7$.  The agreement with the standard
FD-SOC model is consistent with an incoherent process for gamma-ray
production, where the gamma-ray flux is proportional to the emitting
volume of a TGF.

\subsubsection{		Lunar Craters and Meteorites		}

An amazingly straight powerlaw size distribution has been found for
the sizes of lunar craters (Fig.~25), with a cumulative powerlaw slope of 
$\alpha_L^{cum}=2.0$ over a size range of $L=0.65 - 69,000$ m,
which covers 5 orders of magnitude (Cross 1966), derived from 
crater statistics measured in pictures of the lunar probes 
{\sl Ranger 7, 8, 9} combined with a lunar map of Wilkins (1946).
Since a cumulative size distribution is flatter than
a differential size distribution (by a value of one), this corresponds
to a powerlaw slope of $\alpha_L = \alpha_L^{cum} + 1 = 3.0$.
A similar powerlaw index of $\alpha_L = 2.75$ was found for the
size distribution of meteorites and space debris from man-made
rockets and satellites in the range of L=10 $\mu$m$-$10 cm
(Fig.~3.11 in Sornette 2004).   

Given the ubiquitous powerlaw shape of these size distributions, 
it is not far-fetched to consider the possibility of an interpretation 
in terms of a SOC model. Since lunar craters are believed to be
produced by meteorite impacts, the directly observed meteorites
and the lunar impact craters have the same origin in the solar system,
although they cover different length scale ranges. We find that
these observed powerlaw slopes of $\alpha_L \approx 2.75-3.0$ agree
remarkably well with the scale-free probability conjecture, which
predicts in 3D space a universal scaling exponent of $\alpha_L=3$ 
(Eq.~1). The reservoir of meteorites is the slow driver and
small bodies that orbit in the solar system and provides projectiles
for lunar or planetary impacts. The dissipated energy is essengially
the kinetic energy of the projectiles, given by the relative velocity
of the projectile ($v_{proj}$) and the target ($v_{target}$),
\begin{equation}
  E_{kin} = {1 \over 2} m_{proj} (v_{proj} - v_{target} )^2 
          \ge  {1 \over 2} m_{proj} v_{inel}^2 \ ,
\end{equation} 
which has to exceed the critical threshold $v_{inel}$ that is given
by the limit between elastic and inelastic collisions. If the 
projectile hits the target below this threshold, it will just bounce
back by conservation of momentum, without producing an impact crater. 
If it hits the target with a
larger velocity, the impact will produce a fractal-diffusive
pattern of cracks on the projectile and target, similar to the
rupture area during the energy release of an earthquake. In this
analogy, lunar impact craters have much in common with earthquake
``damage areas'', which is considered as a SOC process.

The size distribution of meteorites and planetesimals may also
be generated by a SOC process in the first place. The slow driver
that provides the trickling of sand grains is the gravity-driven
formation process of the solar system itself, which clumps the local
molecular cloud into meteorites and planets.The aspect of
self-organized criticality, which is a balance between the gravity 
and the frictional force that controls the critical angle of 
repose in Bak's sandpile, can be understood as a critical point 
between the condensation rate of planetesimals or meteorites 
(by self-gravity) and the diffusion rate (driven by thermal
pressure and external gravitational disturbances). This critical 
threshold given by the balance of the condensation rate and 
the diffusion rate has to be exceeded in order to initiate the  
gravitational collapse that forms a solar system body. The
gravitational collapse is the underlying instability 
in a physical SOC concept (Fig.~1, right frame).

Hence, from such a generalized point of view, we might consider 
the meteorite formation as a SOC pr
ocess and the resulting lunar
cratering as the imprint of this process.
The main benefit of the FD-SOC framework is the direct
prediction of the scale-free size distribution of crater sizes,
i.e., $N(L) \propto L^{-3}$ (Eq.~1), which can also be used
as a prediction for any other targets in the solar system, such
as cratering on Earth, Mars, or Mercury. This allows us, for
instance, to predict the collisional probability of an asteroid
hitting our Earth, although we have to take into account the 
variability of the impact rate, which varied drastically during the
lifetime of our solar system. Both the Moon and the Earth
were subject of intese bormardment between 4.0 and 3.7 billion
years ago, which was the final stage of the sweep-up of debris 
left over from the formation of the solar system (Bottke et al.~2012). 
The impact rate at that time was thousands of times higher 
than it is today.

\subsubsection{		The Asteroid Belt 		}

The asteroid belt is a large accumulation of irregular small 
solar system bodies orbiting the Sun between the orbits of
Mars and Jupiter. The largest of these small bodies is
Ceres, with a diameter of 1,020 km, followed by Pallas
(538 km), Vesta (549 km), Juno (248 km), and extends down to the size
of dust particles. While most planetesimals from the primoridal
solar nebula formed larger planets under the influence of
self-gravity, the gravitational perturbations from the giant
planets Jupiter and Saturn prevented a stable conglomeration
of planetesimals in the zone between Mars and Jupiter.
This fragmented soup of primordial planetesimals makes up the
asteroid belt. The larger asteorids ($\ge 120$ km) are believed
to be primordial, while the smaller ones are likely to be a
byproduct of fragmentation events (Bottke et al.~2005).

Statistics of the sizes of asteroids has been carried out in the
{\sl Palomar Leiden Survey} (Van Houten et al.~1970), in the
{\sl Spacewatch Surveys} (Jedicke and Metcalfe 1998), in a
{\sl Sloan Sky Survey} (Ivezic et al.~2001), 
and in the {\sl Subaru Main-Belt Asteroid Survey} (Yoshida
et al.~2003; Yoshida and Nakamura 2007). Most of these
statistics yield a powerlaw-like function for the cumulative 
size distribution (Fig.~25). From these values $\alpha_L^{cum}$
we can estimate the powerlaw slopes of the differential size
distributions $\alpha_L=\alpha_L^{cum}+1$, which yield
$\alpha_L=2.8$ (Jedicke and Metcalfe 1998; Jedicke et al.~2002), 
a double powerlaw of $\alpha_L=2.3-4.0$ (Ivezic et al.~2001),
and $\alpha_L=2.3$ (Yoshida et al.~2003; Yoshida and
Nakamura 2007), see compilation in Table 12. 
Observational selection effects in asteroid surveys, of course,
affect the reported powerlaw slopes, as discussed in
Jedicke et al.~(2002).

If the small bodies in the asteroid belt are formed by a
SOC process, the scale-free probability conjecture predicts a
size distribution of $N(L) \propto L^{-3}$, which is indeed
close to what is observed (Fig.~25). However, there are slight deviations
from a single powerlaw distribution for small and large
bodies, which indicate some additional effects. Nevertheless,
an almost scale-free behavior is observed for a range of
$L\approx 0.4-50$ km, which makes it appropriate to consider 
the formation process in terms of a SOC system. As we discussed 
for the formation of meteorites above (Section 3.3.3), the aspect 
of self-organized criticality can be understood as a critical point 
between the condensation rate of planetesimals or meteorites 
by self-gravity, and the diffusion rate driven by 
external gravitational disturbances, mostly from the giant
planets Jupiter and Saturn. If this critical threshold of the
ratio of the condensation rate to the diffusion rate exceeds the
value of unity, the self-gravity force takes over and forms 
a small solar system body, which represents an avalanche
process with a well-defined instability threshold. 

\subsubsection{		Mars				}

It has also been suggested to apply SOC dynamics to Martian
fluvial systems (Rosenshein 2003). The motivation was 
that complexity theory provides powerful methods to
analyze, interpret, and model terrestrial fluvial systems,
inlcuding the fractal structure of meandering, sediment
dynamics, bedrock incision, and braiding.

Another application of SOC systems to Mars is the statistics
of dust storms, especially the interannual variability of Mars 
global dust storms (Pankine and Ingersoll 2004a,b). Previously
it was thought that the threshold for wind speed for starting
saltation and lifting dust from the Martian surface was a finely
tuned process. In the study of Pankine and Ingersoll (2004a,b),
however, 
it was shown that the fine-tuning of this parameter could be the
result of a negative feeback mechanism that lowers the threshold
of the wind speed.
In this way, the Martian atmosphere/dust system could organize
itself as a SOC system, and no fine-tuning of a critical
threshold is required. 

\begin{figure}[t]
\centerline{\includegraphics[width=1.0\textwidth]{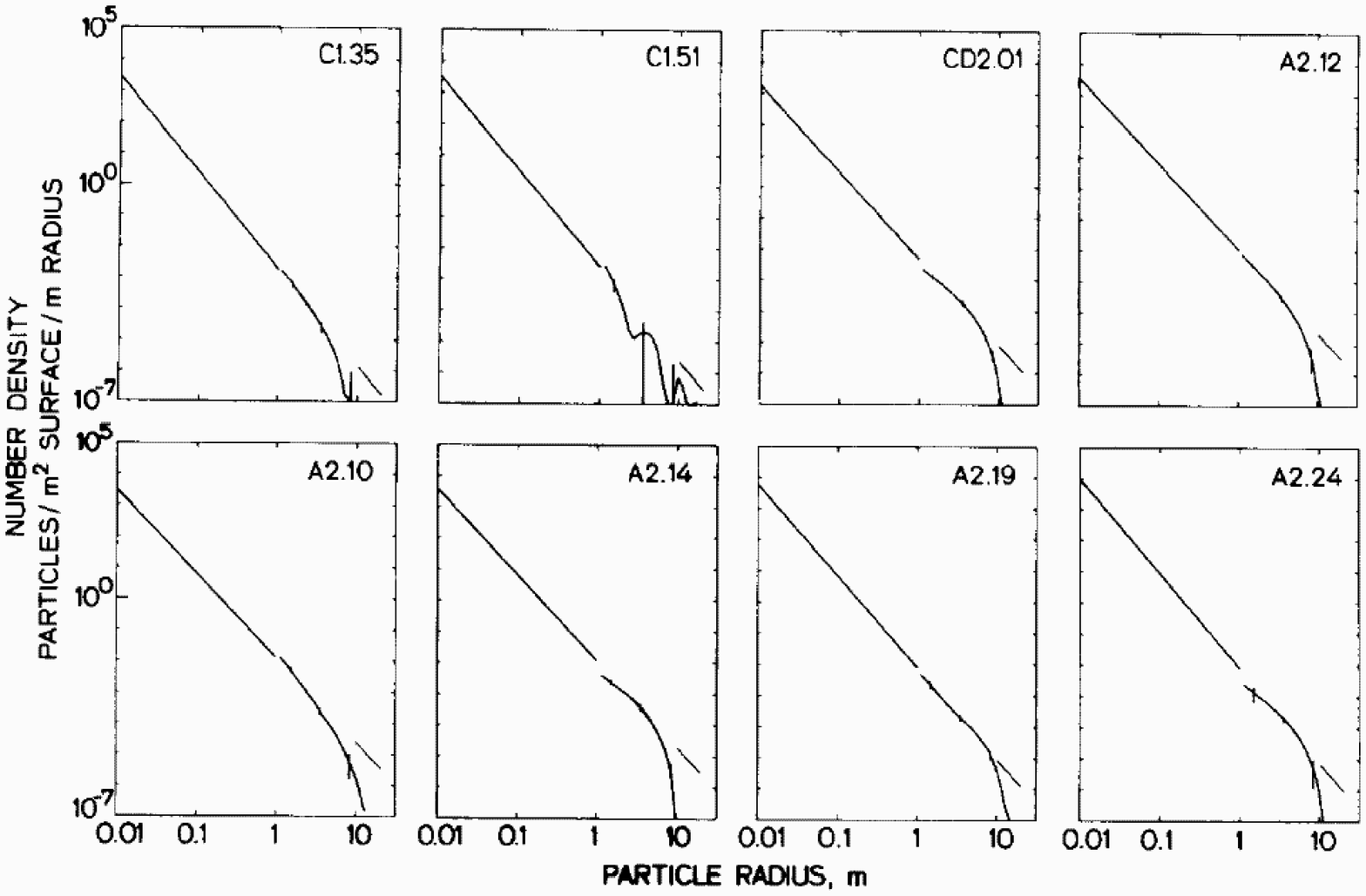}}
\captio{Measurements of the particle size distribution
functions for 8 ring regions with {\sl Voyager I} radio occultation
measurements (Ring C: C1.35, C1.51; Cassini division:
CD2.01; Ring A: A2.12, A2.10, A2.14, A2.19, A2.24). The slopes of the
fitted powerlaw functions in these 8 regions are: $\alpha_L=$
3.11, 3.05, 2.79, 2.74, 2.70, 2.75, 2.93, 3.03. The range of
particle sizes is $L=0.01-10$ m (Zebker et al.~1985).}
\end{figure}

\subsubsection{		Saturn's Ring System 		}

Saturn and Jupiter are the most massive planets in our solar system
with a gravity that is sufficiently strong to keep 
numerous moons, rings, and ringlets in their strong gravitational
field. The Saturn ring extends from 7,000 km to 80,000 km above
Saturn's equator, consisting of particles ranging from 1 cm to
10 m, with a total mass of $3 \times 10^{19}$ kg, which is comparable
with the mass of its moon Mimas. Theories about the origin of Saturn's
ring range from nebular material left over from the formation of Saturn
itself to the tidal disruption of a former moon.

When {\sl Voyager 1} passed the orbit of Saturn, it carried out
radio occultation observations, which were analyzed with a scattering
model and yielded the size distribution of ring particles in the range
of $L=0.01-10$ m, being a powerlaw distribution of 
$N(L) \propto L^{-3}$ (Fig.~25). The results of 8 size distributions obtained
from 8 different locations in Ring A, C, and the Cassini division
are shown in Fig.~24, which all are found to exhibit powerlaw indices in the range
of $\alpha_L=2.74-3.11$ (Zebker et al.~1985). Related to this
is a wavelet transform analysis of the Encke gap ringlets in
Saturn's ring system (Bendjoya et al.~1993).

The coincidence of predicted and observed powerlaw distributions 
for meteorites, lunar craters, asteroids, and Saturn ring particles
may have all the same explanation, namely SOC systems, although
operating in different locations in our solar system, and in
different ranges of length scales (Fig.~25). The critical threshold 
in all these systems is apparently given by the balance between
the local self-gravity force and external gravity disturbances.
All the gaps between the Saturn rings have been explained by mechanical
resonances of Saturn's moons, which orbit outside the ring and
amplify gravitational disturbances whenever two moons have an
integer ratio of their orbital periods. Thus, we have all parts of
a physical SOC system: the driver, the instability, and the avalanches.
Saturn's moons are the driver of the system, because they provide 
random/periodic disturbances that lead to chaotic orbits of the ring 
particles. The instability is given by amplification of resonant orbits
that leads to avalanches of particles, which clump in zones of non-resonant
orbits. The appeal of SOC models is the simple way to predict the final size
distribution of ringlets that result in the end, which cannot easily
be predicted by celestial mechanics or chaos theory.

\begin{figure}[t]
\centerline{\includegraphics[width=1.0\textwidth]{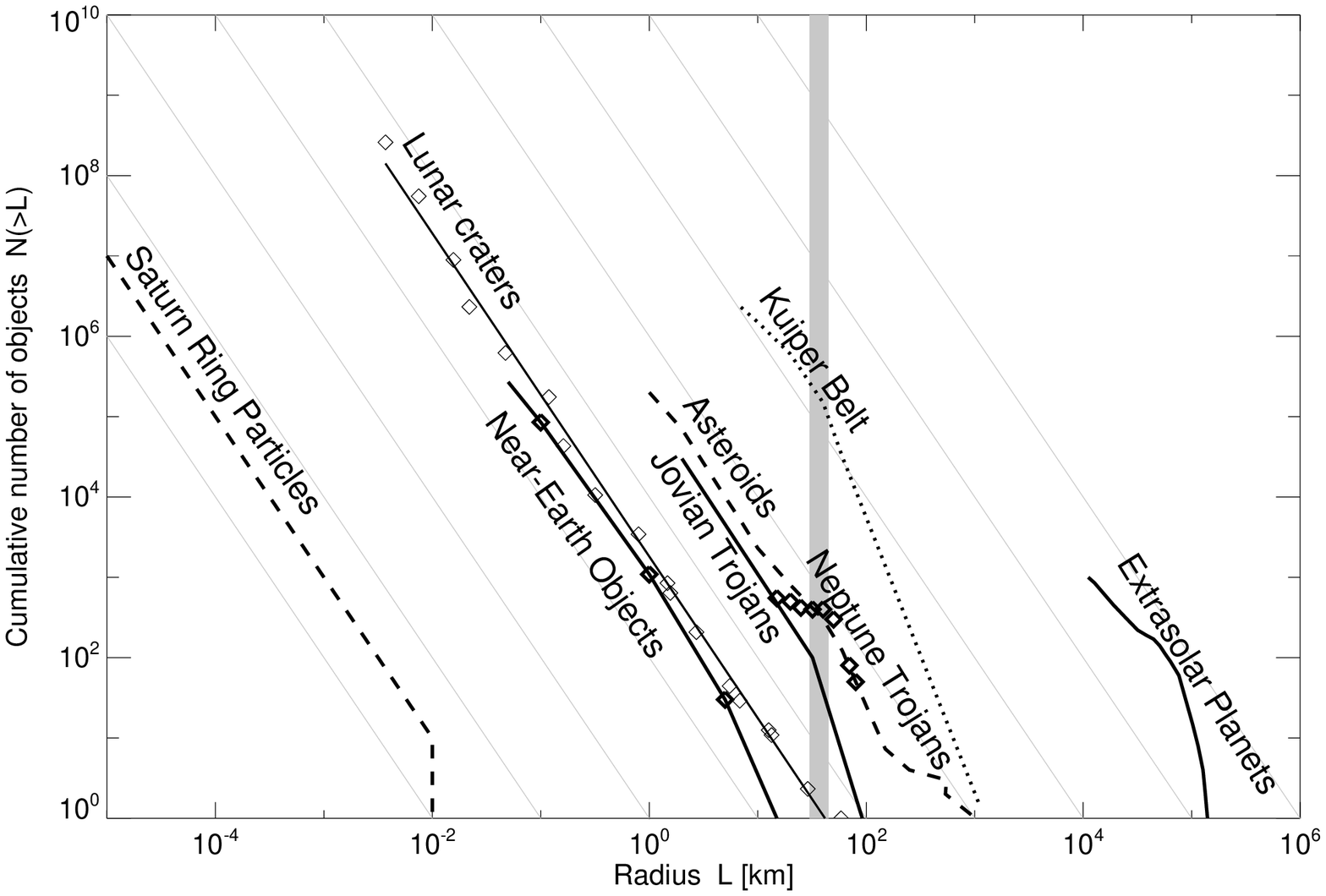}}
\captio{Cumulative size distribution of 
Saturn ring particles (Zebker et al.~1985),
near-Earth objects (McFadden and Binzel 2007),
Jovian Trojans (Jewitt et al.~2000), 
asteroids (Jedicke et al.~2002), 
Neptunian Trojans (Sheppard and Trujillo 2010),
lunar craters (Cross 1966),
Kuiper Belt objects (Fraser \& Kavelaars 2008; Fuendes \& Holman 2008),
and Earth-sized extrasolar planets (Catanzarite and Shao 2011).
The grey diagonal lines indicates the prediction of the FD-SOC model,
with a powerlaw slope of $\alpha_L^{cum} = 2$ for the cumulative size
distribution, corresponding to a powerlaw slope of $\alpha_L
=\alpha_L^{cum}+1=3$ for the differential occurrence frequency distribution.
A zone of paucity is indicated at a size range of $L=30-45$ km
identified from Neptunian Trojans (Sheppard and Trujillo 2010).} 
\end{figure}

\subsubsection{		Jovian and Neptunian Trojans 		}

The Jovian Trojans are two swarms of asteroids, which lead or trail
Jupiter by $\pm60^\circ$ on its orbit, known as the Lagrangian L4 and L5 point.
The Jovian Trojans contain some 250 members. Their origin has been
interpreted in terms of trapping of asteroidal fragments. 
A statistical analysis yielded a differential size distribution of 
$N(L) \propto L^{-3.0\pm0.3}$ in the size range of $L=2-30$ km,
and $N(L) \propto L^{-5.5\pm0.9}$ in the size range of $L=50-84$ km
(Jewitt et al.~2000).

Similarly, Trojans have been detected in the L4 and L5 regions of 
the planet Neptune, with a size distribution that approaches
a powerlaw slope of $\alpha_L = 5 \pm 1$ at the upper end
(Sheppard and Trujillo 2010), while a flatter slope is found at 
the lower end. The scarcity of intermediate- and smaller-sized
Neptune Trojans ($\le 45$ km), which is also found for other
objects in the Kuiper Belt, Jovian Trojans, and main belt asteroids,
was interpreted in terms of a primordial origin, rather than a
collisional or fragmentational origin, for which a size distribution
of $N(L) \propto L^{-3}$ is expected in the SOC model. 
However, the smaller bodies of the Neptunian Trojans in the range of
$L=2-30$ km could still be consistent with a SOC origin, 
if they have the same distribution as Jovian Trojans
(with $N(L) \propto L^{-3.0\pm0.3}$; Jewitt and Trujillo 2000).
Their size range and distribution is close to that of asteroids (Fig.~25).

\subsubsection{		Kuijper Belt Objects 		}

The Kuijper belt is a region of our solar system beyond the orbit
of Neptune (at 30 AU) out to $\approx 50$ AU, consisting of many
small bodies. A size distribution of $N(L) \propto L^{-4.3}$
was found for objects with $L \gapprox 100$ km (Fraser et al.~2008;
Fraser and Kavelaars 2008; Fuentes and Holman 2008). A comparison
of the cumulative size distributions of Kuiper Belt objects,
Neptunian Trojans, Jovian Trojans, and asteroids is shown in
Fig.~25. Obviously, there is a paucity of objects in the zone
of $L\approx 30-45$ km that shows up in the Neptunian Trojans
and in the Kuiper belt objects (Fig.~25). The data seem to be 
consistent with the predicted powerlaw slope of $\alpha_L \approx 3$ 
only for small length scales of $L\approx 1-30$ km. 

\subsubsection{		Extrasolar Planets		}

The oligarchic growth of protoplanets has been brought into the context
of a self-organized protoplanet-planetesimal system (Kokubo and Ida 1998).
The growth and orbital evolution of protoplanets embedded in a swarm of
planetesimals has been simulated with a 3D N-body code, which shows the
relative distribution of large planets that grow oligarchically, while
most of the planetesimals remain small (Kokubo and Ida 1998). 

Using the {\sl Kepler} space telescope for search of Sun-like stars
and (extrasolar) planets, a sample of over 150,000 stars was measured
during the first 4 months of the mission. The Kepler science team determined
sizes, surface temperatures, orbit sizes, and periods for over
a thousand new planet candidates. From a size distribution of 1176
Earth-sized planet candidates within a range of $L=2,...,20$ Earth
radii, a powerlaw distribution was found in the range of $L\approx 2-10$
Earth radii (Fig.~25), with a powerlaw slope of $\alpha_L = \alpha_L^{cum} 
+ 1 = 1.48 + 1 = 2.48$, while the relatively narrow distribution falls 
of steeply between $L\approx 10-20$ Earth radii (Catanzarite and Shao 2011). 

This sample from 1176 different stars can be considered as a galactic
SOC system, in which case a size distribution of $N(L) \approx L^{-3}$
is predicted by the FD-SOC model, which is close to the observed value of 
$N(L) \approx L^{-2.5}$
for a subset of Earth-like planets. The accretion of an Earth-like planet
represents then an avalanche event, triggered by a gravitational instability
in each stellar system. 

\clearpage %%%%%%%%%%%%%%%%%%%%%%%%%%%%%%%%%%%%%%%%%%%%%%%%%%%%%%%%%%%%%%%%%%%%%%%

\subsection{		STARS AND GALAXIES 		}

We can obtain information on spatial scales and spatio-temporal scaling laws
from SOC phenomena in our solar system (i.e., from the Sun, the planets, the 
magnetosphere), while such information from the rest of the universe is 
concealed by distance and cosmological time scales. Nevertheless, a number 
of stellar phenomena have been attributed to
SOC phenomena. The observables are mostly time durations $T$, peak fluxes $P$,
and fluences $E$ of electromagnetic emission in some wavelength range, measured
with some automated event detection algorithm from time series of a stellar object.
We will compile such observations from stellar flares, pulsars, soft gamma-ray
repeaters, blazars, and black-hole objects in the following, and compare them
with the predictions of the FD-SOC model.

\subsubsection{		Stellar Flares			}

Time series with rapidly fluctuating emission in soft X-rays, EUV, and visible
light from individual stars have been gathered with EXOSAT (Collura et al.~1988;
Pallavicini et al.~1990), the {\sl Hubble Space Telescope (HST)}
(Robinson et al.~1999), the {\sl Extreme Ultraviolet Explorer (EUVE)}
(Osten and Brown 1999; Audard et al.~1999, 2000; Kashyap et al.~2002; 
G\"udel et al.~2003; Arzner
et al.~2007), the {\sl X-ray Multi-Mirror Mission (XMM)} or {\sl Newton}
(Stelzer et al.~2007), and most recently with the surveys of the {\sl Kepler} 
mission (Walkowicz et al.~2011; Maehara et al.~2012; Shibayama et al.~2013). 
Impulsive bursts detected in the time series in excess of the noise level
have been interpreted as stellar flares, because they show
similar temporal and wavelength characteristics as solar flares, except that
they exceed solar flares in their luminosity by several orders of magnitude
(Aschwanden et al.~2008c).
Therefore, they should be considered as ``giant flares'' by solar standards.
These stellar flares have been observed mostly in solar-like G-type stars
(Notsu et al.~2013; Mahara et al.~2012; Shibayama et al.~2013), and in cool
dwarf (dMe) stars (Robinson et al.~1999; Audard et al.~2000; Kashyap et al.~2002; 
G\"udel et al.~2003; Arzner et al.~2007; Stelzer et al.~2007; Walcowicz 
et al.~2011; Maehara et al.~2012). From soft X-ray and EUV spectroscopy,
flare temperatures of $T_e \approx 10-100$ MK have been determined in some
of the stellar flares, exceeding solar flare temperatures ($T_e \approx
5-35$ MK). Consequently, the same physical interpretation
in terms of magnetic reconnection with subsequent heating of chromospheric
plasma has been proposed for stellar flares, in analogy to their solar
analogs, although their total emission measure is a few orders of
magnitude larger than for solar flares (Aschwanden et al.~2008c). 

\begin{table}[t]
\begin{center}
\captio{Frequency distributions measured from stellar flares.
The predictions (marked in boldface) are based on the FD-SOC model (Aschwanden 2012a).}
\medskip
\begin{tabular}{llrlll}
\hline
Star	        &Instrument     &Number          &Powerlaw      &Powerlaw   	&References:\\
                &               &of events       &slope of   	&slope of   	&\\
                &               &                &peak flux  	&fluences   	&\\
                &               &                &$\alpha_P$ 	&$\alpha_E$ 	&\\
\hline
\hline
13 M dwarfs	&EXOSAT         &17              &              &1.52$\pm$0.08 &Collura et al.~(1988)\\
22 M dwarfs 	&EXOSAT         &20              &              &1.7$\pm$0.1   &Pallavicini et al.~(1990)\\
RS CVn          &EUVE           &25              &              &1.5$-$1.7     &Osten and Brown (1999)\\
47 Cas, EK Dra	&EUVE		&28		 &              &1.8$-$2.3     &Audard et al.~(1999)\\
YZ Cmi          &HSP/HST        &54              &              &2.25$\pm$0.10 &Robinson et al.~(1999)\\
HD 2726         &EUVE           &15              &              &1.9$-$2.6     &Audard et al.~(2000)\\
47 Cas          &EUVE           &12              &              &2.0$-$2.6     &Audard et al.~(2000)\\
EK Dra          &EUVE           &16              &              &1.8$-$2.3     &Audard et al.~(2000)\\
$\kappa$ Cet 1994 &EUVE         &5               &              &1.9$-$2.6     &Audard et al.~(2000)\\
$\kappa$ Cet 1995 &EUVE         &10              &              &2.2$-$2.5     &Audard et al.~(2000)\\
AB Dor          &EUVE           &16              &              &1.8$-$2.0     &Audard et al.~(2000)\\
$\epsilon$ Eri  &EUVE           &15              &              &2.4$-$2.5     &Audard et al.~(2000)\\
GJ 411          &EUVE           &15              &              &1.6$-$2.0     &Audard et al.~(2000)\\
AD Leo          &EUVE           &12              &              &1.7$-$2.0     &Audard et al.~(2000)\\
EV Lac          &EUVE           &12              &              &1.8$-$1.9     &Audard et al.~(2000)\\
CN Leo 1994     &EUVE           &14              &              &1.9$-$2.2     &Audard et al.~(2000)\\
CN Leo 1995     &EUVE           &14              &              &1.5$-$2.1     &Audard et al.~(2000)\\
FK Aqr          &EUVE           &50              &              &2.60$\pm$0.34 &Kashyap et al.~(2002)\\
V1054 Oph       &EUVE           &70              &              &2.74$\pm$0.35 &Kashyap et al.~(2002)\\
AD Leo          &EUVE           &145             &              &2.1$-$2.3     &Kashyap et al.~(2002)\\
AD Leo          &EUVE           &261             &              &2.0$-$2.5     &G\"udel et al.~(2003)\\
AD Leo          &EUVE           &                &              &2.3$\pm$0.1   &Arzner \& G\"udel (2004)\\
HD 31305        &XMM            &22              &              &1.9$-$2.5     &Arzner et al.~(2007)\\
TMC             &XMM            &126             &              &2.4$\pm$0.5   &Stelzer et al.~(2007)\\
G5-stars        &Kepler         &1538            &1.88$\pm$0.09 &2.04$\pm$0.13 &Shibayama et al.~(2013)\\
\hline
FD-SOC prediction &             &                &{\bf 1.67}   &{\bf 1.50}    &Aschwanden (2012a)\\
\hline
\end{tabular}
\end{center}
\end{table}

Let us have a look at the obtained size distributions of flare durations $T$,
peak fluxes $P$, and fluences $E$ that have been sampled from flares on 
individual stars, which are compiled in Table 13. Most powerlaw slopes of
fluences are found in the range of $\alpha_E \approx 1.9-2.3$, which is
significantly higher than measured in solar flares, where we found $\alpha_E 
\approx 1.4-1.9$ in soft X-rays (Table 3) and $\alpha_E \approx 1.4-2.3$
in EUV (Table 4), while the FD-SOC model predicts a value of $\alpha_E
= 1.5$, which is matched indeed by solar flare observations in hard X-rays,
i.e., $\alpha_E \approx 1.4-1.7$. However, several observations found
powerlaw slopes of $\alpha_E \approx 1.5-1.7$ (Collura et al.~1988;
Pallavicini et al.~1990; Osten and Brown 1999) that are consistent with
the predictions of the FD-SOC model ($\alpha_E \approx 1.5$).
Almost all size distributions of stellar flares have been characterized
as powerlaw functions. The only exception (with an exponential size
distribution) has been reported from optical flares of low-mass young 
stellar objects in the Orion nebula (Akopian 2012a) and from the region
region of $\rho$ Ophiuchi (Akopian 2012b).

This raises the question why most of the stellar (and a few solar) flare 
samples appear to have a different (steeper) size distribution than expected? 
Part of the explanation is probably the difference in luminosity, which puts the
stellar flares at the upper end of the size distribution of solar flares,
where size distributions tend to fall off steeper due to finite observing
time and finite system-size effects. Moreover, cumulative size distributions,
$N^{cum}(>x)$, as they generally are obtained in small samples of stellar flares 
(from inverse rank-order plots), show
an exponential-like fall-off towards the largest event. This is a mathematical
consequence of the integration of a powerlaw function that extends over a 
finite range $[x_1,x_2]$, i.e., the differential frequency distribution,
\begin{equation}
  	N(x) \propto (\alpha - 1) x^{-\alpha} \ , \qquad x_1 \le x \le x_2 \ ,
\end{equation}
which yields the cumulative frequency distribution,
\begin{equation}
  	N^{cum}(>x) = n {{\int_{x}^{x_2}} N(x') dx' \over 
                   {\int_{x_1}^{x_2} N(x') dx'}}
                   = n {(x^{1-\alpha} - x_2^{1-\alpha}) \over 
                       (x_1^{1-\alpha} - x_2^{1-\alpha})} \ .
\end{equation}
The powerlaw slope $\alpha^{cum}$ of a cumulative size distribution needs to 
be fitted with this expression (Eq.~94), in order to obtain the exact value
of the powerlaw slope $\alpha = \alpha^{cum}+1$ of the differential size 
distribution.
Applying this method yields somewhat smaller slopes for stellar flare
size distributions, in the order of $\alpha_E \approx 1.8-2.1$ (Table 7.7
in Aschwanden 2011a), but does not completely explain the difference
between solar and stellar flares. The average of flare star observations
with EUVE yields $\alpha_E = 2.2 \pm 0.3$, while the optical observations
with Kepler exhibit a similar value ($\alpha_E=2.0 \pm 0.1$), which are
both steeper than predicted by the FD-SOC model.

\begin{figure}[tpbh]
\centerline{\includegraphics[width=1.0\textwidth]{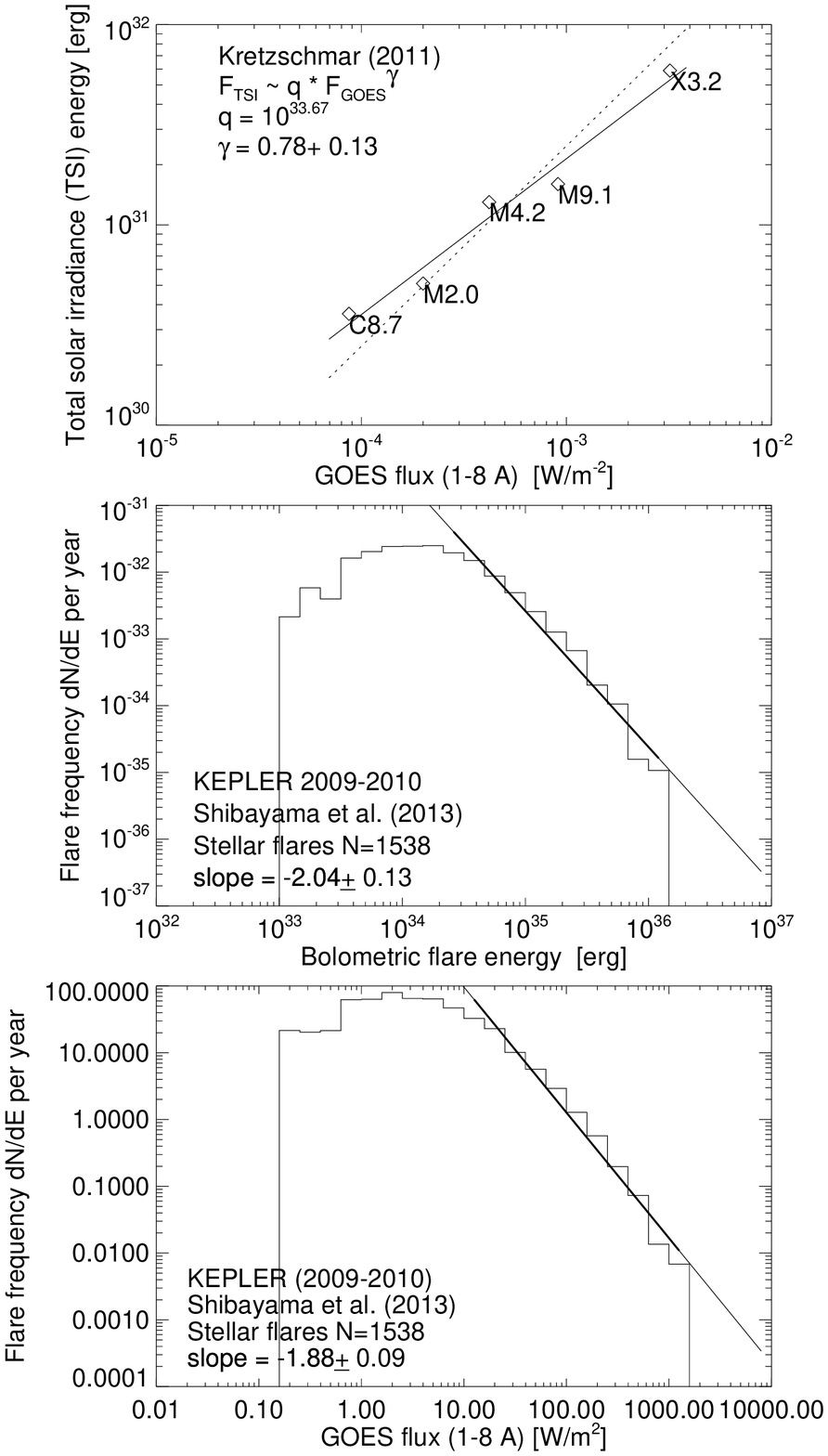}}
\captio{{\sl Top:} The scaling law of the total solar irradiance (TSI) and 
the GOES 1-8 \ang\ flux based on a linear regression fit (solid line)
to data from Kretzschmar (2011) is shown, i.e., $E_b \propto P_x^{0.78}$.
A linear relationship is indicated with a dotted line. {\sl Middle:}
The bolometric flare energy of 1538 stellar flares observed with
{\sl Kepler} is histogrammed, yielding a size distribution with a powerlaw
slope of $\alpha_E=2.04\pm0.13$. {\sl Bottom:} The size distribution of
GOES fluxes inferred from the scaling law of Kretzschmar (2011) yields 
a powerlaw slope of $\alpha_P=1.88\pm0.09$ (Aschwanden 2014).}
\end{figure}

Another explanation for the steeper powerlaw slopes of stellar flare
distributions is the nonlinear scaling between the observed
bolometric energy $E_b$ and the soft X-ray peak flux $P$. From Kretzschmar
(2011, Table 1 therein) we derive a scaling law between the bolometric
fluence (total solar irradiance), which is equivalent to the
bolometric energy $E_b$, and the soft X-ray GOES 1-8 \ang\ peak flux 
$P_x$ (Fig.~26, top),
\begin{equation}
	E_b \propto P_x^{0.78\pm0.13} \ .
\end{equation}
Using this scaling law and the observed size distribution of bolometric
energies measured with Kepler (Shibayama et al.~2013), 
i.e., $N(E_b) \propto E_b^{2.04\pm0.13}$,
we can derive the distribution of equivalent GOES peak fluxes of stellar 
flares,
\begin{equation}
	N(P_x)\ dP_x \propto N(E_b[P_x])
	{dE_b \over dP_x}\ dP_x \propto P_x^{-1.81\pm0.12} \  dP_x \ ,
\end{equation}
which is indeed more consistent with the size distribution of observed
solar GOES peak fluxes, $\alpha_P=1.88\pm0.09$ (Fig.~26 bottom and Table 3) 
and with the predictions of the FD-SOC
model ($\alpha_P = 1.67$). The scaling law (Eq.~95) and the size 
distributions of the bolometric flare energies and corresponding
soft X-ray GOES peak fluxes obtained from 1538 stellar flares observed 
with Kepler (Shibayama et al.~2013) are shown in Fig.~26.

Since solar flares
show the trend of a steeper powerlaw slope $\alpha_E$ in the fluences
measured in soft X-rays and EUV, compared to hard X-rays, we
suspect also that the prolonged thermal emission in soft X-rays and EUV,
due to plasma cooling, boosts the time-integrated fluence so that the 
total dissipated energy is overestimated, unlike the fluences in hard 
X-rays, where thermal emission is completely negligible at electron 
energies $E \ge 25$ keV. Unfortunately, current hard X-ray detectors
are not sensitive enough to detect hard X-ray emission from stellar flares.

Thus, we conclude that hard X-rays provide the most accurate measurements
of dissipated energies during flares, which are also consistent with the 
predictions of the FD-SOC model, while soft X-rays, EUV emission, and
white-light (bolometric) emission exhibits a nonlinear scaling with the
emitted energy. The fluence measured in soft X-rays and EUV emission
are boosted due to plasma heating and cooling processes. The reconciliation
of measurement methods of the total dissipated energy in hard X-rays,
soft X-rays, and EUV is still an open problem, which could be resolved
with multi-wavelength statistics of solar data, and by modeling the
scaling laws between dissipated energies and the fluxes in different 
wavelengths. Apparently the bias
in the soft X-ray and EUV wavelengths affects the energy distributions
measured from (giant) stellar flares to a larger degree than those of
solar flares.

\subsubsection{Star Formation}

The formation of stars is initiated by gravitational
collapses of molecular clouds. Such a gravitational collapse can be
triggered by collisions of two molecular clouds, by the explosion
of a nearby supernova, which ejects shocked matter, or even by
galactic collisions, which cause compression of matter and tidal
forces. If there is a critical mass reached, which is quantified
by the {\sl Jeans mass criterion}, which mostly depends on the initial
size of the unstable galactic fragment, the collapsing molecular cloud
will build up a dense core by self-gravity, which forms a star with
nuclear burning. Smaller sizes develop into non-radiating brown
dwarfs.

Considering star formation as a SOC process, the situation is similar
to the formation of planetesimals and planets, where a critical
condition is given by the balance between the forces of self-gravity and
diffusion. A collapsing molecular cloud gains kinetic energy from the
gravitational potential according to the conservation of angular
momentum. However, tidal forces, external gravitational disturbances,
and thermal pressure represent forces that contribute to the local
diffusion of the molecular cloud. Therefore there is a threshold
for the instability of a gravitational collapse, which is self-organizing
by the given balance between the opposing forces of contraction and 
diffusion. This process could possibly also be modeled in terms of 
a percolation model.

SOC avalanches have a fractal structure, and hence fractals are expected for
star-forming regions also. Indeed, fractal and self-fimilar patterns have
been observed in the Milky Way from dense cores to giant molecular clouds
in a range of $0.1 < L < 100$ pc (Elmegreen and Scalo 2004;
Bergin and Tafalla 2007), as well as in star-forming regions in the 
Andromeda nebula M33 (Sanchez et al.~2010). The fractal dimension
in the interstellar medium has a value of $D_3 \approx 2.3 \pm 0.3$
(Elmegreen and Falgarone 1996), in bright young stars and molecular gas
is $D_2 \approx 1.9$, and in fainter stars and HII regions is
$D_3 \approx 2.2-2.5$. The predictions of the FD-SOC model is
$D_3 \approx 2.0$. The fractal structure has generally been attributed
to interstellar turbulence, which however does not exclude a
generalized description in terms of a SOC process. It has been argued
that the interstellar mass function (IMF) of starbursts is 
independent of local processes governing star formation and thus
can be considered as a universal self-organized criticality process
(Melnick and Selman 2000). 

\subsubsection{Pulsars}

A pulsar is a highly magnetized, rapidly-rotating neutron star that
emits a beam of electromagnetic radiation. Since the beamed emission is  
aligned with the magnetic axis, we observe rotationally modulated
pulses whenever the beam axis points to the Earth (line-of-sight direction)
during each period of its rapid rotation. Besides these regular periodic 
pulses on time scales of milliseconds, which are measured with high 
accuracy, there occur sporadic glitches in pulse amplitudes and 
frequency shifts, probably caused by sporadic unpinning of vortices
that transfer momentum to the crust (Warzawski and Melatos 2008). 
Conservation of angular momentum produces then a tiny increase of 
the angular rotation rate, called {\sl ``positive spin-ups''} of the 
neutron star. 

\begin{table}[t]
\begin{center}
\captio{Frequency distributions measured from giant pulses of
pulsars (Crab, Vela, PSR), soft gamma-ray repeaters (SGR),
black-hole objects (Cygnus X-1, Sgr A$^*$), and a blazar (GC 0109+224).
The size distributions were reported in units of (cumulative) pulse 
energies (Argyle and Gower 1972), in radio flux densities (Lundgren
et al.~1995), (cumulative) pulse amplitudes (Cognard et al.~1996),
electric fields (Cairns 2004), fractional increase of the spin 
frequency $(\Delta \nu/\nu)$ (Melatos et al.~2008), or peak fluxes
(Ciprini et al.~2003). Powerlaw slopes of peak fluxes are marked
with parentheses.  Uncertainties (standard deviations) are quoted 
in brackets.}
\medskip
\begin{tabular}{lllll}
\hline
Object          &Waveband       &Number          &Powerlaw   	   &References:\\
                &               &of events       &slope of   	   &\\
		&		&		 &energies	   &\\
                &               &                &$\alpha_S, (\alpha_P)$ &\\
\hline
\hline
Crab pulsar	& 146 MHz	& 440		 & 3.5 		   &Argyle and Gower (1972)\\
Crab pulsar	& 813-1330 MHz	& 3$\times 10^4$ & 3.06$-$3.36     &Lundgren et al.~(1995)\\
PSR B1937+21	& 430 MHz	& 60		 & 2.8$\pm$0.1	   &Cognard et al.~(1996)\\
PSR B1706-44	& 1.5 GHz 	&		 & 6.4$\pm$0.6     &Cairns (2004)\\
Vela pulsar	& 2.3 GHz 	&		 & 6.7$\pm$0.6     &Cairns (2004)\\
PSR B0950+08	& 0.4 GHz 	&		 & 6.2$\pm$0.5     &Cairns (2004)\\
Crab pulsar	& 0.8 GHz	&		 & 5.6$\pm$0.6     &Cairns (2004)\\
PSR B1937+214	& 0.4 GHz	&		 & 4.6$\pm$0.2     &Cairns (2004)\\
PSR B1821-24	& 1.5 GHz 	&		 & 9.0$\pm$2.0     &Cairns (2004)\\
PSR 0358+5413	& 		& 6		 & 2.4 [1.5,5.2]   &Melatos et al.~(2008)\\
PSR 0534+2200	& 		& 26		 & 1.2 [1.1,1.4]   &Melatos et al.~(2008)\\
PSR 0537+6910	& 		& 23		 & 0.42 [0.39,0.43]&Melatos et al.~(2008)\\
PSR 0631+1036 	&		& 9		 & 1.8 [1.2,2.7]   &Melatos et al.~(2008)\\
PSR 0835+4510	& 		& 17		 & -0.13 [-0.20,+0.18]&Melatos et al.~(2008)\\
PSR 1341+6220	& 		& 12		 & 1.4 [1.2,2.1]   &Melatos et al.~(2008)\\
PSR 1740+3015	& 		& 30		 & 1.1 [0.98,1.3]  &Melatos et al.~(2008)\\
PSR 1801+2304	& 		& 9		 & 0.57 [0.092,1.1]&Melatos et al.~(2008)\\
PSR 1825+0935	& 		& 8		 & 0.36 [-0.30,1.0]&Melatos et al.~(2008)\\
\hline
SGR 1806-20     &  		&		 & 1.6             &Chang et al.~(1996)\\
SGR 1900+14	& $>25$ keV	&	 	 & 1.66		   &Gogus et al.~(1999)\\
SGR 1806-20	& $>21$ keV	&	 	 & 1.43, 1.76, 1.67&Gogus et al.~(2000)\\
Gamma-ray bursts&               & 83             & 1.06$\pm$0.15   &Wang \& Dai (2013)\\
\hline
GC 0109+224	& optical	&		 & (1.55)	   &Ciprini et al.~(2003)\\
\hline
Cygnus X-1	& $1.2-58.4$ keV&		 & (7.1)	   &Mineshige and Negoro (1999)\\
Sgr A$^*$	& $2-8$ keV     &		 & 1.5, (1.0)	   &Neilsen et al.~(2013)\\
\hline
\hline
FD-SOC prediction & 		&		 & {\bf 1.50, (1.67)} &Aschwanden (2012a)\\
\hline
\end{tabular}
\end{center}
\end{table}

Statistics of these sporadic glitches (Table 14) exhibit 
powerlaw distributions of the pulsar peak fluxes or fluences, such as
observed from the Crab pulsar and other pulsars in radio wavelengths 
(Argyle and Gower 1972; Lundgren et al.~1995; Cognard et al.~1996;
Cairns (2004); Melatos et al.~2008), and thus were interpreted in 
terms of a SOC system (Young and Kenny 1996). While early 
measurements with extensive statistics exhibit powerlaw distributions 
with relatively steep slopes of $\alpha_P \approx 3.0$ (Argyle 
and Gower 1972; Lundgren et al.~1995; Cognard et al.~1996), 
more recent observations with
smaller samples yield a large scatter of powerlaw slopes in the
range of $-0.13 \le \alpha_P \le 2.4$ (Melatos et al.~2008) and
$\alpha_E \approx 4.6-9.0$ (Cairns 2004). Other recent studies of 
giant micropulses from pulsars report a log-normal distribution 
of energies (Johnston and Romani 2002; Cairns et al.~2004), 
which is only consistent with 
a powerlaw function as an asymptotic limit in the tail of a 
log-normal function. Therefore, the distributions of giant pulses
from pulsar glitches do not give rise to a narrow range of powerlaw
slopes, and thus are not easy to explain in terms of a simple SOC
model. Part of the large uncertainties of powerlaw slope measurements
is clearly attributable to the small-number statistics in small
samples (i.e., 6-30 pulses in the data sets of Melatos et al.~2008).
The unusual steepness of reported powerlaw slopes may be associated
with finite-size effects in a SOC system, which can cause an
exponential-like cutoff at the upper end of the size distribution,
as it is suspected for giant stellar flares (Section 3.4.1).  
Furthermore, the size distributions listed in Table 14, have
been reported in different physical units (i.e., flux, pulse energy, 
electric field, frequency decrease ratios), and in form of both 
differential and cumulative size distributions, which need to be
converted into the same energy units in order to make them directly 
comparable. 

A cellular automaton model has been developed for pulsar glitches,
based on the superfluid vortex unpinning paradigm (Warszawski and
Melatos 2008, 2012; Melatos and Warszawski 2008). 
The lattice grid in this model simulates the
collective behavior of up to $10^{16}$ vortices in the interior
of the pulsar. The cellular automaton generates scale-free
avalanche distributions with powerlaw slopes of $\alpha_S =
2.0-4.3$ for avalanche sizes, and $\alpha_T=2.2-5.5$ for avalanche
durations. This numerical model produces size distributions that
are not too far off the predictions of the FD-SOC model ($\alpha_E
\approx 1.5$, $\alpha_T=2.0$), but covers an intermediate range
between the flatter slopes reported by Melatos et al.~(2008) and
the steeper slopes observed in radio wavelengths earlier.
Larger observational statistics and a consistent definition of
avalanche energies is needed to settle the pulsar SOC problem. 

\subsubsection{Soft Gamma Ray Repeaters}

A class of gamma-ray bursts that were detected with the {\sl Compton
Gamma Ray Observatory (CGRO)}, the {\sl Rossi X-ray Timing
Explorer (RXTE)}, and {\sl International Cometary Explorer (ICE)} 
in hard X-rays $\approx 20-40$ keV (a wavelength regime 
that is also called soft gamma-rays), with repeated detections 
from the same source location, has been dubbed {\sl Soft Gamma Ray
Repeaters (SGR)}. These gamma-ray bursts are believed to originate
from slowly rotating, extremely magnetized neutron stars 
(magnetars) that
are located in supernova remnants (Kouveliotou et al.~1998, 1999),
where neutron star crust fractures occur, driven by the stress
of an evolving, ultrastrong magnetic field in the order of
$B \gapprox 10^{14}$ G (Thompson and Duncan 1996). We should
be aware that repeated bursts from the same source are the
exception rather than the rule for gamma-ray bursts.

The size distributions of the fluences of sources SGR 1900+14
and SGR 1806-20 were found to exhibit powerlaw distributions with
slopes of $\alpha_E=1.66$ (Gogus et al.~1999) and
$\alpha_E$=1.43, 1.76, and 1.67 (Gogus et al.~2000),
extending over a range of about 4 orders of magnitude in fluence.
The waiting time distributions were found to be consistent with
a log-normal distribution (which is approximately a powerlaw 
function in the upper tail). Based on these observational
statistics, SGR bursts have been interpreted in terms of
a SOC process (Gogus et al.~1999; 2000). 
Since the source location is identical for
an object that produces SGR bursts, we can identify it with
a single SOC system, an assumption that cannot be made for
other gamma-ray bursts, which are non-repetitive and often
do not have an unambiguous source identification with known
distance.  Moreover we find that the fluence or energy 
distribution of the bursts matches the prediction of the
fractal-diffusive SOC model, with $\alpha_E = 1.5$. In the
magnetar model, the triggering mechanism for SGR bursts is
a hybrid of stress-induced starquakes and magnetically 
powered flares (Thompson and Duncan 1996), and thus has 
some similarity with the physical process of earthquakes.

A recent study was carried out with data from the {\sl Swift}
satellite, which has a rapid response, suitable for detecting
afterglows
of gamma-ray bursts. In a sample with 83 localized sources for
which the redshift was known (and thus the distance), a
size distribution of (distance-corrected) energies could
be constructed, and a powerlaw distribution with slope of
$\alpha_E = 1.06\pm0.15$ was found (Wang and Dai 2013).  
The size distribution of time duration was found to have
a slope of $\alpha_T=1.10 \pm 0.15$. These results were
interpreted in terms of a 1D SOC system (Wang and Dai 2013), 
for which the
FD-SOC model predicts $\alpha_E=1$ and $\alpha_T=1$. This 1D
interpretation for gamma-ray bursts with afterglows appears
to be different from soft gamma-ray repeaters, which  are
consistent with a 3D SOC system.

\subsubsection{Blazars}

Blazars are very compact objects associated with 
super-massive black holes in the center of active, giant elliptical
galaxies. They represent a subgroup of {\sl active galactic
nuclei (AGN)}, which emit a relativistic beam or jet 
that is aligned or nearly-aligned with the line-of-sight 
direction to Earth. Due to this particular geometry, blazars
exhibit highly variable and highly polarized emission in radio 
and X-ray emission. Optically violent variable (OVV) quasars
are a subclass of blazars. 

The optical variability of blazar GC 0109+224 was monitored
from 1994 onwards and the light curve exhibited a power spectrum 
$P(\nu) \approx \nu^{-p}$,
with $1.57 < p < 2.05$ (Ciprini et al.~2003), which is consistent
with the 1/f or flicker noise characteristics of SOC avalanches
in the BTW model (Bak et al.~1987; Hufnagel and Bregman 1992). 
The frequency distribution of radio peak fluxes of flaring events
from blazar GC 0109+224 was found to be a powerlaw distribution
(over about one order of magnitude), $N(P) \propto P^{-1.55}$
(Ciprini et al.~2003), which is consistent with the prediction 
of the FD-SOC model, i.e., $N(P) \propto P^{-1.67}$, within the 
uncertainties of the measurements. Interpreting blazars as a
SOC phenomenon, the critical threshold for a pulse is given by
the geometric coalignment condition between the emitted beam
direction (of accelerated particles producing gyrosynchrotron
emission) and the observer's line-of-sight direction from Earth.
The intermittency of blazar bursts observed on Earth is believed
to be caused by sporadic bursts of energy releases, 
created by internal shocks that occur within AGN jets.

\subsubsection{Black Holes and Accretion Disks}

The first Galactic X-ray source that has been identified as a
black-hole candidate, Cygnus X-1, emits hard X-ray pulses with
a time variability down to 1 ms. These hard X-ray pulses are
attributed to inverse Compton scattering of soft photons by hot
electrons heading toward the event horizon within the blak hole's
accretion disk. 

Statistics of the fluctuations in the light curve from Cygnus X-1,
observed in hard X-rays with {\sl Ginga} and {\sl Chandra}, exhibit
complex 1/f noise spectra and size distributions of peak fluxes with
very steep powerlaw slopes of $\alpha_P \approx 7.1$ (Negoro et al.~1995;
Mineshige and Negoro 1999), which have been interpreted in terms of
SOC models applied to accretion disks (Mineshige et al.~1994a,b;
Takeuchi et al.~1995; Mineshige and Negoro 1999). A SOC interpretation
was also suggested for the VY Scl-type cataclysmic variable KR Aurigae
(Kato et al.~2002; Dobrotka et al.~2012), UU Aqr (Dobrotka et al.~2012),
and for the broad-line radio galaxy 3C-390.3 (Leighly and O'Brien 1997),
for the Seyfert I MCG-6-30-15 (Sivron and Goralski 1998; Sivron 1998),
or for the extreme narrow-line Seyfert 1 galaxy IRAS 13224-3909
(Gaskell 2004), which all exhibit a highly intermittent variability 
on top of a shot noise background like Cygnus X-1.

In contrast, a total of 39 X-ray flares observed with {\sl Chandra}
from Sgr A$^*$, the $4 \times 10^6$ M$_{\odot}$ black hole at the 
center of our Galaxy, revealed powerlaw distributions with slopes
of $\alpha_P=1.9\pm0.4$ for the peak luminosity (of the 2-8 keV flux)
and $\alpha_E=1.5\pm 0.2$ for the fluence (Neilsen et al.~2013), 
which is perfectly consistent with the predictions of the FD-SOC model 
($\alpha_P=1.67$ and $\alpha_E=1.5$).

Cellular automaton models were constructed to mimic mass accretion
by avalanches that are triggered when the mass density of the disk
exceeds some critical value, which could reproduce the 1/f power
spectra $N(\nu) \propto \nu^{-1.6}$ and produced size distributions 
with powerlaw slopes of $\alpha_E=2.8$ for energies and $\alpha_T=1.4$ 
for durations (Mineshige et al.~1994a,b; Yonehara et al.~1997). 
A BTW-related model produced
an energy distribution of $\alpha_E = 1.35$ (Mineshige et al.~1994a,b)
that is closer to the FD-SOC prediction ($\alpha_E=1.5$). 
Adding gradual diffusion to the SOC avalanches in the cellular
automaton simulations produced a steeper (exponential) energy size 
distribution that was closer to the observations (Takeuchi et al.~1995).
Further modified cellular automaton models were developed that include 
reservoirs of different capacities (Negoro et al.~1995),
hydrodynamic models of advection-dominated accretion disks
(Takeuchi and Mineshige 1997), relativistic effects
(Xiong et al. 2000), non-local transport of angular momentum
in terms of the kinematic viscosity of magnetic loops in the
accretion disk corona (Pavlidou et al.~2001), and boson clouds
around black holes (Mocanu and Grumiller 2012).

Most of the various cellular automaton models designed to mimic
a physical mechanism operating in black-hole objects have difficulty 
to reproduce the observed steep size distributions, while most of
them seem to produce 1/f power spectra without special assumptions.
The observed steep size distributions may represent deviations of
the accretion disk system from a pure SOC system. The notion of
of SOC may still be useful to understand the observations,
but it cannot explan all properties of the fluctuations.

\subsubsection{		Galactic Structures		}

What physical mechanism produces galactic structures? 
A nonlinear theory was proposed in which the structure of spiral
galaxies arises from percolation phase transition (Schulman and
Seiden 1986; Seiden and Schulman 1990). The differential
rotation of the galaxy triggers propagating patterns of star formation.
This scenario is very similar to a SOC model, since it has a critical
point at the second-order phase transition associated with the
percolation threshold, which causes avalanches of star formations. 
Percolation processes, however, require fine-tuning, in contrast
to SOC systems. The process of stochastic self-propagating star formation 
was simulated with a cellular automaton model that provides a
representation of the percolation process operating in spiral galaxies
(Seiden and Schulman 1990).

The formation of galaxies has been modeled with two opposite scenarios,
the {\sl top-down scenario} that starts with a monolithic collapse
of a large cloud (Eggen, Lynden-Bell, and Sandage 1962; Zeldovich 1970), 
versus the now more widely accepted
{\sl bottom-up scenario}, where smaller objects merge and form larger
structures that ultimately turn into galaxies (Searle and Zinn 1978;
Peebles 1980).
The second scenario is more widely accepted now and corresponds also
closer to a SOC-driven avalanching scenario. In most models of galaxy
formation, thin, rotating galactic disks result as a consequence
of clustering of dark matter halos, gravitational forces and
disturbances, and conservation of angular momentum. The fractal-like
patterns of the universe from galactic down to solar system scales is
thought to be a consequence of the gravitational self-organization
of matter (Da Rocha and Nottale 2003). Fractal structures are observed
throughout the universe (Baryshev and Teerikorpi 2002). It is conceivable
that gravitational forces in an expanding universe lead to sporadic
density fluctuations or waves that initiate a local instability of
self-gravitating matter like an avalanche in a sandpile SOC model,
in case a critical threshold exists without need of fine-tuning.

\begin{figure}[t]
\centerline{\includegraphics[width=1.0\textwidth]{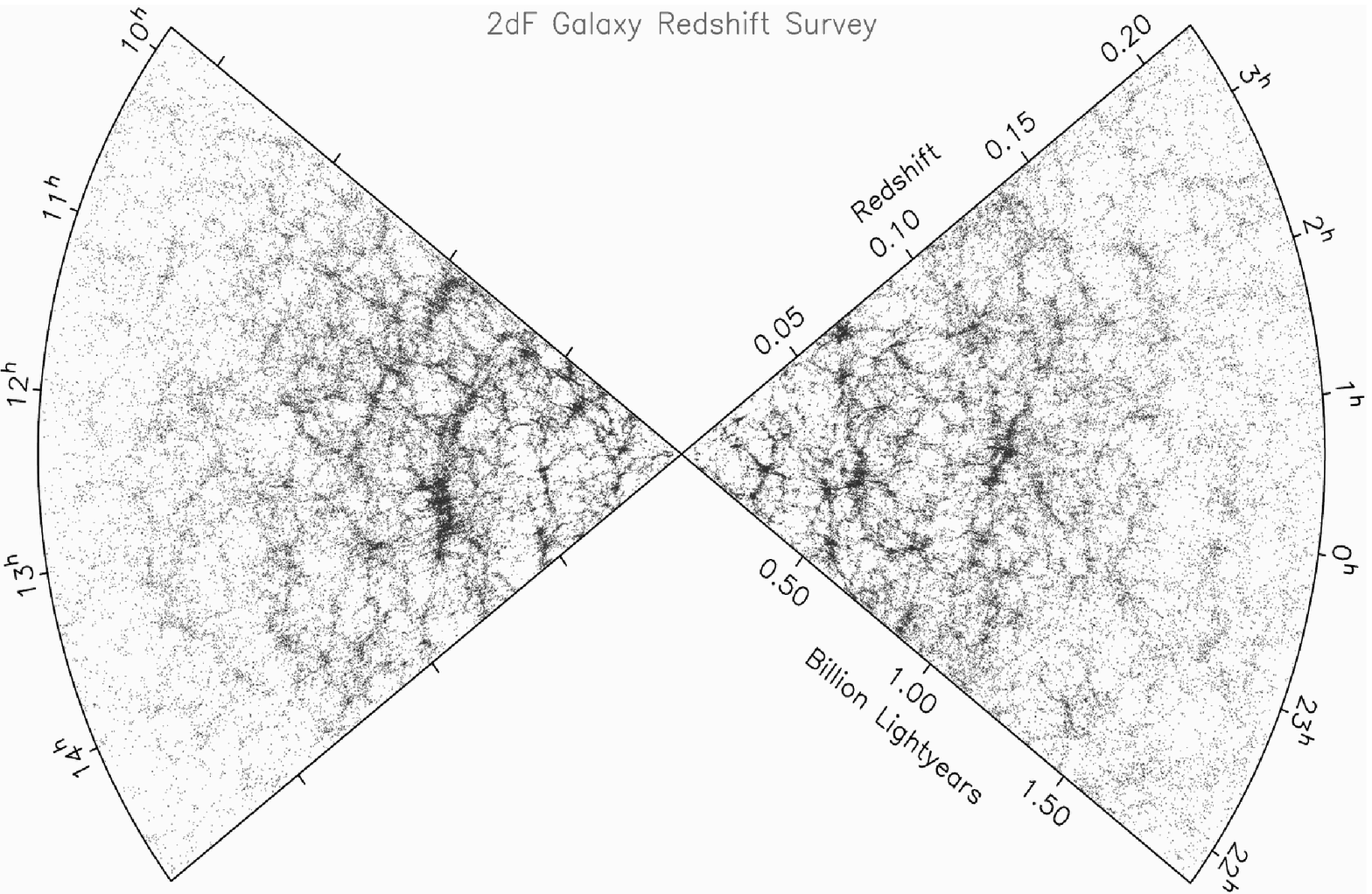}}
\captio{The 2dF galaxy redshift survey (2dFGRS), conducted at the
Anglo-Australian Observatory, shows a map of the galaxy distribution
out to redshifts of $z=0.23$ or approximately 2 billion lightyears,
which includes approximately 250,000 galaxies. Note the fractal
large-scale structure of the universe that makes up the galaxy
density (Colless et al.~2001).}
\end{figure}

\subsubsection{		Cosmology		}

The spatial structure of the universe exhibits fractal structures
of galaxy clusters out to a redshift of $z=0.23$ (Fig.~27), but
becomes very homogeneous and isotropic at cosmological scales
of the microwave background, with inhomogeneities of
$\lapprox 10^{-4}$ according to the latest results of the COBE
and WMAP missions. Self-organization and fractal scaling has
been applied to some large-scale structures in our universe,
such as to the galactic spiral structure (Nozakura and Ikeuchi 1988), 
the formation of the interstellar medium (Tainaka et al.~1993), 
the initial mass function of starbursts (Melnick and Selman 2000), 
the stellar dynamics in elliptical galaxy formation (Kalapotharakos 
et al.~2004), or to the gravitational structure formation in general 
on many scales (Da Rocha and Nottale 2003). 

The spatial flatness, homogeneity, and isotropy of the universe
at cosmological scales can be considered as a critical point that
would require an extreme fine-tuning, unless there is a self-organizing
principle that creates such a special state in a natural way.
Moffat (1997) proposes that the universe evolves as a SOC system
(in the sense of a BTW model), where the Hubble expansion undergoes 
{\sl ``punctuated equilibria''} like the SOC scenario of
intermittent evolution (Bak and Sneppen 1993). The inflationary
scenario, which predicts a rapid expansion of the early universe
to explain the flatness and the horizon problem, could be the
manifestation of a major SOC avalanche, while a SOC scenario
would predict many intermittent inflationary phases (Moffat 1997).
The critical point of a cosmological system would be the
critical density $\Omega=1$ that discriminates between an
open $(\Omega < 1)$ and a closed ($\Omega > 1$) universe, independent
of the initial conditions and without fine tuning of the parameters.
A related SOC concept has also been applied to quantum gravity
(Ansari and Smolin 2008). With the recent advent of string theory and
multi-verses, we might even consider our universe being only one
single avalanche episode in a multi-verse SOC scenario. 

\begin{figure}[tpbh]
\centerline{\includegraphics[width=0.8\textwidth]{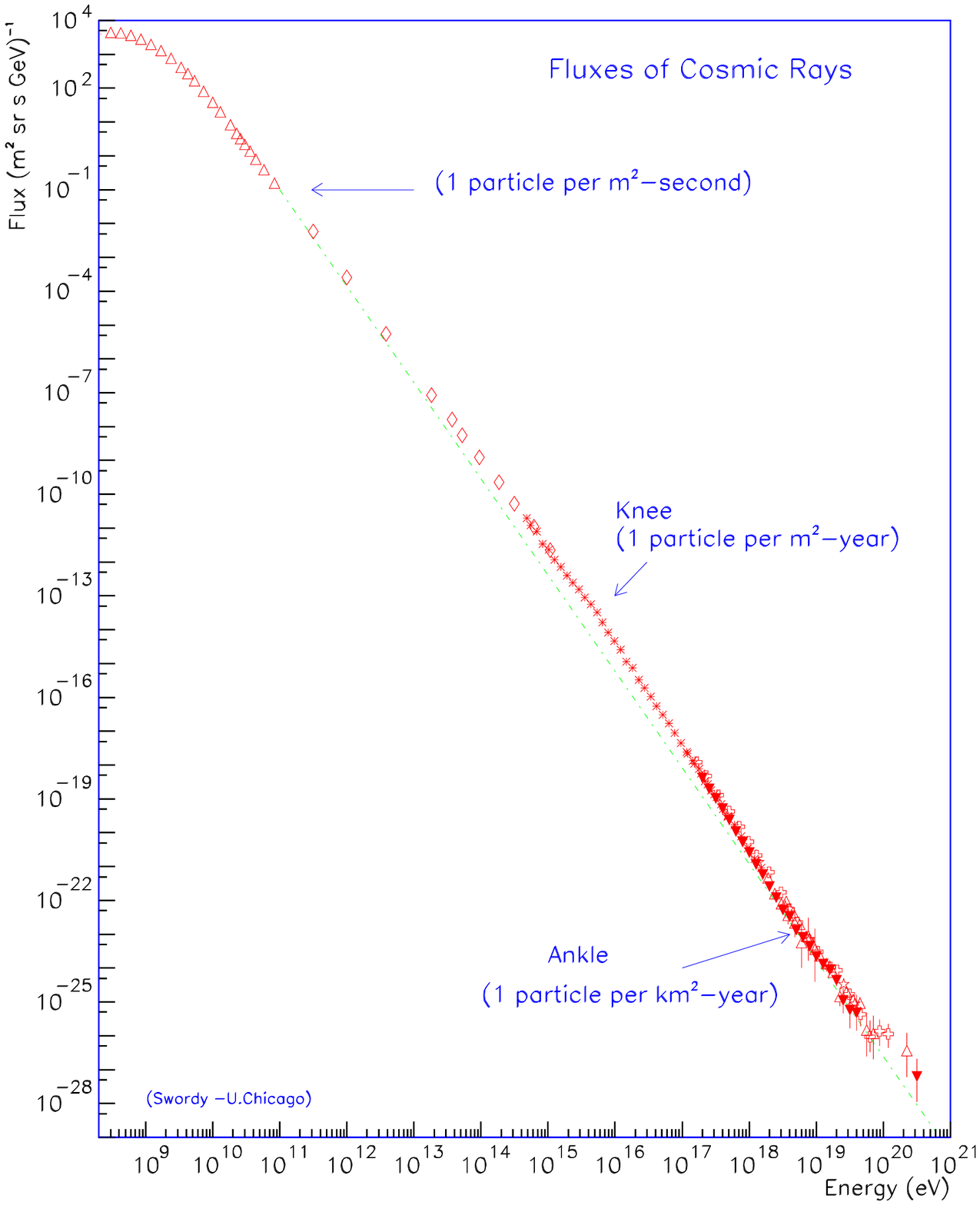}}
\captio{Cosmic ray spectrum in the energy range of $E=10^9-10^{21}$ eV,
covering 12 orders of magnitude. There is a ``knee'' in the spectrum
around $E \approx 10^{16}$ eV, which separates cosmic rays originating
within our galaxy (at lower energies) and those from outside the galaxy
(at higher energies) (Credit: Simon Swordy, University of Chicago).}
\end{figure}

\subsubsection{		Cosmic Rays		}

Cosmic rays are high-energetic particles (protons, helium nuclei,
or electrons) that originate from within our Milky Way, as well as
from extragalactic space, and are detected when they hit the
Earth's atmosphere and produce a shower of high-energy particles
(muons). The energy spectrum of cosmic rays extends over a large
range of $10^9$ eV $\lapprox E \lapprox 10^{21}$ eV, with an
approximate powerlaw slope of $\alpha_E \approx 3.0$ (Fig.~28).
A closer inspection reveals a broken powerlaw with a ``knee''
at $E_{knee} \approx 10^{16}$ eV, which separates the cosmic rays
accelerated inside our Milky Way (with a spectral slope of
$\alpha_{E1} \approx 2.7$) and in extragalactic space (with a
slope of $\alpha_{E2} \approx 3.3$). The sources of cosmic rays
are believed to be supernova remnants, pulsars, pulsar-wind
nebulae, and gamma-ray burst sources. The particles with higher
energies ($E \gapprox E_{knee}$) have a uniform and isotropic 
distribution over the sky and are believed to originate mostly
from active galactic nuclei (AGN).

High-energy particles can be accelerated by a number of physical
mechanisms, e.g., by electric fields, by shock waves, or by 
stochastic wave-particle interactions, such as by cyclotron
resonance, which requires magnetic fields. In Section 3.2.6
we discussed how a first-order Fermi process as well as
a fractal reconnection model can produce the observed powerlaw
spectra of high-energy particles (Nishizuka and Shibata 2013). 
Cosmic rays, which
travel through a large part of the universe, probably undergo
many local acceleration processes, and thus their trajectories
may look like a diffusive random walk. The acceleration process
of cosmic rays has been interpreted in terms of a SOC process
(Aschwanden 2014). The critical threshold is the ``runaway regime''
(e.g., Holman 1985) of a charged particle in a thermal distribution,
which is a critical velocity, i.e., $v_{crit} \gapprox 4 v_{th}$,
that is necessary to enable efficient acceleration out of the
thermal distribution. Considering the subsequent acceleration
process as a SOC avalanche, which can be achieved by an arbitrary
number of localized acceleration steps, the particles are likely
to undergo a diffusive random walk, as it is characterized by
the fractal-diffusive SOC model. The FD-SOC model predicts than
a powerlaw distribution for the energy spectrum of accelerated
particles, which is approximately fullfilled for cosmic rays
(as well as for nonthermal particles in solar flares). The
FD-SOC model predicts an energy spectrum of $N(E) \approx E^{-1.5}$,
which is however different from the observed cosmic ray spectrum
with $N(E) \approx E^{-3.0}$. This discrepancy has been interpreted
in terms of an incomplete sampling effect of cosmic-ray avalanches
(Aschwanden 2014). Since cosmic rays are in-situ measurements in a 
very localized target region (i.e., the Earth surface), only a
small 1-D cone of an isotropic cosmic-ray avalanche is sampled,
leading to an energy gain that is proportional to the traveled
length scale, i.e., $L \propto E$, and thus to an energy spectrum
$N(E) \propto N(L) \propto L^{-3} \propto E^{-3}$. Solar flare
observations, in contrast, provide remote-sensing of a complete
SOC avalanche of accelerated particles, and thus are expected
to have an energy spectrum of $N(E) \propto E^{-1.5}$, which is
indeed an asymptotic limit for the hardest solar flare spectra
(e.g., Dennis 1985; Miller et al.~1997).

\clearpage
%%%%%%%%%%%%%%%%%%%%%%%%%%%%%%%%%%%%%%%%%%%%%%%%%%%%%%%%%%%%%%%%%%%%%%%%

\section{    DISCUSSION: SOC CONCEPTS, CRITIQUES, NEW TRENDS, AND OPEN PROBLEMS   }

\subsection{ A Dual Approach of Self-Organized Criticality Systems }

A theory or a physical model is only useful (or acceptable) if it can 
make quantitative predictions, and if these predictions can be tested 
by observations, and hence the theory is falsifyable. What is the
current status of a SOC theory or a SOC model? In this review we
stress the dual nature of SOC models, in the sense that they include 
(i) universal statistical aspects that apply to all SOC systems, 
and (ii) special physical mechanisms that are idiosyncratic to a
particular SOC phenomenon. There is a consensus that the powerlaw 
function of the size distribution of a SOC observable is a universal
statistical aspect that is common to all SOC systems, regardless
whether we sample statistics of solar flares or earthquakes, while
the underlying physical mechanisms are completely different, such as
magnetic reconnection in solar flares, or mechanical stressing in
earthquakes. If we accept this dichotomy, we should be able to build
a generalized SOC theory that predicts the universal statistical properties,
which should be purely of ``mathematical nature'' and ``physics-free'',
while the nonlinear energy dissipation process of a SOC event still
can be described with (single or multiple) specific physical SOC models 
that are different for every SOC manifestation. In this spirit we reviewed 
the basic elements of a generalized SOC theory in Section 2, while we 
touched on possible interpretations in terms of particular physical 
mechanisms that produce a SOC phenomenon in Section 3.

Let us review how the definition of a generalized SOC theory evolved
over the last 25 years. The BTW model essentially defined
a SOC process by simulating a cellular automaton, which demonstrated
that a powerlaw size distribution resulted for avalanche sizes and
durations. Since 1/f noise has a power spectrum in the form of a 
powerlaw function, the claim was made that both phenomena may be related.
Many of the subsequent studies came up with different cellular automaton
models, which produced a range of powerlaw slopes (see Table 1 and
Pruessner 2012), some of them produced exact powerlaw size distributions 
over many orders of magnitudes (which demonstrated ``universality'' with 
regard to the scale-free size range, such as the Manna and Oslo model), 
while others exhibited significant deviations from exact powerlaw 
distributions (and thus cannot claim universality).
The next important insight concerned the relationships between the
powerlaw slopes of different SOC parameters, which depend on the nonlinear
scaling laws between the SOC parameters. Further progress was made by
predicting the statistical probability distributions of SOC parameters,
using branching theory, percolation theory, discretized diffusion models,
or renormalization group theory. A more detailed review on these 
theoretical and mathematical efforts is given in the article by Watkins
et al. in this volume. A very simple theoretical framework that unifies
many features of previous SOC models is the fractal-diffusive SOC model,
based on the scale-free probability conjecture (Eq.~1), which is able to
predict probability distributions of observable SOC parameters and the
underlying scaling laws between the SOC parameters. This basic SOC model
has no free parameters for the most common case of 3D Euclidean space and
classical diffusion transport, and offers a prediction for most of the astrophysical
observations of SOC systems reviewed here. The model can also be adjusted to
a different space dimension, fractal dimension, and type of diffusive 
transport. The FD-SOC model should be considered as a macroscopic 
approximation of the complex micro-dynamic processes in a SOC system.

Is this SOC theory complete? By no means, there is still a lot of
statistics and data analysis required to pin down the scaling laws,
statistical truncation bias, event selection bias, and other unknown
effects for those SOC phenomena where the generic FD-SOC model yields
a different prediction than what is observed. In addition, there are
a number open questions in SOC models that try to reproduce real-world
data, such as the time variabiltiy of the driver, effects that cause
deviations from ideal powerlaw distributions, predictive capabilities,
alternative SOC-related processes, which are discussed in the following 
sections.

\subsection{	Universal Aspects of SOC Systems			}

The universal aspects that are common to all SOC phenomena define 
a SOC theory. In Fig.~1 we sketched the basic characteristics of
a SOC system: (i) a critical threshold for instabilities, (ii) a 
statistically slow driver that continuously nudges the system toward
a critical 
point, and (iii) a nonlinear energy dissipation process when an 
avalanche is triggered. The most crucial and testable predictions
of a SOC theory are the statistical probability distributions.
The central key feature of the FD-SOC theory is the statistical
probability argument for geometric length scales, the so-called
scale-free probability conjecture, $N(L) dL \propto L^{-d} dL$.
This statistical argument is derived by the same principle as a
binomial distribution is derived for a stochastic process by 
enumerating all possible outcomes of dice combinations.
This conjecture can easily be tested by extensive statistics of
length scales, such as we demonstrated for lunar craters, asteroid
sizes, Saturn ring particle sizes, magnetospheric aurora sizes,
solar flare sizes, and can be done in the same way for other SOC 
processes, such as earthquake rupture areas, for instance. 
An additional assumption of the FD-SOC model is the fractal-diffusive
transport, which involves random walk statistics for the avalanche
transport, i.e., $L \propto T^{1/2}$ for classical diffusion, and
a fractal geometry of the instantaneous avalanche size $V_f \propto
L^{D_d}$, where the mean fractal dimension can be estimated from
the mean-value dimension $D_d \approx (1+d)/2$. Integrating such
a fractal-diffusive avalanche in time yields then the total size $S$
of an avalanche. If energy dissipation of an avalanche is proportional
to the time-integrated size of an avalanche, we obtain the total
dissipated energy $E \propto S$, the energy dissipation rate
$F \propto V_f$, and the peak energy dissipation rate $P \propto V$.
In astrophysical applications, the energy dissipation rate $F$ is generally
measured by the flux or intensity of electromagnetic radiation in some
wavelength, but the universal meaning of the energy dissipation rate
is simply the instantaneous avalanche size during a snapshot, while
the total energy is the time-integrated avalanche volume.
Thus this generalized SOC concept is still universally applicable
to every SOC system, regardless if it is observed by an astronomical
instrument, by a geophysical monitor, by financial statistics,
or by computer lattice simulations.

\subsection{	Physical Aspects of SOC Systems				}

The physics comes in once we identify the avalanche, the threshold,
and the dissipated energy with a particular instability in the real world
(Table 16). For solar flares, for instance, the threshold may be 
given by a critical stressing angle between a potential and 
non-potential magnetic field line in an active region, the avalanche 
may be manifested by a solar flare emitting in all wavelengths, 
triggered by a magnetic reconnection process
of the over-stressed magnetic field lines, and the dissipated energy
can be measured by the change of magnetic energy before and after the
flare, or by the thermal energy of the heated plasma, or by the total
kinetic energy of accelerated particles. If we consider an earthquake,
the threshold may be given by the limit of elastic stressing of
tectonic plates, the instability is the slip-stick motion of the
tectonic plates, the avalanche is the spatio-temporal pattern of the
rupture area on the Earth's surface, and the measured energy is the
magnitude indicated by the vibrations detected by a seismometer.
The advantage of separating the universal aspects from the physical
aspects of a SOC system is that we can understand the statistics of
SOC parameters independently of the physical model of a SOC phenomenon.
For instance, we have very vague ideas about the exact physical process
that occurs in pulsar glitches, in giant pulses from black-hole candidates,
or in the bursts from soft gamma ray repeaters, but the FD-SOC model
can predict the distributions and basic scaling laws between spatial
and temporal parameters. In the case of imaging observations, where
we can measure both spatial and temporal scales, the FD-SOC 
model can place absolute values on the diffusion coefficients, which
may help to identify the physical transport process that occurs during
an avalanche. We should also be aware that the universal FD-SOC model
assumes a proportionality between the avalanche size $S$ and total
dissipated energy $E$, which may not always be the case, such as
for coherent emission mechanisms (e.g., laser or maser emission),
which requires a specific physical model.

\subsection{	Powerlaws and Deviations 		}

The functional shape of size distributions of SOC parameters
is generally expected to be a powerlaw function, i.e.,
$N(x) dx \propto x^{-\alpha} dx$, which is a consequence of the
scale-free nature of SOC processes. Numerical simulations of
cellular automaton models were indeed capable to reproduce an
exact powerlaw probability distribution function for avalanche
sizes over many orders of magnitude, such as the Manna model 
(Manna 1991) or the Oslo model (Christensen et al.~1996), 
while substantial deviations from ideal powerlaw functions
have been found in real-world observations, which raises the
question how well the ideal powerlaw distributions predicted
by standard SOC models characterizes real-world data.
Taken to the extreme, sceptics doubt whether powerlaws
have any relevance at all (Stumpf and Porter 2012).

Starting from first principles, a powerlaw function of length
scales is predicted from the scale-free probability conjecture
(Eq.~1) in our generic standard model, which is fundamentally
based on the principle of statistical maximum likelihood,
and does not depend on any other assumption. However, this
ideal distribution function is always limited within a finite range
of spatial sizes $[x_1, x_2]$, given by the spatial resolution
limit or lower limit of complete sampling $x_1$, and the 
finite system size or maximum avalanche size $x_2$ that happened 
during the observed time interval $x_2$. So, the powerlaw function 
is expected only over this limited range $[x_1, x_2]$, while there 
is generally a rollover at the lower end and an exponential-like
drop-off at the upper end. However, this range can be enlarged
by lowering the lower limit $x_1$ by more sensitive instruments,
and by increasing the upper limit $x_2$ by extending the total
observing time (in case the largest avalanche does not exceed
the finite system size).

Starting from the powerlaw function of the length scale 
distribution $N(L) \propto L^{-d}$, the FD-SOC model predicts
powerlaw distribution functions for all other parameters,
such as the area $A$, the volume $V$, the fractal area $A_f$,
the fractal volume $V_f$, the flux $F$, the peak flux $P$, 
the fluence $S$, and energy $E$, because these SOC parameters
are all related to each other by powerlaw relationships,
such as by the definition of the Hausdorff dimension $D_d$,
or the diffusive transport with spreading exponent $\beta$.
Even the intermittent waiting times are predicted to be a 
powerlaw for contiguous flaring periods, with the only 
exception of quiescent time intervals, which may follow
an exponential distribution (if they are produced randomly).

Then we should also be aware of the different predictions for
a differential and for a cumulative size distribution. Even when
a perfect powerlaw function exists for the differential size
distribution over some range $[x_1, x_2]$,
\begin{equation}
  	N(x) \propto (\alpha - 1) x^{-\alpha} \ , \qquad x_1 \le x \le x_2 \ ,
\end{equation}
the cumulative frequency distribution is not a perfect powerlaw
function, with a slope that is flatter by one, but exhibits an
exponential-like drop-off at the upper end, because the
distribution goes to zero at the upper end $x_2$ by definition,
\begin{equation}
  	N^{cum}(>x) = n {{\int_{x}^{x_2}} N(x') dx' \over 
                   {\int_{x_1}^{x_2} N(x') dx'}}
                   = n {(x^{1-\alpha} - x_2^{1-\alpha}) \over 
                       (x_1^{1-\alpha} - x_2^{1-\alpha})} \ .
\end{equation}
where $n$ is the total number of events.
Therefore, cumulative size distributions show always a steeper
distribution at the upper end $x_2$ (i.e., the bin containing
the largest event) than $(1-\alpha)$, for a differential 
size distributions with slope $\alpha$,
an effect that partially explains why (small) stellar flare samples have
steeper powerlaw slopes than (large) solar flare samples (Section 3.4.1).

There are a number of additional effects that cause deviations from
ideal powerlaw distribution functions. The most obvious deviations
occur from the truncation of distribution functions. If we have
statistics over many orders of magnitude, the truncation effects
are less severe, but are crucial for small samples. Let us explain
this with an example that is illustrated in Fig.~20. Solar flare
statistics is usually limited by a peak count threshold, i.e.,
complete sampling is only achieved for $P \ge P_{thresh}$. In this
case we expect a perfect powerlaw for the differential size distribution
of peak fluxes in the range of $P_{thresh} \le P \le P_{max}$, where
$P_{max}$ is the count rate of the largest observed flare (for instance
see Fig.~20e). However, if we sample the statistics of a related
parameter, such as the thermal energy (Fig.~20f), the peak count
threshold causes a truncation effect that extends over the lower half
(logarithmic) range of energies, where sampling is not complete in
energy and thus produces a broken powerlaw with a flatter slope
in the lower half (logarithmic) range. The same truncation effect
affects also linear regression fits. Nevertheless, these truncation
effects can be numerically simulated or analytically calculated
(e.g., see example in Aschwanden and Shimizu 2013, Appendix A),
and this way can be taken into account in the prediction of the
probability distribution functions of SOC parameters.

\subsection{	The Meaning of Self-Organized Criticality	}

After we have reviewed a large number of astrophysical observations
(Section 3) with powerlaw behavior, the question arises whether all
of these observed phenomena are SOC systems, and which are not
consistent with a SOC interpretation. To answer this question we
remind again our pragmatic generalized definition of a SOC system:
{\sl SOC is a critical state of a nonlinear energy dissipation
system that is slowly and continuously driven towards a critical value
of a system-wide instability threshold, producing scale-free,
fractal-diffusive, and intermittent avalanches with powerlaw-like 
size distributions} (Aschwanden 2014). This definition is
independent of any particular physical mechanism, but describes
only some universal system behavior that is common to virtually all 
threshold-operated nonlinear energy dissipation processes, 
in the limit of slow driving. Given this much larger perspective
of a SOC definition, we can ask whether the term ``self-organized
crtiticality'' is still justified in this context, which includes also
``critical points'' now that define a threshold for an instability.
The terms ``self-organizing'', ``self-tuning'', or ``self-adjusting''
mean in this context only that the system is continuoulsy driven 
towards a critical threshold, without necessity of external control.
If we have a self-sustaining slow driver, the continuous pushing
of the system towards an instability threshold is automatically
organized. In case the driver stops, the triggering of
instabilities stops too, and the system becomes static. For instance,
the cratering of the Moon has almost stopped, and thus the observed
craters are only remnants of a dynamical state. However, during the
times of heavy lunar meteorite bombardment, a SOC
system with a critical (relative-velocity) threshold that triggered
impacts on the Moon like earthquakes, or a scale-free
distribution of meteorites could be an alternative source. 
The same is true for solar
flares: there are quiescent static periods during the solar cycle
minimum when nearly no magnetic flux is generated by the solar dynamo,
while flaring during the maximum of the solar cycle constitutes 
a highly dynamic period of a continuously driven SOC system. 

\subsection{		SOC and Turbulence				}

What is the relationship between a SOC system and a turbulent system?
Because both systems exhibit powerlaw functions in the power spectrum,
scale-free size distributions, and many degrees of freedom,
there are commonalities that make their distinction difficult.
Vortices are in turbulence what avalanches are in SOC. 
A first difference was noted in the predicted waiting time distribution.
The originial BTW model considered SOC avalanches as statistically 
independent and thus predicted an exponential waiting time distribution,
while turbulent media exhibit long correlation times and predict 
powerlaw-like waiting time distributions (Boffetta et al. 1999;
Giuliani et al.~1999; Freeman et al.~2000a).
This argument, however, is alleviated by alternative SOC models 
(Section 2.12), such as non-stationary flaring rates (Wheatland et al.~1998),
the fractal-diffusive SOC model (Aschwanden 2014), or models with
persistence and memory as modeled with the Weibull distribution
(Telloni et al.~2014). 

How is SOC different from turbulence? Both processes may produce
similar statistics for slow driving, but start to differ when we 
move from slow to intermediate driving, when the smallest avalanches 
are ``swamped'', but the large avalanches persist, so that
intermittent turbulence shows only finite-range powerlaw scaling 
(Chapman et al.~2009; Chapman and Watkins 2009; Chapman and Nicol 2009).

Instead of considering single vortices in fully developed turbulence as the
equivalent of a SOC avalanche, a more satisfactory concept may be the notion 
of SOC in the state of near-critical turbulence, which is in the transition 
between the laminar state and the fully developed turbulence state. In this 
regime, the system profiles that store the free energy exciting the 
turbulence (i.e., pressure or temperature gradients, in a fusion plasma 
for instance) are very close to their local threshold values for the onset 
of instability. As a result, the local perturbations excited when these 
thresholds are overcome (giving rise to local eddies) may propagate to 
nearby locations (other eddies) as the former are relaxed, and the local 
profiles are brought back below critical. This is similar to the BTW sandpile. 
This regime is considered to be important in tokamak plasmas, because the local 
turbulent fluxes that bring the profiles back below a marginal state are strongest 
at higher temperatures. Thus, the equivalent to an avalanche is not a 
single eddy, but a chain of eddies at different locations connected in 
time, very much as a sand avalanche would happen in a sandpile. And the 
system is still in a turbulence-dominated regime, although turbulence 
is fully-developed only locally, not globally. These ideas were proposed 
in the mid-90s in the lab fusion community (Carreras et al.~1996;
Newman and Carreras 1996;
Mier et al.~2008; Sanchez et al.~2009), and have been given rise to a 
large body of work in the area of SOC, dealing with self-similarity, 
long-temporal correlations, and non-diffusive transport. 

Intermittent turbulence (IT) and self-organized criticality (SOC)
seem to co-exist in the magnetic field fluctuations of the solar wind 
at time scales of $T=10-10^3$ s (Podesta et al. 2006a,b, 2007), 
and in the solar corona (Uritsky et al.~2007).
It was proposed that the coexistence of SOC and IT may be a generic feature
of astrophysical plasmas, although the explicit complementarity
between SOC and IT in astrophysical observations has not been
demonstrated (Uritsky et al.~2007), IT phenomena can be
explained without invoking SOC (Watkins et al.~2009a),
and may need multi-fractal scaling (Macek and Wawrzasek 2009),
or three turbulence regimes (Meyrand and Galtier 2010).

The extent to which SOC and turbulence phenomena are really separable 
in complex systems is subject to a few conditions and topological 
constraints, also involving the ambient dimensionality. In two 
embedding dimensions, there is a theoretical possibility that SOC 
couples to turbulence via the inverse cascade of the energy, giving 
rise to large-amplitude events beyond the range of applicability of 
the conventional SOC (Milovanov and Rasmussen 2013). It has been 
discussed that the phenomenon occurs universally in two-dimensional 
fluid (as well as fluid-like, such as the drift-wave and drift-Alfven)
turbulence and requires time scale separation in that the Rhines 
time of the vortical system must be small compared with the instability 
growth time. Then the typical avalanching behavior associable with 
SOC will be amplified by the inverse cascade, which acts as to fuel 
the SOC avalanches "on-the-fly" with the energy. The energy reservoir 
for this behavior is only limited to the finite size of the system. 
It has been suggested that this new complexity phenomenon, the 
SOC-turbulence coupling, has serious implications for operational 
stability of big fusion confinement devices such as for instance 
the future power plants, where it may trigger transport events of 
potentially a catastrophic character (Milovanov and Rasmussen 2013). 
In this regard, it was argued that SOC was not really an alternative 
to the notion of turbulence and that there is kind of SOC-turbulence 
duality instead, coming along with the condition for time scale 
separation. A hybrid SOC-turbulence model has also been developed 
based on statistical arguments, using nonlocal transport and the 
formalism of a space-fractional Fokker-Planck equation 
(Milovanov and Rasmussen 2013). According to the hybrid model, 
the processes of amplification taking place will manifest 
themselves in the form of algebraic tails on top of the typical 
log-normal behavior of the probability distribution function of 
the flux-surface averaged transport. This suggestion finds further 
justification in the general properties of log-normal behavior in 
hierarchical systems with subordination (Montroll and Shlesinger 1982). 
In the realm of solar physics, SOC predicts scale-free distributions 
for large avalanche events (e.g., in solar flaring active regions) 
down to the smallest avalanche events (e.g., in nano-flaring or 
non-flaring active regions), which implies also the same turbulence 
characteristics for flaaring and non-flaring active regions, as it 
has been observationally verified (Georgoulis 2012). Related 
uniïfications of SOC processes, intermittent turbulence, and 
chaos theory include analysis of dynamical complexity via 
nonextensive Tsallis entropy (Milovanov and Zelenyi 2000; 
Balasis et al. 2011; Pavlos et al. 2012), fractional transport 
models (Zelenyi and Milovanov 2004; del-Castillo-Negrete 2006), 
and the formalism of fractional Ginzburg-Landau equation 
(Milovanov and Rasmussen 2005; Milovanov 2013).

\subsection{SOC and Percolation}

A recent discussion of the SOC concept versus the percolation problem 
is given in Milovanov (2013). Both SOC and percolation systems share 
the implications of threshold behavior, the spatial self-similarity, 
and fractality. One essential difference is that percolation is a purely 
geometrical model, while SOC involves also the temporal fractality, 
i.e., the 1/f noise. Another difference is the role of fine-tuning, 
which needs an externally manipulated control parameter in a 
percolation system, while it is automatically self-organizing 
in a SOC system. However, some nonlinear phenomena have been modeled 
with both SOC and percolation models, such as the spread of diseases 
or forest fires, which indicates a strong commonality between the 
two models, as well as some ambiguity in the choice of the most suitable 
model for a given observed phenomenon (e.g., Grassberger and Zhang 1996). 
Regarding numerical simulations, both models can be represented with 
iterative lattice-grid simulations, using similar mathematical 
re-distribution rules in each iterative step. It has been discussed 
that SOC and percolation systems can be both represented with cellular 
automation models, but having different re-distribution rules. In the 
basic theoretical perspective, though, this lattice-grid approach 
seems to overly simplify the integral picture of the self-organization, 
as it tends to disregard the peculiar role of nonlinearity behind the 
phenomena of SOC. Generally, standard percolation processes can be 
made self-organized by including a feedback loop generating 
self-organization in a marginally stable state. Then marginal dynamical 
stability of systems with spatio-temporal coupling will also require 
marginal topological connectedness (Milovanov 2013), so that in the 
presence of many dynamical degrees of freedom the operation of 
nonlinear feedback will automatically lead the system into a state 
of critical percolation. This general theoretical framework has been 
demonstrated on a lattice model using random walks to represent the 
microscopic re-distribution rules and the idea of "holes" or missing 
occupied sites which by themselves could participate to the random walk 
and dynamically generate a feedback (Milovanov 2010; 2011).

From a practical (or observational) perspective, the question arises 
whether the percolation and SOC models predict the same, or different, 
size distributions, after adjustment of the optimum control parameters. 
However, since any automation model is only an idealized representation 
of microscopic physics in complex systems, none of the two models is 
expected to mimic microscopic transport to an accurate level, but may 
rather approximate the microscopic size distributions. In this regard, 
the advantage of the random walk approach once again lies in a 
theoretically consistent picture of the dynamics, making it possible 
to obtain non-Markovian kinetic equations at criticality in terms of 
fractional calculus (Milovanov 2009; 2011). The main idea here is that 
fractional generalizations of the diffusion and Fokker-Planck equations 
(e.g., Metzler and Klafter 2000 for review) incorporate via a Laplace 
convolution the key signatures of non-Gaussianity and long-time 
dependence characteristic of the dynamical systems at or near SOC. 
One by-product of the the fractional model is the prediction that the 
relaxation of a super-critical system to SOC is of Mittag-Leffler type
(similar to the Cole-Cole behavior in glassy systems and polymers: 
see Milovanov 2011). The Mittag-Leffler relaxation implies that the 
behavior is multi-scale with a broad distribution of durations of 
relaxation events consistently with a description in terms of 
fractional relaxation equation (e.g., Metzler and Klafter 2000; 
Sokolov et al. 2002) and at odds with a single-exponential relaxation 
dynamics of the Debye type (Coffey 2004 for an overview; references 
therein). We should stress that the notion of feedback plays a very 
important role in the phenomena of SOC, as it ensures a steady state, 
where the system is marginally stable against a disturbance 
(Kadanoff 1991). For instance, in sandpiles, the unstable sand slides 
off to decrease the slope and reinstall stability, thus providing a 
feedback of the particle loss process on the dynamical state of 
the pile. Following Sornette (1992), we also note that, using the 
idea of feedback, it is possible to convert the standard critical 
phenomena into self-organized criticality dynamics, thereby extending 
considerably the span of models exhibiting SOC. One example of this 
conversion is localization-delocalization transition on a separatrix 
system of nonlinear Schr{\"o}dinger equation with disorder and 
self-adjusting nonlinearity, giving rise to a percolation structure 
in wave-number space, which is critical and self-organized (Milovanov 
and Iomin 2012; 2014).

In solar and astrophysics, percolation models have been applied to 
the formation of galaxies (Schulman and Seiden 1986; Seiden and 
Schulman 1990), to magnetotail current systems and the phenomena 
of tail current disruption (Milovanov et al. 1996; Milovanov 
et al. 2001; 2013; Arzner et al. 2002), to the solar dynamo 
(Schatten 2007), to photospheric magnetic flux concentrations 
(Balke et al. 1993), and to the emergence of solar active regions 
(Wentzel and Seiden 1992; Seiden and Wentzel 1996). Each of these 
phenomena can also be modeled with a threshold-operated instability 
in a SOC system. Hence, the jury is still out which model describes 
the real-world observations better.

\subsection{SOC and Branching Theory}

A branching process is a Markov process (i.e., a memory-less process)
that models a population with a random distribution at time step $n$
to predict the number of individuals in the next generation or time
step $n+1$ according to some probability distributions. To some
degree, the branching process during a single time step has the  
same purpose as the re-distribution rule in a cellular automaton
simulation. The question is whether the two processes have the same
probability distributions for the spatio-temporal evolution of an
avalanche event. The branching theory was mostly applied to the
evolution of a population, which ended either in infinite growth
or in global extinction. SOC avalanches end always after a finite
time interval, and thus can only evolve as a branching process with
final extinction. What is common to both processes is a critical
threshold or critical probability for next-neighbor or next-generation
propagation. Therefore, a self-organized branching process with
critical probabilities (Zapperi et al.~1995; Corral and Font-Clos 2013) 
has much in common with a SOC system of the BTW-type. Again, for
practical purposes to model observations, we may ask whether the
two models predict equal or different size distributions, using
some suitable critical probabilities.

In astrophysics, self-organizing branching theory has been applied
to magnetotail current systems (Milovanov et al.~2001),
to solar soft X-rays (Martin et al.~2010), and to solar flares
(MacKinnon and Macpherson 1997; Macpherson and MacKinnon 1999; Litvinenko 1998). 
The branching theory applied to solar flares 
(MacKinnon and Macpherson 1997; Macpherson and MacKinnon 1999; Litvinenko 1998a) 
as well as the self-organized branching process (SOBP) model (Zapperi et al.~1995;
Hergarten 2012) predict both a size distribution of $N(S) \propto
S^{-3/2}$, which is also predicted by the fractal-diffusive
self-organized criticality (FD-SOC) model, and thus indicates
an equivalent description of the multiplicative avalanche growth 
characteristics, and makes these two models indistinguishable 
with regard to their size distributions.

\subsection{Challenges and Open Questions in Solar SOC Models}

Attempting to connect the idealized analytical or numerical SOC 
models with real-world (astro)physical systems, one faces a host 
of questions that remain unanswered. We briefly touch on a few 
issues that arised from solar SOC models. 

{\bf Evolving SOC Drivers:} The driver of a SOC system may
naturally evolve in and out of a SOC state, vary cyclically or
intermittently, or oscillate between low and high states. 
Well-known examples are the solar dynamo that cyclically modulates the 
solar flare rate, or the variability of low and high states in the 
black-hole object Cygnus X-1. Another example is a time-variable
driver of solar active regions, as decribed in McAteer \etal~(2014),
which emulates how a dissipative, nonlinear dynamical system 
enters a SOC state. Standard SOC models, such as the BTW model, 
assume a steady driver and do not take into account the particular
system behavior of variable SOC drivers. Real-world SOC systems 
are operated by time-variable drivers that are
never exactly constant, which may alter the statistical
distributions that are predicted from a constant driver.
During the decay phase of a SOC driver, the dissipative 
properties of the system may possibly diffuse the available energy 
in a gradual (non-intermittent) fashion, and this way reduce the 
system's control parameter to a value below the critical threshold, 
and this way inhibit intermittent instabilities (avalanches). 

In a tectonic system, for example, earthquakes in an area would stop when 
inter-plate stresses are somehow mollified by repelling mantle motions
below. While this is a hypothetical and hardly observable fact, 
at least within reasonable geological timescales, a solar active region, 
even a fiercely flaring/eruptive one, emerges, evolves, and disappears 
within weeks. If this system evolves into a SOC state, as amply argued 
in this review, then the discontinuation of magnetic-flux emergence 
from the solar interior signals the start of this active region's 
demise. It is both, the time-dependent proper motions within the region 
(e.g., shear, sunspot rotation, outflows), as well as the overall 
solar differential rotation, which apparently quench the 
SOC-decaying driver in a gradual and non-intermittent manner
by exhausting the region's free magnetic energy. 
Flux emergence and related motions, on the other hand, become the 
realization of the classical {\it SOC-building} driver. The 
implementation of distinct SOC-building and SOC-decaying drivers, 
with the first being dominant during the SOC phase of the system,
but weak or absent during the system's decay, can characterize the 
finite lifetime of SOC states.

In the limit of a statistically slow SOC-building driver, 
another conceivable way for a system to exit SOC is by a ``catastrophic''
quenching of the SOC state by a single, system-wide instability that dissipates 
a substantial part of the system's available energy. While this is in 
principle not prohibited in a SOC system and could occur 
when the entire system becomes a network of marginally stable 
configurations, observations suggests otherwise: earthquake clusters 
(e.g., Corral 2004) reveal that only a relatively small portion of 
the stress-accumulated free energy is released, regardless how powerful
the earthquake is.
In other words, a seismic fault does not disappear after any 
single earthquake. The same is qualitatively the case for solar 
active regions, where the total available free energy can be 
calculated or estimated (e.g., Tziotziou et al. 2013). This rule
of thumb seems to indicate that SOC states fade away gradually only,
rather than by a catastrophic event at once.

{\bf Hidden, Anisotropic, and Composite SOC States:} 
A potentially interesting 
finding on the evolution of a time-variable SOC driver was obtained 
from experiments with a static data-driven (S-IFM) and a dynamically 
driven (D-IFM) flare model of Dimitropoulou et al. (2011; 2013), 
described in McAteer et al.~(2014). A given nonlinear force-free 
extrapolated magnetic field of an observed solar active 
region is first evolved into a SOC state, yielding a random but valid 
divergence-free magnetic configuration due to the S-IFM's random 
driving. Next, the system is evolved back to the initial extrapolated 
state via D-IFM, while monitoring tests confirmed that the SOC state
has not been destroyed. Therefore, one cannot rule out that 
the initial extrapolated-field state, despite being a 
force-free-equilibrium state, may in fact be a SOC state. 
The subject active region for the test happened to be an eruptive one; 
however, the eruptive property was not used in the test. Therefore, 
unless eruptive solar active regions have a topologically or 
otherwise distinct magnetic structure compared to non-eruptive ones, 
the same test might possibly work equally well with a non-eruptive 
active region. Should this be confirmed, it would be evidence that 
solar active regions, regardless of an eruptive or non-eruptive nature, 
may be in a SOC state. The question then arises, besides
active regions, whether the quiet-Sun (or global stellar) magnetic 
field is in a SOC state also? 
This remains to be assessed. The lack of major flares and eruptions 
from non-eruptive active regions and the quiet Sun may be due to 
the lack of available free-energy density accumulation, a much weaker 
SOC-building driver, or a critical threshold of a different nature, 
heuristically proposed as an ``anisotropic'' SOC threshold by 
Vlahos et al. (1995) and subsequent works. It is now observed from 
exceptionally high-resolution solar observations that small-scale 
energy-release events resembling the hypothesized nanoflares occur 
in the active and the quiet solar corona (Cirtain et al. 2013; 
Winebarger et al. 2013). If the entire solar corona is in a ``composite''
SOC state, albeit with different critical thresholds and drivers in 
different regions, then there is a possibility to extend SOC validity 
over the global magnetic configurations of magnetically active, 
main-sequence stars. 

{\bf Robustness of Power Laws:} Probability distributions of sizes
and durations exhibit generally a powerlaw function with a specific
slope for a given observable (such as the peak count rate, fluence,
rise time, or decay time). The value of the powerlaw slope becomes
the more robust, the larger the statistics is, gathered over 
sampling times as long as possible. Even for small statistics and
short sampling times, the value of the powerlaw slope may be robust,
as long as the driver is constant and the sample is statistically
representative. However, this robustness is lost when subsets of
data are histogrammed that contain some selection bias.
This loss of robustness has been demonstrated in a study by
Crosby et al.~(1998), using a sample of some 1500 X-ray flares from 
the WATCH/GRANAT satellite, when subsets were selected by groups with
different event durations: the power laws were found to be steeper 
for subsets with short duration, while they progressively flattened 
for longer events.  A similar result was found for total-count 
distribution functions of these flares by Georgoulis et al.~(2001), 
which was also used for a ``statistical flare SOC cellular automaton 
model'' (Georgoulis and Vlahos 1998).

The effect of a selection bias in time durations $T$ on the size 
distribution function of an observable, such as the peak flux $P$,
can easiest be understood from a scatterplot between the parameters
$P$ and $T$. If there would be an exact correlation with a 
correlation $P \propto T^a$, the distribution $N(P; T=T_i)$
of a subset with duration $T_i$ would be a $\delta$-function
$N(P=P_i)$
at the value $P_i \propto T_i^a$. In reality, the correlations
have a substantial scatter, which broadens the size distributions
of each subset, but the trend that they are clustered around the
value $P_i \propto T_i^a$ persists. A consequence of this scatter
between correlated parameters is also that the threshold in an 
observable (say in the peak count rate, $P \ge P_0$), causes 
a truncation bias in the correlated parameter 
(say $T \gapprox T_0$). Therefore, even when the peak rate
distribution $N(P)$  exhibits an exact powerlaw down to the threshold
value $P_0$ of the sample, the correlated time duration
distribution $N(T)$ will have a smooth rollover, which is
a significant deviation from an ideal powerlaw. At the upper end of
size distributions, finite-size effects cause an addtional
fall-off, which is another deviation from an ideal powerlaw
distribution. These well-understood effects should be taken into account
in arguments countering power laws and their validity and interpretation, 
as expressed by Stumpf and Porter (2012). 

{\bf Hybrid SOC Models and Multi-Fractal Effects:} 
There is also a controversy about the hypothesized 
``soft'' nanoflare population (Parker 1988) that must be 
sufficiently steep ($\alpha _E >2$) to allow the bulk of the 
dissipated energy to originate from the lower end of the distribution, 
via a mostly thermal energy release, thus balancing the coronal 
energy losses and maintaining a hot corona (Hudson 1991; 
see also Section 3.2.8).  This review presents evidence that 
nanoflares share the same powerlaw distribution of energies
as microflares and large flares do.
Therefore, the bulk of the released energy stems from large flares 
in the upper end of the distribution, which is debated by some 
studies to be insufficient to maintain the corona at its observed 
temperature. Indeed, statistical properties of small-scale 
events have been revisited to correct for multiple selection 
biases and have been shown to obey flatter power laws than 
originally found. Given the ever-improving but always finite 
observational sensitivity, however, it is conceivable that 
such a soft population, if existing, may still be eluding  
observation or may be partially suppressed by the better 
sampled intermediate and large events, as it appears to be the 
case with the results of Crosby \etal~(1998) and 
Georgoulis \etal~(2001). In addition, the prediction of the 
statistical flare model (Georgoulis and Vlahos 1998) 
for a dual population of instabilities 
and a ``knee'' between them, moving from a steeper (softer) 
to a flatter (harder) power law (Georgoulis and Vlahos 1996), 
has yet to be  confirmed or ruled out. The statistical flare 
model remains the only SOC model that produces double scaling 
owning to a double instability criterion featuring ``isotropic''
and ``anisotropic'', directional relaxation (see, however, 
Figure 4 of Hughes et al. (2003) and relevant discussion).

Hybrid models can explain broken-powerlaw distributions,
which imply also multi-fractality, a property that has been
measured in a number of solar active region studies on the
magnetic flux distributions (Lawrence~\etal 1993;
Cadavid \etal~1994; Gallagher et al.~1998; McAteer \etal~2005; 
Conlon et al.~2008, 2010; Hewett \etal.~2008). 

{\bf Predictability in a SOC System:} Are large events resulting 
from a SOC system predictable? This remains a widely open 
question with profound geophysical (i.e., earthquake prediction) 
and space-weather (i.e., solar-flare/eruption prediction) 
implications. The question can naturally be linked to the 
question of inter-event, or waiting times. Extensive 
discussion on waiting times and their distribution in this 
review (Section 2.12 and references therein) has established 
that the form of the SOC waiting-time distribution is not an 
invariant SOC property such as the power-law distribution 
functions of event size. The degree of memory, intrinsic and 
different in each SOC system, determines the form of the 
waiting-time distribution. The opposite is not true generally, 
because the form of the waiting-time distribution cannot 
uniquely specify the degree of memory of the SOC system that 
created it. 
In addition, an instability -- regardless how intense -- tends 
to release only a small fraction of the system's available energy, 
hence always imposing a finite degree of stochasticity that is 
complementary to the finite memory of the system. In case of 
no memory, that gives rise to a classical BTW exponential 
waiting-time distribution, events are purely random and 
cannot be predicted. In particular cases -- such as, e.g., 
deterministically driven models -- Strugarek and Charbonneau 2014 
showed that the memory of the SOC system could be raised up to 
a level where large events can be forecasted systematically. 
Though, it must be noted that predictions from a SOC system 
necessarily rely on different realizations of the stochastic 
process and by such are intrinsically probabilistic. Achieving 
the most significant prediction probabilities then depends on 
the memory level of the model and is a matter of the specific 
physics of the SOC system in question (discussion below).

{\bf Helictiy Conservation in Solar SOC Models:}  
What physical quantity is conserved in a SOC system?
Two of the telltale SOC features are {\it metastability} 
and {\it marginal stability}. Metastability typically involves 
a conservative property of the system in the course of driving 
as it occurs in the original BTW concept, while marginal stability 
reflects the mere result of an upper accumulation limit 
for the conserved parameter, hence defining the critical 
threshold. Perturbing a low-beta, magnetized environment 
of a solar active region, for instance, one builds electric 
currents while conserving magnetic flux. Using a flux critical 
threshold, however, would be misleading, as large, severely 
flux-imbalanced active regions (e.g., a single compact sunspot 
surrounded by scattered opposite-polarity flux) do not flare 
or erupt in general. Electric current density could constitute 
a critical threshold for magnetic reconnection and hence for 
an instability, but it is not a conserved quantity: when 
stopping the SOC-building driver, the free magnetic energy 
due to electric currents will be gradually dissipated via a 
SOC-decaying driver, returning the system to eruption-free 
stability reflected in a current-free, potential state 
(e.g., Contopoulos et al. 2011). Although a few non-conservative 
SOC models have been proposed (Vespignani and Zapperi 1998; 
Pruessner and Jensen 2002 and references therein), the greatly 
larger number of conservative SOC models implies that one 
should perhaps look into a conservative control parameter 
first to identify a critical threshold: an attractive concept 
is that of magnetic helicity, a physical quantity that is 
roughly conserved in high magnetic Reynolds-number plasmas 
even during reconnection (e.g., Berger 1999). Magnetic helicity 
could indeed provide a critical threshold, complemented by 
a minimum free magnetic energy necessary to keep in pace 
with the accumulated helicity (Tziotziou et al. 2012). This may 
lead to an unbiased interpretation of eruptions as instabilities 
occur not because of magnetic reconnection primarily, but 
because a part or the entire magnetic structure reached its 
limit in terms of accumulated helicity. Uncovering the crucial 
physical details of this and similar mechanisms, including how 
the control quantity of the system (magnetic helicity in this 
example) consistently accumulates until the system becomes 
unstable, may potentially achieve closure between physical
models and statistical interpretations of complexity systems
governed by SOC.

\clearpage

\section{		SUMMARY AND CONCLUSIONS			}

The literature on self-organized criticality (SOC) models counts over 
3000 refereed publications at the time of writing, with about 
500 papers dedicated to solar and astrophysics. Given the relatively 
short time interval of 25 years since the SOC concept was born
(Bak et al.~1987), the productivity in this interdisciplinary and 
innovative field speaks for the generality, versatility, and inspirational
power of this new scientific theory. Although there exist some previous
similar concepts in complexity theory, such as phase transitions,
turbulence, percolation, or branching theory, the SOC concept seems 
to have the broadest scope and the most general applicability to 
phenomena with nonlinear energy dissipation in complex systems
with many degrees of freedom. Of course there is no such thing as
a single ``SOC theory'', but we rather deal with various SOC concepts
(that are more qualitative rather than quantitative), which in some cases
have been developed into more rigorous quantitative SOC models
that can be tested with real-world data. Computer simulations of the
BTW type provide toy models that can mimic complexity phenomena,
but they generally lack the physics of real-world SOC phenomenona,
because their discretized lattice grids do not reflect in any way
the microscopic atomic or subatomic structure of real-world physical 
systems. 

In this review we focus on the astrophysical applications only,
including solar physics, magnetospheric, planetary, stellar,
and galactic physics. We summarize first some basic concepts 
of a generalized SOC theory, covering different SOC definitions,
the driver, instability and criticality, avalanches, microscopic
structures, basic spatio-temporal scaling laws and derivations of basic
occurrence frequency or size distributions, waiting time distributions,
and a comparison of basic numerical cellular automaton simulations.
Most of these aspects are the ingredients of a generalized
fractal-diffusive self-organized criticality (FD-SOC) model 
(Aschwanden 2014),
which we use as a standard model for the macroscopic description
of a SOC system, bearing in mind that it represents only a 
first-order approximation to the statistics of the microphysics 
of SOC avalanches. This standard model is based on the scale-free
probability conjecture, fractal geometry, and diffusive transport.
This model can explain most of the astrophysical observations
and enables us to discriminate which SOC-related observations 
can be explained with standard scaling laws, and which phenomena
represent mavericks that need either a special model, an
improved data analysis, or better statistical completeness.
We summarize the major findings of this review in the following:

\begin{enumerate}
\item{A general working definition of a SOC system that can be
	applied to the majority of the observed astrophysical
	phenomena interpreted as SOC phenomena can be formulated as:
	{\sl SOC is a critical state of a nonlinear energy dissipation
	system that is slowly and continuously driven towards a critical value
	of a system-wide instability threshold, producing scale-free,
	fractal-diffusive, and intermittent avalanches with powerlaw-like 
	size distributions} (Aschwanden 2014). This generalized
	definition expands the original meaning of self-organized
	``criticality'' to a wider class of critical points and
	instability thresholds that have a similar (nonlinear)
	dynamical behavior and produce similar (powerlaw-like)
	statistical size distributions.}

\item{A generalized (macroscopic description of a) SOC model can be 
	formulated as a function of the Euclidean space dimension $d$, 
	the spatio-temporal spreading exponent $\beta$, a fractal 
	dimension $D_d$, and a volume-flux scaling (or radiation 
	coherency) exponent $\gamma$. For standard conditions
	[$d=3$, $D_d \approx (1+d)/2$, $\beta=1$, and $\gamma=1$],
	this SOC model predicts (with no free parameters) powerlaw
	distributions for all SOC parameters, namely $\alpha_L=3$
	for length scales, $\alpha_A=2$ for areas, $\alpha_V=5/3$
	for volumes,
	$\alpha_F=2$ for fluxes or energy dissipation rates,
	$\alpha_F=5/3$ for peak fluxes or peak energy dissipation rates,
	and $\alpha_E=3/2$ for time-integrated fluences or energies 
	of SOC avalanches.}

\item{The underlying correlations or scaling laws are:
	$A \propto L^2$ for the maximum avalanche area, 
	$A_f \propto L^{D_d}$ for the fractal avalanche area, 
	$V \propto L^3$ for the maximum avalanche volume, 
	$V_f \propto L^{D_d}$ for the fractal avalanche volume, 
	$T \propto L^{(2/\beta)}$ for the avalanche duration,
	$F \propto L^{(\gamma D_d)}$ for the flux or energy dissipation rate, 
	$P \propto L^{(\gamma d)}$ for the peak flux or peak energy dissipation rate, 
	$E \propto L^{(\gamma D_d+2/\beta)}$ for the fluence or total energy.}

\item{Moreover, the FD-SOC model predicts a waiting time distribution with a slope of
	$\alpha_{\Delta t}=2$ for short waiting times, and an exponential
	drop-off for long waiting times, where the two waiting time
	regimes are attributed to intermittently active periods,
	and to randomly distributed quiescent periods. The contiguous
	activity periods are predicted to have persistence and memory.}

\item{Among the astrophysical applications we find agreement between
	the predicted and observed size distribution for 10 out of 14
	reported phenomena, including lunar craters, meteorites, 
	asteroid belts, Saturn ring particles, auroral events during
	magnetospheric substorms, outer radiation belt electron events,
	solar flares, soft gamma-ray repeaters, blazars, and black-hole
        objects.}

\item{Discrepancies between the predicted and observed size distributions
	are found for solar energetic particle (SEP) events, stellar flares, 
	pulsar glitches, the Cygnus X-1 black hole, and cosmic rays, which require a
	modification of the standard FD-SOC model or improved data analysis.
	The disagreement for SEP events is believed to be due to a
	selection bias for large events, or could alternatively be
	modeled with a different dimensionality of the SOC system.
	For stellar flares we conclude that the bolometric fluence
	is not proportional to the dissipated energy and flaring volume.
	Pulsar glitches are subject to small-number statistics.
	Black hole pulses from Cygnus X-1 have an extremely steep size 
	distribution that could be explained by a suppression of large 
	pulses for a certain period after a large pulse. For cosmic rays, the
	energy distribution appears to be subject to incomplete
	uni-directional sampling by in-situ observations, rather than 
	omni-directional sampling by remote-sensing methods.}

\item{Some of the SOC-associated phenomena have also been modeled with
	alternative models regarding their size or waiting time distributions
	and were found to be commensurable, such as in terms of turbulence,
	percolation, branching theory, or phase transitions. All these
	theories have some commonalities in their concept and can often not be
	discriminated based on their observed size distributions alone. 
	Some of the
	physical processes may coexist and not exclude each other, 
	such as SOC and turbulence in the solar wind.}
\end{enumerate}

A summary of theoretically predicted and observed powerlaw indices of
selected astrophysical SOC phenomena is listed in Table 15, while
more complete compilations for each phenomenon are given in Tables 2 to 14. 
The variation of powerlaw values among the same phenomena indicates 
incompatible data analysis methods or statistically irreconcilable samples. 
Improved data analysis, larger statistics, and more detailed complexity
models are called for in future studies, which should reconcile existing
discrepancies and answer the existing open questions and challenges. 
Besides the statistical improvements, also physical models (Table 16) 
that reproduce the underlying scaling
laws are expected in future work. All these tasks present a rich and
rewarding activity of future research in the field of complex systems.
The SOC concept has clearly stimulated a new way of thinking and
analyzing the dynamics and statistics of complex systems.

\acknowledgments
The author team acknowledges the hospitality and partial support for two
workshops on ``Self-Organized Criticality and Turbulence'' at the
{\sl International Space Science Institute (ISSI)} at Bern, Switzerland,
during October 15-19, 2012, and September 16-20, 2013, as well as 
constructive and stimulating discussions (in alphabetical order)
with Sandra Chapman, Paul Charbonneau, Henrik Jeldtoft Jensen, 
Maya Paczuski, John Rundle, Loukas Vlahos, and Nick Watkins. 
This work was partially supported by NASA contract NNX11A099G 
``Self-organized criticality in solar physics'' and NASA contract 
NNG04EA00C of the SDO/AIA instrument to LMSAL. MKG acknowledges
partial support by the EU Seventh Framework Marie-Curie Programme
under grant agreement No. PIRG07-GA-2010-268245.

\clearpage

\clearpage
\begin{table}[t]
\begin{center}
\normalsize
\captio{Summary of theoretically predicted and observed powerlaw
indices of size distributions in astrophysical systems.}
\medskip
\begin{tabular}{|l|l|l|l|l|l|l|}
\hline
                                 & Length     & Area        & Duration    & Peak flux   & Energy     & Waiting \\
                                 & $\alpha_L$ & $\alpha_A$,
                                                $\alpha_{th,A}$ & $\alpha_T$  & $\alpha_P$  & $\alpha_E$ 
											& time $\alpha_{\Delta t}$\\
\hline
FD-SOC prediction                & {\bf 3.0}  & {\bf 2.33}  & {\bf 2.0}   & {\bf 1.67}  & {\bf 1.50} &{\bf 2.0}\\
\hline
\underbar{Lunar craters:}        &            &             &             &             &            &\\
Mare Tranquillitatis $^1)$       & 3.0        &             &             &             &            &\\
Meteorites and debris $^2)$      & 2.75       &             &             &             &            &\\
\hline
\underbar{Asteroid belt:}        &            &             &             &             &            &\\
{\sl Spacewatch Surveys}$^3)$    & 2.8        &             &             &             &            &\\
{\sl Sloan Survey}$^4)$          & 2.3-4.0    &             &             &             &            &\\
{\sl Subaru Survey}$^5)$         & 2.3        &             &             &             &            &\\
\hline
\underbar{Saturn ring:}          &            &             &             &             &            &\\
Voyager 1$^6)$                   & 2.74-3.11  &             &             &             &            &\\
\hline
\underbar{Magnetosphere:}        &            &             &             &             &            &\\
EUV auroral events$^7$           &            &$1.73-1.92$  & $2.08-2.39$ &$1.66-1.82$  & $1.39-1.61$&\\
EUV auroral events$^8$           &            &$1.85-1.98$  & $2.25-2.53$ &$1.71-2.02$  & $1.50-1.74$&\\
Outer radiation belt$^{9})$      &            &             &             & 1.5-2.1     &            &\\
\hline
\underbar{Solar Flares:}         &            &             &             &             &            &\\
HXR, ISEE-3$^{10}$               &            &             & 1.88-2.73   & 1.75-1.86   & 1.51-1.62  &\\
HXR, HXRBS/SMM$^{11}$            &            &             &$2.17\pm0.05$&$1.73\pm0.01$&$1.53\pm0.02$&2.0$^a$\\
HXR, BATSE/CGRO$^{12}$           &            &             & 2.20-2.42   & 1.67-1.69   & 1.56-1.58  & 2.14$\pm$0.01$^b$\\
HXR, RHESSI$^{13}$               &            &             & 1.8-2.2     & 1.58-1.77   & 1.65-1.77  & 2.0$^a$\\
SXR, Yohkoh$^{14}$               & 1.96-2.41  & 1.77-1.94   &             & 1.64-1.89   & 1.4-1.6    & \\
SXR, GOES$^{15}$                 &            &             & 2.0-5.0     & 1.86-1.98   & 1.88       & 1.8$-$2.4$^c$\\
EUV, SOHO/EIT$^{16}$             &            & 2.3-2.6     & 1.4-2.0     &             &            &\\
EUV, TRACE$^{17}$                & 2.50-2.75  & 2.4-2.6     &             & 1.52-2.35   & 1.41-2.06  &\\
EUV, AIA/SDO$^{18}$              &$3.2\pm0.7$ &$2.1\pm0.3$  &$2.10\pm0.18$&$2.0\pm0.1$  &$1.6\pm0.2$ &\\
EUV, EIT/SOHO$^{19}$             &$3.15\pm0.18$ &$2.52\pm0.05$ &$1.79\pm0.03$&          &$1.47\pm0.03$ &\\
Radio microwave bursts$^{20}$    &            &             &             & 1.2-2.5     &            &\\
Radio type III bursts$^{21}$     &            &             &             & 1.26-1.91   &            &\\
Solar energetic particles$^{22}$ &            &             &             & 1.10-2.42   & 1.27-1.32  &\\
\hline
\underbar{Stellar Flares:}       &            &             &             &             &            &\\
EUVE flare stars$^{23}$          &            &             &             &             &$2.17\pm0.25$&\\
KEPLER flare stars$^{24}$        &            &             &             &$1.88\pm0.09$&$2.04\pm0.13$&\\
\hline
\underbar{Astrophysical Objects:}&            &             &             &             &            &\\
Crab pulsar$^{25}$               &            &             &             & 3.06$-$3.50 &            &\\
PSR B1937+21$^{26}$              &            &             &             & $2.8\pm0.1$ &            &\\
Soft Gamma-Ray repeaters$^{27}$  &            &             &             &             &$1.43-1.76$ &\\
Cygnus X-1 black hole$^{28}$     &            &             &             & 7.1         &            &\\
Sgr A$^*$ black hole$^{29}$      &            &             &             & $1.9\pm0.4$ &$1.5\pm0.2$ &\\
Blazar GC 0109+224$^{30}$        &            &             &             & 1.55        &            &\\
Cosmic rays$^{31}$               &            &             &             &             &$2.7-3.3$   &\\
\hline
\end{tabular}
\end{center}
\end{table}

\clearpage
{ %\footnotesize 
\underbar{References to Table 15:}
$^1)$ Cross (1966);
$^2)$ Sornette (2004);
$^3)$ Jedicke and Metcalfe (1998);
$^4)$ Ivezic et al.~(2001);
$^5)$ Yoshida et al.~(2003), Yoshida and Nakamura (2007);
$^6)$ Zebker et al.~(1985), French and Nicholson (2000);
$^7)$ Uritsky et al.~(2002);
$^8)$ Kozelov et al.~(2004);
$^{9})$ Crosby et al.~(2005)
$^{10})$ Lu et al.~(1993), Lee et al.~(1993);
$^{11})$ Crosby et al.~(1993);
$^{12})$ Aschwanden ~(2011b);
$^{13})$ Christe et al.~(2008), Lin et al.~(2001), Aschwanden ~(2011a);
$^{14})$ Shimizu (1995), Aschwanden and Parnell~(2002);
$^{15})$ Lee et al.~(1995), Feldman et al.~(1997), Veronig et al.~(2002a,b),
         Aschwanden and Freeland (2012);
$^{16})$ Krucker and Benz (1998), McIntosh and Gurman (2005);
$^{17})$ Parnell and Jupp (2000), Aschwanden et al. (2000b), Benz and Krucker (2002),
         Aschwanden and Parnell (2002), Georgoulis et al.~(2002);
$^{18})$ Aschwanden and Shimizu (2013), Aschwanden et al.~(2013a); 
$^{19})$ Uritsky et al.~(2002);
$^{20})$ Akabane (1956), Kundu (1965), Kakinuma et al.~(1969), 
	 Das et al.~(1997), Nita et al.~(2002);
$^{21})$ Fitzenreiter et al.~(1976), Aschwanden et al.~(1995), Das et al.~(1997), Nita et al.~(2002);
$^{22})$ Van Hollebeke et al.~(1975), Belovsky and Ochelkov (1979), Cliver et al. (1991),
         Gabriel and Feynman (1996), Smart and Shea (1997), Mendoza et al.~(1997),
         Miroshnichenko et al.~(2001), Gerontidou et al.~(2002);
$^{23})$ Robinson et al.~(1999). Audard et al.~(2000), Kashyap et al.~(2002),
         G\"udel et al.~(2003), Arzner and G\"udel (2004), Arzner et al.~(2007), 
	 Stelzer et al.~(2007), Maehara et al. 2012; Shibayama et al.~(2013);
$^{24})$ Maehara et al.~(2012); Shibayama et al.~(2013);
$^{25})$ Argyle and Gower (1972), Lundgren et al.~(1995);
$^{26})$ Cognard et al.~(1996);
$^{27})$ Gogus et al.~(1999, 2000);
$^{28})$ Negoro et al.~(1995), Mineshige and Negoro (1999);
$^{29})$ Neilsen et al.~(2013);
$^{30})$ Ciprini et al.~(2003);
$^{31})$ e.g., Fig.~28 (courtesy of Simon Swordy, Univ.Chicago);
$^{a})$  Aschwanden and McTiernan (2010);
$^{b})$  Grigolini et al.~(2002);
$^{c})$  Wheatland (2001, 2003), Boffetta et al.~(1999), Lepreti et al.~(2001).}

\begin{table}
\begin{center}
\small
\captio{Physical mechanisms operating in self-organized criticality systems.}
\medskip
\begin{tabular}{|l|l|l|l|}
\hline
Phenomenon	& Energy Input 		& Instability threshold & Energy output   \\
		& (steady driver)	& (criticality)	   & (intermittent avalanches) \\	
\hline
SOC-related Systems:&			&		   &		     \\	
Sandpile 	& gravity (dripping sand)& angle of repose & sand avalanches \\
Superconductor	& magnetic field change & phase transition & vortex avalanches\\
Ising model	& temperature increase  & phase transition & atomic spin-flip\\ 
Tea kettle	& temperature increase  & boiling point	   & vapour bubbles  \\	
Earthquakes	& tectonic stressing	& dynamical friction & rupture area    \\
Forest fire	& tree growth           & fire ignition point & burned area     \\
BTW cellular automaton & input at random nodes& critical threshold &next-neighbor redistribution \\ 
\hline
ASTROPHYSICS:	&			&		   &		     \\	
Lunar craters	& meteorite production	& lunar collision  & lunar impact craters\\
Asteroid belt	& planetesimals		& critical mass density & asteroids       \\
Saturn ring	& gravitational disturbances & collision rate  & Saturn ring particles\\
Magnetospheric substorm& solar wind	& magnetic reconnection & auroral bursts \\
Radiation belt  & solar wind            & magnetic trapping/untrapping & electron bursts	\\
Solar flares	& magnetic stressing	& magnetic reconnection & nonthermal particles\\
Stellar flares	& magnetic stressing	& magnetic reconnection & nonthermal particles\\
Pulsar glitches & neutron star spin-up  & vortex unpinning  &neutron starquakes\\
Soft gamma-ray repeaters& magnetic stressing &star crust fracture &neutron starquakes\\
Black-hole objects & gravity            & accretion and inflow   &X-ray bremsstrahlung pulses\\ 
Blazars		& quasar jets		& jet direction jitter & optical radiation pulses\\
Cosmic rays	& galactic magnetic fields& (run-away) acceleration threshold & high-energy particles\\
\hline
\end{tabular}
\end{center}
\end{table}

\clearpage

\section{ACRONYMS}

\begin{tabular}{ll}
1D, 2D, 3D & 1-, 2-, 3-dimensional \\
ACE	& Advanced Composition Explorer (spacecraft) \\
AE	& Auroral Electron jet index \\
AGILE	& Astro-Rivelatore Gamma a Immagini LEggero (spacecraft) \\
AGN	& Active Galactic Nuclei \\
AIA	& Atmospheric Imaging Assembly (on SDO) \\
AlMg	& Aluminium-Magnesium filter (on Yohkoh spacecraft) \\
AU	& Auroral Upper geomagnetic index \\ 
AU	& Astronomical Unit (Sun-Earth distance) \\
BATSE	& Burst And Transient Source Experiment (on CGRO) \\
BC	& Box Counting (fractal dimension) \\
BCS	& Bent Crystal Spectrograph (on SMM) \\
BTW	& Bak, Tang, and Wiesenfeld (SOC model) \\
COBE	& COsmic Background Explorer (spacecraft) \\
CGRO	& Compton Gamma Ray Observatory (spacecraft) \\
CLUSTER & magnetospheric mission with 4 spacecraft \\
CME	& Coronal Mass Ejection \\
DCIM-P  & Decimetric pulsation radio burst \\
DCIM-S  & Decimetric spike radio burst \\
DNL	& Distant Neutral Line (in geotail) \\
EIT	& Extreme ultra-violet Imager Telescope (on SOHO) \\
EM	& Emission Measure \\
ESA	& European Space Agency \\
EUV	& Extreme Ultra-Violet \\
EUVE	& Extreme Ultra-Violet Explorer (spacecraft) \\
EXOSAT	& European X-ray Observatory SATellite \\
FD-SOC  & Fractal-Diffusive Self-Organized Criticality model \\
Fermi	& hard X-ray spacecraft \\
FSOC	& Forced Self-Organized Criticality \\
GEOTAIL & magnetospheric mission (spacecraft) \\
GOES	& Geostationary Orbiting Earth Satellite (spacecraft) \\
GRANAT	& Gamma Ray Astronomical observatory (Russian spacecraft) \\
HSP	& High Speed Photometer (instrument on HST) \\
HST	& Hubble Space Telescope \\
HXR	& Hard X-Rays \\
HXRBS	& Hard X-Ray Burst Spectrometer (on SMM) \\
ICE	& International Cometary Explorer (spacecraft) \\
IMAGE   & magnetospheric spacecraft \\
IMF	& Interplanetary Magnetic Field \\
IMP	& Interplanetary Monitoring Platform (spacecraft) \\
ISEE-3	& International Cometary Explorer (spacecraft) \\
ISSI	& International Space Science Institute, Bern, Switzerland \\
IT	& Intermittent Turbulence \\
keV	& kilo electron Volt \\
LA	& Linear size vs. Area method (fractal dimension) \\
LASClO	& Large-Angle Solar COronagraph (on SOHO) \\
LIM	& Local Intermittency Measure (method) \\
\end{tabular}

\begin{tabular}{ll}
MHD	& Magneto-HydroDynamics	\\
MeV	& Mega electron Volt \\
MG	& magnetogram \\
MK	& Mega Kelvin \\
MW	& Microwave Burst \\
MW-S	& Microwave Burst Synchrotron emission \\
NASA	& National Aeronautics and Space Administration \\
NOAA	& National Oceanic and Atmospheric Administration \\ 
OFC	& Olami-Feder-Christensen (SOC model) \\
OGO	& Orbiting Geophysical Observatory (spacecraft) \\
OSO  	& Orbiting Solar Observatory (spacecraft) \\
PA	& Perimeter versus Area (fractal dimension) \\
PDF	& Probability Distribution Function \\
PHEBUS  & Gamma ray burst instrument (on GRANAT) \\
POLAR	& magnetospheric mission (spacecraft) \\
PSR	& pulsar \\
RHESSI	& Ramaty High Energy Solar Spectroscopic Imager (spacecraft) \\
RTV	& Rosner-Tucker-Vaiana model (coronal heating) \\
RXTE	& Rossi X-ray Timing Explorer (spacecraft) \\
SDF	& Surviving Distribution Function \\
SGR	& Soft Gamma-ray Repeaters \\
SDO	& Solar Dynamics Observatory (spacecraft) \\
SEP	& Solar Energetic Particle event \\
SMM	& Solar Maximum Mission (spacecraft) \\
SOBP	& Self-Organized Branching Process \\
SOC	& Self-Organized Criticality \\		
SOHO	& SOlar and Heliospheric Observatory (spacecraft) \\
SST	& Swedish Solar Telescope (observatory) \\
STEREO	& Sun TErrestrial RElations Observatory (spacecraft) \\
SuperDARN & Super Dual Auroral Radar Network \\
SXR	& Soft X-Rays \\
SXT	& Soft X-ray Telescope (on Yohkoh spacecraft) \\
TGF	& Terrestrial Gamma-ray Flashes \\
TRACE	& TRAnsition region and Coronal Explorer (spacecraft) \\
TV	& TeleVision (camera) \\
UCB	& University of California, Berkeley \\
ULYSSES & interplanetary mission (spacecraft) \\
UV	& Ultra-Violet \\
UVI	& Ultra-Violet Image (on POLAR spacecraft) \\
WATCH	& Wide Angle Telescope of Cosmic Hard X-rays (on GRANAT) \\	
XMM	& X-ray Multi-Mirror mission (spacecraft) \\
WIC	& Wideband Imaging Camera (on IMAGE spacecraft) \\
WL	& White Light \\
WMAP	& Wilkinson Microwave Anisotropy Probe (spacecraft) \\
WTD	& Waiting Time Distribution function \\
Yohkoh	& Japanese Solar-A mission (spacecraft) \\
\end{tabular}

\end{document}